\documentclass{aastex631}

\usepackage{amsmath}
\usepackage{changepage}
\usepackage{tasks}
\usepackage{booktabs}
\usepackage{hyperref}
\usepackage{diagbox}
\usepackage{graphicx}
\usepackage{hhline}
\usepackage{capt-of}
\usepackage{xspace}
\usepackage{csvsimple}
\usepackage{upgreek}
\usepackage{makecell}

\shorttitle{\projname II}
\shortauthors{Huang, Inchausti, Storfer, Tabares-Tarquinio, Moustakas, Schlegel et al.}

\graphicspath{{./}{figures/}}
\newcommand{\ang}{\AA\xspace}
\newcommand{\ho}{\ensuremath{H_0}\xspace}
\newcommand{\hst}{\emph{HST}\xspace}
\newcommand{\HST}{\emph{Hubble Space Telescope}\xspace}
\newcommand{\projname}{DESI Strong Lens Foundry\xspace}
\newcommand{\srccircle}{cyan\xspace}
\newcommand{\lenscircle}{magenta\xspace}
\newcommand{\progname}{Strong Lensing Secondary Target Program\xspace}
\newcommand{\nfujisys}{73\xspace}
\newcommand{\nspecshown}{91\space}
\newcommand{\noursys}{17\xspace}
\newcommand{\nothersys}{56\xspace}

\newcommand{\ntotsys}{2157\xspace}
\newcommand{\ntotspec}{4383\xspace}
\newcommand{\nourtotsys}{1653}
\newcommand{\nothertotsys}{504\xspace}
\newcommand{\nL}{20\xspace}

\newcommand{\ha}{H\ensuremath{\alpha}\xspace}
\newcommand{\hb}{H\ensuremath{\beta}\xspace}
\newcommand{\hc}{H\ensuremath{\gamma}\xspace}
\newcommand{\hd}{H\ensuremath{\delta}\xspace}
\newcommand{\oii}{[\ion{O}{2}]\xspace}

\newcommand{\lya}{Ly\ensuremath{\alpha}\xspace}

\newcommand{\lam}{\ensuremath{\uplambda}\xspace}
\newcommand{\lamlam}{\ensuremath{\uplambda\uplambda}\xspace}
\newcommand{\phz}{photo-\emph{z}\xspace}
\newcommand{\rr}{\emph{Redrock}\xspace}
\newcommand{\sigv}{\ensuremath{\sigma_v}\xspace}

\newcommand{\thetaE}{\ensuremath{\theta_E}\xspace}
\newcommand{\zs}{\ensuremath{z_s}\xspace}
\newcommand{\zd}{\ensuremath{z_d}\xspace}

\newcommand{\twopr}{^{\prime \prime}}

\newcommand{\cwr}{\textcolor{black}}
\newcommand{\cwrcolor}{\color{black}}

\newcommand{\hsc}{HSC~DR2\xspace}
\newcommand{\ls}{LS~DR9\xspace}
\newcommand{\nfujispec}{106}
\usepackage{enumitem}
\usepackage{array}
\setlist[itemize]{noitemsep}

\begin{document}

\title{\projname~II:\\
DESI Spectroscopy for Strong Lens Candidates
}

\correspondingauthor{Xiaosheng Huang}
\email{xhuang22@usfca.edu}

\author[0000-0001-8156-0330]{Xiaosheng~Huang}
\affiliation{Department of Physics \& Astronomy, University of San Francisco, San Francisco, CA 94117, USA}
\affiliation{Physics Division, Lawrence Berkeley National Laboratory, 1 Cyclotron Road, Berkeley, CA 94720, USA}

\author[0009-0009-8667-763X]{Jose Carlos Inchausti}
\affiliation{Department of Physics \& Astronomy, University of San Francisco, San Francisco, CA 94117, USA}

\author[0000-0002-0385-0014]{Christopher~J.~Storfer}
\affiliation{Institute for Astronomy, University of Hawai'i, 
Honolulu, HI 96822-1897, USA}
\affiliation{Physics Division, Lawrence Berkeley National Laboratory, 1 Cyclotron Road, Berkeley, CA 94720, USA}

\author[0000-0003-3533-5890]{S. Tabares-Tarquinio}
\affiliation{Physics Division, Lawrence Berkeley National Laboratory, 1 Cyclotron Road, Berkeley, CA 94720, USA}
\affiliation{Department of Physics \& Astronomy, University of San Francisco, San Francisco, CA 94117, USA}

\author[0000-0002-2733-4559]{J.~Moustakas}
\affiliation{Department of Physics and Astronomy, 
Siena College, 515 Loudon Road, Loudonville, NY 12211, USA}

\author[0000-0003-1889-0227]{W.~Sheu}
\affiliation{Department of Physics \& Astronomy, 
University of California, Los Angeles, Los Angeles, CA 90095, USA}
\affiliation{Physics Division, Lawrence Berkeley National Laboratory, 
1 Cyclotron Road, Berkeley, CA 94720, USA}

\author[0000-0002-2350-4610]{S.~Agarwal}
\affiliation{University of Chicago, Department of Astronomy, Chicago, IL 60615, USA}
\affiliation{Physics Division, Lawrence Berkeley National Laboratory, 1 Cyclotron Road, Berkeley, CA 94720, USA}

\author[0009-0008-0518-8045]{M.~Tamargo-Arizmendi}
\affiliation{Physics Division, Lawrence Berkeley National Laboratory, 
1 Cyclotron Road, Berkeley, CA 94720, USA}
\affiliation{Department of Physics \& Astronomy, University of Pittsburgh, Pittsburgh, PA 15260, USA}

\author[0000-0002-5042-5088]{D.J.~Schlegel}
\affiliation{Physics Division, Lawrence Berkeley National Laboratory, 
1 Cyclotron Road, Berkeley, CA 94720, USA}

\author{J.~Aguilar}
\affiliation{Physics Division, Lawrence Berkeley National Laboratory, 1 Cyclotron Road, Berkeley, CA 94720, USA}

\author[0000-0001-6098-7247]{S.~Ahlen}
\affiliation{Physics Dept., Boston University, 590 Commonwealth Avenue, Boston, MA 02215, USA}

\author{G.~Aldering}
\affiliation{Physics Division, Lawrence Berkeley National Laboratory, 1 Cyclotron Road, Berkeley, CA 94720, USA}

\author[0000-0003-4162-6619]{S.~Bailey}
\affiliation{Physics Division, Lawrence Berkeley National Laboratory, 1 Cyclotron Road, Berkeley, CA 94720, USA}

\author{S.~Banka}
\affiliation{Department of Electrical Engineering \& Computer Sciences, 
University of California, Berkeley, Berkeley, CA 94720}

\author[0000-0001-5537-4710]{S.~BenZvi}
\affiliation{Department of Physics \& Astronomy, University of Rochester, 
206 Bausch and Lomb Hall, P.O. Box 270171, 
Rochester, NY 14627-0171, USA}

\author[0000-0001-9712-0006]{D.~Bianchi}
\affiliation{Dipartimento di Fisica ``Aldo Pontremoli'', Universit\`a degli Studi di Milano, 
Via Celoria 16, I-20133 Milano, Italy}

\author[0000-0002-9836-603X]{A.~Bolton}
\affiliation{NSF's National Optical-Infrared Astronomy Research Laboratory, Tucson, AZ 85719, USA}

\author{D.~Brooks}
\affiliation{Department of Physics \& Astronomy, University College London, 
Gower Street, London, WC1E 6BT, UK}

\author[0000-0001-7101-9831]{A.~Cikota}
\affiliation{Gemini Observatory / NSF's NOIRLab, Casilla 603, La Serena, Chile}

\author{T.~Claybaugh}
\affiliation{Physics Division, Lawrence Berkeley National Laboratory, 
1 Cyclotron Road, Berkeley, CA 94720, USA}

\author[0000-0002-0553-3805]{K.~S.~Dawson}
\affiliation{Department of Physics and Astronomy, The University of Utah,
115 South 1400 East, Salt Lake City, UT 84112, USA}

\author[0000-0002-1769-1640]{A.~de~la~Macorra}
\affiliation{Instituto de F\'{\i}sica, Universidad Nacional Aut\'{o}noma de M\'{e}xico, 
Circuito de la Investigaci\'{o}n Cient\'{\i}fica, Ciudad Universitaria, 
Cd. de M\'{e}xico  C.~P.~04510,  M\'{e}xico}

\author[0000-0002-4928-4003]{A.~Dey}
\affiliation{NSF's National Optical-Infrared Astronomy Research Laboratory, Tucson, AZ 85719, USA}

\author{P.~Doel}
\affiliation{Department of Physics \& Astronomy, University College London, 
Gower Street, London, WC1E 6BT, UK}

\author{J.~Edelstein}
\affiliation{Space Sciences Laboratory, University of California, Berkeley, 7 Gauss Way, Berkeley, CA 94720, USA}

\author[0000-0002-2890-3725]{J.~E.~Forero-Romero}
\affiliation{Departamento de F\'isica, Universidad de los Andes, Cra. 1 No. 18A-10, Edificio Ip, CP 111711, Bogot\'a, Colombia}
\affiliation{Observatorio Astron\'omico, Universidad de los Andes, 
Cra. 1 No. 18A-10, Edificio H, CP 111711 Bogot\'a, Colombia}

\author{E.~Gazta\~{n}aga}
\affiliation{Institut d'Estudis Espacials de Catalunya (IEEC), 
c/ Esteve Terradas 1, Edifici RDIT, Campus PMT-UPC, 08860 Castelldefels, Spain}
\affiliation{Institute of Cosmology and Gravitation, 
University of Portsmouth, Dennis Sciama Building, Portsmouth, PO1 3FX, UK}
\affiliation{Institute of Space Sciences, ICE-CSIC, Campus UAB, 
Carrer de Can Magrans s/n, 08913 Bellaterra, Barcelona, Spain}

\author[0000-0003-3142-233X]{S.~Gontcho~A~Gontcho}
\affiliation{Physics Division, Lawrence Berkeley National Laboratory, 
1 Cyclotron Road, Berkeley, CA 94720, USA}

\author[0000-0003-4089-6924]{A.~X.~Gonzalez-Morales}
\affiliation{Departamento de F\'{\i}sica, DCI-Campus Le\'{o}n, Universidad de Guanajuato, Loma del Bosque 103, Le\'{o}n, Guanajuato C.~P.~37150, M\'{e}xico}

\author[0000-0003-2748-7333]{A.~Gu}
\affiliation{Quantum Science and Engineering, 
Harvard University, Cambridge, MA 02138, USA}

\author[0000-0002-6550-2023]{K.~Honscheid}
\affiliation{Center for Cosmology and AstroParticle Physics, 
The Ohio State University, 191 West Woodruff Avenue, Columbus, OH 43210, USA}
\affiliation{Department of Physics, The Ohio State University, 
191 West Woodruff Avenue, Columbus, OH 43210, USA}
\affiliation{The Ohio State University, Columbus, 43210 OH, USA}


\author[0000-0002-6024-466X]{M.~Ishak}
\affiliation{Department of Physics, The University of Texas at Dallas, 800 W. Campbell Rd., Richardson, TX 75080, USA}

\author[0000-0002-0000-2394]{S.~Juneau}
\affiliation{NSF NOIRLab, 950 N. Cherry Ave., Tucson, AZ 85719, USA}

\author{R.~Kehoe}
\affiliation{Department of Physics, Southern Methodist University, 
3215 Daniel Avenue, Dallas, TX 75275, USA}

\author[0000-0003-3510-7134]{T.~Kisner}
\affiliation{Physics Division, Lawrence Berkeley National Laboratory, 
1 Cyclotron Road, Berkeley, CA 94720, USA}

\author[0000-0003-2644-135X]{S.~E.~Koposov}
\affiliation{Institute for Astronomy, University of Edinburgh, Royal Observatory, Blackford Hill, Edinburgh EH9 3HJ, UK}
\affiliation{Institute of Astronomy, University of Cambridge, Madingley Road, Cambridge CB3 0HA, UK}

\author[0000-0001-9802-362X]{K.J.~Kwon}
\affiliation{Department of Physics, University of California, Santa Barbara, 
Santa Barbara, CA 93106, USA}

\author{A.~Lambert}
\affiliation{Physics Division, Lawrence Berkeley National Laboratory, 
1 Cyclotron Road, Berkeley, CA 94720, USA}

\author[0000-0003-1838-8528]{M.~Landriau}
\affiliation{Physics Division, Lawrence Berkeley National Laboratory, 
1 Cyclotron Road, Berkeley, CA 94720, USA}

\author[0000-0002-1172-0754]{D.~Lang}
\affiliation{Perimeter Institute for Theoretical Physics, 
Waterloo, ON N2L 2Y5, Canada}

\author[0000-0001-7178-8868]{L.~Le~Guillou}
\affiliation{Sorbonne Universit\'{e}, CNRS/IN2P3, 
Laboratoire de Physique Nucl\'{e}aire et de Hautes Energies (LPNHE), 
FR-75005 Paris, France}

\author[0000-0003-1887-1018]{M.~E.~Levi}
\affiliation{Lawrence Berkeley National Laboratory, 1 Cyclotron Road, Berkeley, CA 94720, USA}

\author{J.~Liu}
\affiliation{Department of Physics, Rose-Hulman Institute of Technology, 
Terre Haute, IN 47803, USA}

\author[0000-0002-1125-7384]{A.~Meisner}
\affiliation{NSF's National Optical-Infrared Astronomy Research Laboratory, Tucson, AZ 85719, USA}

\author{R.~Miquel}
\affiliation{Instituci\'{o} Catalana de Recerca i Estudis Avan\c{c}ats, 
Passeig de Llu\'{\i}s Companys, 23, 08010 Barcelona, Spain}
\affiliation{Institut de F\'{i}sica d’Altes Energies (IFAE), 
The Barcelona Institute of Science and Technology, Edifici Cn, Campus UAB, 
08193, Bellaterra (Barcelona), Spain}

\author{A.~D.~Myers}
\affiliation{Department of Physics \& Astronomy, 
University of Wyoming, 1000 E. University, Dept.~3905, 
Laramie, WY 82071, USA}

\author[0000-0002-4436-4661]{S.~Perlmutter}
\affiliation{Physics Division, Lawrence Berkeley National Laboratory, 
1 Cyclotron Road, Berkeley, CA 94720, USA}
\affiliation{Department of Physics, University of California, 
Berkeley, CA 94720, USA}

\author[0000-0003-3188-784X]{N.~Palanque-Delabrouille}
\affiliation{IRFU, CEA, Universit\'{e} Paris-Saclay, F-91191 Gif-sur-Yvette, France}
\affiliation{Lawrence Berkeley National Laboratory, 1 Cyclotron Road, Berkeley, CA 94720, USA}

\author[0000-0001-6979-0125]{I.~P\'erez-R\`afols}
\affiliation{Departament de F\'isica, EEBE, Universitat Polit\`ecnica de Catalunya, 
c/Eduard Maristany 10, 08930 Barcelona, Spain}

\author{C.~Poppett}
\affiliation{Lawrence Berkeley National Laboratory, 1 Cyclotron Road, Berkeley, CA 94720, USA}
\affiliation{Space Sciences Laboratory, University of California, Berkeley, 7 Gauss Way, Berkeley, CA  94720, USA}
\affiliation{University of California, Berkeley, 110 Sproul Hall \#5800 Berkeley, CA 94720, USA}

\author[0000-0001-7145-8674]{F.~Prada}
\affiliation{Instituto de Astrof\'{i}sica de Andaluc\'{i}a (CSIC), 
Glorieta de la Astronom\'{i}a, s/n, E-18008 Granada, Spain}

\author{G.~Rossi}
\affiliation{Department of Physics and Astronomy, 
Sejong University, 209 Neungdong-ro, Gwangjin-gu, Seoul 05006, Republic of Korea}

\author[0000-0001-5402-4647]{D.~Rubin}
\affiliation{Department of Physics \& Astronomy, 
University of Hawai'i at M\~{a}noa, Honolulu, HI 96822, USA}
\affiliation{Physics Division, Lawrence Berkeley National Laboratory, 
1 Cyclotron Road, Berkeley, CA 94720, USA}

\author[0000-0002-9646-8198]{E.~Sanchez}
\affiliation{CIEMAT, Avenida Complutense 40, E-28040 Madrid, Spain}

\author{M.~Schubnell}
\affiliation{Department of Physics, University of Michigan, 
450 Church Street, Ann Arbor, MI 48109, USA}
\affiliation{University of Michigan, 500 S. State Street, 
Ann Arbor, MI 48109, USA}

\author{Y.~Shu}
\affiliation{Purple Mountain Observatory, Chinese Academy of Sciences, 
Nanjing 210023, People's Republic of China}

\author{E.~Silver}
\affiliation{Department of Astronomy, University of California, 
Berkeley, CA 94720, USA}

\author{D.~Sprayberry}
\affiliation{NSF's National Optical-Infrared Astronomy Research Laboratory, Tucson, AZ 85719, USA}

\author[0000-0001-7266-930X]{N.~Suzuki}
\affiliation{Physics Division, Lawrence Berkeley National Laboratory, 
1 Cyclotron Road, Berkeley, CA 94720, USA}

\author[0000-0003-1704-0781]{G.~Tarl\'{e}}
\affiliation{University of Michigan, 500 S. State Street, 
Ann Arbor, MI 48109, USA}

\author{B.A.~Weaver}
\affiliation{NSF's National Optical-Infrared Astronomy Research Laboratory, Tucson, AZ 85719, USA}

\author[0000-0002-6684-3997]{H.~Zou}
\affiliation{National Astronomical Observatories, Chinese Academy of Sciences, 
A20 Datun Rd., Chaoyang District, Beijing, 100012, P.R. China}


\begin{abstract}
We present the Dark Energy Spectroscopic Instrument (DESI) \progname.
This is a spectroscopic follow-up program for strong gravitational lens candidates found in the DESI Legacy Imaging Surveys footprint.  
Spectroscopic redshifts for the lenses and lensed source are crucial for lens modeling to obtain physical parameters.
The spectroscopic catalog in this paper consists of \nfujisys candidate systems from the DESI Early Data Release (EDR).
We have confirmed \nL strong lensing systems and determined four to not be lenses. 
For the remaining systems, more spectroscopic data from ongoing and future observations will be presented in future publications.
We discuss the implications of our results for lens searches with neural networks in existing and future imaging surveys as well as for lens modeling.
This \progname is part of the \projname project, and this is Paper II of a series on this project. 
\end{abstract}


\section{Introduction} \label{sec:intro}
Strong gravitational lensing \citep{einstein1936a, zwicky1937a} systems are a powerful tool for astrophysics and cosmology. 
They provide an effective way to measure dark matter (DM) mass distribution in the central regions of galaxies and galaxy clusters \citep[e.g.,][]{kochanek1991a, koopmans2002a, bolton2006a, koopmans2006a, bolton2008a, 
bradac2008a, huang2009a, jullo2010a, grillo2015a, tessore2016a, shu2017a}
and the only way to detect DM substructure at cosmological distances 
 or line-of-sight low-mass halos \citep[e.g.,][]{vegetti2010a, hezaveh2016a, cagan-sengul2022a}. 
Moreover, time-delay $H_{0}$ measurements from multiply imaged supernovae \citep[e.g.,][]{pierel2019a, kelly2023a, suyu2023a} and lensed quasars \citep[e.g.,][]{wong2020a}, 
combined with measurements from distance ladders \citep[e.g.,][]{freedman2020a, riess2022a},
provide consistency checks and competitive constraints on $H_{0}$.

We found $\sim 3500$ new strong gravitational lensing candidates in the DESI Legacy Imaging Surveys.\footnote{The entire catalog of these candidates can be found on our project website \url{https://sites.google.com/usfca.edu/neuralens/}.} 
These consist of $\sim 3000$ new lenses candidates identified using residual neural networks \citep[ResNet;][hereafter H20, H21, and S24, respectivley]{huang2020a, huang2021a, storfer2024a}
and 436 new lensed quasar candidates identified using an autocorrelation algorithm \citep{dawes2023a}.
In the same dataset, we also found lensed supernova \citep{sheu2023a} and lensed quasar candidates in targeted lensed transient searches using differencing image techniques \citep{sheu2024a}.

For high resolution imaging, our \HST SNAP program (Go-15867, PI: Huang) 
observed 51 of our most promising candidates. 
We report the results of this program in Paper~I of this series \citep{huang2025a}.  
Spectroscopic observations are
also being carried out. 
The Dark Energy Spectroscopic Instrument  \citep[DESI;][]{desi2016a, desi2016b}
is uniquely suited for spectroscopic observations of such a large number of lensing systems.
In this paper --- Paper~II in the \projname series --- we present the DESI \progname \citep[see also][]{desi2023a}.
For a subset of the candidate systems in this program \cwr{(preliminarily, $\sim 30\%$)} the source redshifts are too high and the typical emission features (e.g., the [\ion{O}{2}] doublet at \lamlam 3726, 3729~\ang) are beyond the optical range.
We target these with an ongoing near-IR  spectroscopic program on the Keck~2 Telescope. 
The first results will be reported in Paper~III in this series (Agarwal et al., in prep).

Here, we describe the DESI \progname
in \S~\ref{sec:desi-imaging_secondary} and present lensing spectroscopic results from DESI Early Data Release (EDR, or equivalently, the Fuji data production) in \S~\ref{sec:spect-data},
including the velocity dispersion of the lensing galaxies (confirmed or putative) measured using \texttt{FastSpecFit}(\citealt{moustakas2023a}; J.\ Moustakas et al., in prep).\footnote{\url{https://fastspecfit.readthedocs.org}}
We discuss our results in \S~\ref{sec:discuss} and conclude in \S~\ref{sec:conclusion}.

\section{DESI \progname}
\label{sec:desi-imaging_secondary}
We briefly describe the DESI Legacy Imaging Surveys
and our lens discoveries in \S~\ref{sec:desi-imaging} and
introduce the \progname 
in \S~\ref{sec:secondary}.

\subsection{Lens Discoveries in the DESI Legacy Imaging Surveys}
\label{sec:desi-imaging}
The Legacy Imaging Surveys (LS) was carried out to provide targets for the DESI spectroscopic observations.
It is composed of three surveys: 
the Dark Energy Camera Legacy Survey (DECaLS), the Beijing-Arizona Sky Survey (BASS), and the Mayall $z$-band Legacy Survey (MzLS). Altogether, these surveys cover a total footprint of $\sim19,000$~deg$^2$ \citep{flaugher2015a,abbott2018a,williams2004a} divided into two contiguous regions separated by the Galactic plane. A more detailed description of the Legacy Surveys can be found on the Legacy Surveys website.\footnote{\href{https://www.legacysurvey.org/dr9/description/}{https://www.legacysurvey.org/dr9/description/}}

We performed three lens searches on this dataset and presented them in H20, H21, S24, totaling $\sim 3500$ new lens candidates.
We used residual neural network architectures \citep[][H21]{lanusse2018a} trained on images of both lenses and non-lenses.
Applying an autocorrelation algorithm to the DESI quasar sample \citep{yeche2020a}, we also found 436 new lensed quasar candidates \citep{dawes2023a}.

\subsection{DESI Strong Lensing Spectroscopic Program}
\label{sec:secondary}
DESI, on the 4-meter Mayall Telescope at Kitt Peak, began observations in 2021 and will measure spectroscopic redshifts of approximately 40 million extragalactic targets over five years, 
including Emission Line Galaxies (ELGs), Luminous Red Galaxies (LRGs) and quasars (QSOs) \citep{Abareshi2022a,desi2023a, guy2023a}.
\cwr{Such an ambitious project is made possible through the wide-field optical corrector installed on the Mayall Telescope \citep{miller2023a} and highly multiplexed fiber spectroscopy \citep{poppett2024a}.}
\cwr{Based on observing conditions, DESI dynamically adjusts the raw exposure time (typically 1000-4000~sec) to achieve a consistent signal-to-noise equivalent to a 1000 sec effective exposure time under nominal observation conditions.}

In addition to its primary science goals, the DESI survey incorporates a range of ``secondary'' targets to pursue bespoke science goals, which includes the spectroscopic follow-up of strong gravitational lenses found in LS.
This is the DESI 
\progname.\footnote{Note that the lensed quasar candidates found in \citet{dawes2023a} will be observed in the DESI main survey. We will present the spectroscopic data for those candidates in a separate publication.}
Spectroscopic redshifts for the lenses and lensed sources are crucial for determining physical parameters from lens modeling.
Furthermore, the combination of velocity dispersion and lens modeling 
can break the mass-sheet degeneracy and constrain the mass profile via Jean's Equation 
\citep[e.g.,][]{grogin1996a, romanowsky1999a, treu2002a,  koopmans2006a, auger2010a, suyu2017a}.

The DESI pipeline converts the observed raw data into sky-subtracted, 
flux-calibrated spectra, classifications, and redshifts. 
Spectral classifications and redshifts are measured using the \rr software package\footnote{\url{https://github.com/desihub/redrock/releases/tag/0.15.4}} (Bailey et al., in prep) via $\chi^2$ fitting of PCA templates.
This procedure is similar to the method used in SDSS/BOSS \citep{bolton2012a}.
Redshift failure is indicated in the pipeline with the ZWARN flag.
ZWARN=0 means no errors due to instrument or data and no flags from spectral fitting issues \citep[for more detail, see Bailey et al. in prep;][\S~5.3.1]{schlafly2023a}.
The most common flag is ZWARN=4.
\cwr{This indicates that bit 2 is set, which is ``DELTACHI2'', meaning that the difference in $\chi^2$ between the best model fit and the second best model is too small to be statistically robust ($< 3\sigma$ significant)}.
Note that the \rr best-fit redshift may not be correct even if ZWARN=0 and visual inspection is necessary.

This program includes all the candidates reported in H20 and H21 that are in the DESI footprint ($\delta > -20^\circ$), or 1324 systems.
We did not impose any grade cut in order to fully evaluate the results of our first two ResNet searches.
We also include 504 systems from other lens searches. Those that were discovered prior to the year 2017 are selected from the compilation of \citet{moustakas2012a}\footnote{ 
Among the system in the The Master Lens Database, those that are included in this work are from \citet{belokurov2007a,hennawi2008a,belokurov2009a,diehl2009a,kubo2010a,bayliss2011a,more2012a,oguri2012a,sonnenfeld2013a,stark2013a,gavazzi2014a,more2016a,shu2016a,tanaka2016a}.}, and those since, from 
\citet{diehl2017a,sonnenfeld2018a,wong2018a,jacobs2019b,petrillo2019a,sonnenfeld2019b,chan2020a,jaelani2020a, li2020a,sonnenfeld2020a,canameras2021a,
odonnell2022a,rojas2022a,savary2022a,shu2022a,wong2022a}.
In these cases, we select only the best candidates.
This is because the methodologies (from search methods to human grading schemes) were highly varied, and it would be difficult to used DESI data to perform meaningful evaluation. 
At the time of target submission, 
we just completed the lens search in LS DR9. 
We included in the target list the best 306 systems from the search based on preliminary human grading (the search results were later reported in S24.
We also included a small number of systems (with a total of 102 fibers) that did not make the quality threshold to be included in S24 but their nature were worth exploring for the purpose of improving the training sample for further lens searches.\footnote{
A reminder to the reader: we use real observed images in our training sample for our lens searches.}
The total number of systems is \ntotsys (Table~\ref{tab:summary}). 
Each system has at least one fiber for the lens and one fiber for the arc, 
and in a small percentage of cases, multiple fibers for source images in a single system.
Thus the total number of targets is slightly more than twice the number of systems, for a total of \ntotspec unique targets.\footnote{Note that this number has been updated from 3558 in \citet{desi2023a}.}


In this paper, we will present results from this program in DESI EDR.
There are a total of \nfujispec\ unique spectra, 
for {\nfujisys} systems.
For 15 spectra (14 systems), the redshifts are inconclusive based on DESI spectra from fibers targeting the lensed sources (14) or the lensing galaxy (1).
We do not include these spectra in this paper.
One spectrum (for DESI-239.561000+42.6188, targeting the lens) has a 120~sec exposure, with ZWARN=1570.
This corresponds to four ZWARN Bits (1, 5, 9, and 10), the most significant of which is 9, or ``NO DATA''.
However, it was observed by another DESI fiber with a different TARGETID, with an exposure time of 938~sec. 
This spectrum is included in this paper.
But note that the DESI TARGET bitmask
does not match the strong lens program \citep[34, from Table~16 of][]{desi2023a}.\footnote{See also, \citet{myers2023a}.} 
We therefore show the \nspecshown spectra that have yielded convincing redshift results.
Of the \nfujisys systems, \noursys are from our searches (H20 and H21),
with the overwhelming majority of the rest from a series of lens searches in HyperSprime-Cam Subaru Strategic Survey \citep[HSC SSP;][]{aihara2018a}.
This is due to a high degree of overlap between DESI Survey Validation 3 (SV3) --- which constitutes a significant portion of the EDR --- and HSC~SSP \citep{desi2023a}.

\medskip 
\begin{adjustwidth}{-2cm}{0cm}
\begin{minipage}{\linewidth}
    \centering
    \captionof{table}{\progname}
    \label{tab:summary}
        \scriptsize
        \hspace{-0.5 in}\begin{tabular}{cccc|c}
        \toprule
        Dataset  & Systems & H20, H21 \& S24 & Other Sources &  Targets \\
        (1) & (2) & (3) & (4) & (5) \\
        \midrule
        EDR & \nfujisys & \textbf{\noursys} & \nothersys  & \nfujispec\\
        Total  & \ntotsys  & \textbf{\nourtotsys$^a$} & \nothertotsys & \ntotspec \\
        \bottomrule
        \end{tabular} \\[5 pt]  
    \begin{adjustwidth}{6 cm}{5 cm}
    \textsc{Note} -- 
    Column~1: DESI EDR (this paper) and the total dataset (future papers in this series).
    Column~2: The number of systems.
    Column~3: The number of systems from our searches (highlighted with boldface).
    Column~4: The number of systems from other search papers.
    Column~5: The number of DESI fibers.\\
    $^a$These also include 102 spectra for systems that are from the LS DR9 lens search but not included in S24 (see text).
    \end{adjustwidth}
\end{minipage}
\end{adjustwidth}



\section{Strong Lens Candidates in DESI EDR} 
\label{sec:spect-data}
We present the spectroscopic results from DESI EDR for this program below.
\cwr{Seventy-three strong lens candidate systems are in the EDR, summarized in Table~\ref{tab:master}.
We inspect all the observed spectra and determine if the \rr redshift is reliable, and make corrections when necessary.}
Overall, for lens targets, confirmed or putative, DESI obtains redshifts for 72 out of \nfujisys observed spectra.
For lensed source galaxy targets, confirmed or putative, DESI obtains redshifts for 22 out of 36 observed spectra.
\cwr{We present the \nL confirmed systems in \S\,\ref{sec:confirmed-lenses}, ordered in ascending RA.
One previously known system is shown \S\,\ref{sec:known-lens}. A small number of such systems are chosen for this program for the purpose of comparison and additional validation.}
\cwr{All remaining spectra, which include  candidates with confirmation pending and confirmed nonlenses, are briefly mentioned in \S\,\ref{sec:other-sys} and presented in the Appendix.}

\begingroup
\cwrcolor
\begin{adjustwidth}{1.5cm}{0cm}
\centering
\captionof{table}{All Systems}\label{tab:master}
    \scriptsize
    \hspace{-2 cm}\begin{longtable}{lcccccc}
    \toprule
    System Name & Discovery Paper & Discovery Name & \zd & \zs & $\sigma_v$ [km s$^{-1}$] & VI \\
    (1) & (2) & (3) & (4) & (5) & (6) & (7) \\
    \midrule
    \multicolumn{7}{c}{\textbf{Confirmed systems (20)}} \\
    \midrule
   DESI~J149.8209+01.0331 & 1 & HSCJ095917+010159 & 0.4461 & 0.8100 & 152 $\pm$ 100 & \\
     &  &  &   & 1.4794 &  & \zs \\
    DESI~J179.0547-01.0339 & 1 & HSCJ115613-010158 & 0.4243 & 1.2600 & 328 $\pm$ 35 &  \\
    DESI~J183.0990-01.5510 & 1, 2 & HSCJ121223-013304 & 0.4039 & 0.9887 & 220 $\pm$ 19 &  \\
    DESI~J183.6001-00.5351 & $3^a$ &  & 0.2481 & 1.1159 & 306 $\pm$ 19 & \zs \\
    DESI~J188.2847+60.1921$^b$ & 4 & SDSS J1233+6011 & 0.5526 & 1.6185 & 261 $\pm$ 29 & \zs \\
    DESI~J212.9021-01.0377 & 5 & HSCJ141136-010215 & 0.9477 & 3.0230 & 168 $\pm$ 47 & \zs \\
    DESI~J215.1311+00.3392 & 5 & HSCJ142031+002021 & 0.4750 & 3.1103 & 227 $\pm$ 18 & \zs \\
    DESI~J216.9515+00.1663 & 6 & HSCJ142748+000958 & 0.5890 & 1.5701 & 253 $\pm$ 24 &  \\
    DESI~J217.0936+03.3000 & 2, 7 & HSCJ142822+031759 & 0.5887 & 1.3072 & 379 $\pm$ 36 &  \\
    DESI~J218.4780+03.0038 & 2, 8 & PS1J1433+0300 & 0.5866 & 1.5721 & 264 $\pm$ 23 &  \\
    DESI~J218.5371-00.2199 & 5 & HSCJ143408-001311 & 0.7135 & 1.4421 & 230 $\pm$ 22 & \zd \\
    DESI~J218.8286-01.4239 & 5 & HSCJ143518-012526 & 0.7553 & 1.2253 & 152 $\pm$ 100 &  \\
    DESI~J220.3619-00.3166 & 1 & HSCJ144126-001859 & 0.2875 & 1.2124 & 338 $\pm$ 13 &  \\
    DESI~J220.3862-00.8995$^d$ & 1, 2 & HSCJ144132-005359 & 0.5372 &  & 417 $\pm$ 27 &  \\
    DESI~J239.5610+42.6188$^b$ & 5 & HSCJ155815-423708 & 0.6772 & 1.3801 & 199 $\pm$ 24 & \zd, \zs \\
    DESI~J239.9897+44.2621 & 6 & HSCJ155957+441543 & 0.5977 & 1.5290 & 394 $\pm$ 37 &  \\
    DESI~J241.7346+42.1102$^b$ & 2, 5 & HSCJ160656+420636 & 0.7727 & 1.5480 & 271 $\pm$ 53 & \zs \\
    DESI~J245.3616+42.7620 & 1 & HSCJ162126+424540 & 0.1354 & 0.9700 & 348 $\pm$ 19 &  \\
    DESI~J215.2654+00.3719 & 9 &  & 0.6566 & 2.2066$^c$ & 331 $\pm$ 33 &  \\
    DESI~J254.4235+34.8162$^d$ & 2 &  & 0.8496 &  & 437 $\pm$ 97 &  \\
    \midrule
    \multicolumn{7}{c}{\textbf{Known system with lens observed by DESI (1)}}\\
    \midrule
    DESI~J216.2042-00.8893 & 13 &  HSCJ142449-005322 & 0.7940 &   & 427 $\pm$ 42 \\
    \midrule
    \multicolumn{7}{c}{\textbf{Confirmation pending with source not yet observed (34)}} \\
    \midrule
    DESI-149.6312+01.1603 & 1 & HSCJ095831+010937 & 0.5505 &  & 274 $\pm$ 22 &  \\
    DESI-150.4045+02.5544 & 9 &  & 0.2476 &  & 208 $\pm$ 21 &  \\
    DESI-151.7664+02.1430 & 9, 10 & HSCJ1007+0208 & 0.3708 &  & 291 $\pm$ 15 &  \\
    DESI-179.0412-00.3258 & 1 & HSCJ115609-001932 & 0.2601 &  & 418 $\pm$ 14 &  \\
    DESI-179.2209-00.6635 & 6 & HSCJ115653-003948 & 0.5081 &  & 286 $\pm$ 21 &  \\
    DESI-182.5944-01.1998 & 1 & HSCJ121022-011201 & 0.5744 &  & 211 $\pm$ 22 &  \\
    DESI-211.9741-00.4708 & 2, 5 & HSCJ140753-002815 & 0.4661 &  & 419 $\pm$ 93 &  \\
    DESI-213.9925+52.6654 & 5 & HSCJ141558+523955 & 0.5275 &  & 375 $\pm$ 28 &  \\
    DESI-214.1941-01.3037 & 5 & HSCJ141646-011813 & 0.7055 &  & 413 $\pm$ 40 &  \\
    DESI-214.8006+53.4366 & 2 &  & 0.6376 &  & 420 $\pm$ 49 &  \\
    DESI-215.2019+00.1259 & 1 & HSCJ142048+000733 & 0.5451 &  & 269 $\pm$ 20 &  \\
    DESI-215.3634+00.2015 & 1 & HSCJ142127+001205 & 0.5261 &  & 258 $\pm$ 23 &  \\
    DESI-215.9039-00.6763 & 5 & HSCJ142336-004034 & 0.8983 &  & 392 $\pm$ 52 &  \\
    DESI-216.6693+00.1663 & 5 & HSCJ142640+000958 & 0.5312 &  & 194 $\pm$ 29 &  \\
    DESI-216.7775+00.7208 & 3$^a$, 1 & HSCJ142706+004314 & 0.2951 &  & 306 $\pm$ 30 &  \\
    DESI-216.9786+00.6624 & 5 & HSCJ142754+003944 & 0.8535 &  & 392 $\pm$ 66 &  \\
    DESI-217.7996+03.0093 & 2 &  & 0.2631 &  & 182 $\pm$ 15 &  \\
    DESI-218.1831-00.7649 & 5 & HSCJ143243-004553 & 0.4662 &  & 373 $\pm$ 47 &  \\
    DESI-218.7266-00.9496 & 6, 1 & HSCJ143454-005658 & 0.7285 &  & 459 $\pm$ 38 &  \\
    DESI-219.0374-01.3295 & 9 &  & 0.5269 &  & 488 $\pm$ 84 &  \\
    DESI-219.8219-00.6550 & 5 & HSCJ143917-003917 & 0.7181 &  & 290 $\pm$ 27 &  \\
    DESI-220.9791-00.1253 & 1 & HSCJ144354-000731 & 0.8053 &  & 282 $\pm$ 48 &  \\
    DESI-236.2225+44.4637 & 1 & HSCJ154453+442749 & 0.6514 &  & 208 $\pm$ 29 &  \\
    DESI-238.3307+43.3068 & 6 & HSCJ155319+431824 & 0.6288 &  & 417 $\pm$ 90 &  \\
    DESI-239.1453+43.4038 & 5 & HSCJ155634+432413 & 0.7159 &  & 315 $\pm$ 42 &  \\
    DESI-239.6111+43.4752 & 6 & HSCJ155826+432830 & 0.4441 &  & 280 $\pm$ 21 &  \\
    DESI-239.6258+44.5610 & 5 & HSCJ155830+443339 & 0.6076 &  & 336 $\pm$ 34 &  \\
    DESI-240.7108+43.5847 & 1 & HSCJ160250+433505 & 0.4140 &  & 355 $\pm$ 21 &  \\
    DESI-242.0652+42.0026 & 1 & HSCJ160815+420009 & 0.6145 &  & 302 $\pm$ 40 &  \\
    DESI-244.5857+54.5052 & 1, 2 & HSCJ161820+543018 & 0.7940 &  & 326 $\pm$ 40 &  \\
    DESI-244.6470+54.8231 & 1 & HSCJ161835+544923 & 0.7905 &  & 258 $\pm$ 42 &  \\
    DESI-246.0930+43.6661 & 5 & HSCJ162422+433958 & 0.8290 &  & 609 $\pm$ 184 &  \\
    DESI-247.4262+43.8280 & 1 & HSCJ162942+434940 & 0.5282 &  & 285 $\pm$ 28 &  \\
    DESI-247.7865+42.5782 & 1 & HSCJ163108+423441 & 0.7300 &  & -- &  \\
    \midrule
    \multicolumn{7}{c}{\textbf{Confirmation pending with \zs not confirmed (13)}} \\
    \midrule
    DESI-149.4588+01.7521 & 11 & HSCJ095750+014507 & 0.5736 &  & 190$\pm$19 &  \\
    DESI-178.8726-00.7156 & 5 & HSCJ115529-004255 & 0.8148 &  & 410$\pm$67 &  \\
    DESI-180.1889-00.9614 & 12 & HSCJ120045-005740 & 0.6367 &  & 328$\pm$107 &  \\
    DESI-182.9875+00.3483 & 1 & HSCJ121156+002054 & 0.7748 &  & 326$\pm$32 &  \\
    DESI-216.2076+00.7006 & 1 & HSCJ142449+004202 & 0.4766 &  & 311$\pm$30 &  \\
    DESI-217.0485-00.8392 & 5 & HSCJ142811-005021 & 0.6318 &  & 349$\pm$22 &  \\
    DESI-217.9619+01.5055 & 5 & HSCJ143150+013019 & 0.7541 &  & 292$\pm$26 &  \\
    DESI-219.7558+00.8550 & 5 & HSCJ143901+005117 & 0.7847 &  & 257$\pm$22 &  \\
    DESI-239.5256+43.2839 & 5 & HSCJ155806+431702 & 0.8129 &  & 320$\pm$38 &  \\
    DESI-239.9599+42.8307 & 5 & HSCJ155950+424950 & 0.7375 &  & 397$\pm$49 &  \\
    DESI-240.5990+43.7727 & 1 & HSCJ160223+434621 & 0.4305 &  & 229$\pm$29 &  \\
    DESI-240.6045+43.7711 & 1 & HSCJ160225+434615 & 0.4264 &  & 278$\pm$40 &  \\
    DESI-240.7213+43.5837 & 1 & HSCJ160253+433501 & 0.4238 &  & 205$\pm$29 &  \\
    \midrule
    \multicolumn{7}{c}{\textbf{Confirmation pending with \zd and \zs not confirmed (1)}} \\
    \midrule
    DESI-243.4004+54.1155 & 1 & HSCJ161336+540656 &  &  &  &  \\
    \midrule
    \multicolumn{7}{c}{\textbf{Confirmed Nonlenses (4)}} \\
    \midrule
    DESI-150.2022+01.6538 & 9 &  & 0.2178 & 0.0805 & 248$\pm$19 &  \\
    DESI-217.2090+51.6467 & 2 &  & 0.7262 & 0.2436 & 215$\pm$26 &  \\
    DESI-218.8713+00.3034 & 5 & HSCJ143529+001812 & 0.5720 & 0.5722 & 240$\pm$46 &  \\
    DESI-219.9227+00.5073 & 9 &  & 0.1378 & 0.1369 & 132$\pm$5 &  \\
\bottomrule \\[-20pt] 
\end{longtable}  
    \begin{adjustwidth}{-.5 cm}{1 cm}
    \scriptsize
    \textsc{Note} -- 
    Column~1: system name used in this paper. 
    Column~2: the discovery paper. 
    Column~3: discover name (the name used in the discovery paper); for system reported in H20 or H21, the discovery name is essentially the same as the system name, but without the ``J" (see text). 
    Column~4 \& 5: lens and source redshifts, \zd and \zs, respectively, from DESI (unless otherwise noted). 
    Note that the first system has two confirmed sources. 
    Column~6: velocity dispersion (see \S~\ref{sec:vdisp}).
    Column~7: redshifts (with ZWARN~$> 0$) corrected or confirmed by human visual inspection (VI; for more details, see text).
    For Column~2, the citation legend is as follows, 
    1:~\citet{jaelani2020a},
    2: H21, 
    3:~\citet{petrillo2019a},
    4:~\citet{talbot2021a},
    5:~\citet{sonnenfeld2020a},
    6:~\citet{sonnenfeld2018a}, 
    7:~\citet{wong2022a},
    8:~\citet{canameras2020a},
    9:~H20,
    10:~\citet{canameras2021a},
    11:~\citet{wong2018a},
    12:~\citet{chan2020a},
    13:~\citet{tanaka2016a}\\
    $^a$This was also reported in Petrillo et al. (2019). From the LinKS website, \url{https://www.astro.rug.nl/lensesinkids/}, this system was given the ID 2187.\\
    $^b$The lens and source redshifts are measured from one spectrum.\\
    $^c$From Keck NIRES (Argarwal et al., in prep, Paper~III)\\
    $^d$Confirmed with \zd from DESI and imaging from \hst.
    \end{adjustwidth}

\end{adjustwidth}
\endgroup

\subsection{Confirmed Strong Lensing Systems}\label{sec:confirmed-lenses}
In this section, we present the \nL confirmed lenses in DESI EDR, \cwr{ordered in ascending RA and summarized in Table~\ref{tab:master}}.
We also present the velocity dispersion, \sigv, from \texttt{FastSpecFit}. 
\cwr{
For most (11) of these systems, \rr successfully measured both the lens and source redshifts.
For the lens spectrum, the 4000~\ang break and the Ca~H\&K feature (\lamlam 3933.66 \& 3968.47  \ang) are always present, and often, the G, \ion{Mg}{1}, and Na~D absorption lines, as well.
For the source spectrum, usually from a fiber centered on the brightest lensed image, the [\ion{O}{2}] doublet emission (\lamlam 3727.092, 3729.87 \ang) is typically the strongest feature by which we determine the redshift.
There are a few cases where corrections need to be made during visual inspection (VI). 
For redshifts above 1.6, the \oii emission feature will be redshifted outside of the optical range of DESI.
If \lya is present, it will appear in the DESI spectrum if the redshift is higher than 2.0.
In fact, two of the confirmed lensing systems exhibit \lya emission in the source spectrum (\S~\ref{sec:lya}), with $\zs > 3$.
However there is a redshift ``desert'' between 1.6 - 2.0 where no prominent emission feature exists in the optical range of DESI spectroscopy (3600 - 9800 \ang).
In addition, \lya is not always present in a spectrum.
That is why for one system in this set, near-IR spectroscopy was necessary for confirmation.
Finally, the lensing nature of two systems were confirmed by DESI redshift for the lens and high resolution \hst images that clearly reveal the lensed arcs.
}

\cwr{Note that in our discovery paper series (H20, H21, S24), the lens candidates are named following this convention: ``DESI'' followed with RA and Dec, both in digital format with four decimal places. 
In 2024, DESI established an official naming convention that is very similar to that for our candidates, with a small difference, the inclusion of ``J''. 
Thus,
in our strong lens discovery paper series (including a fourth one, J.~Inchausti et al. in prep) and the DESI Strong Lens Foundry series (including this paper), 
all candidates will continue to be named without the ``J'' whereas fully confirmed systems will include the ``J''. }

\begin{figure}[h]
  \centering
  \includegraphics[width=0.7\textwidth]{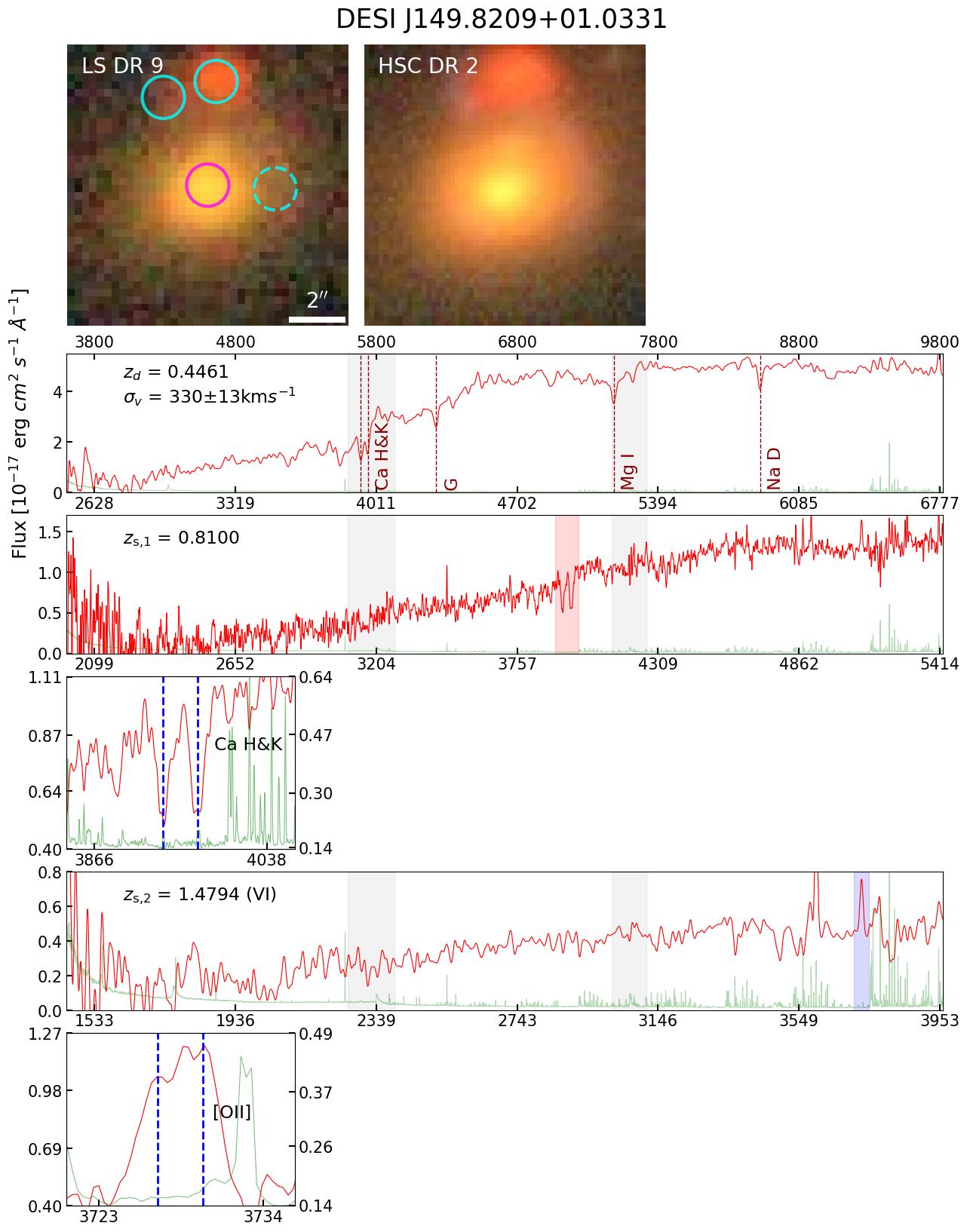}
  \caption{DESI~J149.8209+01.0331.
   \cwr{The following description generally applies to all confirmed systems in this subsection:
in the first row, we show the \ls image (left), and when available, also the \hsc image (right).
North is up, and east to the left.
In the LS image, the \lenscircle and \srccircle circles mark fiber positions for the lens and source(s), respectively, with dashed circles indicating a DESI targets not yet observed in EDR.
The size of the circles corresponds to the fiber diameter ($1.5''$).
The second row shows the lensing galaxy spectrum (red), together with the noise spectrum (green).
The top and bottom horizontal axes show the restframe and the observed wavelengths, respectively.
For the lens spectrum we mark the locations of the Ca~H\&K feature along with the G, Mg~I, and Na~D absorptions (dashed lines).
The third row shows a source spectrum, with the zoom-in of the detected  Ca~H\&K feature shown in the fourth row.
The fifth row shows the spectrum for a second source.
We highlight the [\ion{O}{2}] doublet emission feature with a blue band, and show its zoom-in in the sixth row.
In the respective panels, we indicate the redshifts of the lens and sources, and the velocity dispersion, \sigv, for the lens.
On both the lens and source spectra, the gray bands indicate the two dichroics separating the V, R and Z DESI spectrographs (see text).}
  }
  \label{fig:eg-lens-fig}
\end{figure}

\href{https://www.legacysurvey.org/viewer/?ra=149.8209&dec=01.0331&layer=hsc-dr2&pixscale=0.262&zoom=16}{\emph{DESI~J149.8209+01.0331}}\footnote{For the first mention of a system preceding the presentation of its figure,
we hyperlink the name to its Legacy Surveys DR9 image, or the \hsc image if available.} \, 
This system has has a lens redshift at $z_d = 0.4461$ (Figure~\ref{fig:eg-lens-fig}).
It has are two lensed sources at two different redshifts.
One source, the bright red object almost directly above (and slightly W of) the lens, is at $z_{s,1} = 0.8100$.
The object was targeted as the DESI LRG program.
It appears to be slightly elongated (more clearly in the HSC image) along the tangential direction, as expected for a lensed arc.
This indicates there may be a counterarc on the other side of the lens, likely very close to it.
The second source is part of a faint purplish arc that extends from 11 o'clock to 4 o'clock around the lens --- it is visible both in the \ls and \hsc images.
Note that on the sky the images of these two sources overlap.
One fiber for this source is centered on the NE end (11 o'clock) of the arc (solid \srccircle circle, Figure~\ref{fig:eg-lens-fig}).
Its spectrum has a strong emission line at 9245 \ang. 
The \rr pipeline misidentified it as \ha (\lam 6562 \ang), and therefore assigned an incorrect redshift of 0.4082.
From VI this feature appears to be doubly peaked with the separation matching that of the \oii doublet. 
Given that this feature is detected at high SNR, if it were \ha,  one could reasonably expect the presence of the weaker [\ion{N}{2}] line at \lam 6585.27 \ang and the [\ion{S}{2}] doublet at \lamlam~6718.29,~6732.67~\ang
(see, e.g., Paper \cwr{III}, Agarwal et al. in prep).
Their absence further weakens the case for this line being \ha.
Moreover, from the color, this object appears to be an elliptical galaxy and 
we note the downward trend starting immediately blueward of this feature.
This is consistent with this feature being \oii
(see, e.g., the downward trend blueward of \lam~3727~\ang in the second row spectrum, for the lensing galaxy, which is an elliptical galaxy);
but it is insistent with this emission being \ha, in which case one would expect a $\sim 1000$~\ang wide portion (restframe) of the spectrum blueward of this feature to be essentially flat.
We thus identify this feature as \oii and determine the redshift to be $z_{s,2} = 1.4794$. 
There is another DESI target that is not in the EDR (thus indicated as a dashed \srccircle circle, first row, Figure~\ref{fig:eg-lens-fig}).
The spectrum for this object is in the DESI Year~1 data.
Based on the presence of the same emission line, again identified as \oii, 
the preliminary redshift is determined to be the same as $z_{s,2}$.
This shows that the two parts of the purple arc-like object indeed belong to the same gravitationally lensed arc.
Therefore we confirm that this is a rare lensing systems with two sources at different redshifts, with the possibility of constraining cosmology \citep[e.g.,][]{sharma2022a}.

\begin{figure}[h]
  \centering
  \includegraphics[width=0.7\textwidth]{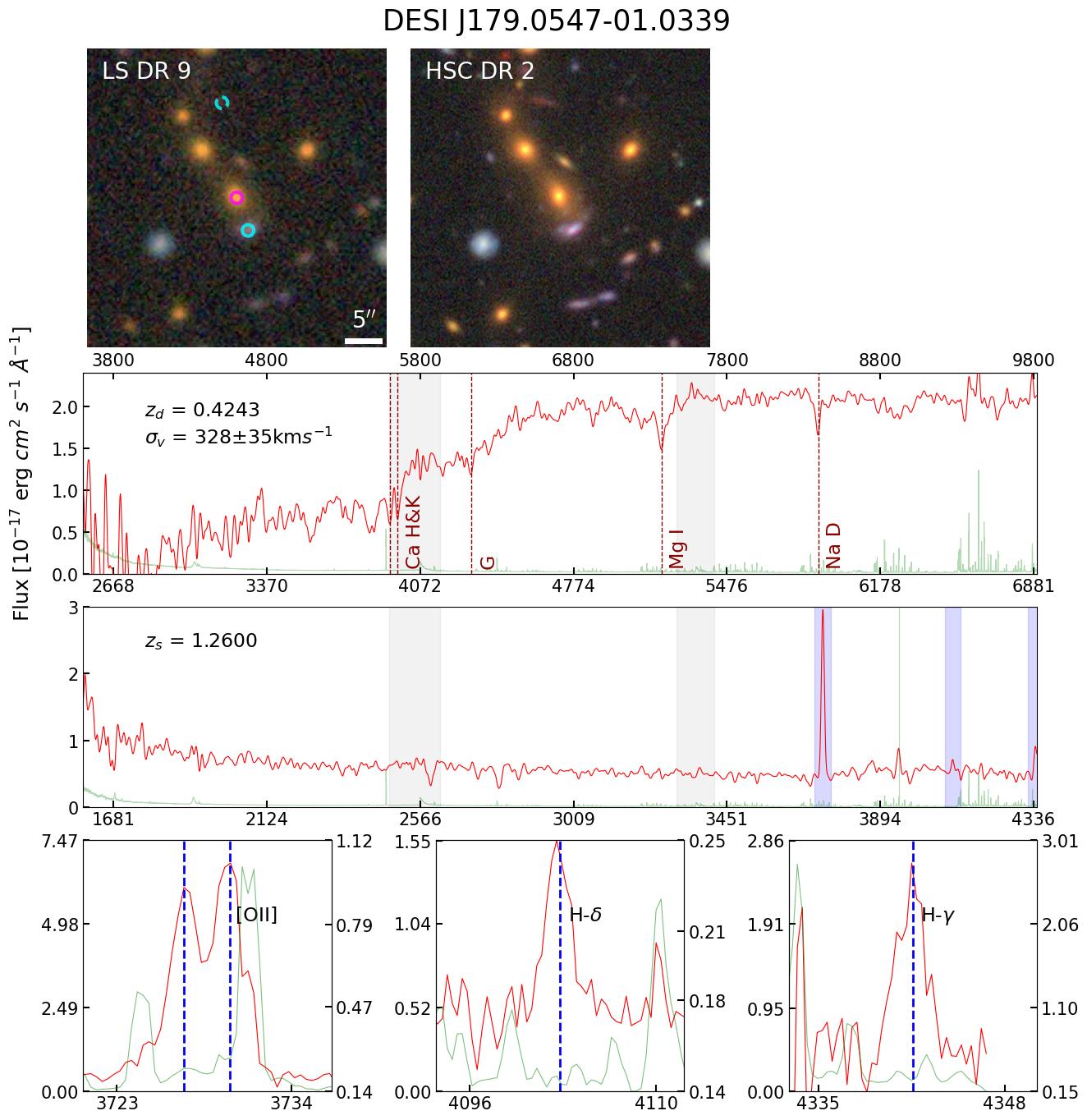}
  \caption{DESI~J179.0547-01.0339. For the arrangement of the panels, see Figure~\ref{fig:eg-lens-fig} caption. 
}
  \label{fig:desi179-01}
\end{figure}

\newpage
\href{https://www.legacysurvey.org/viewer/?ra=179.0547&dec=-1.0339&layer=hsc-dr2&pixscale=0.262&zoom=16}{\emph{DESI~J179.0547-01.0339}} \,
The lens for this system, at \zd = 0.4243 (Figure~\ref{fig:desi179-01}), appears to be the brightest galaxy in a group or a small cluster.
The purple lensed arc, at \zs = 1.2600, is approximately 5$''$ away.

\newpage
\href{https://www.legacysurvey.org/viewer/?ra=83.0990&-01.5510&layer=hsc-dr2&pixscale=0.262&zoom=16}{\emph{DESI~J183.0990-01.5510}} \, 
It is interesting to contrast this system (Figure~\ref{fig:desi183-01}) with DESI~J150.2022+01.6538 (Figure~\ref{fig:desi150+01}) in Appendix~\ref{sec:nonlenses}.
While having similar appearance in the imaging data, especially in \ls, 
one is a lensing system and the other is not (for more discussion on this comparison, see \S~\ref{sec:discuss}).
\begin{figure}[h]
  \centering
  \includegraphics[width=0.7\textwidth]{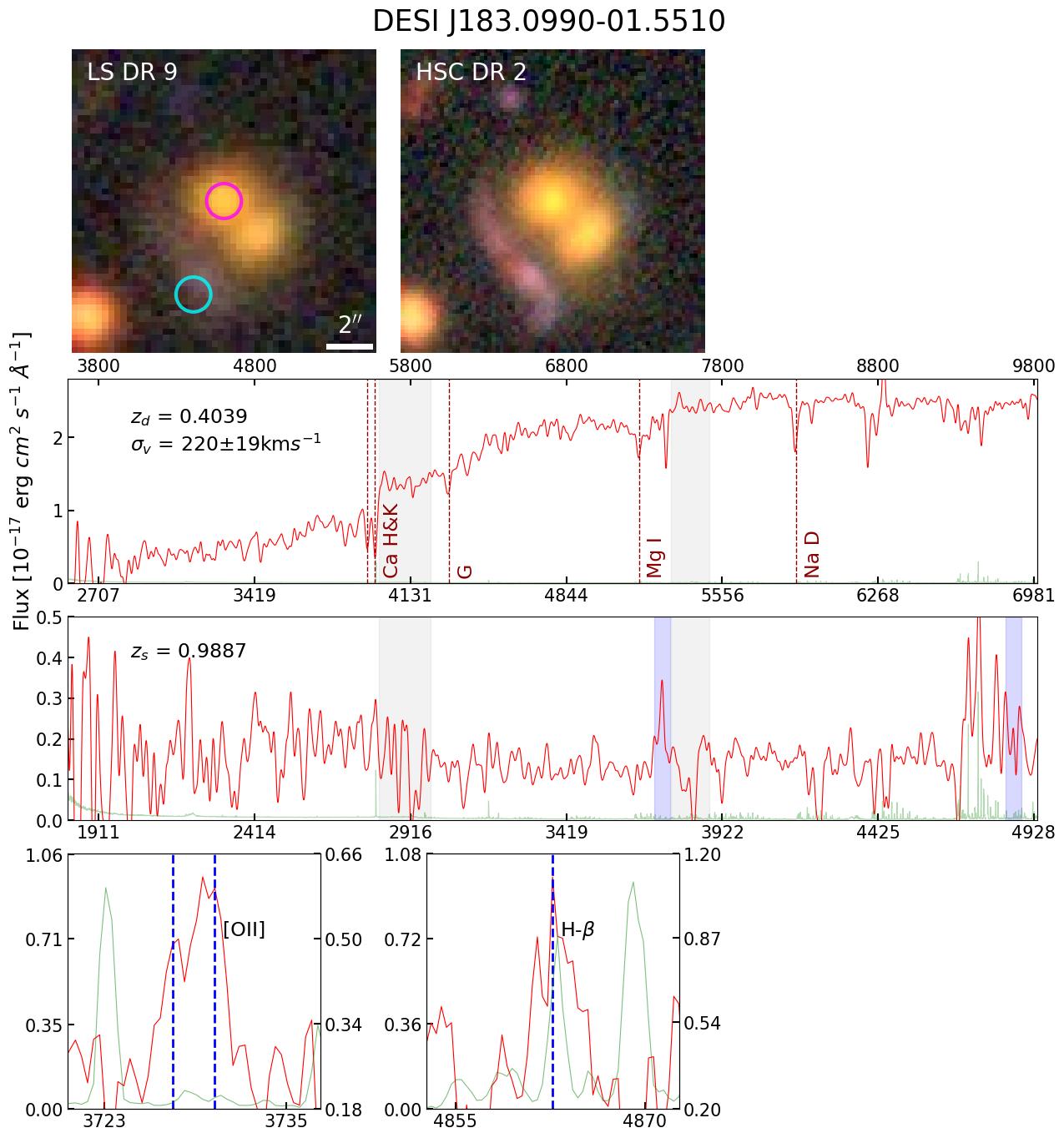}
  \caption{DESI~J183.0990-01.5510.  
  For the arrangement of the panels, see Figure~\ref{fig:eg-lens-fig} caption. }
  \label{fig:desi183-01}
\end{figure}

\begin{figure}[h]
\centering
\includegraphics[width=0.7\linewidth]{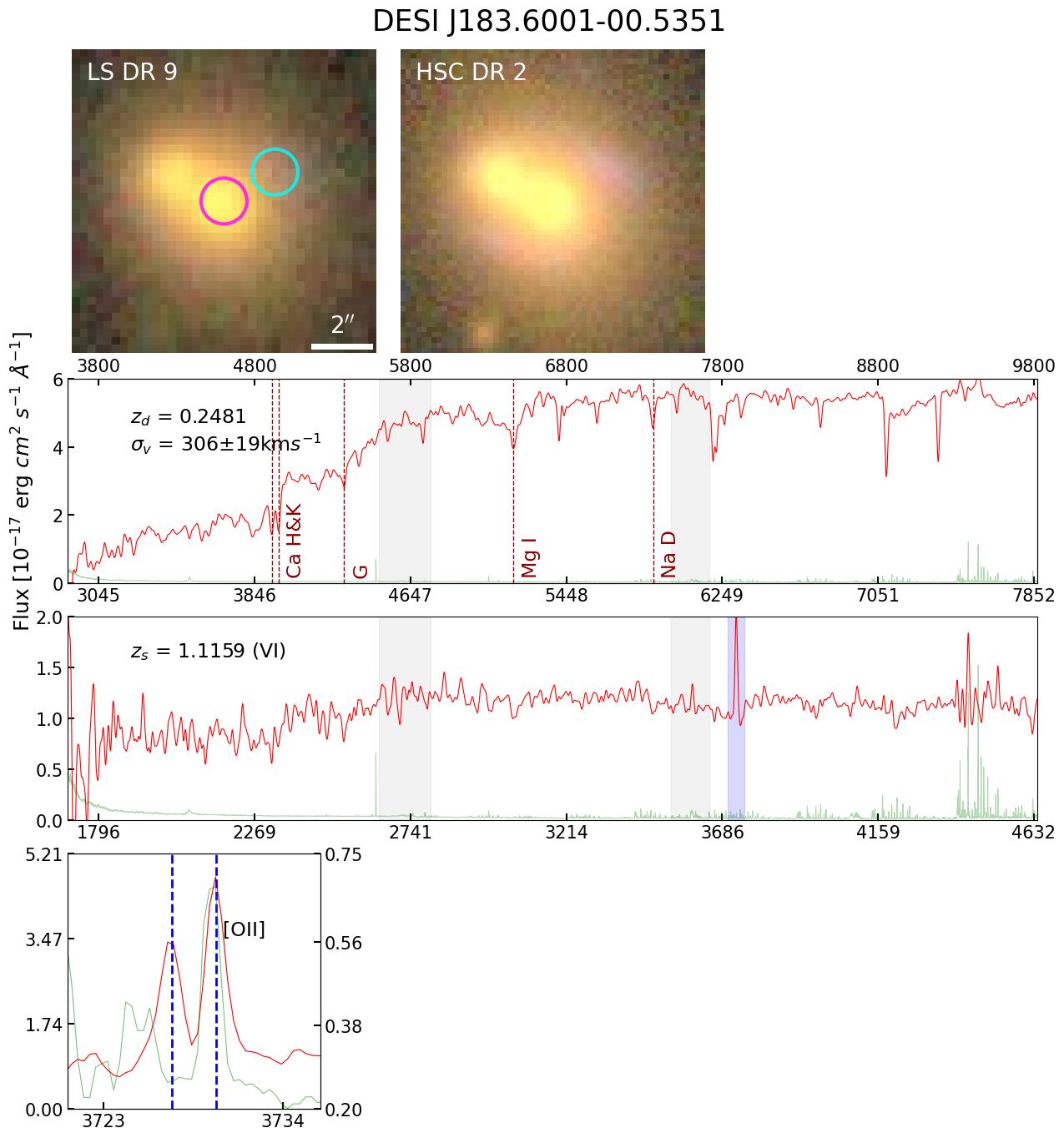}
\caption{
DESI~J183.6001-00.5351. For the arrangement of the panels, 
see Figure~\ref{fig:eg-lens-fig} caption.
}
\label{fig:desi183-00}
\end{figure}

\newpage
\href{https://www.legacysurvey.org/viewer/?ra=183.6001&-00.5351&layer=hsc-dr2&pixscale=0.262&zoom=16}{\emph{DESI~J183.6001-00.5351}} 
The lens for this system (Figure~\ref{fig:desi183-00}) comprises two massive elliptical galaxies, with nearly identical colors and very similar \phz's (0.233$\pm0.072$ and 0.241$\pm0.021$ for the northern and southern objects, respectively).
From the DESI spectrum shown in Figure~\ref{fig:desi183-00}, \rr determines the redshift of the southern elliptical to be 0.2481.
An bluish arc-like object is barely visible in the \ls images, but can be more clearly seen in the \hsc image (to the NW of the pair of ellipticals). 
In the fiber targeting this arc-like object (\srccircle circle, Figure~\ref{fig:desi183-00}),
there is a strong emission emission feature at 7888 \ang.
\rr identifies this as \ha (\lam 6562 \ang), yielding a redshift of 0.2013, but with ZWARN=4.
VI shows this feature is doubly peaked, with the separation between the peaks matching that of the \oii doublet. 
Thus, we identify this emission feature as \oii, resulting in a redshift of 1.1159 and confirming the lensing nature of this system.

\begin{figure}[h]
\centering
\includegraphics[width=0.7\linewidth]{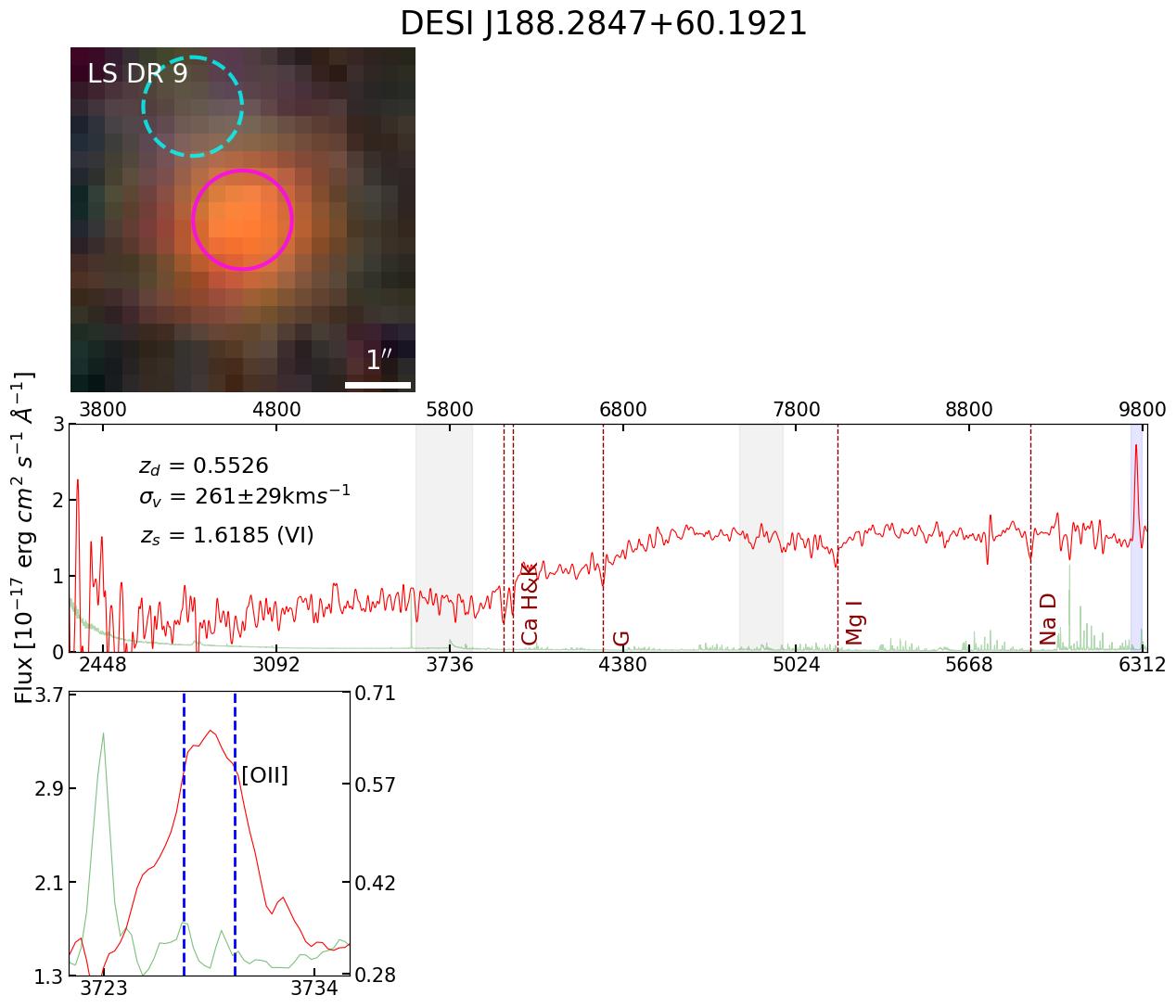}
\caption{DESI~J188.2847+60.1921. 
Note that for this system there is only one DESI spectrum (one target ID) from a fiber centered on the lens.
But given that it is a small Einstein radius system,
from VI we recognize that it contains spectral features of both the lens and the source.
In the second row panel, we label the absorption features for the lensing galaxy with red dashed lines
and highlight the \oii emission from the source with a blue band. 
}
\label{fig:desi188+60}
\end{figure}

\newpage
\href{https://www.legacysurvey.org/viewer/?ra=188.2847&dec=+60.1921&layer=ls-dr9&pixscale=0.262&zoom=16}{\emph{DESI~J188.2847+60.1921}}\,  
This spectrum for this system (Figure~\ref{fig:desi188+60}) is from a fiber centered on the lensing galaxy.  
\rr determined its redshift to be \zd = 0.5526, based on clear absorption lines.
Through VI, we identified another feature, a prominent emission line at 9763~\ang.
Given its width and shape, 
it is most likely the \oii doublet, which implies there is a background object, at \zs = 1.6185. 
This system was first identified by \citet{talbot2021a} in SDSS DR16, also from a fiber centered on the lens. 
They made the same identification for this emission.
DESI will separately target a second lensed source (to the NE of the foreground galaxy) in future observations.

\begin{figure}[h]
\centering
\includegraphics[width=.7\linewidth]{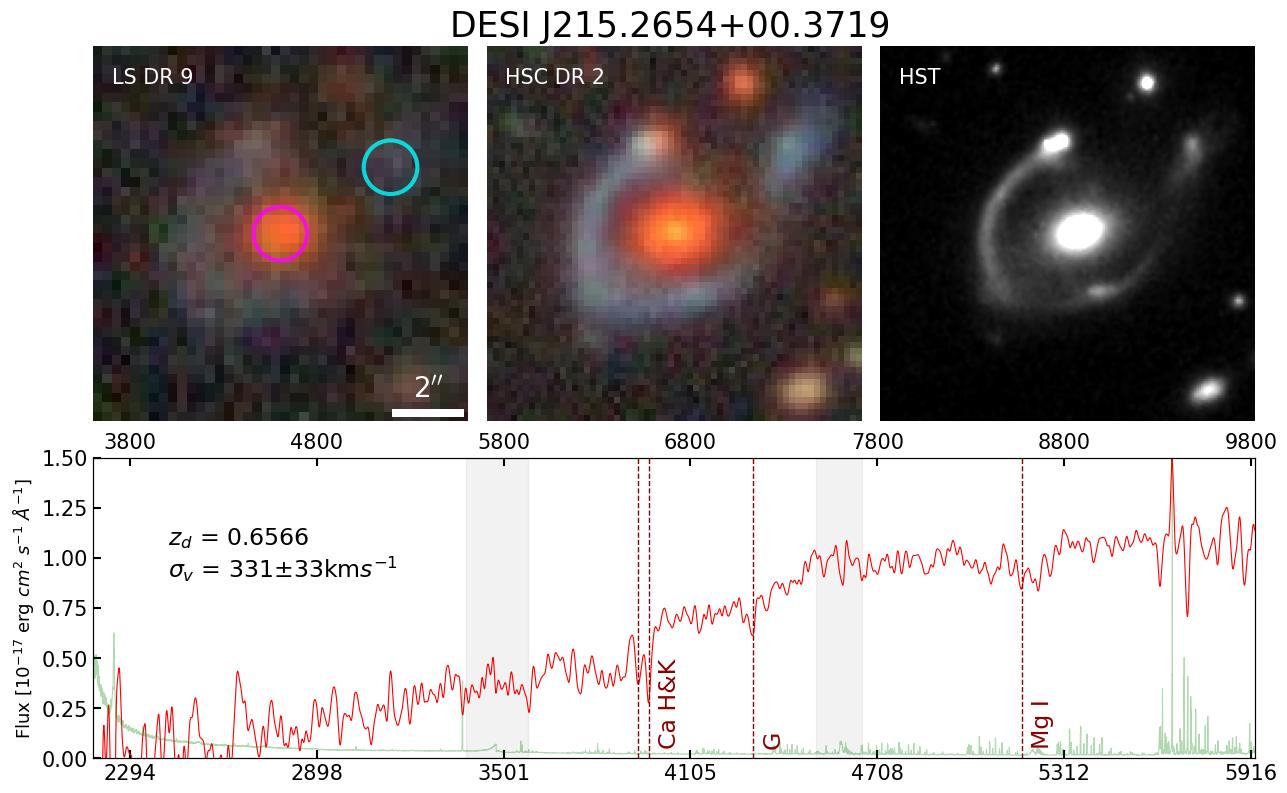}
\caption{DESI~J215.2654+00.3719.
In the first row, we show the \ls (left), the \hsc (middle), and \hst (right) images.
The lens spectrum is shown in the second row.
We determined the source redshift to be \zs = 2.2066 from Keck NIRES (see text).
}
\label{fig:desi215.2+00}
\end{figure}

\newpage
\href{https://www.legacysurvey.org/viewer/?ra=215.2654&dec=+00.3719&layer=hsc-dr2&pixscale=0.262&zoom=16}{\emph{DESI~J215.2654+00.3719}}\, 
This system (Figure~\ref{fig:desi215.2+00}) was confirmed by DESI and Keck spectra to be a strong lensing system.
It has additional confirmation from \hst observation (GO-15867; Paper~I).
The lens galaxy has $\zd = 0.6566$ from its DESI spectrum.
The lensed source was observed by DESI, too, but \rr flagged the spectrum (not shown) with ZWARN=4 (see \S~\ref{sec:secondary}), because it does not appear to have identifiable features.
Our Keck NIRES program targeted the source and determined its redshift to be \zs = 2.2066,
based on \ha and [\ion{O}{3}] emission (Paper~III; Argawal et al. in prep.).
For such a redshift value, without \lya being present, which does not seem to be detected in the DESI spectrum, it would be difficult for DESI to measure its redshift in the optical range.

\begin{figure}[h]
\centering
\includegraphics[width=0.7\linewidth]{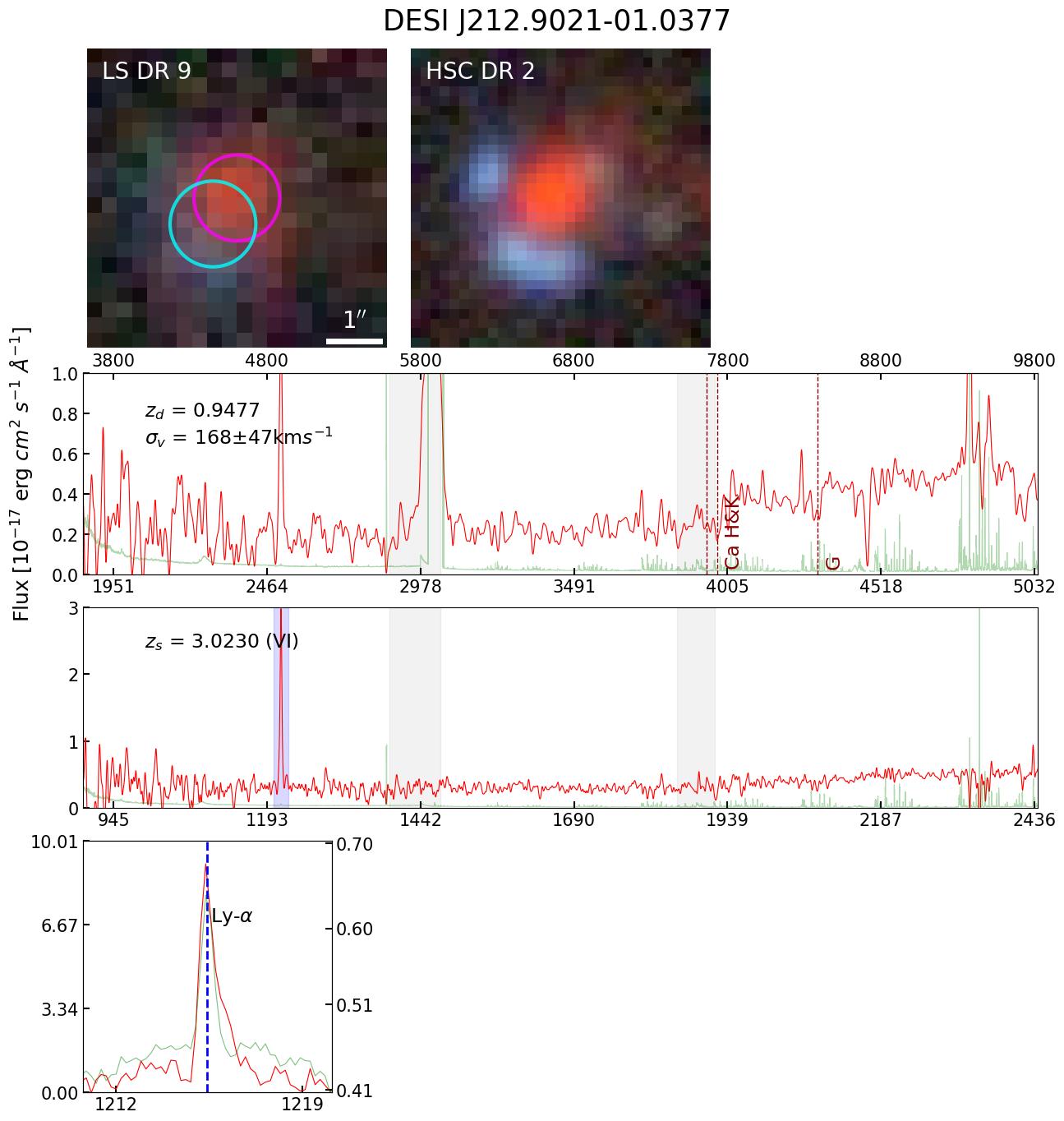}
\caption{DESI~J212.9021-01.0377. 
  For the arrangement of the panels, see Figure~\ref{fig:eg-lens-fig}.
It is interesting to note that, given that this is a small Einstein radius system, the same \lya emission can be seen in the lens and source spectra (shown in the second and third row, respectively).
} 
\label{fig:desi212-01}
\end{figure}

\newpage
\href{https://www.legacysurvey.org/viewer/?ra=212.9021&dec=-01.0377&layer=hsc-dr2&pixscale=0.262&zoom=16}{\emph{DESI~J212.9021-01.0377}}. 
The \rr pipeline determined the lens redshift to be \zd = 0.9477 (Figure~\ref{fig:desi212-01}).
\citet{koopmans2002a} reported a doubly imaged quasar system with a comparable lens redshift at $z_d = 1.004$.
They noted the significant overdensity of galaxies around the lens that possibly indicated 
this system might be a protocluster.
It is interesting to note that DESI~J212.9021-01.0377 also appears to be in an overdense region.
For the source spectrum, through VI, we identify a prominent emission feature as \lya and determine its redshift to be \zs = 3.0230,
whereas the \rr pipeline failed because the current version does not fit for the \lya emission.
The last row of Figure~\ref{fig:desi212-01} shows the zoom-in of this feature, which displays the characteristic ``blue edge, red rail'' profile expected for \lya emission \citep[e.g.,][]{shu2016a}.

\begin{figure}[h]
\centering
\includegraphics[width=0.7\linewidth]{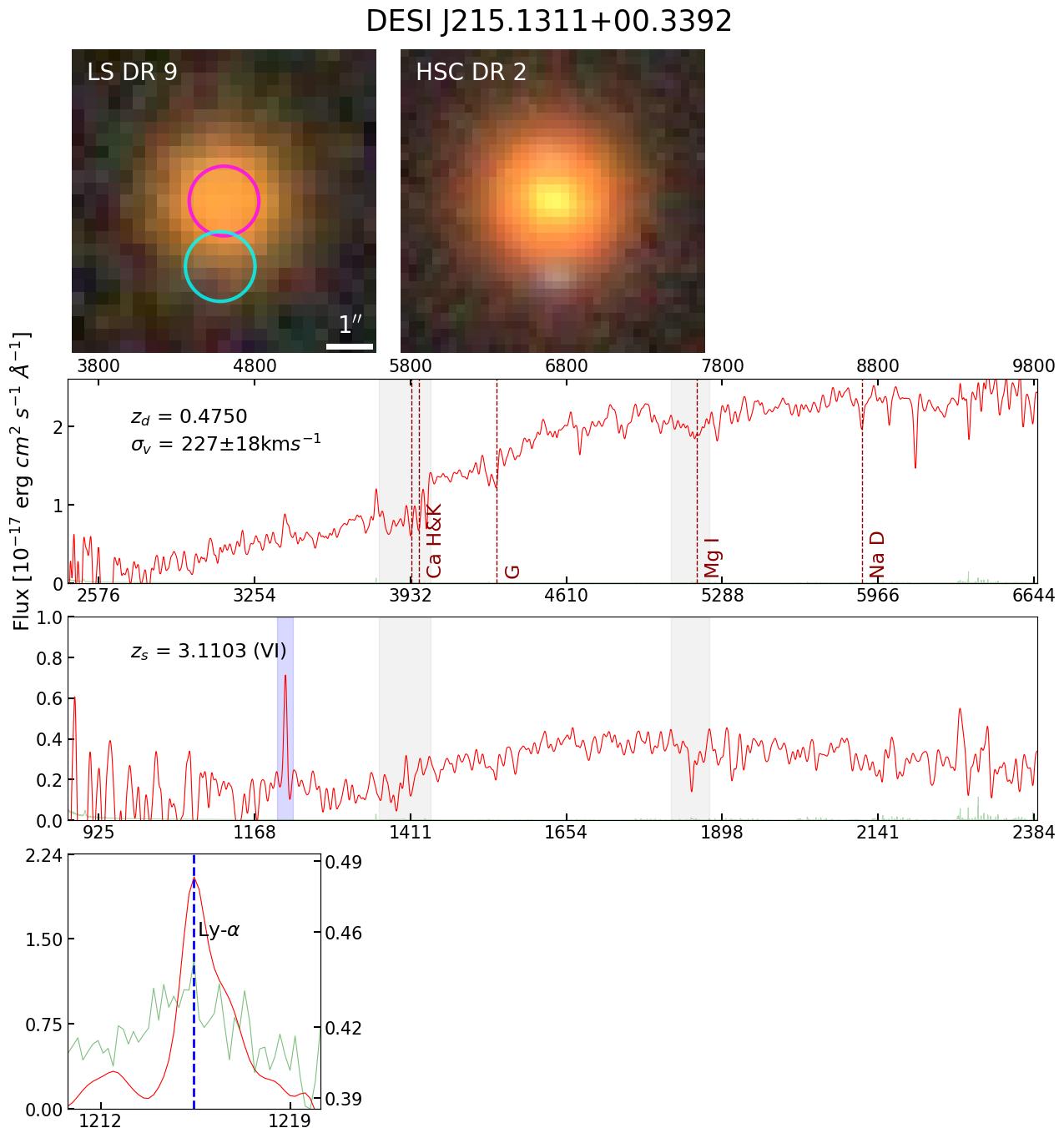}
\caption{
DESI~J215.1311+00.3392.
For the arrangement of the panels, see Figure~\ref{fig:eg-lens-fig}.
The fiber intended to measure the source redshift is centered on the blue object below the lens (only visible in the HSC image). 
Just as for DESI~J212.9021-01.0377 (Figure~\ref{fig:desi212-01}), VI identifies the presence of the \lya emission (middle panel, blue band) in the source spectrum.}
\label{fig:desi215+00}
\end{figure}

\newpage
\href{https://www.legacysurvey.org/viewer/?ra=215.1311&dec=+00.3392&layer=hsc-dr2&pixscale=0.262&zoom=16}{\emph{DESI~J215.1311+00.3392}}\,  DESI targeted  both the lens and the source of this system.
\rr determined the lens redshift to be \zd = 0.4750 (Figrure~\ref{fig:desi215+00}).
For the source spectrum, VI identifies \lya emission, with the characteristic profile of ``blue edge, red tail'',
and determines \zs = 3.1103. 
\rr failed to obtain the source redshift because it does not fit for \lya.
Along with DESI~J212.9021-01.0377 (Figure~\ref{fig:desi212-01}), this is the second $\zs > 3$ galaxy-scale strong lensing system confirmed in this work.

\begin{figure}[h]
  \centering
  \includegraphics[width=0.7\textwidth]{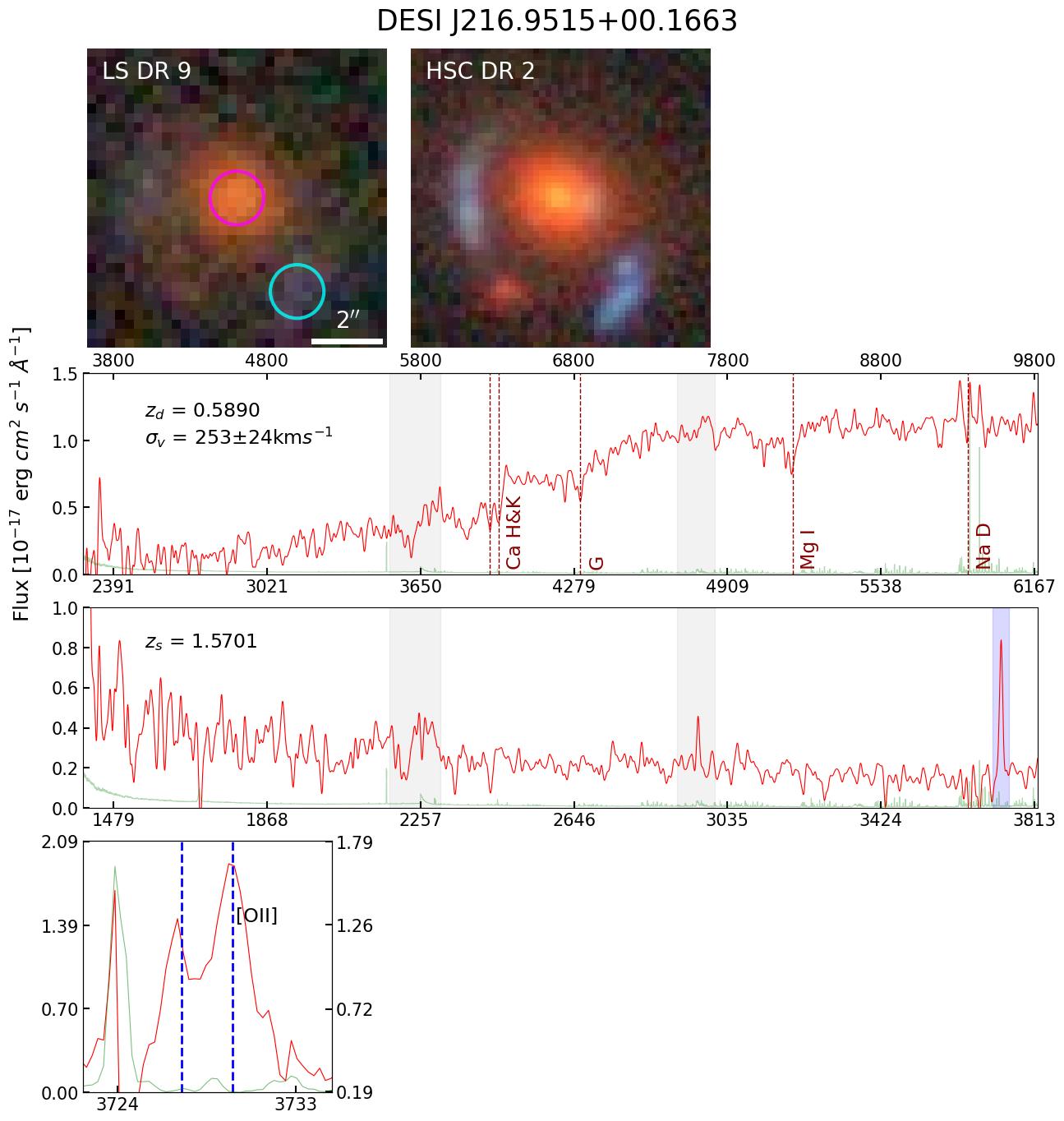}
  \caption{DESI~J216.9515+00.1663. For the arrangement of the panels, see Figure~\ref{fig:eg-lens-fig} caption.}
  \label{fig:desi216.9+00.1}
\end{figure}

\newpage
\href{https://www.legacysurvey.org/viewer/?ra=216.9515&dec=+00.1663&layer=hsc-dr2&pixscale=0.262&zoom=16}{\emph{DESI~J216.9515+00.1663}} \,
The lensing galaxy has \zd = 0.5890 (Figure~\ref{fig:desi216.9+00.1}).
the fiber for the source is centered on the blue arc to the SW of the lensing galaxy.
For the lensed arc, the \oii feature is detected with high SNR, with \zs = 1.5701.
\cwr{There appears to be an arc-counterarc pair closer to the lens, likely corresponding to a second source.}

\begin{figure}[h]
  \centering
  \includegraphics[width=0.7\textwidth]{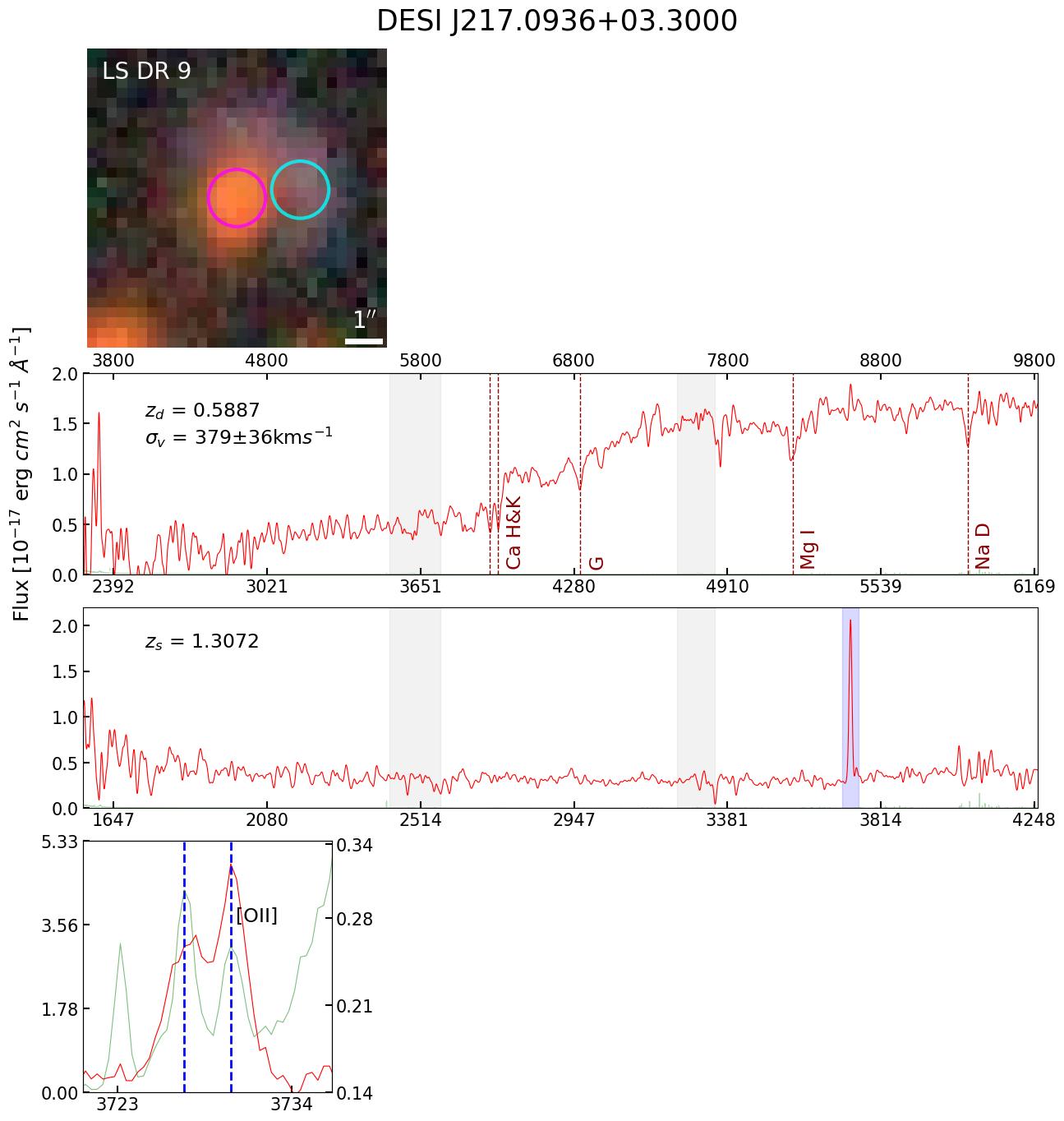}
  \caption{DESI~J217.0936+03.3000. For the arrangement of the panels, see Figure~\ref{fig:eg-lens-fig} caption.}
  \label{fig:desi217+03}
\end{figure}

\newpage
\href{https://www.legacysurvey.org/viewer/?ra=217.0936&dec=+03.3000&layer=ls-dr9&pixscale=0.262&zoom=16}{\emph{DESI~J217.0936+03.3000}} \, 
The lensing galaxy, at \zd = 0.5887 (Figure~\ref{fig:desi217+03}), appears to be part of a galaxy group.
A prominent blue arc to the NW is approximately 
3$''$ away.
The \oii  feature is detected with high SNR, with \zs = 1.3072.
This system was a target for \hst GO-15867, but was not successfully observed due to the loss of a guide star.

\begin{figure}[h]
  \centering
  \includegraphics[width=0.7\textwidth]{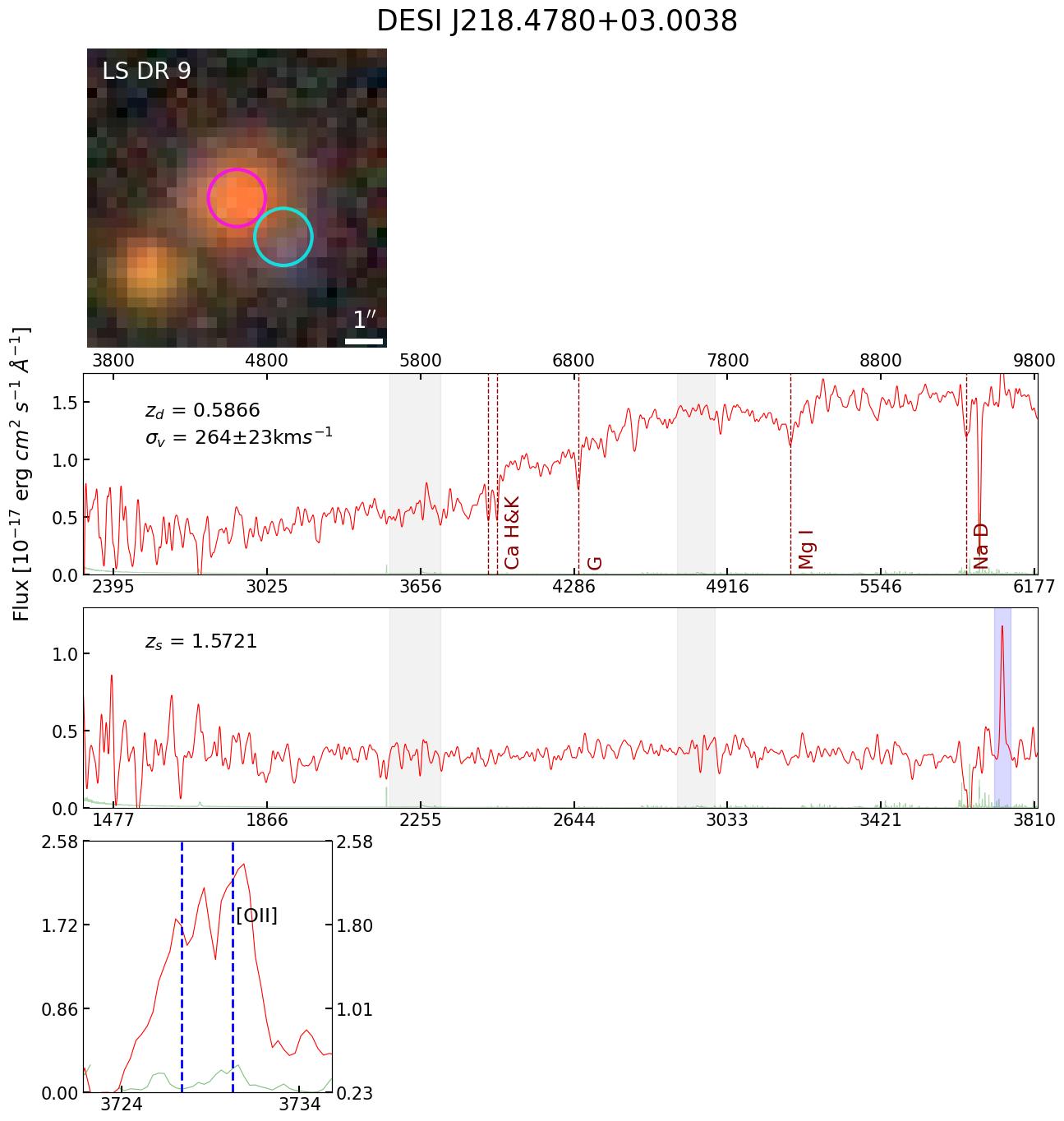}
  \caption{DESI~J218.4780+03.0038. For the arrangement of the panels, see Figure~\ref{fig:eg-lens-fig} caption.}
  \label{fig:desi218+03}
\end{figure}

\newpage
\href{https://www.legacysurvey.org/viewer/?ra=218.4780&dec=+03.0038&layer=ls-dr9&pixscale=0.262&zoom=16}{\emph{DESI~J218.4780+03.0038}}\, The lensing galaxy is at \zd = 0.5866 (Figure~\ref{fig:desi218+03}). The lensed source forms a nearly 3/4 ring with a radius of approximately 
1.7$''$. 
The \oii emission is clearly detected, with \zs = 1.5721.

\begin{figure}[h]
\centering
\includegraphics[width=0.7\linewidth]{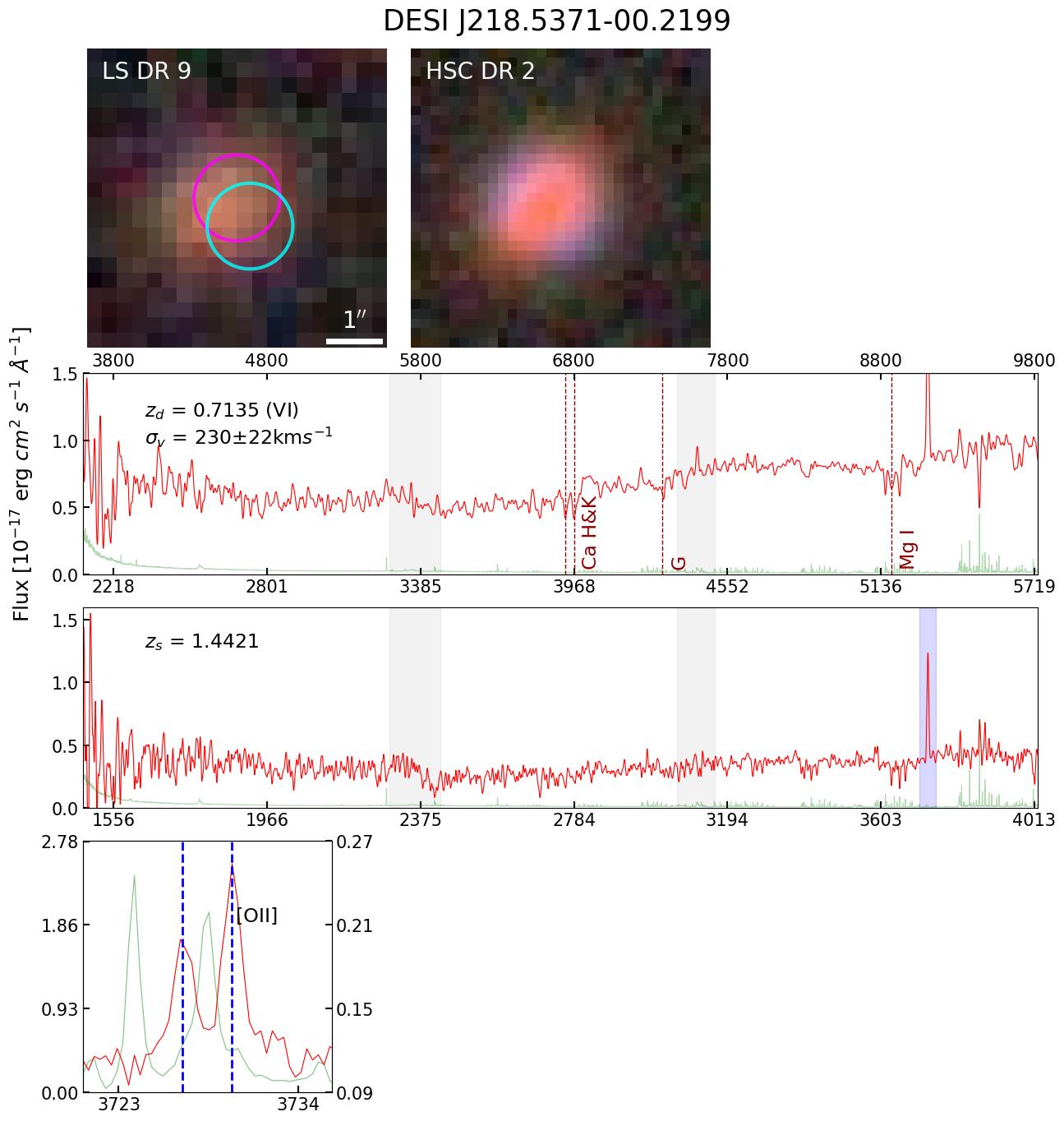}
\caption{DESI~J218.5371-00.2199.
For the arrangement of the panels, see Figure~\ref{fig:eg-lens-fig} caption.}.
\label{fig:desi218-00}
\end{figure}

\newpage 
\href{https://www.legacysurvey.org/viewer/?ra=218.5371&dec=-00.2199&layer=hsc-dr2&pixscale=0.262&zoom=16}{\emph{DESI~J218.5371-00.2199}} \, This system (Figure~\ref{fig:desi218-00}) appears to have a small Einstein radius ($\thetaE \sim 1''$).
Thus, the light from the lensed source is abundantly present in the fiber centered on the lensing galaxy.
The source also happens to have strong \oii emission.
The emission feature of the source appears in the spectrum coming from the fiber centered on the lens. 
The \rr pipeline correctly identified it as \oii, and therefore incorrectly assigned the source redshift to the lensing galaxy.
This probably also due to the fact that the 4000 \ang break is relative weak for the lensing galaxy, although the Ca H\&K absorption is very clear.
We determine the lens redshift from VI to be \zd = 0.7135.

\begin{figure}[h]
  \centering
  \includegraphics[width=0.7\textwidth]{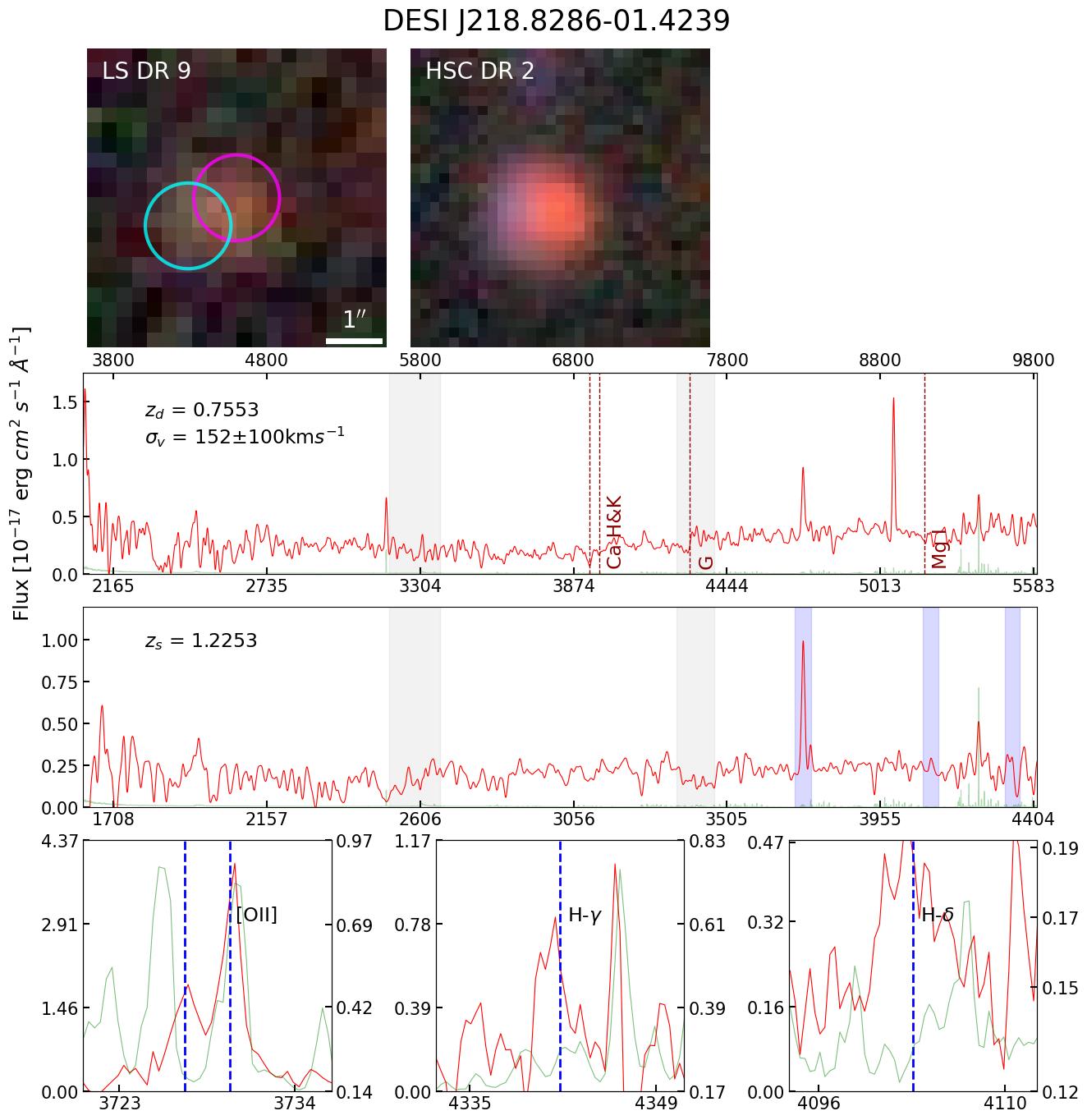}
  \caption{DESI~J218.8286-01.4239. 
  For the arrangement of the panels, see Figure~\ref{fig:eg-lens-fig} caption. Note that the \oii feature for the lensed source appears in the lens spectrum as well, given the proximity of the two objects on the sky.}
  \label{fig:desi218-01}
\end{figure}

\newpage
\href{https://www.legacysurvey.org/viewer/?ra=218.8286&dec=-01.4239&zoom=16&layer=hsc-dr3&mark=178.872100,-0.715200}{\emph{DESI~J218.8286-01.4239}}\, 
This system has \zd = 0.7553 and \zs = 1.2253 (Figure~\ref{fig:desi218-01}).
Using imaging data alone, even in the \hsc image, which is deeper and has higher resolution than the \ls observations,
a human expert would likely assign a B grade but not higher.
This system has a similar appearance as two confirmed non-lenses, DESI-217-2090+51.6467 (Figure~\ref{fig:desi217+51}) and DESI-218.87813+00.3034 (Figure~\ref{fig:desi218+00}), with a lower and comparable human grade, respectively.
In \S~\ref{sec:human-vs-spect}, we will discuss the implications of this comparison for visual inspection of lens candidates selected from imaging surveys \cwr{Given AB's comment, possibly include the discussion here?).}

\begin{figure}[h]
  \centering
  \includegraphics[width=0.7\textwidth]{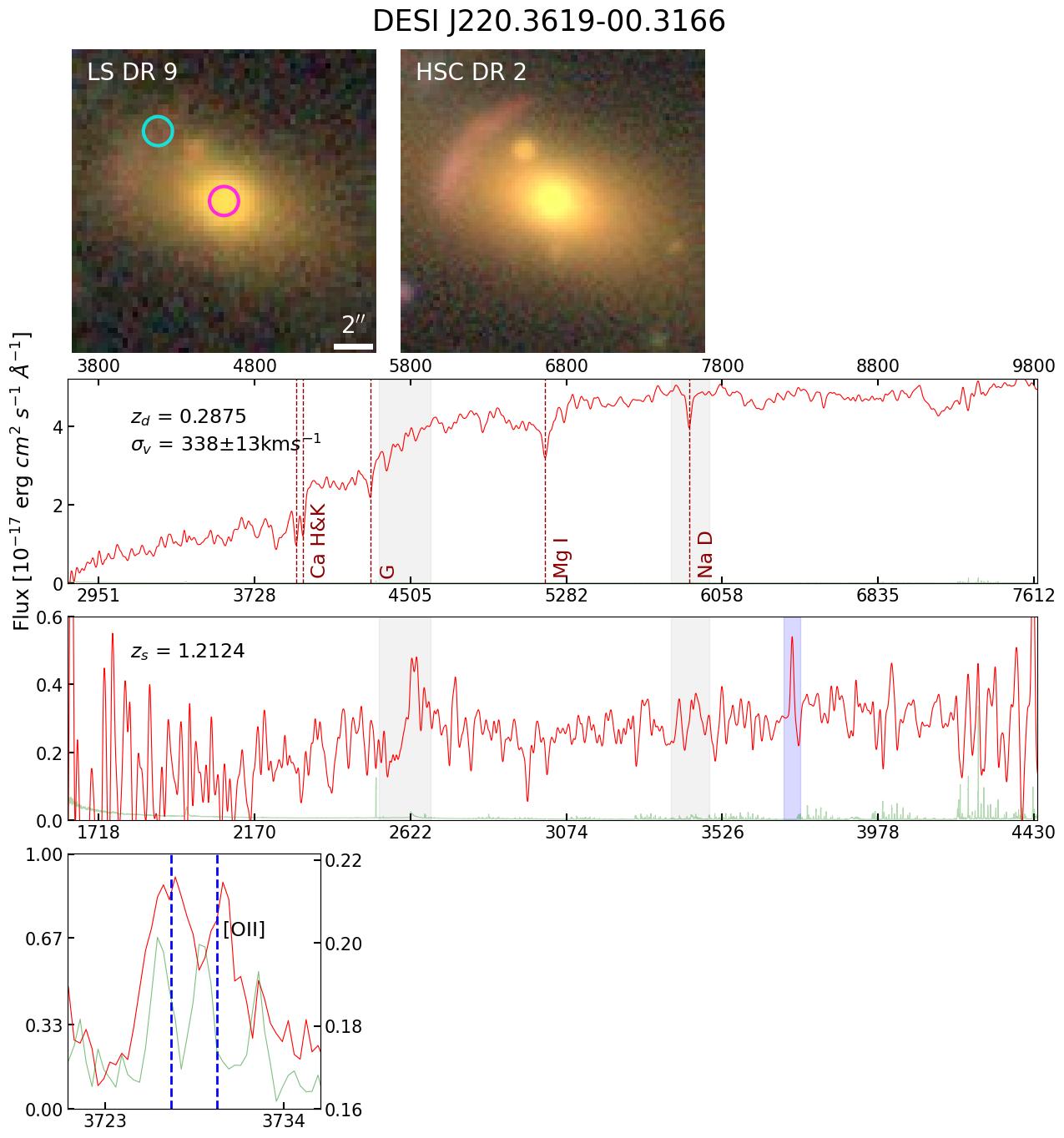}
  \caption{DESI~J220.3619-00.3166. 
  For the arrangement of the panels, 
see Figure~\ref{fig:eg-lens-fig} caption.}
  \label{fig:desi220.3-00}
\end{figure}

\newpage
\href{https://www.legacysurvey.org/viewer/?ra=220.3619&dec=-00.316&layer=ls-dr10-grz&pixscale=0.262&zoom=16}{\emph{DESI~J220.3619-00.3166}}\,
The lens for this system, at $\zd = 0.2875$ (Figure~\ref{fig:desi220.3-00}), appears to be the brightest member of a galaxy group, 
based on photometric redshifts of nearby galaxies with similar colors.\footnote{All photo-$z$ values in this work are from \citet{zhou2020a}.}
The reddish lensed arc, at $\zs = 1.2600$, is approximately 5$''$ away.

\begin{figure}[h]
  \centering
  \includegraphics[width=.7\textwidth]{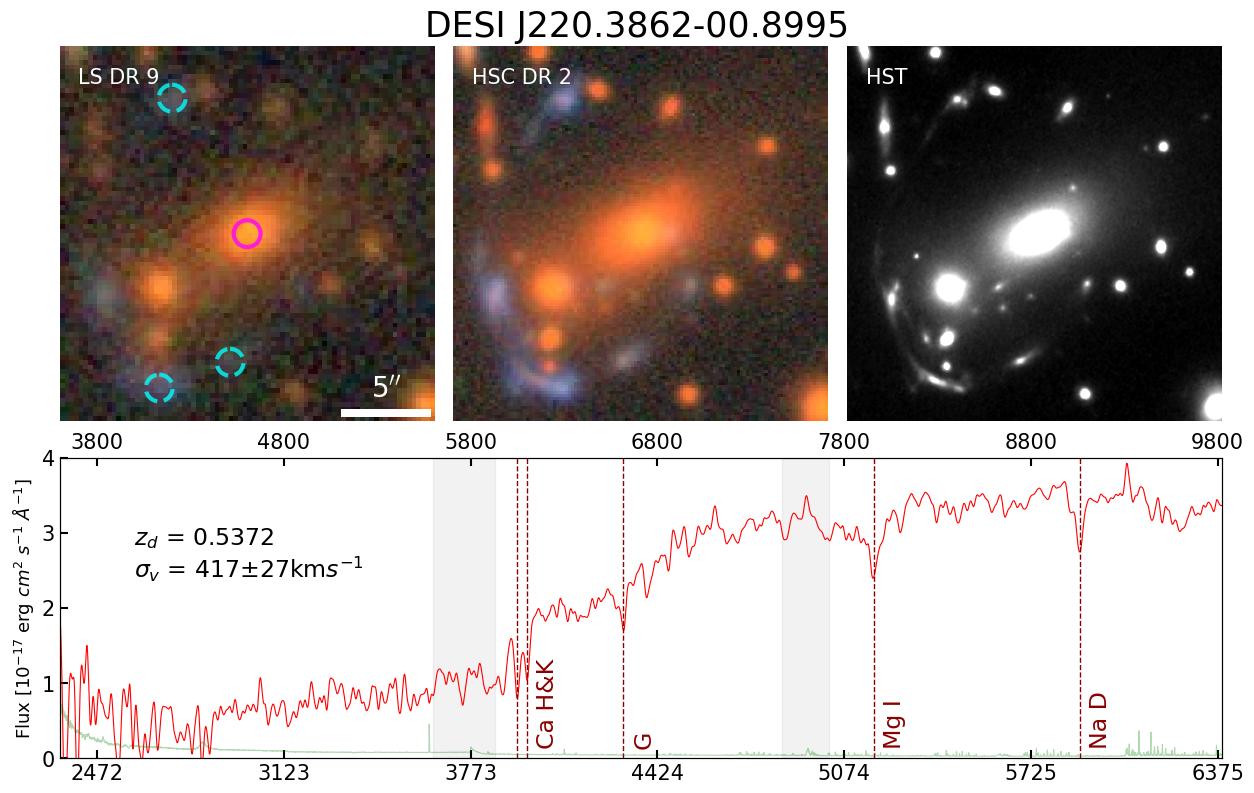}
  \caption{DESI~J220.3862-00.8995.
  For the arrangement of the panels, see Figure~\ref{fig:desi215.2+00}.} 
  \label{fig:desi220-00}
\end{figure}

\newpage
\href{https://www.legacysurvey.org/viewer/?ra=220.3862&dec=-00.8995&layer=ls-dr10-grz&pixscale=0.262&zoom=16}{\emph{DESI~J220.3862-00.8995}} 
This system (Figure\,\ref{fig:desi220-00}) was discovered in H20. 
DESI has obtained the redshift for the massive elliptical galaxy at the center of the cutout image, $\zd = 0.5372$.
It appears to be the brightest galaxy in a group.
From both the LS DR9 and HSC DR2 images, multiple blue arcs can be seen around the center of the foreground group of galaxies, 
confirmed by \hst observation (GO-15867; Paper I). 
Three of these will be observed by DESI.

\begin{figure}[h]
  \centering
  \includegraphics[width=0.7\textwidth]{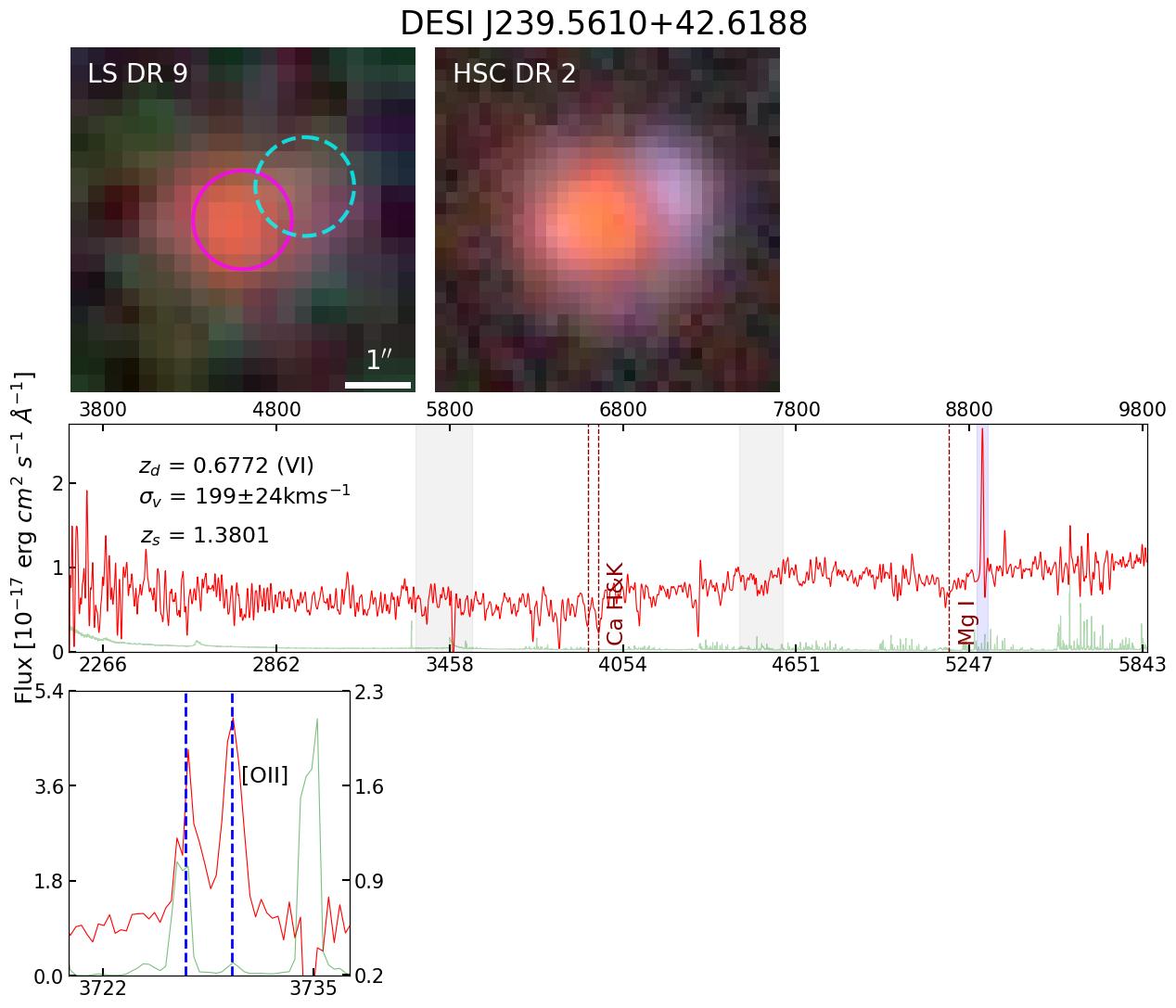}
  \caption{DESI~J239.5610+42.6188. For the arrangement of the panels, see Figure~\ref{fig:eg-lens-fig} caption.
Note that for this system there is only one DESI spectrum (one target ID) from a fiber centered on the lens.
But given that it is a small Einstein radius system,
from VI we recognize that it contains spectral features of both the lens and the source.
In the second row panel, we label the absorption features for the lensing galaxy with red dashed lines
and highlight the \oii emission from the source with a blue band.}
  \label{fig:desi239.6+42}
\end{figure}

\newpage
\href{https://www.legacysurvey.org/viewer/?ra=239.5610&dec=+42.6188&zoom=16&layer=hsc-dr3&mark=178.872100,-0.715200}{\emph{DESI~J239.5610+42.6188}}\, 
This system (Figure~\ref{fig:desi239.6+42}) has a small Einstein radius system ($\thetaE \lesssim 1\twopr$), with the \hsc image showing a blue arc and a possible counterarc.
The spectral features of the lens and the source appear in same spectrum from a fiber targeting the lens. 
\rr assigned a flag of ZWARN=2048. 
This corresponds to the ZWARN bit of 11, with the bit name, \texttt{POORDATA}, and a description of 
``Poor input data quality but try fitting anyway''.
VI reveals that the data has sufficient SNR, and the \rr pipeline fit for the source redshift (by identifying the \oii doublet) is correct, $\zs = 1.3801$.
From VI, we are also able to determine the lens redshift (based on clear Ca~H\&K, as well as G, Mg, \hb, \hc, and \hd absorption lines) to be $\zd = 0.6772$.
We mentioned in \S~\ref{sec:secondary} that the bitmask for this system does not match the strong lens secondary target bitmask value.

\begin{figure}[h]
  \centering
  \includegraphics[width=0.7\textwidth]{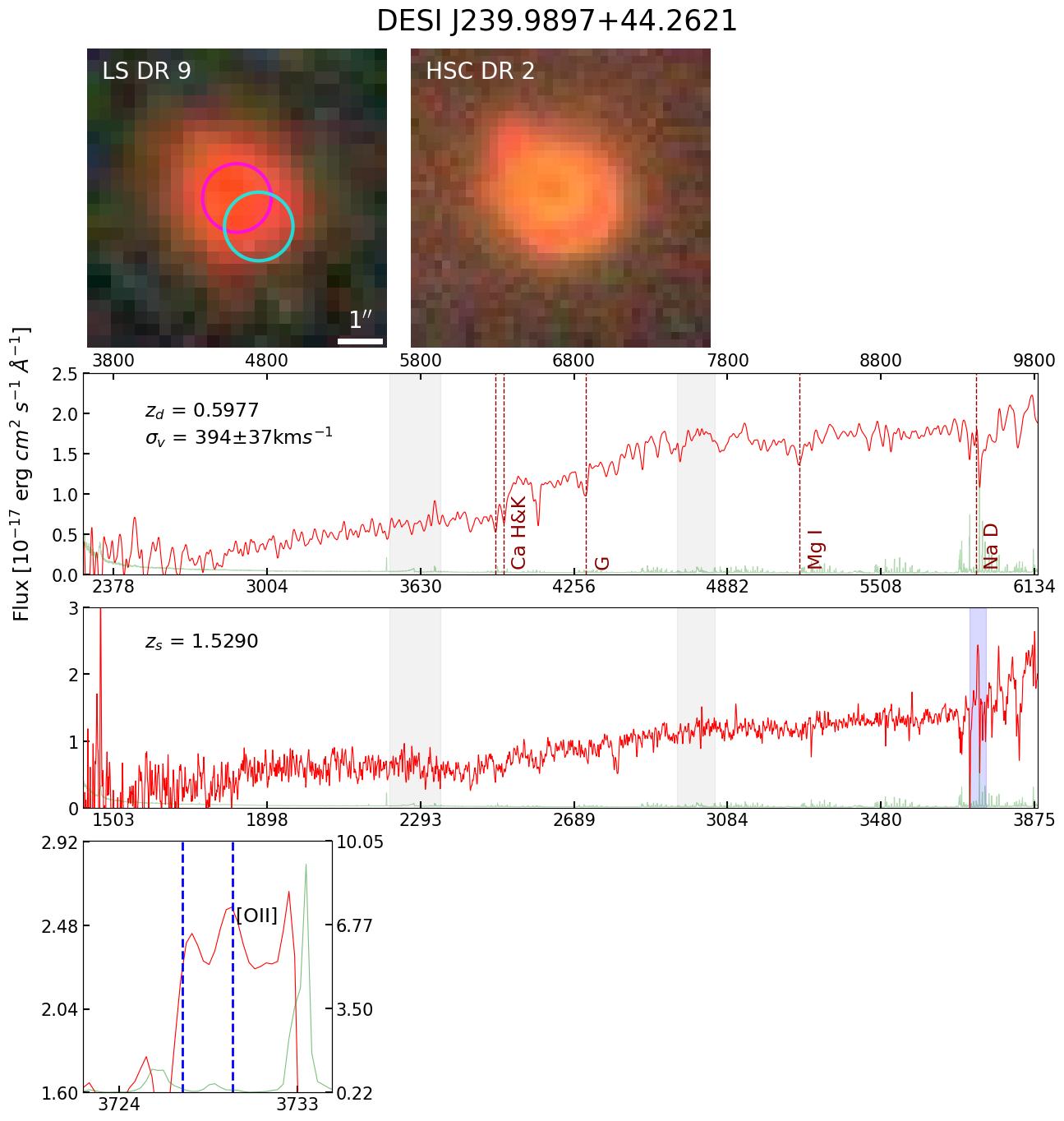}
  \caption{DESI~J239.9897+44.2621.
  For the arrangement of the panels, see Figure~\ref{fig:eg-lens-fig} caption.}
  \label{fig:desi239.9+44}
\end{figure}

\newpage
\href{https://www.legacysurvey.org/viewer/?ra=239.9898&dec=44.2621&layer=hsc-dr2&pixscale=0.262&zoom=16}{\emph{DESI~J239.9897+44.2621}}\,
The \hsc image of this system (Figure~\ref{fig:desi239.9+44})
shows the presence of an arc and counterarc pair around an elliptical galaxy.
The spectrum for the lensed source galaxy has a clear emission line.
Based on the double peak appearance, we identify it as the \oii emission, placing the redshift at $\zs = 1.5290$.

\begin{figure}[h]
  \centering
  \includegraphics[width=0.7\textwidth]{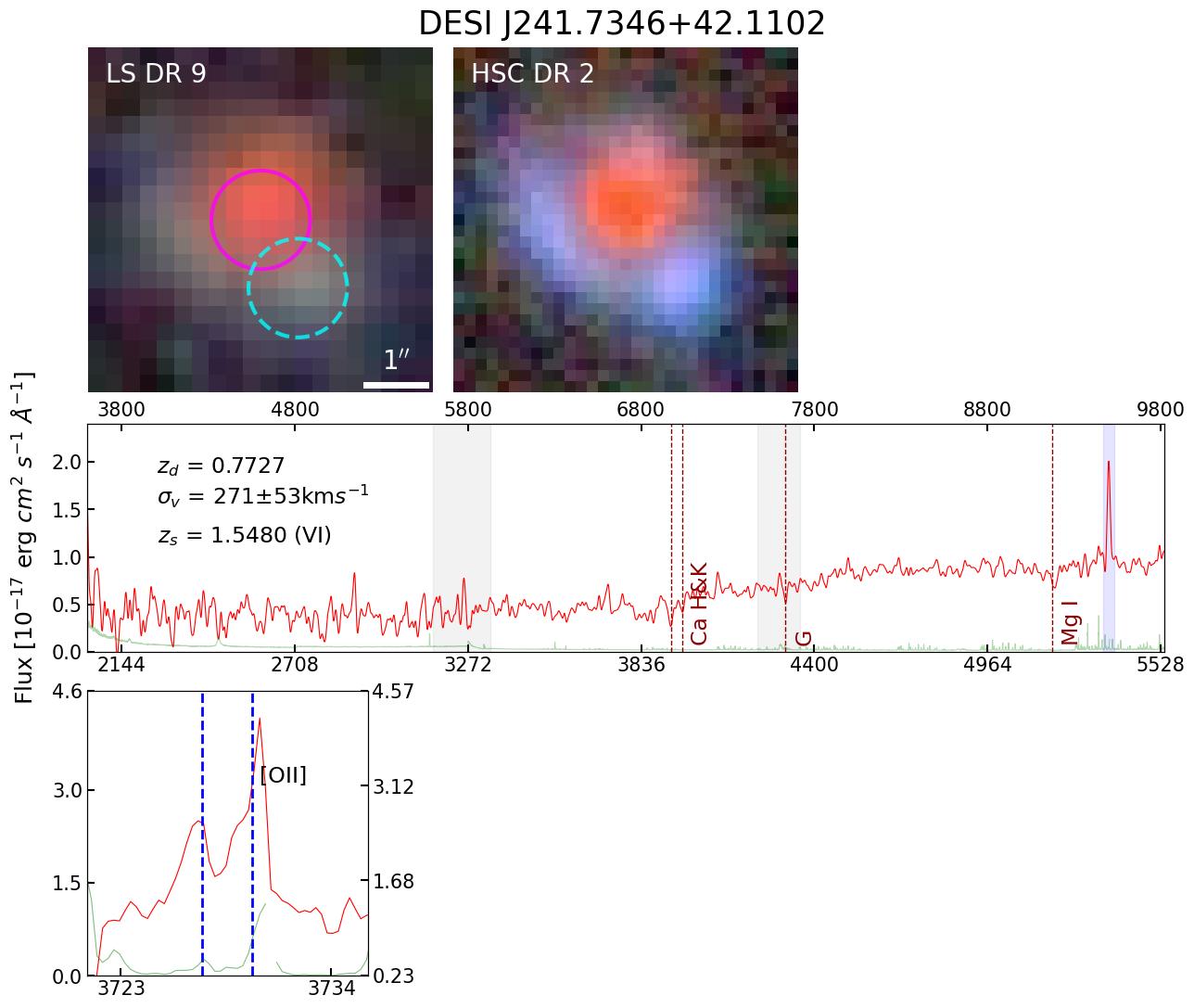}
  \caption{DESI~J241.7346+42.1102. 
  For the arrangement of the panels, see Figure~\ref{fig:eg-lens-fig} caption.
  Note that for this system there is only one DESI spectrum (one target ID) from a fiber centered on the lens.
But given that it is a small Einstein radius system,
from VI we recognize that it contains spectral features of both the lens and the source.
In the second row panel, we label the absorption features for the lensing galaxy with red dashed lines
and highlight the \oii emission from the source with a blue band.
  }
  \label{fig:desi241+42}
\end{figure}

\newpage
\href{https://www.legacysurvey.org/viewer/?ra=241.7346&dec=+42.1102&layer=hsc-dr2&pixscale=0.262&zoom=16}{\emph{DESI~J241.7346+42.1102}}  The lensing galaxy has \zd = 0.7727 (Figure~\ref{fig:desi241+42})
For the lensed arc, the \oii  feature is detected with high SNR, with \zs = 1.5480.

\begin{figure}[h]
  \centering
  \includegraphics[width=0.7\textwidth]{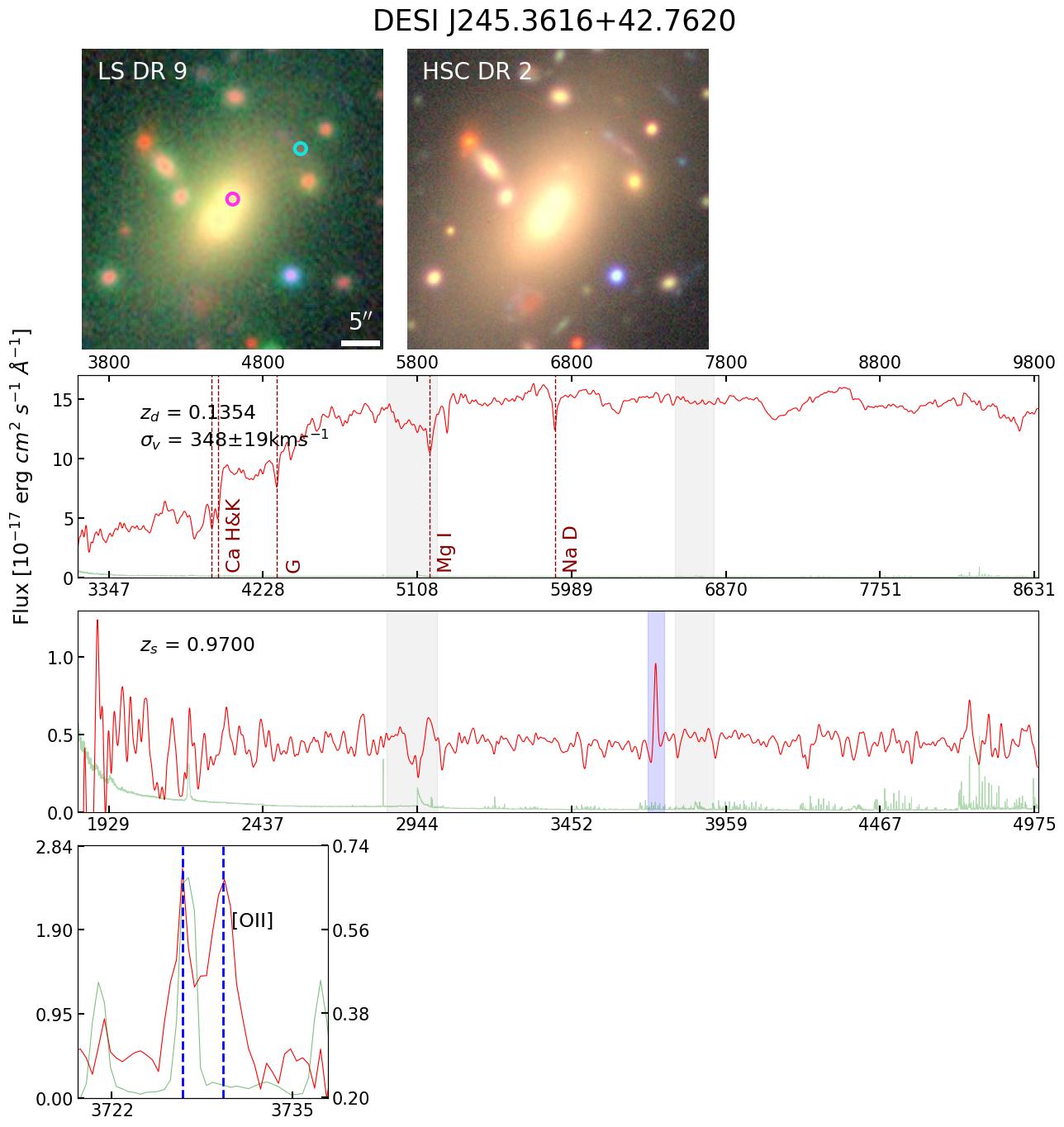}
  \caption{DESI~J245.3616+42.7620. For the arrangement of the panels, see Figure~\ref{fig:eg-lens-fig} caption.
  }
  \label{fig:desi245+42}
\end{figure}

\newpage
\href{https://www.legacysurvey.org/viewer/?ra=245.3616&dec=+42.7620&zoom=14&layer=hsc-dr3&mark=178.872100,-0.715200}{\emph{DESI~J245.3616+42.7620}} This is a cluster or group lens (Figure~\ref{fig:desi245+42}).
The fiber for the lens was centered on one of the two central galaxies.
The fiber for the source is centered on the arc approximately $12''$ NW of the lens center.
This is the lowest redshift lens in this dataset, at \zd = 0.1354.
For the lensed arc, the \oii feature is detected with high significance, with \zs = 0.9700.


\newpage 
\href{https://www.legacysurvey.org/viewer/?ra=254.4235&dec=+34.8162&layer=ls-dr10-grz&pixscale=0.262&zoom=16}{\emph{DESI~J254.4235+34.8162}}\, This system has a high lens redshift of \zd = 0.8496 (Figure~\ref{fig:desi215.2+00}).
The main lens appears to be at the center of a galaxy group, based on \phz.
From the \ls image, a large blue arc can be seen around the lens.
This arc will be observed by DESI.
The \hst image reveals that there is possible radial counterarc on the other side of the lens.
DESI will also target a small perturber \citep[see, e.g.,][]{vegetti2010a}.
It is very faint on the \ls image, with a \phz of $0.881\pm0.213$, indicating that it is likely a low-mass group member.
\begin{figure}[h]
  \centering
  \includegraphics[width=.7\textwidth]{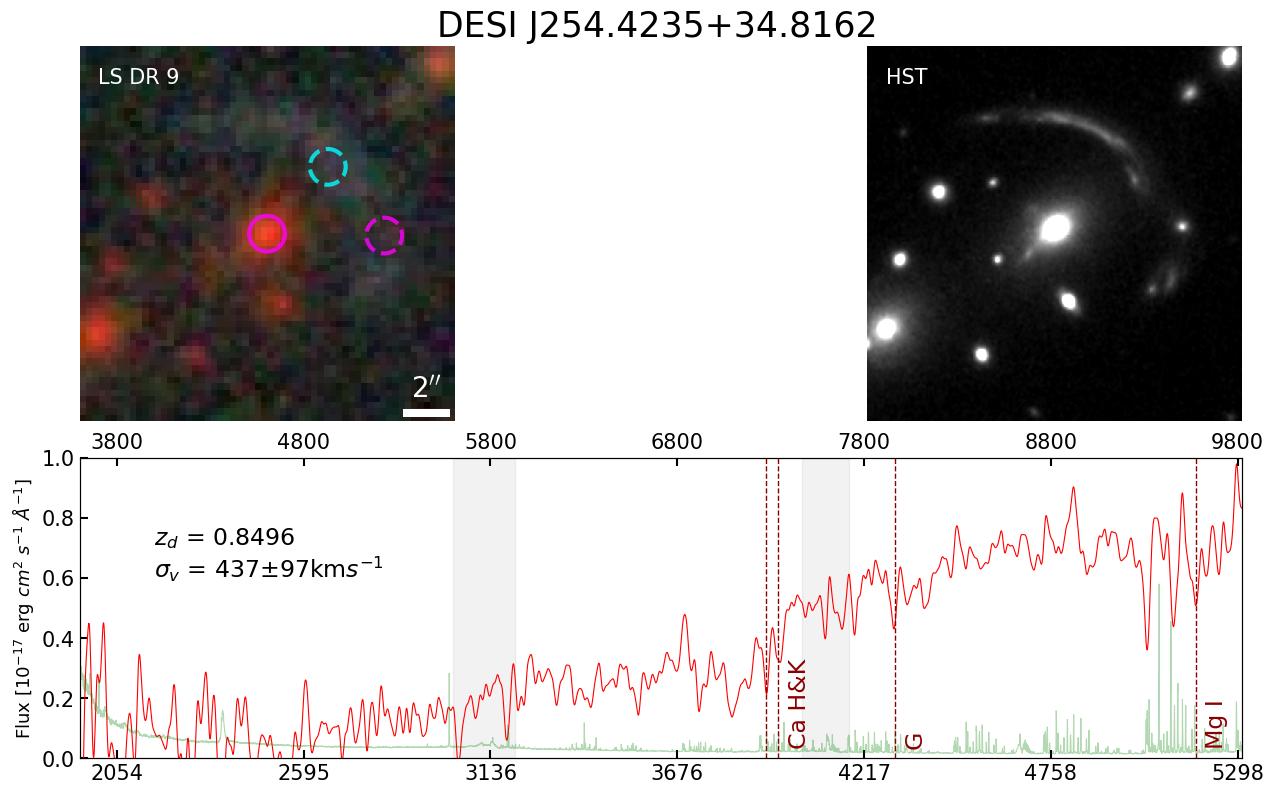}
  \caption{DESI~J254.4235+34.8162. 
  For the arrangement of the panels, see Figure~\ref{fig:desi215.2+00} caption.}
  \label{fig:desi236}
\end{figure}

\subsection{A Known System}\label{sec:known-lens}
\begin{figure}
  \centering
  \includegraphics[width=.7\textwidth]{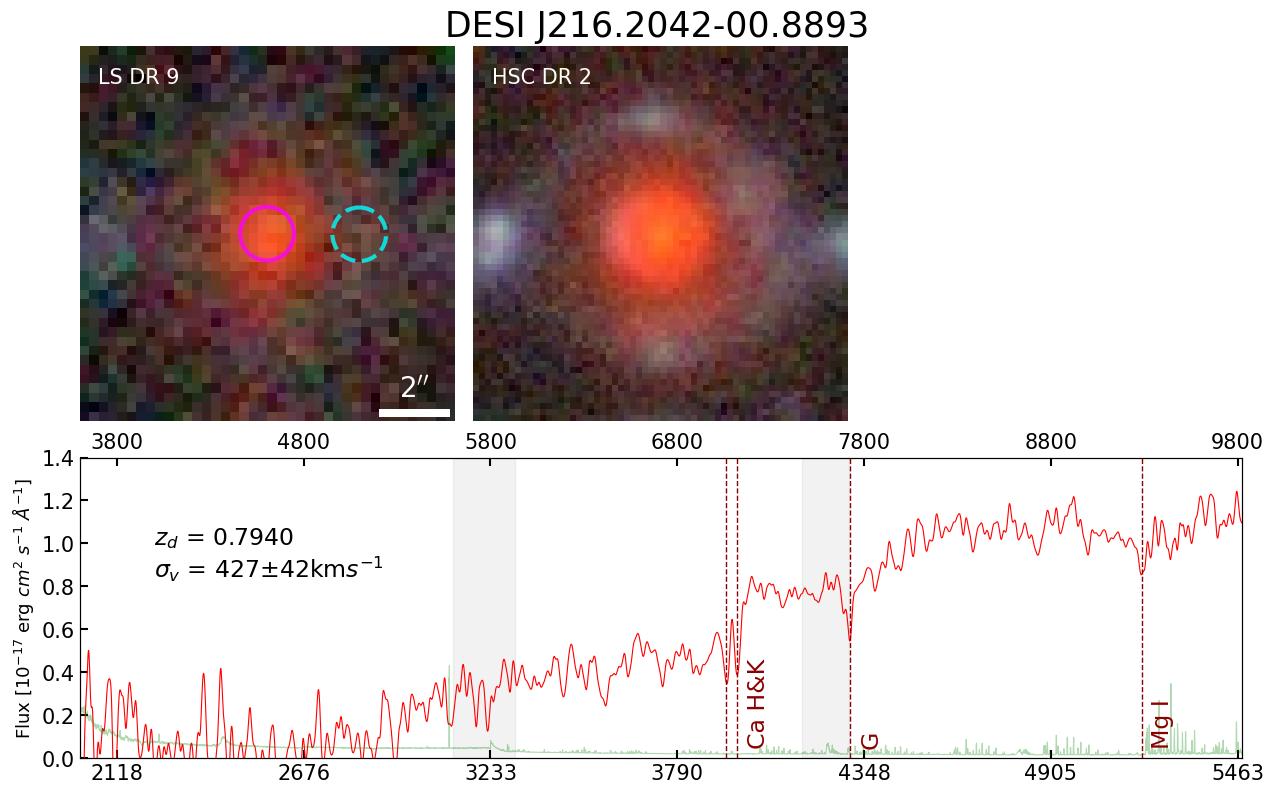}
  \caption{DESI-216.2042-00.8893. 
  We show the \ls and \hsc images, and the DESI lens spectrum.}
  \label{fig:desi216-00}
\end{figure}

We present the one previously known system observed in 
DESI EDR.
\href{https://www.legacysurvey.org/viewer/?ra=216.2042&dec=-00.8893&layer=hsc-dr2&pixscale=0.262&zoom=16}{DESI~J216.2042-00.8893}
was serendipitously discovered and spectroscopically confirmed to be a double-source lensing system by \citet{tanaka2016a}, consisting of a quadruply lensed source ($z_{s1} = 1.998$) and a doubly lensed source ($z_{s2} = 1.302$).
DESI measured its lens redshift, \zd = 0.7940 (Figure~\ref{fig:desi216-00}).
The lens redshift from SDSS agrees with the \zd value from DESI.
The velocity dispersion was also provided by SDSS, $\sigv =  334.8 \pm 83.2$~km/s
\citep{bolton2012a, shu2012a}.\footnote{See also, \url{https://www.sdss4.org/dr17/algorithms/redshifts/}}
This is marginally consistent with the \sigv value from \texttt{FastSpecFit} of $427\pm42$~km/s.

\subsection{All Other Systems}\label{sec:other-sys}
\cwr{For all other systems, those that remain candidates (48 in total) are shown in Appendix~A
and those that are confirmed to be not lenses (4 in total) are shown in the Appendix B.}




\section{Discussion} \label{sec:discuss}
We discuss the efficacy of the \rr pipeline in \S~\ref{sec:rr-vi}, 
the quality of lens candidates found with ResNet in imaging surveys in \S~\ref{sec:cand-quality}, 
redshift distributions in \S~\ref{sec:z-distr},
and velocity dispersion in \S~\ref{sec:vdisp}.

 \subsection{The Redrock pipeline fit and Visual Inspection}\label{sec:rr-vi}
 
The \rr redshift fitting pipeline overall performs very well, as expected.
In all except two cases \cwr{(one each in the categories of confirmed systems and candidates pending confirmation)}, for the lensing galaxies, 
the \rr fit is successful with no warning flags raised.
On the other hand, 
for objects targeted as source galaxies (confirmed or putative), 
only 21 out of 35 spectra (or 60\%) have sufficient signal-to-noise ratios for redshift determination.
Furthermore, \cwr{in the confirmed category}, VI corrected a total of six \zs values given by the \rr pipeline (Table~\ref{tab:master}).

In total, for the \nfujispec\ spectra in this paper, VI identified eight cases (Table~\ref{tab:master}) where either the \rr redshift fit is incorrect or needs confirmation (where ZWARN $> 0$).
This is not representative of the failure rate of the \rr pipeline as these lensing system present unique challenges.
As we will demonstrate, in the context of \rr's main goal, these eight systems clearly comprise a very small fraction, albeit belonging to a class of edge cases that is of high value for a different science goal.

Below, we break down these eight cases into three categories and discuss them separately. 
For two of them, the \oii emission in the background lensed galaxy was  misidentified as \ha (\S~\ref{sec:oii-ha}).
For the next two, VI identified \lya emission from the lensed source galaxy.
Currently \rr does not fit for \lya for galaxy target (which it does for quasar targets).
For the last four cases, the lens and source spectral features are present in the same spectrum (\S~\ref{sec:single-fiber}). 
This causes \rr to fit for the features of one of the two.
Each of these corrections has been noted in the  descriptions for that system.



\subsubsection{\texorpdfstring{\oii}{[O II]} mistaken as H\texorpdfstring{$\alpha$}{alpha} by Redrock}\label{sec:oii-ha}

For DESI~J149.8209+01.0331 (Figure~\ref{fig:eg-lens-fig}) and DESI~J183.6001-00.5351 (Figure~\ref{fig:desi183-00}), \rr 
identified a strong emission feature in the lensed source spectrum to be \ha, with ZWARN=4.
From VI, we determine the respective feature in both spectra to be \oii and confirm the lensing nature of these two systems.

\subsubsection{\textit{Ly}\texorpdfstring{$\alpha$}{-alpha} not fit by Redrock}\label{sec:lya}
For DESI~J212.9021-01.0377 (Figure~\ref{fig:desi212-01})
and DESI~J215.1311+00.3392 (Figure~\ref{fig:desi215+00}),
the lensed sources are \lya emitters, with high redshift values at 
 \zs = 3.023 and 3.1103, respectively.
 \cwr{\rr galaxy templates do not cover restframe \lya and thus are only used to fit for galaxies with redshifts $z<1.6$. Although the \rr quasar templates include \lya coverage and thus can get to higher redshifts, these templates are inefficient for modeling \lya emitters such as these two systems.}
Hence the redshift values from the pipeline are incorrect in both cases \cwr{and were determined by VI}.
Furthermore given the small Einstein radii ($\lesssim 1''$) in both cases and the $0\farcs75$ DESI fiber radius,
the spectra for the two source are contaminated by the light of the nearby lensing galaxy.
Thus it is not a surprise that the \rr best-fit redshift is that of the lens with ZWARN=0.

\subsubsection{Lens and source spectra in the same fiber}\label{sec:single-fiber}

There are three confirmed systems for which there is only one DESI spectrum \cwr{each}. 
\cwr{Each spectrum is from a DESI fiber that targeted the lens of the respective system}.
For DESI~J188.2847+60.1921 (Figure~\ref{fig:desi188+60}), 
and DESI~J241.7346+42.1102 (Figure~\ref{fig:desi241+42}), 
\rr correctly determined the redshift of the lens, without raising a warning flag.
For the third case, DESI~J239.5610+42.6188 (Figure~\ref{fig:desi239.6+42}), 
the \rr best-fit redshift had a ZWARN flag.
But VI confirms that it correctly determined the source redshift. 
For the lensed sources, through VI, we identified an emission feature (\oii in all three cases) in the same spectrum.
This of course was the approach pioneered by the SLACS \citep{bolton2006a} program \citep[see also][]{brownstein2012a, shu2016a}. 
In fact, one of them, DESI-188.2847+60.192, was previously identified by \citet{talbot2021a}, by applying the same technique to the final SDSS data release.
It is the only candidate in this  paper that was first found in spectroscopic data.
For a fourth system, DESI~J218.5371-00.2199 (Figure~\ref{fig:desi218-00}), DESI observed the lens and the source, but the \rr best-fit redshift for both spectra is that of the source, given its strong \oii emission.
From VI, we identify absorption features of the foreground galaxy and determine its redshift.
For all four systems, there is at least one arc that is  $\lesssim 0\farcs75$ (the DESI fiber radius) away from the lens.

\subsubsection{Overall Performance by Redrock}\label{sec:redrock-performance}

Seventy-three lensing galaxies (confirmed or putative) were observed by DESI.
We obtain redshifts for 72 of these,
with two redshifts requiring verification by VI (\rr provided the correct redshift, but with a ZWARN flag).
Only one system has an inconclusive redshift (the one system with the C3 status).
Thirty-seven lensed source galaxies (confirmed or putative) were observed by DESI.
We obtain redshifts for 21 of these, with
seven redshifts requiring verification or correction by VI.

\subsection{Quality of Lens Candidates Found in Imaging Surveys}\label{sec:cand-quality}

We first provide a summary in \S~\ref{sec:lens-cands-quality} of the quality of the lens candidates found in imaging surveys. 
We then make two sets of side-by-side comparisons.
1) In light of DESI spectral results, we discuss 
the effect of imaging data quality on human grades (\S~\ref{sec:img-quality}) via the comparison of the \ls and \hsc datasets.
2) The comparison between human judgment based on the imaging data alone and the spectroscopic determination (\S~\ref{sec:human-vs-spect}) of the natures of these systems (whether they are lenses or not).
Finally, we briefly discuss systems that remain candidates in \S~\ref{sec:dicuss-C-status}.
The comparison of lens search algorithms in imaging data is clearly important, and will be carried out when the spectroscopic sample becomes significantly larger.

\subsubsection{Summary of Lens Candidate Quality}\label{sec:lens-cands-quality}
A strong majority of the candidate systems in this work are from lens searches in HSC~SSP. 
The image quality of HSC~SSP is significantly better,
in terms of seeing and depth, than the dataset we used (LS)
to find lens candidates.
This is especially true for those in the MzLS/BASS \citep[the northern footprint; for more details, see][\S~2.2]{dey2019a}. 
In addition, we included all of our candidates, without grade cut, 
from H20 and H21 (none of the systems from S24 are in EDR), 
in order to better understand the selection function from our ResNet-based search.
Whereas for candidates found in other surveys (including HSC~SSP), we selected the best to include in the target list.
Both factors contribute to the higher confirmation rate for systems found in the HSC~SSP.
Of the \noursys candidate systems from H20 and H21 that are in EDR, 
seven are confirmed to be lenses and three confirmed to be nonlenses, with the remaining four awaiting more observations.
It is also worth keeping in mind that without spectroscopy, the three non-lenses could not have been ruled out as lensing systems as they appeared as plausible lensing candidates in the LS imaging data.
Even though this is still a relatively small sample, this is an encouraging indication that the ResNet-based search yielded high quality candidates, even with comparatively moderate quality imaging data.
Given that the image quality of LSST will be much better than LS (and comparable to HSC~SSP), we expect a ResNet-based search will be very effective.

At this time, for candidates found in the HSC~SSP (as well as KiDS and PanSTARRS) search papers, we cannot do such an evaluation, because we selected the best candidates to be included as DESI targets. 
We nevertheless report that for the 61 lens candidates found in HSC~SSP that were observed, 16 have been confirmed as lenses,
and one as nonlenses.
Forty four remain candidates.
For the two observed candidates found in KiDS, one was confirmed to be a lensing system, and one remains a candidate. 
The one candidate found in Pan-STARRS has been confirmed to be a lensing system.
Note that some systems were found in multiple datasets (Table~\ref{tab:master}).

\subsubsection{Imaging Data Quality}\label{sec:img-quality}
To concretely demonstrate the effect of having higher quality imaging data, 
we first compare between  DESI-150.2022+01.6538 (Figure~\ref{fig:desi150+01}) and DESI~J183.0990-01.5510 (Figure~\ref{fig:desi183-01}), 
identified in H20 and H21, respectively, both given a C~grade (Table~\ref{tab:image-search-compare}).
For the latter system, H21 also provided a human numerical score, 2.5 out of 4, indicating our confidence in it being a lens is higher than grade C candidates with a score of 2.0 (see \S~4.3 of H21).
Both systems were observed by the DECam \citep{dey2019a}.
The grades are based on the images in DECaLS, which is part of LS. 

\medskip

\begin{adjustwidth}{2cm}{1cm}
\begin{minipage}{\linewidth}
    \centering
    \scriptsize
    \captionof{table}{Classification from Imaging Data Comparison\label{tab:image-search-compare}}
        \hspace{-1.8 cm}\begin{tabular}{lcccc}
        \toprule
        System Name &  
        \multicolumn{2}{c}{Human Grade} 
        & Status & Figure\\
        \cline{2-3}
           &     \ls & \hsc  & &\\
        \midrule
        DESI~J150.2022+01.6538  &  C & B & Lens & \ref{fig:desi150+01} \\
        DESI-183.0990-01.5510 &  C & C & non-lens & \ref{fig:desi183-01} \\
        \bottomrule
        \end{tabular}   
    \parbox{0.65\linewidth}{
    \tablecomments{
    DESI~J150.2022+01.6538 and DESI-183.0990-01.5510 were identified in H20 and H21, respectively. 
    The later was also found in HSC~SSP by \citet[][HSCJ121223-013304]{jaelani2020a}.} 
    }
\end{minipage}
\end{adjustwidth}

\medskip

In Table~\ref{tab:image-search-compare}, we also show the human grades for these two systems if we use the \hsc images. 
For DESI-183.0990-01.5510, the arc nature of the purple object to the SE is clearer. 
In addition, a small image with a similar color to the arc now can be seen more clearly to the NE; this is possibly another lensed image of the same source.
Thus we now assign a B grade based on the \hsc image.
However, relying on the \ls images alone, human inspection will likely give these two systems similar grades.
Since HSC~SSP covers a much smaller footprint than the LS, for most candidates at this grade level, the images in the Legacy Surveys are all that we can rely on.
This will be the case for most of such candidates, 
which are not reachable by the LSST,
for the foreseeable future.
However, not surprisingly, this comparison does show a promising future for lens searches in the LSST imaging data:
for candidate systems with this level of evidence, we expect to be able to judge their quality notably better, as shown here.

\subsubsection{Lensing Confirmation That Requires Spectroscopy}\label{sec:human-vs-spect}
The comparison below reinforces that spectroscopy is indispensable for the confirmation of at least some systems, even if they were observed under the best (non-AO assisted) conditions, comparable to what is expected for LSST.
We compare between two confirmed non-lenses that appear as plausible lens candidates with a confirmed lens (Table~\ref{tab:image-search-compare2}).
All three images have a faint blue arc-like object next to a red galaxy.

\begin{adjustwidth}{2cm}{1cm}
\begin{minipage}{\linewidth}
    \centering
    \scriptsize
    \captionof{table}{Necessity for Spectroscopy\label{tab:image-search-compare2}}
        \hspace{-1.8 cm}\begin{tabular}{lccccc}
        \toprule
        System Name &   \multicolumn{2}{c}{Human Grade} & Status & Figure\\
        \cline{2-3}
           &     \ls & \hsc  & &\\
        \midrule
        DESI-217.2090+51.6467  &  C & - & non-lens & \ref{fig:desi217+51} \\
        DESI-218.8713+00.3034 &  - & B & non-lens & \ref{fig:desi218+00} \\
        DESI~J218.8286-01.4239 &  - & B & Lens  & \ref{fig:desi218-01} \\
        \bottomrule
        \end{tabular}   
    \parbox{0.65\linewidth}{
    \tablecomments{
    DESI-217.2090+51.6467 was identified in H20, and DESI-218.8713+00.3034 and DESI-218.8286-01.4239 in \citet{sonnenfeld2020a}.}}
    \end{minipage}
\end{adjustwidth}

\medskip

The human grade, C, for DESI-217.2090+51.6467 is from H20.
The other two systems were identified in the HSC~SSP program, HSCJ143529+001812 and HSCJ143518-012526, or DESI-218.8713+00.3034 and  DESI~J218.8286-01.4239, respectively, in this paper.
Both of them were assigned a B grade by the discovery paper \citep{sonnenfeld2020a}, which we agree.
For these three systems, the grades are mainly based on how arc-like a blue object around an elliptical galaxy appears to be, 
and to a lesser extent, on the surface brightness of this object, with a lower surface brightness being the more likely  appearance of a lensed arc. 
It is a very small sample, but this comparison is the beginning of understanding the performance ``ceiling'' for strong lens searches in imaging surveys, 
even using the best (non-AO assisted) ground-based imaging data, such as those from the HSC.
The comparison between the two systems found in the HSC~SSP program therefore is especially useful.
Not only were they found in the same HSC~SSP imaging data, 
but \hsc data are comparable to the future LSST observations, in resolution and in depth.


In these two comparisons, we concretely demonstrate that given the imaging quality of DECaLS (observed by DECam), 
there are candidates systems that need spectroscopy for confirmation.
This need is even greater in the northern MzLS/BASS region of the LS footprint, where the seeing is worse.
For a subset of these cases, images with higher resolution and greater depth as provided by the HSC~SSP observation can help (see Table~\ref{tab:image-search-compare}).
However, not surprisingly, using non-AO assisted ground-based imaging data alone, there still is a ceiling for how good 
human judgment can be (see Table~\ref{tab:image-search-compare2}), which would also be the ceiling for the performance of search algorithms.

\subsubsection{Systems that remain candidates}\label{sec:dicuss-C-status}

The lens candidates from HSC~SSP constitute a strong majority of the systems \cwr{that remain candidates} (of the \cwr{48}, 7 are from our searches, one each from KiDS and PanSTARRS, and the rest from HSC~SSP, 
with some systems found in multiple searches).
Therefore, as more DESI data become available for these systems, 
we expect the confirmation rate to be high.
This is not only due to the better seeing and greater depth for the HSC~SSP data
(compared with the LS),
but also because, unlike the target selection from H20 and H21,
we selected only the best candidates 
from those found in the HSC~SSP search papers (and lens search results from other surveys).


\subsection{Redshift Distributions}\label{sec:z-distr}

Below we discuss the redshift distributions of the confirmed lensing systems and those that remain candidates (\S~\ref{sec:z-distr-lens-cands}) 
and the redshift distribution of the confirmed non-lenses (\S~\ref{sec:z-distr-non-lenses}).

\subsubsection{Redshift Distributions of Confirmed Lenses and Candidates}\label{sec:z-distr-lens-cands}
\begin{figure}[h]
  \centering
  \includegraphics[width=0.45\textwidth]{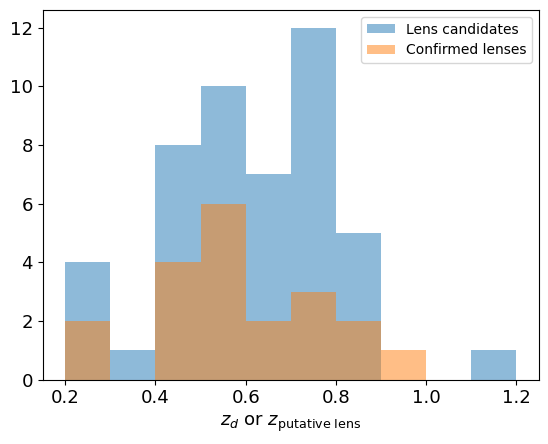}
  \caption{Redshift distributions of lens targets in DESI EDR, both confirmed and putative.}
  \label{fig:z_d_hist}
\end{figure}

The redshift distributions for confirmed lenses and putative lenses  from systems that remain candidates are shown in Figure~\ref{fig:z_d_hist}. 
The distributions for these two categories appear to be consistent with each other.
Both distributions appear approximately Gaussian and peak around a redshift of 0.6, which is significantly higher than the SLACS lenses, but comparable to the lenses found in the BELLS programs \cite[e.g.,][]{brownstein2012a, shu2016a}. However, the high end of the lens redshift for the systems in this work extends to higher values. 
As noted in \S~\ref{sec:confirmed-lenses}, DESI~J212.9021-01.0377 with $\zd = 0.9477$ is among the highest redshift galaxy-scale strong lenses.
These systems are especially valuable for constraining cosmological parameters \citep[e.g.,][]{chen2019a}{}{}.

\begin{figure}[h]
  \centering
  \includegraphics[width=0.45\textwidth]{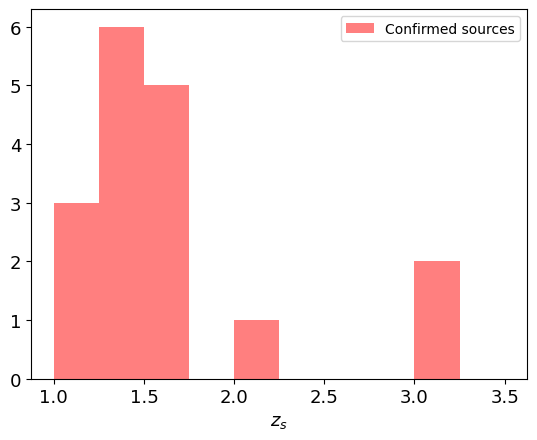}
  \caption{Source redshift distributions of confirmed lensing systems in EDR.}
  \label{fig:z_s_hist}
\end{figure}

The redshift distributions for the sources of confirmed lensing systems are shown in Figure~\ref{fig:z_s_hist}. 
The sample size is relatively small.
The mode of the distribution is around 1.5, close to the edge of what DESI is able to measure.
This, combined with the long tail toward beyond 3.0, is likely an early indication that the source redshift distribution from the DESI~\progname will likely be significantly higher than in SLACS.
Even for the candidates from the BELLS GALLERY program \citep{shu2016a}, which were found in SDSS~III (BOSS) with \lya emission in the lensed sources, 
the \zs values for all 21 systems were $<3$.
Whereas in the relatively small sample of confirmed lenses in this paper, we found two systems with $\zs > 3$ (we identify a third system from Paper III in this series).
These are rare.
Prior to the availability of deep and wide imaging surveys (e.g., DES, LS), among galaxy-scale lenses, in the compilation by \citet{chen2019a}, there are seven such systems.\footnote{Among group and cluster scale lensing systems, seven systems with 
$3 < \zs < 4$ were discovered \citep{smail2007a, gilbank2008a, huang2009a, christensen2012a, oguri2012a, dessauges-zavadsky2017a, rigby2018a}.
Recently, \citet{zhang2023a} added five more.}
In recent years, the AGEL survey \citep{tran2022a} report two systems with \zs in this range.
Both of them were first discovered in H21 (\href{https://www.legacysurvey.org/viewer/cutout.?ra=278.8338&dec=+46.1076&layer=ls-dr9&pixscale=0.262&zoom=15}{\cwr{DESI~J278.8338+46.1076}} and \href{https://www.legacysurvey.org/viewer/cutout.?ra=293.9927&dec=+58.1525&layer=ls-dr9&pixscale=0.262&zoom=15}{\cwr{DESI~J293.9927+58.1525}}), the first of which  was observed by our \hst program (GO-15867).
 These highly magnified galaxies allow us to investigate much smaller physical scales than would be possible without strong lensing and provide an unparalleled opportunity to understand star formation processes at high redshift. 
Given that the lenses in EDR represents a small fraction of the targets in the \progname, we expect many more $\zs > 3$ systems and high \zd systems to be discovered.

\subsubsection{Redshift Distributions of Confirmed Non-Lenses}\label{sec:z-distr-non-lenses}

\begin{figure}[h]
\centering  
\includegraphics[width=0.42\textwidth]{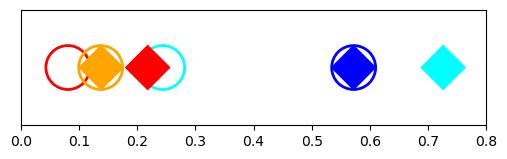}
\caption{Redshifts of the four \cwr{false positive} candidate systems that are confirmed to \cwr{not be lenses}. For each pair, with the same color coding, the diamond and the circle show the redshifts of the ``lens target" and ``source target", respectively.
}
\label{fig:z_nonlens}
\end{figure}

Redshifts of the four candidate systems that are confirmed to be not lenses are shown in Figure~\ref{fig:z_nonlens}.
For two of the four pairs, the  ``lens target'' and the ``source target'' redshifts are very similar.
In the case of DESI-218.87813+00.3034, it appears that these two targets at $z = 0.5770$ and $z = 0.5772$, respectively, are two parts of the same galaxy.
In the other case, DESI-219.9227+0.5073, they are possibly members of the same galaxy group.
In the remaining two cases, what appeared to be possible lensed arcs turn out to be relatively faint foreground galaxies.

\subsection{Velocity Dispersion}\label{sec:vdisp}

We estimate the velocity dispersion (\sigv) of each lens using \texttt{FastSpecFit}, a spectral synthesis and emission-line modeling code developed specifically to analyze DESI spectroscopy (\citealt{moustakas2023a}; J.\ Moustakas et~al., in prep). Briefly, \texttt{FastSpecFit} masks pixels which may contain emission lines, and then it fits an inverse-variance weighted, non-negative linear combination of stellar population synthesis (SPS) models to the data, carefully accounting for the resolution matrix (or instrumental line-spread function) of each DESI camera. The SPS models are generated using the theoretical \texttt{C3K\_a} spectral grid (with $\mathcal{R}\approx10,000$) from the Flexible Stellar Population Synthesis \texttt{FSPS}) library\footnote{\url{https://github.com/cconroy20/fsps}}, and span a range of age and dust attenuation values. \texttt{FastSpecFit} carries out a simple $\chi^{2}$(\sigv) scan using models convolved with \sigv{} ranging between 50 and 700~km/s with a uniform spacing of 10~km/s; the maximum-likelihood \sigv{} is then determined by fitting a quadratic polynomial to the three velocity dispersions with the lowest $\chi^{2}$, and the uncertainty is given by the \sigv{} interval corresponding to $\Delta\chi^2=1$. If the minimum $\chi^2$ value lands on either end of the scan in \sigv, then we adopt a nominal (default) velocity dispersion of 350~km/s and set the corresponding inverse variance to zero.\footnote{This applies to only one spectrum in this work: DESI-247.7865+42.5782 (Table~\ref{tab:master}).}

The velocity dispersion of the lensing galaxies can be used to break the mass-sheet degeneracy in lens modeling.
This is a subject we will delve into in conjunction  with lens modeling in future work.
Here we discuss briefly how this information can begin to help us understand the selection function of lensing systems found in imaging surveys.

\begin{figure}
  \centering
  \includegraphics[width=0.5\textwidth]{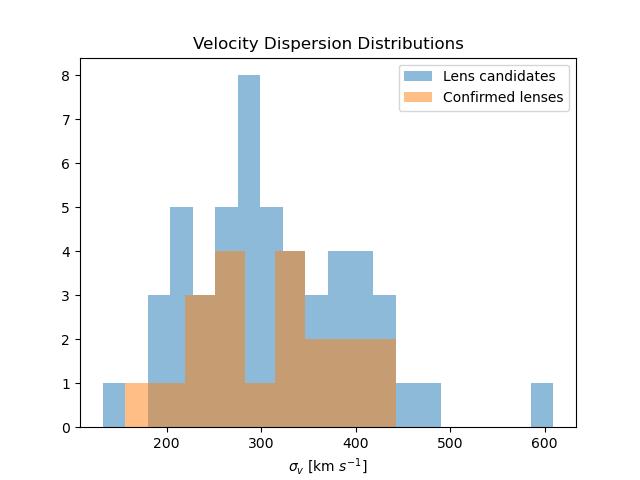}
  \caption{Distribution of velocity distribution for lens targets, both confirmed and putative.}
  \label{fig:vdisp_hist}
\end{figure}

\citet{collett2015a} forecast lens discoveries for several major imaging surveys, including DES and LSST. 
Their Figure~1 showed the forecast for all galaxy-scale strong lenses with a velocity dispersion of $\sigv > 100$ km/s.
In the third panel of their Figure~1, the \sigv distribution has a mode $\sim$220~km/s and tapers off $\sim$400~km/s. 
The contrast with the results in this paper is striking.
In Figure~\ref{fig:vdisp_hist}, we show the distribution of \sigv for both confirmed lenses and those that remain candidates. 
These two distributions are broadly consistent with each other.
It is clear that the mode is $\sim$300~km/s, and the highest values extend to 450~km/s and beyond.
This is not a surprise. 
In lens searches in imaging surveys, even for HSC~SSP, large Einstein radius (\thetaE) systems are comparatively easier to find, whether by a ``traditional'' algorithmic approach or a neural network based approach.

The lensing galaxy for \cwr{such large Einstein radius} systems \cwr{often} appears to be at the center of a galaxy group.
Finding an abundance of such systems from imaging surveys is not a surprise.
The fact that so far the two prevailing categories of strong lenses are galaxy-scaled vs.\ cluster-scaled is a result of the selection functions of the search methods before $\sim 2017$.
Prior to that time, the cluster-scale lenses are typically found by human eye, as they easily stand out in imaging data,
and the galaxy-scale lenses are found dozens at time through searches in SDSS spectroscopic data. 
SDSS I\&II had a fiber diameter of 3$''$, whereas III \& IV had a fiber diameter of $2''$. 
This approximately translates to finding systems with Einstein radii $\lesssim 1.5''$ and $\lesssim 1''$, respectively.
Even more massive galaxies, sometimes situated at the center of a group, likely will produce a set of arcs with Einstein radii $1.5''  \lesssim  \thetaE \lesssim 5''$, typically from one background source.
These systems are essentially out of the reach of the SDSS spectroscopic search method.
They also do not stand out in imaging data nearly as much as a typical cluster-scale strong lens with $\thetaE \gtrsim 5''$, often extending to much greater values, with numerous arcs.
Thus, to the best of our knowledge, hitherto, it is not clear how abundant such strong lenses are.
This is a new mass range that bridges galaxy-scale lensing systems with \thetaE $\lesssim1.5''$ and cluster-scale strong lenses. 
Lens searches in large imaging surveys, especially those using neural networks, now begin to fill this void. 
Although a thorough study of the completeness of neural network based searches have yet to be carried out, a large number of recent results show preliminarily that these searches are very effective.
They already indicate that there are many such systems in the universe.
These systems will likely open another window of using strong lensing to address astrophysical (e.g., galaxy evolution, structure formation, stellar mass profile) 
and cosmological (e.g., testing CDM and measuring \ho using lensed quasars and supernovae).
Current search results show that with deeper observations and wider coverage in the southern sky,
we will find many more such systems in LSST.



\section{Conclusions} \label{sec:conclusion}

We introduce the DESI Strong Lens Secondary Target Program as part of the \projname series.
This program will spectroscopically follow up \ntotsys strong lens candidate systems found in the DESI Legacy Imaging Surveys footprint, with a total of \ntotspec unique targets.
A highly multiplexed fiber spectroscopic instrument like DESI is uniquely suited to obtain the spectra, requisite for strong lensing science goals, for such a large number of lens candidates.

In this paper, Paper~II of the series, we present
spectra from DESI Early Data Release (EDR) for \nfujisys candidate systems.
DESI successfully measured the redshifts of the putative lens for 72 of them,  or 98.6\%.
Of these, 20 have been spectroscopically confirmed as strong lensing systems, i.e., their source redshifts have also been successfully measured.
Of the confirmed systems, 15 are from searches in HSC~SSP, 
7 from our lens searches in the DESI Legacy Imaging Surveys, 1 each from KiDS and Pan-STARRS.
Note that some systems were found in multiple imaging surveys.
Additionally, the DESI observation provides stronger evidence for a system identified in the spectroscopic search of \citet{talbot2021a}.
There are only four confirmed non-lenses. 
\cwr{DESI EDR also contains a previously known lens.}
The remaining 48 systems remain candidates.

For the candidates from our lens searches (H20 and H21), without grade cuts,
of the \noursys included in this paper, 7 
have been confirmed with DESI spectroscopy alone, DESI and Keck NIRES, and/or \hst. 
Three have been confirmed to be non-lenses, and the remaining 7 candidates await additional observations (i.e., the putative lensed sources have not been observed by DESI).
This is still a relatively small sample, but it shows promising signs that lens candidates from our neural net based searches in imaging surveys with $\sim 1''$ seeing (DECam), even up to $\sim 1.7''$ (which is the case for the $gr$ bands in the northern region of the Legacy Surveys footprint), will likely have a high confirmation rate.


We also note that the HSC~SSP data are comparable to what LSST observations will obtain, in resolution and depth.
As the spectroscopic data grow, more detailed characterization of these systems can shed significant light on the purity and completeness of lens candidates discoverable in the LSST data, given a grade cut.
This will be more fully investigated as this series continues.


We also provide velocity dispersions for the lensing galaxies, confirmed and putative, from \texttt{FastSpecFit}.
The distribution has a mode $\sim 300$~km/s, and extends to beyond 450 km/s, much higher than for previously known galaxy-scale strong lenses. 
This is due to the fact that lenses found in imaging surveys have a  different selection function compared with that for lenses found in fiber spectroscopy, which is by far the largest source of discovery for previously known lenses. 
In particular, it is clear that the lensing galaxies for some of the systems reported here are at the center of a group or a cluster.
Such lensing galaxies, in terms of their mass and environment, bridge the gap between galaxy and cluster scales.
This is a hitherto relatively under-studied regime.

We note the wide range of lens redshifts of these systems (Figure~\ref{fig:z_d_hist}), both those that are confirmed and those that remain candidates: \zd ranging between 0.1354 and 0.9477 for DESI\cwr{~J}212.9021-01.0377, 
which, to our knowledge, is the highest lens redshift among galaxy-scale strong lensing systems.
The mode of the distributions ($\sim 0.6$) is much higher than the SLACS lenses and comparable to systems from the BELLS programs.
Crucially, it matches that of the lensed quasars. 
Detailed modeling for a wide lens redshift range provides important priors for dark matter density profile for time delay $H_{0}$ measurements \citep[e.g.,][]{birrer2020a, shajib2020a}. 
There are five confirmed lenses that have $\zd > 0.75$.
These systems are highly valuable in understanding the evolution of the mass profiles of elliptical galaxies \citep[e.g.,][]{filipp2023a}.
Based on photometric redshifts of the targets in the DESI \progname, there will be a large number of high lens redshift systems. 
DESI\cwr{~J}212.9021-01.0377 is also one of two systems with $\zs > 3$. 
The other is DESI\cwr{~J}215.1311+00.3392.
We confirmed a third system, with $\zs > 3$ using both DESI and Keck NIRES spectroscopy (DESI\cwr{~J}094.5639+50.305, Paper III, Agarwal et al. in prep).
To our knowledge, these are the first three galaxy-scale strong lensing systems with $\zs > 3$. 
We expect there will be many more such systems from this project.

The implications and applications of this dataset will be far and wide. 
Future publications in this series will report even larger spectroscopic samples of strong lenses from this project.

\begin{acknowledgments}
This research used resources of the National Energy Research Scientific Computing Center (NERSC), a U.S. Department of Energy Office of Science User Facility operated under Contract No. DE-AC02-05CH11231 and the Computational HEP program in The Department of Energy's Science Office of High Energy Physics provided resources through the ``Cosmology Data Repository" project (Grant \#KA2401022).
X.H. acknowledges the University of San Francisco Faculty Development Fund. 
A.D.'s research is supported by National Science Foundation's National Optical-Infrared Astronomy Research Laboratory, 
which is operated by the Association of Universities for Research in Astronomy (AURA) under cooperative agreement with the National Science Foundation.

\cwr{This material is based upon work supported by the U.S. Department of Energy (DOE), Office of Science, Office of High-Energy Physics, under Contract No. DE–AC02–05CH11231, and by the National Energy Research Scientific Computing Center, a DOE Office of Science User Facility under the same contract. Additional support for DESI was provided by the U.S. National Science Foundation (NSF), Division of Astronomical Sciences under Contract No. AST-0950945 to the NSF’s National Optical-Infrared Astronomy Research Laboratory; the Science and Technology Facilities Council of the United Kingdom; the Gordon and Betty Moore Foundation; the Heising-Simons Foundation; the French Alternative Energies and Atomic Energy Commission (CEA); the National Council of Humanities, Science and Technology of Mexico (CONAHCYT); the Ministry of Science, Innovation and Universities of Spain (MICIU/AEI/10.13039/501100011033), and by the DESI Member Institutions: \url{https://www.desi.lbl.gov/collaborating-institutions}.}

\cwr{The DESI Legacy Imaging Surveys consist of three individual and complementary projects: the
Dark Energy Camera Legacy Survey (DECaLS), the Beijing-Arizona Sky Survey (BASS), and the
Mayall z-band Legacy Survey (MzLS). DECaLS, BASS and MzLS together include data obtained,
respectively, at the Blanco telescope, Cerro Tololo Inter-American Observatory, NSF’s NOIRLab; the
Bok telescope, Steward Observatory, University of Arizona; and the Mayall telescope, Kitt Peak
National Observatory, NOIRLab. NOIRLab is operated by the Association of Universities for Research
in Astronomy (AURA) under a cooperative agreement with the National Science Foundation.
Pipeline processing and analyses of the data were supported by NOIRLab and the Lawrence
Berkeley National Laboratory. Legacy Surveys also uses data products from the Near-Earth Object
Wide-field Infrared Survey Explorer (NEOWISE), a project of the Jet Propulsion
Laboratory/California Institute of Technology, funded by the National Aeronautics and Space
Administration. Legacy Surveys was supported by: the Director, Office of Science, Office of High
Energy Physics of the U.S. Department of Energy; the National Energy Research Scientific
Computing Center, a DOE Office of Science User Facility; the U.S. National Science Foundation,
Division of Astronomical Sciences; the National Astronomical Observatories of China, the Chinese
Academy of Sciences and the Chinese National Natural Science Foundation. LBNL is managed by
the Regents of the University of California under contract to the U.S. Department of Energy. The
complete acknowledgments can be found at \url{https://www.legacysurvey.org/}.}

\cwr{Any opinions, findings, and conclusions or recommendations expressed in this material are those of the author(s) and do not necessarily reflect the views of the U. S. National Science Foundation, the U. S. Department of Energy, or any of the listed funding agencies.}

\cwr{The authors are honored to be permitted to conduct scientific research on I'oligam Du'ag (Kitt Peak), a mountain with particular significance to the Tohono O’odham Nation.}

\end{acknowledgments}




\newpage
\bibliography{dustarchive}{}

\begin{thebibliography}{}
\expandafter\ifx\csname natexlab\endcsname\relax\def\natexlab#1{#1}\fi
\providecommand{\url}[1]{\href{#1}{#1}}
\providecommand{\dodoi}[1]{doi:~\href{http://doi.org/#1}{\nolinkurl{#1}}}
\providecommand{\doeprint}[1]{\href{http://ascl.net/#1}{\nolinkurl{http://ascl.net/#1}}}
\providecommand{\doarXiv}[1]{\href{https://arxiv.org/abs/#1}{\nolinkurl{https://arxiv.org/abs/#1}}}

\bibitem[{{Abbott} {et~al.}(2018){Abbott}, {Abdalla}, {Allam}, {Amara},
  {Annis}, {Asorey}, {Avila}, {Ballester}, {Banerji}, {Barkhouse}, {Baruah},
  {Baumer}, {Bechtol}, {Becker}, {Benoit-L{\'e}vy}, {Bernstein}, {Bertin},
  {Blazek}, {Bocquet}, {Brooks}, {Brout}, {Buckley-Geer}, {Burke}, {Busti},
  {Campisano}, {Cardiel-Sas}, {Carnero Rosell}, {Carrasco Kind}, {Carretero},
  {Castander}, {Cawthon}, {Chang}, {Chen}, {Conselice}, {Costa}, {Crocce},
  {Cunha}, {D'Andrea}, {da Costa}, {Das}, {Daues}, {Davis}, {Davis}, {De
  Vicente}, {DePoy}, {DeRose}, {Desai}, {Diehl}, {Dietrich}, {Dodelson},
  {Doel}, {Drlica-Wagner}, {Eifler}, {Elliott}, {Evrard}, {Farahi}, {Fausti
  Neto}, {Fernandez}, {Finley}, {Flaugher}, {Foley}, {Fosalba}, {Friedel},
  {Frieman}, {Garc{\'\i}a-Bellido}, {Gaztanaga}, {Gerdes}, {Giannantonio},
  {Gill}, {Glazebrook}, {Goldstein}, {Gower}, {Gruen}, {Gruendl}, {Gschwend},
  {Gupta}, {Gutierrez}, {Hamilton}, {Hartley}, {Hinton}, {Hislop}, {Hollowood},
  {Honscheid}, {Hoyle}, {Huterer}, {Jain}, {James}, {Jeltema}, {Johnson},
  {Johnson}, {Kacprzak}, {Kent}, {Khullar}, {Klein}, {Kovacs}, {Koziol},
  {Krause}, {Kremin}, {Kron}, {Kuehn}, {Kuhlmann}, {Kuropatkin}, {Lahav},
  {Lasker}, {Li}, {Li}, {Liddle}, {Lima}, {Lin}, {L{\'o}pez-Reyes}, {MacCrann},
  {Maia}, {Maloney}, {Manera}, {March}, {Marriner}, {Marshall}, {Martini},
  {McClintock}, {McKay}, {McMahon}, {Melchior}, {Menanteau}, {Miller},
  {Miquel}, {Mohr}, {Morganson}, {Mould}, {Neilsen}, {Nichol}, {Nogueira},
  {Nord}, {Nugent}, {Nunes}, {Ogand o}, {Old}, {Pace}, {Palmese},
  {Paz-Chinch{\'o}n}, {Peiris}, {Percival}, {Petravick}, {Plazas}, {Poh},
  {Pond}, {Porredon}, {Pujol}, {Refregier}, {Reil}, {Ricker}, {Rollins},
  {Romer}, {Roodman}, {Rooney}, {Ross}, {Rykoff}, {Sako}, {Sanchez}, {Sanchez},
  {Santiago}, {Saro}, {Scarpine}, {Scolnic}, {Serrano}, {Sevilla-Noarbe},
  {Sheldon}, {Shipp}, {Silveira}, {Smith}, {Smith}, {Smith}, {Soares-Santos},
  {Sobreira}, {Song}, {Stebbins}, {Suchyta}, {Sullivan}, {Swanson}, {Tarle},
  {Thaler}, {Thomas}, {Thomas}, {Troxel}, {Tucker}, {Vikram}, {Vivas},
  {Walker}, {Wechsler}, {Weller}, {Wester}, {Wolf}, {Wu}, {Yanny}, {Zenteno},
  {Zhang}, {Zuntz}, {DES Collaboration}, {Juneau}, {Fitzpatrick}, {Nikutta},
  {Nidever}, {Olsen}, {Scott}, \& {NOAO Data Lab}}]{abbott2018a}
{Abbott}, T.~M.~C., {Abdalla}, F.~B., {Allam}, S., {et~al.} 2018, \apjs, 239,
  18, \dodoi{10.3847/1538-4365/aae9f0}

\bibitem[{Adame {et~al.}(2024)Adame, Aguilar, Ahlen, Alam, Aldering, Alexander,
  Alfarsy, Allende~Prieto, Alvarez, Alves, Anand, Andrade-Oliveira, Armengaud,
  Asorey, Avila, Aviles, Bailey, Balaguera-Antolínez, Ballester, Baltay,
  Bault, Bautista, Behera, Beltran, BenZvi, Beraldo~e Silva, Bermejo-Climent,
  Berti, Besuner, Beutler, Bianchi, Blake, Blum, Bolton, Brieden, Brodzeller,
  Brooks, Brown, Buckley-Geer, Burtin, Cabayol-Garcia, Cai, Canning,
  Cardiel-Sas, Carnero~Rosell, Castander, Cervantes-Cota, Chabanier,
  Chaussidon, Chaves-Montero, Chen, Chen, Chuang, Claybaugh, Cole, Cooper,
  Cuceu, Davis, Dawson, de~Belsunce, de~la Cruz, de~la Macorra, Della~Costa,
  de~Mattia, Demina, Demirbozan, DeRose, Dey, Dey, Dhungana, Ding, Ding, Doel,
  Doshi, Douglass, Edge, Eftekharzadeh, Eisenstein, Elliott, Ereza, Escoffier,
  Fagrelius, Fan, Fanning, Fawcett, Ferraro, Flaugher, Font-Ribera,
  Forero-Romero, Forero-Sánchez, Frenk, Gänsicke, García, García-Bellido,
  Garcia-Quintero, Garrison, Gil-Marín, Golden-Marx, Gontcho A~Gontcho,
  Gonzalez-Morales, Gonzalez-Perez, Gordon, Graur, Green, Gruen, Guy,
  Hadzhiyska, Hahn, Han, Hanif, Herrera-Alcantar, Honscheid, Hou, Howlett,
  Huterer, Iršič, Ishak, Jacques, Jana, Jiang, Jimenez, Jing, Joudaki, Joyce,
  Jullo, Juneau, Karaçaylı, Karim, Kehoe, Kent, Khederlarian, Kim, Kirkby,
  Kisner, Kitaura, Kizhuprakkat, Kneib, Koposov, Kovács, Kremin, Krolewski,
  L’Huillier, Lahav, Lambert, Lamman, Lan, Landriau, Lang, Lange, Lasker,
  Leauthaud, Le~Guillou, Levi, Li, Linder, Lyons, Magneville, Manera, Manser,
  Margala, Martini, McDonald, Medina, Medina-Varela, Meisner, Mena-Fernández,
  Meneses-Rizo, Mezcua, Miquel, Montero-Camacho, Moon, Moore, Moustakas,
  Mueller, Mundet, Muñoz-Gutiérrez, Myers, Nadathur, Napolitano, Neveux,
  Newman, Nie, Nikutta, Niz, Norberg, Noriega, Paillas, Palanque-Delabrouille,
  Palmese, Pan, Parkinson, Penmetsa, Percival, Pérez-Fernández,
  Pérez-Ràfols, Pieri, Poppett, Porredon, Pothier, Prada, Pucha, Raichoor,
  Ramírez-Pérez, Ramirez-Solano, Rashkovetskyi, Ravoux, Rocher, Rockosi,
  Ross, Rossi, Ruggeri, Ruhlmann-Kleider, Sabiu, Said, Saintonge, Samushia,
  Sanchez, Saulder, Schaan, Schlafly, Schlegel, Scholte, Schubnell, Seo,
  Shafieloo, Sharples, Sheu, Silber, Sinigaglia, Siudek, Slepian, Smith,
  Soumagnac, Sprayberry, Stephey, Suárez-Pérez, Sun, Tan, Tarlé, Tojeiro,
  Ureña-López, Vaisakh, Valcin, Valdes, Valluri, Vargas-Magaña, Variu,
  Verde, Walther, Wang, Wang, Weaver, Weaverdyck, Wechsler, White, Xie, Yang,
  Yèche, Yu, Yuan, Zhang, Zhang, Zhao, Zheng, Zhou, Zhou, Zou, Zou, \&
  Zu}]{desi2023a}
Adame, A.~G., Aguilar, J., Ahlen, S., {et~al.} 2024, The Astronomical Journal,
  168, 58, \dodoi{10.3847/1538-3881/ad3217}

\bibitem[{{Aihara} {et~al.}(2018){Aihara}, {Armstrong}, {Bickerton}, {Bosch},
  {Coupon}, {Furusawa}, {Hayashi}, {Ikeda}, {Kamata}, {Karoji}, {Kawanomoto},
  {Koike}, {Komiyama}, {Lang}, {Lupton}, {Mineo}, {Miyatake}, {Miyazaki},
  {Morokuma}, {Obuchi}, {Oishi}, {Okura}, {Price}, {Takata}, {Tanaka},
  {Tanaka}, {Tanaka}, {Uchida}, {Uraguchi}, {Utsumi}, {Wang}, {Yamada},
  {Yamanoi}, {Yasuda}, {Arimoto}, {Chiba}, {Finet}, {Fujimori}, {Fujimoto},
  {Furusawa}, {Goto}, {Goulding}, {Gunn}, {Harikane}, {Hattori}, {Hayashi},
  {He{\l}miniak}, {Higuchi}, {Hikage}, {Ho}, {Hsieh}, {Huang}, {Huang},
  {Imanishi}, {Iwata}, {Jaelani}, {Jian}, {Kashikawa}, {Katayama}, {Kojima},
  {Konno}, {Koshida}, {Kusakabe}, {Leauthaud}, {Lee}, {Lin}, {Lin},
  {Mandelbaum}, {Matsuoka}, {Medezinski}, {Miyama}, {Momose}, {More}, {More},
  {Mukae}, {Murata}, {Murayama}, {Nagao}, {Nakata}, {Niida}, {Niikura},
  {Nishizawa}, {Oguri}, {Okabe}, {Ono}, {Onodera}, {Onoue}, {Ouchi}, {Pyo},
  {Shibuya}, {Shimasaku}, {Simet}, {Speagle}, {Spergel}, {Strauss}, {Sugahara},
  {Sugiyama}, {Suto}, {Suzuki}, {Tait}, {Takada}, {Terai}, {Toba}, {Turner},
  {Uchiyama}, {Umetsu}, {Urata}, {Usuda}, {Yeh}, \& {Yuma}}]{aihara2018a}
{Aihara}, H., {Armstrong}, R., {Bickerton}, S., {et~al.} 2018, \pasj, 70, S8,
  \dodoi{10.1093/pasj/psx081}

\bibitem[{{Auger} {et~al.}(2010){Auger}, {Treu}, {Bolton}, {Gavazzi},
  {Koopmans}, {Marshall}, {Moustakas}, \& {Burles}}]{auger2010a}
{Auger}, M.~W., {Treu}, T., {Bolton}, A.~S., {et~al.} 2010, \apj, 724, 511,
  \dodoi{10.1088/0004-637X/724/1/511}

\bibitem[{{Bayliss} {et~al.}(2011){Bayliss}, {Hennawi}, {Gladders}, {Koester},
  {Sharon}, {Dahle}, \& {Oguri}}]{bayliss2011a}
{Bayliss}, M.~B., {Hennawi}, J.~F., {Gladders}, M.~D., {et~al.} 2011, \apjs,
  193, 8, \dodoi{10.1088/0067-0049/193/1/8}

\bibitem[{{Belokurov} {et~al.}(2009){Belokurov}, {Evans}, {Hewett}, {Moiseev},
  {McMahon}, {Sanchez}, \& {King}}]{belokurov2009a}
{Belokurov}, V., {Evans}, N.~W., {Hewett}, P.~C., {et~al.} 2009, \mnras, 392,
  104, \dodoi{10.1111/j.1365-2966.2008.14075.x}

\bibitem[{{Belokurov} {et~al.}(2007){Belokurov}, {Evans}, {Moiseev}, {King},
  {Hewett}, {Pettini}, {Wyrzykowski}, {McMahon}, {Smith}, {Gilmore}, {Sanchez},
  {Udalski}, {Koposov}, {Zucker}, \& {Walcher}}]{belokurov2007a}
{Belokurov}, V., {Evans}, N.~W., {Moiseev}, A., {et~al.} 2007, \apj, 671, L9,
  \dodoi{10.1086/524948}

\bibitem[{Birrer {et~al.}(2020)Birrer, Shajib, Galan, Millon, Treu, Agnello,
  Auger, Chen, Christensen, Collett, Courbin, Fassnacht, Koopmans, Marshall,
  Park, Rusu, Sluse, Spiniello, Suyu, Wagner-Carena, Wong, Barnabè, Bolton,
  Czoske, Ding, Frieman, \& Van~de Vyvere}]{birrer2020a}
Birrer, S., Shajib, A.~J., Galan, A., {et~al.} 2020, Astronomy \&amp;
  Astrophysics, 643, A165, \dodoi{10.1051/0004-6361/202038861}

\bibitem[{{Bolton} {et~al.}(2008){Bolton}, {Burles}, {Koopmans}, {Treu},
  {Gavazzi}, {Moustakas}, {Wayth}, \& {Schlegel}}]{bolton2008a}
{Bolton}, A.~S., {Burles}, S., {Koopmans}, L. V.~E., {et~al.} 2008, \apj, 682,
  964, \dodoi{10.1086/589327}

\bibitem[{{Bolton} {et~al.}(2006){Bolton}, {Burles}, {Koopmans}, {Treu}, \&
  {Moustakas}}]{bolton2006a}
{Bolton}, A.~S., {Burles}, S., {Koopmans}, L.~V.~E., {Treu}, T., \&
  {Moustakas}, L.~A. 2006, \apj, 638, 703, \dodoi{10.1086/498884}

\bibitem[{{Bolton} {et~al.}(2012){Bolton}, {Schlegel}, {Aubourg}, {Bailey},
  {Bhardwaj}, {Brownstein}, {Burles}, {Chen}, {Dawson}, {Eisenstein}, {Gunn},
  {Knapp}, {Loomis}, {Lupton}, {Maraston}, {Muna}, {Myers}, {Olmstead},
  {Padmanabhan}, {P{\^a}ris}, {Percival}, {Petitjean}, {Rockosi}, {Ross},
  {Schneider}, {Shu}, {Strauss}, {Thomas}, {Tremonti}, {Wake}, {Weaver}, \&
  {Wood-Vasey}}]{bolton2012a}
{Bolton}, A.~S., {Schlegel}, D.~J., {Aubourg}, {\'E}., {et~al.} 2012, \aj, 144,
  144, \dodoi{10.1088/0004-6256/144/5/144}

\bibitem[{Bradač {et~al.}(2008)Bradač, Allen, Treu, Ebeling, Massey, Morris,
  von~der Linden, \& Applegate}]{bradac2008a}
Bradač, M., Allen, S.~W., Treu, T., {et~al.} 2008, \apj, 687, 959–967,
  \dodoi{10.1086/591246}

\bibitem[{{Brownstein} {et~al.}(2012){Brownstein}, {Bolton}, {Schlegel},
  {Eisenstein}, {Kochanek}, {Connolly}, {Maraston}, {Pandey}, {Seitz}, {Wake},
  {Wood-Vasey}, {Brinkmann}, {Schneider}, \& {Weaver}}]{brownstein2012a}
{Brownstein}, J.~R., {Bolton}, A.~S., {Schlegel}, D.~J., {et~al.} 2012, \apj,
  744, 41, \dodoi{10.1088/0004-637X/744/1/41}

\bibitem[{{Ca{\~n}ameras} {et~al.}(2021){Ca{\~n}ameras}, {Schuldt}, {Shu},
  {Suyu}, {Taubenberger}, {Meinhardt}, {Leal-Taix{\'e}}, {Chao}, {Inoue},
  {Jaelani}, \& {More}}]{canameras2021a}
{Ca{\~n}ameras}, R., {Schuldt}, S., {Shu}, Y., {et~al.} 2021, \aap, 653, L6,
  \dodoi{10.1051/0004-6361/202141758}

\bibitem[{{Cañameras, R.} {et~al.}(2020){Cañameras, R.}, {Schuldt, S.},
  {Suyu, S. H.}, {Taubenberger, S.}, {Meinhardt, T.}, {Leal-Taixé, L.},
  {Lemon, C.}, {Rojas, K.}, \& {Savary, E.}}]{canameras2020a}
{Cañameras, R.}, {Schuldt, S.}, {Suyu, S. H.}, {et~al.} 2020, A\&A, 644, A163,
  \dodoi{10.1051/0004-6361/202038219}

\bibitem[{{{\c{C}}a{\v{g}}an {\c{S}}eng{\"u}l} {et~al.}(2022){{\c{C}}a{\v{g}}an
  {\c{S}}eng{\"u}l}, {Dvorkin}, {Ostdiek}, \& {Tsang}}]{cagan-sengul2022a}
{{\c{C}}a{\v{g}}an {\c{S}}eng{\"u}l}, A.~{\c{C}}., {Dvorkin}, C., {Ostdiek},
  B., \& {Tsang}, A. 2022, \mnras, 515, 4391, \dodoi{10.1093/mnras/stac1967}

\bibitem[{{Chan} {et~al.}(2020){Chan}, {Suyu}, {Sonnenfeld}, {Jaelani}, {More},
  {Yonehara}, {Kubota}, {Coupon}, {Lee}, {Oguri}, {Rusu}, \&
  {Wong}}]{chan2020a}
{Chan}, J. H.~H., {Suyu}, S.~H., {Sonnenfeld}, A., {et~al.} 2020, \aap, 636,
  A87, \dodoi{10.1051/0004-6361/201937030}

\bibitem[{{Chen} {et~al.}(2019){Chen}, {Li}, {Shu}, \& {Cao}}]{chen2019a}
{Chen}, Y., {Li}, R., {Shu}, Y., \& {Cao}, X. 2019, \mnras, 488, 3745,
  \dodoi{10.1093/mnras/stz1902}

\bibitem[{{Christensen} {et~al.}(2012){Christensen}, {Richard}, {Hjorth},
  {Milvang-Jensen}, {Laursen}, {Limousin}, {Dessauges-Zavadsky}, {Grillo}, \&
  {Ebeling}}]{christensen2012a}
{Christensen}, L., {Richard}, J., {Hjorth}, J., {et~al.} 2012, \mnras, 427,
  1953, \dodoi{10.1111/j.1365-2966.2012.22006.x}

\bibitem[{{Collett}(2015)}]{collett2015a}
{Collett}, T.~E. 2015, \apj, 811, 20, \dodoi{10.1088/0004-637X/811/1/20}

\bibitem[{Dawes {et~al.}(2023)Dawes, Storfer, Huang, Aldering, Cikota, Dey, \&
  Schlegel}]{dawes2023a}
Dawes, C., Storfer, C., Huang, X., {et~al.} 2023, The Astrophysical Journal
  Supplement Series, 269, 61, \dodoi{10.3847/1538-4365/ad015a}

\bibitem[{{DESI Collaboration} {et~al.}(2016{\natexlab{a}}){DESI
  Collaboration}, {Aghamousa}, {Aguilar}, {Ahlen}, {Alam}, {Allen}, {Allende
  Prieto}, {Annis}, {Bailey}, {Balland}, {Ballester}, {Baltay}, {Beaufore},
  {Bebek}, {Beers}, {Bell}, {Bernal}, {Besuner}, {Beutler}, {Blake}, {Bleuler},
  {Blomqvist}, {Blum}, {Bolton}, {Briceno}, {Brooks}, {Brownstein},
  {Buckley-Geer}, {Burden}, {Burtin}, {Busca}, {Cahn}, {Cai}, {Cardiel-Sas},
  {Carlberg}, {Carton}, {Casas}, {Castander}, {Cervantes-Cota}, {Claybaugh},
  {Close}, {Coker}, {Cole}, {Comparat}, {Cooper}, {Cousinou}, {Crocce}, {Cuby},
  {Cunningham}, {Davis}, {Dawson}, {de la Macorra}, {De Vicente}, {Delubac},
  {Derwent}, {Dey}, {Dhungana}, {Ding}, {Doel}, {Duan}, {Ealet}, {Edelstein},
  {Eftekharzadeh}, {Eisenstein}, {Elliott}, {Escoffier}, {Evatt}, {Fagrelius},
  {Fan}, {Fanning}, {Farahi}, {Farihi}, {Favole}, {Feng}, {Fernandez},
  {Findlay}, {Finkbeiner}, {Fitzpatrick}, {Flaugher}, {Flender}, {Font-Ribera},
  {Forero-Romero}, {Fosalba}, {Frenk}, {Fumagalli}, {Gaensicke}, {Gallo},
  {Garcia-Bellido}, {Gaztanaga}, {Pietro Gentile Fusillo}, {Gerard},
  {Gershkovich}, {Giannantonio}, {Gillet}, {Gonzalez-de-Rivera},
  {Gonzalez-Perez}, {Gott}, {Graur}, {Gutierrez}, {Guy}, {Habib}, {Heetderks},
  {Heetderks}, {Heitmann}, {Hellwing}, {Herrera}, {Ho}, {Holland}, {Honscheid},
  {Huff}, {Hutchinson}, {Huterer}, {Hwang}, {Illa Laguna}, {Ishikawa},
  {Jacobs}, {Jeffrey}, {Jelinsky}, {Jennings}, {Jiang}, {Jimenez}, {Johnson},
  {Joyce}, {Jullo}, {Juneau}, {Kama}, {Karcher}, {Karkar}, {Kehoe}, {Kennamer},
  {Kent}, {Kilbinger}, {Kim}, {Kirkby}, {Kisner}, {Kitanidis}, {Kneib},
  {Koposov}, {Kovacs}, {Koyama}, {Kremin}, {Kron}, {Kronig}, {Kueter-Young},
  {Lacey}, {Lafever}, {Lahav}, {Lambert}, {Lampton}, {Landriau}, {Lang},
  {Lauer}, {Le Goff}, {Le Guillou}, {Le Van Suu}, {Lee}, {Lee}, {Leitner},
  {Lesser}, {Levi}, {L'Huillier}, {Li}, {Liang}, {Lin}, {Linder}, {Loebman},
  {Luki{\'c}}, {Ma}, {MacCrann}, {Magneville}, {Makarem}, {Manera}, {Manser},
  {Marshall}, {Martini}, {Massey}, {Matheson}, {McCauley}, {McDonald},
  {McGreer}, {Meisner}, {Metcalfe}, {Miller}, {Miquel}, {Moustakas}, {Myers},
  {Naik}, {Newman}, {Nichol}, {Nicola}, {Nicolati da Costa}, {Nie}, {Niz},
  {Norberg}, {Nord}, {Norman}, {Nugent}, {O'Brien}, {Oh}, {Olsen}, {Padilla},
  {Padmanabhan}, {Padmanabhan}, {Palanque-Delabrouille}, {Palmese},
  {Pappalardo}, {P{\^a}ris}, {Park}, {Patej}, {Peacock}, {Peiris}, {Peng},
  {Percival}, {Perruchot}, {Pieri}, {Pogge}, {Pollack}, {Poppett}, {Prada},
  {Prakash}, {Probst}, {Rabinowitz}, {Raichoor}, {Ree}, {Refregier}, {Regal},
  {Reid}, {Reil}, {Rezaie}, {Rockosi}, {Roe}, {Ronayette}, {Roodman}, {Ross},
  {Ross}, {Rossi}, {Rozo}, {Ruhlmann-Kleider}, {Rykoff}, {Sabiu}, {Samushia},
  {Sanchez}, {Sanchez}, {Schlegel}, {Schneider}, {Schubnell}, {Secroun},
  {Seljak}, {Seo}, {Serrano}, {Shafieloo}, {Shan}, {Sharples}, {Sholl},
  {Shourt}, {Silber}, {Silva}, {Sirk}, {Slosar}, {Smith}, {Smoot}, {Som},
  {Song}, {Sprayberry}, {Staten}, {Stefanik}, {Tarle}, {Sien Tie}, {Tinker},
  {Tojeiro}, {Valdes}, {Valenzuela}, {Valluri}, {Vargas-Magana}, {Verde},
  {Walker}, {Wang}, {Wang}, {Weaver}, {Weaverdyck}, {Wechsler}, {Weinberg},
  {White}, {Yang}, {Yeche}, {Zhang}, {Zhao}, {Zheng}, {Zhou}, {Zhou}, {Zhu},
  {Zou}, \& {Zu}}]{desi2016a}
{DESI Collaboration}, {Aghamousa}, A., {Aguilar}, J., {et~al.}
  2016{\natexlab{a}}, arXiv e-prints, arXiv:1611.00036,
  \dodoi{10.48550/arXiv.1611.00036}

\bibitem[{{DESI Collaboration} {et~al.}(2016{\natexlab{b}}){DESI
  Collaboration}, {Aghamousa}, {Aguilar}, {Ahlen}, {Alam}, {Allen}, {Allende
  Prieto}, {Annis}, {Bailey}, {Balland}, {Ballester}, {Baltay}, {Beaufore},
  {Bebek}, {Beers}, {Bell}, {Bernal}, {Besuner}, {Beutler}, {Blake}, {Bleuler},
  {Blomqvist}, {Blum}, {Bolton}, {Briceno}, {Brooks}, {Brownstein},
  {Buckley-Geer}, {Burden}, {Burtin}, {Busca}, {Cahn}, {Cai}, {Cardiel-Sas},
  {Carlberg}, {Carton}, {Casas}, {Castander}, {Cervantes-Cota}, {Claybaugh},
  {Close}, {Coker}, {Cole}, {Comparat}, {Cooper}, {Cousinou}, {Crocce}, {Cuby},
  {Cunningham}, {Davis}, {Dawson}, {de la Macorra}, {De Vicente}, {Delubac},
  {Derwent}, {Dey}, {Dhungana}, {Ding}, {Doel}, {Duan}, {Ealet}, {Edelstein},
  {Eftekharzadeh}, {Eisenstein}, {Elliott}, {Escoffier}, {Evatt}, {Fagrelius},
  {Fan}, {Fanning}, {Farahi}, {Farihi}, {Favole}, {Feng}, {Fernandez},
  {Findlay}, {Finkbeiner}, {Fitzpatrick}, {Flaugher}, {Flender}, {Font-Ribera},
  {Forero-Romero}, {Fosalba}, {Frenk}, {Fumagalli}, {Gaensicke}, {Gallo},
  {Garcia-Bellido}, {Gaztanaga}, {Pietro Gentile Fusillo}, {Gerard},
  {Gershkovich}, {Giannantonio}, {Gillet}, {Gonzalez-de-Rivera},
  {Gonzalez-Perez}, {Gott}, {Graur}, {Gutierrez}, {Guy}, {Habib}, {Heetderks},
  {Heetderks}, {Heitmann}, {Hellwing}, {Herrera}, {Ho}, {Holland}, {Honscheid},
  {Huff}, {Hutchinson}, {Huterer}, {Hwang}, {Illa Laguna}, {Ishikawa},
  {Jacobs}, {Jeffrey}, {Jelinsky}, {Jennings}, {Jiang}, {Jimenez}, {Johnson},
  {Joyce}, {Jullo}, {Juneau}, {Kama}, {Karcher}, {Karkar}, {Kehoe}, {Kennamer},
  {Kent}, {Kilbinger}, {Kim}, {Kirkby}, {Kisner}, {Kitanidis}, {Kneib},
  {Koposov}, {Kovacs}, {Koyama}, {Kremin}, {Kron}, {Kronig}, {Kueter-Young},
  {Lacey}, {Lafever}, {Lahav}, {Lambert}, {Lampton}, {Landriau}, {Lang},
  {Lauer}, {Le Goff}, {Le Guillou}, {Le Van Suu}, {Lee}, {Lee}, {Leitner},
  {Lesser}, {Levi}, {L'Huillier}, {Li}, {Liang}, {Lin}, {Linder}, {Loebman},
  {Luki{\'c}}, {Ma}, {MacCrann}, {Magneville}, {Makarem}, {Manera}, {Manser},
  {Marshall}, {Martini}, {Massey}, {Matheson}, {McCauley}, {McDonald},
  {McGreer}, {Meisner}, {Metcalfe}, {Miller}, {Miquel}, {Moustakas}, {Myers},
  {Naik}, {Newman}, {Nichol}, {Nicola}, {Nicolati da Costa}, {Nie}, {Niz},
  {Norberg}, {Nord}, {Norman}, {Nugent}, {O'Brien}, {Oh}, {Olsen}, {Padilla},
  {Padmanabhan}, {Padmanabhan}, {Palanque-Delabrouille}, {Palmese},
  {Pappalardo}, {P{\^a}ris}, {Park}, {Patej}, {Peacock}, {Peiris}, {Peng},
  {Percival}, {Perruchot}, {Pieri}, {Pogge}, {Pollack}, {Poppett}, {Prada},
  {Prakash}, {Probst}, {Rabinowitz}, {Raichoor}, {Ree}, {Refregier}, {Regal},
  {Reid}, {Reil}, {Rezaie}, {Rockosi}, {Roe}, {Ronayette}, {Roodman}, {Ross},
  {Ross}, {Rossi}, {Rozo}, {Ruhlmann-Kleider}, {Rykoff}, {Sabiu}, {Samushia},
  {Sanchez}, {Sanchez}, {Schlegel}, {Schneider}, {Schubnell}, {Secroun},
  {Seljak}, {Seo}, {Serrano}, {Shafieloo}, {Shan}, {Sharples}, {Sholl},
  {Shourt}, {Silber}, {Silva}, {Sirk}, {Slosar}, {Smith}, {Smoot}, {Som},
  {Song}, {Sprayberry}, {Staten}, {Stefanik}, {Tarle}, {Sien Tie}, {Tinker},
  {Tojeiro}, {Valdes}, {Valenzuela}, {Valluri}, {Vargas-Magana}, {Verde},
  {Walker}, {Wang}, {Wang}, {Weaver}, {Weaverdyck}, {Wechsler}, {Weinberg},
  {White}, {Yang}, {Yeche}, {Zhang}, {Zhao}, {Zheng}, {Zhou}, {Zhou}, {Zhu},
  {Zou}, \& {Zu}}]{desi2016b}
---. 2016{\natexlab{b}}, arXiv e-prints, arXiv:1611.00037,
  \dodoi{10.48550/arXiv.1611.00037}

\bibitem[{{DESI Collaboration} {et~al.}(2022){DESI Collaboration}, {Abareshi},
  {Aguilar}, {Ahlen}, {Alam}, {Alexander}, {Alfarsy}, {Allen}, {Allende
  Prieto}, {Alves}, {Ameel}, {Armengaud}, {Asorey}, {Aviles}, {Bailey},
  {Balaguera-Antol{\'\i}nez}, {Ballester}, {Baltay}, {Bault}, {Beltran},
  {Benavides}, {BenZvi}, {Berti}, {Besuner}, {Beutler}, {Bianchi}, {Blake},
  {Blanc}, {Blum}, {Bolton}, {Bose}, {Bramall}, {Brieden}, {Brodzeller},
  {Brooks}, {Brownewell}, {Buckley-Geer}, {Cahn}, {Cai}, {Canning}, {Capasso},
  {Carnero Rosell}, {Carton}, {Casas}, {Castander}, {Cervantes-Cota},
  {Chabanier}, {Chaussidon}, {Chuang}, {Circosta}, {Cole}, {Cooper}, {da
  Costa}, {Cousinou}, {Cuceu}, {Davis}, {Dawson}, {de la Cruz-Noriega}, {de la
  Macorra}, {de Mattia}, {Della Costa}, {Demmer}, {Derwent}, {Dey}, {Dey},
  {Dhungana}, {Ding}, {Dobson}, {Doel}, {Donald-McCann}, {Donaldson},
  {Douglass}, {Duan}, {Dunlop}, {Edelstein}, {Eftekharzadeh}, {Eisenstein},
  {Enriquez-Vargas}, {Escoffier}, {Evatt}, {Fagrelius}, {Fan}, {Fanning},
  {Fawcett}, {Ferraro}, {Ereza}, {Flaugher}, {Font-Ribera}, {Forero-Romero},
  {Frenk}, {Fromenteau}, {G{\"a}nsicke}, {Garcia-Quintero}, {Garrison},
  {Gazta{\~n}aga}, {Gerardi}, {Gil-Mar{\'\i}n}, {Gontcho a Gontcho},
  {Gonzalez-Morales}, {Gonzalez-de-Rivera}, {Gonzalez-Perez}, {Gordon},
  {Graur}, {Green}, {Grove}, {Gruen}, {Gutierrez}, {Guy}, {Hahn}, {Harris},
  {Herrera}, {Herrera-Alcantar}, {Honscheid}, {Howlett}, {Huterer},
  {Ir{\v{s}}i{\v{c}}}, {Ishak}, {Jelinsky}, {Jiang}, {Jimenez}, {Jing},
  {Joyce}, {Jullo}, {Juneau}, {Kara{\c{c}}ayl{\i}}, {Karamanis}, {Karcher},
  {Karim}, {Kehoe}, {Kent}, {Kirkby}, {Kisner}, {Kitaura}, {Koposov},
  {Kov{\'a}cs}, {Kremin}, {Krolewski}, {L'Huillier}, {Lahav}, {Lambert},
  {Lamman}, {Lan}, {Landriau}, {Lane}, {Lang}, {Lange}, {Lasker}, {Le Guillou},
  {Leauthaud}, {Le Van Suu}, {Levi}, {Li}, {Magneville}, {Manera}, {Manser},
  {Marshall}, {Martini}, {McCollam}, {McDonald}, {Meisner},
  {Mena-Fern{\'a}ndez}, {Meneses-Rizo}, {Mezcua}, {Miller}, {Miquel},
  {Montero-Camacho}, {Moon}, {Moustakas}, {Mueller}, {Mu{\~n}oz-Guti{\'e}rrez},
  {Myers}, {Nadathur}, {Najita}, {Napolitano}, {Neilsen}, {Newman}, {Nie},
  {Ning}, {Niz}, {Norberg}, {Noriega}, {O'Brien}, {Obuljen},
  {Palanque-Delabrouille}, {Palmese}, {Zhiwei}, {Pappalardo}, {PENG},
  {Percival}, {Perruchot}, {Pogge}, {Poppett}, {Porredon}, {Prada},
  {Prochaska}, {Pucha}, {P{\'e}rez-Fern{\'a}ndez}, {P{\'e}rez-R{\`a}fols},
  {Rabinowitz}, {Raichoor}, {Ramirez-Solano}, {Ram{\'\i}rez-P{\'e}rez},
  {Ravoux}, {Reil}, {Rezaie}, {Rocher}, {Rockosi}, {Roe}, {Roodman}, {Ross},
  {Rossi}, {Ruggeri}, {Ruhlmann-Kleider}, {Sabiu}, {Safonova}, {Said},
  {Saintonge}, {Salas Catonga}, {Samushia}, {Sanchez}, {Saulder}, {Schaan},
  {Schlafly}, {Schlegel}, {Schmoll}, {Scholte}, {Schubnell}, {Secroun}, {Seo},
  {Serrano}, {Sharples}, {Sholl}, {Silber}, {Silva}, {Sirk}, {Siudek}, {Smith},
  {Sprayberry}, {Staten}, {Stupak}, {Tan}, {Tarl{\'e}}, {Tie}, {Tojeiro},
  {Ure{\~n}a-L{\'o}pez}, {Valdes}, {Valenzuela}, {Valluri},
  {Vargas-Maga{\~n}a}, {Verde}, {Walther}, {Wang}, {Wang}, {Weaver},
  {Weaverdyck}, {Wechsler}, {Wilson}, {Yang}, {Yu}, {Yuan}, {Y{\`e}che},
  {Zhang}, {Zhang}, {Zhao}, {Zhou}, {Zhou}, {Zou}, {Zou}, {Zou}, {Zu}, \& {DESI
  Collaboration}}]{Abareshi2022a}
{DESI Collaboration}, {Abareshi}, B., {Aguilar}, J., {et~al.} 2022, \aj, 164,
  207, \dodoi{10.3847/1538-3881/ac882b}

\bibitem[{{Dessauges-Zavadsky} {et~al.}(2017){Dessauges-Zavadsky}, {Zamojski},
  {Rujopakarn}, {Richard}, {Sklias}, {Schaerer}, {Combes}, {Ebeling}, {Rawle},
  {Egami}, {Boone}, {Cl{\'e}ment}, {Kneib}, {Nyland}, \&
  {Walth}}]{dessauges-zavadsky2017a}
{Dessauges-Zavadsky}, M., {Zamojski}, M., {Rujopakarn}, W., {et~al.} 2017,
  \aap, 605, A81, \dodoi{10.1051/0004-6361/201628513}

\bibitem[{{Dey} {et~al.}(2019){Dey}, {Schlegel}, {Lang}, {Blum}, {Burleigh},
  {Fan}, {Findlay}, {Finkbeiner}, {Herrera}, {Juneau}, {Landriau}, {Levi},
  {McGreer}, {Meisner}, {Myers}, {Moustakas}, {Nugent}, {Patej}, {Schlafly},
  {Walker}, {Valdes}, {Weaver}, {Y{\`e}che}, {Zou}, {Zhou}, {Abareshi},
  {Abbott}, {Abolfathi}, {Aguilera}, {Alam}, {Allen}, {Alvarez}, {Annis},
  {Ansarinejad}, {Aubert}, {Beechert}, {Bell}, {BenZvi}, {Beutler}, {Bielby},
  {Bolton}, {Brice{\~n}o}, {Buckley-Geer}, {Butler}, {Calamida}, {Carlberg},
  {Carter}, {Casas}, {Castander}, {Choi}, {Comparat}, {Cukanovaite}, {Delubac},
  {DeVries}, {Dey}, {Dhungana}, {Dickinson}, {Ding}, {Donaldson}, {Duan},
  {Duckworth}, {Eftekharzadeh}, {Eisenstein}, {Etourneau}, {Fagrelius},
  {Farihi}, {Fitzpatrick}, {Font-Ribera}, {Fulmer}, {G{\"a}nsicke},
  {Gaztanaga}, {George}, {Gerdes}, {Gontcho}, {Gorgoni}, {Green}, {Guy},
  {Harmer}, {Hernand ez}, {Honscheid}, {(Wendy Huang}, {James}, {Jannuzi},
  {Jiang}, {Joyce}, {Karcher}, {Karkar}, {Kehoe}, {Jean-Paul}, {Kneib},
  {Kueter-Young}, {Lan}, {Lauer}, {Le Guillou}, {Le Van Suu}, {Lee}, {Lesser},
  {Perreault Levasseur}, {Li}, {Mann}, {Marshall}, {Mart{\'\i}nez-V{\'a}zquez},
  {Martini}, {du Mas des Bourboux}, {McManus}, {Meier}, {M{\'e}nard},
  {Metcalfe}, {Mu{\~n}oz-Guti{\'e}rrez}, {Najita}, {Napier}, {Narayan},
  {Newman}, {Nie}, {Nord}, {Norman}, {Olsen}, {Paat}, {Palanque-Delabrouille},
  {Peng}, {Poppett}, {Poremba}, {Prakash}, {Rabinowitz}, {Raichoor}, {Rezaie},
  {Robertson}, {Roe}, {Ross}, {Ross}, {Rudnick}, {Safonova}, {Saha},
  {S{\'a}nchez}, {Savary}, {Schweiker}, {Scott}, {Seo}, {Shan}, {Silva},
  {Slepian}, {Soto}, {Sprayberry}, {Staten}, {Stillman}, {Stupak}, {Summers},
  {Sien Tie}, {Tirado}, {Vargas-Maga{\~n}a}, {Vivas}, {Wechsler}, {Williams},
  {Yang}, {Yang}, {Yapici}, {Zaritsky}, {Zenteno}, {Zhang}, {Zhang}, {Zhou}, \&
  {Zhou}}]{dey2019a}
{Dey}, A., {Schlegel}, D.~J., {Lang}, D., {et~al.} 2019, \aj, 157, 168,
  \dodoi{10.3847/1538-3881/ab089d}

\bibitem[{{Diehl} {et~al.}(2009){Diehl}, {Allam}, {Annis}, {Buckley-Geer},
  {Frieman}, {Kubik}, {Kubo}, {Lin}, {Tucker}, \& {West}}]{diehl2009a}
{Diehl}, H.~T., {Allam}, S.~S., {Annis}, J., {et~al.} 2009, \apj, 707, 686,
  \dodoi{10.1088/0004-637X/707/1/686}

\bibitem[{{Diehl} {et~al.}(2017){Diehl}, {Buckley-Geer}, {Lindgren}, {Nord},
  {Gaitsch}, {Gaitsch}, {Lin}, {Allam}, {Collett}, {Furlanetto}, {Gill},
  {More}, {Nightingale}, {Odden}, {Pellico}, {Tucker}, {da Costa}, {Fausti
  Neto}, {Kuropatkin}, {Soares-Santos}, {Welch}, {Zhang}, {Frieman}, {Abdalla},
  {Annis}, {Benoit-L{\'e}vy}, {Bertin}, {Brooks}, {Burke}, {Carnero Rosell},
  {Carrasco Kind}, {Carretero}, {Cunha}, {D'Andrea}, {Desai}, {Dietrich},
  {Drlica-Wagner}, {Evrard}, {Finley}, {Flaugher}, {Garc{\'{\i}}a-Bellido},
  {Gerdes}, {Goldstein}, {Gruen}, {Gruendl}, {Gschwend}, {Gutierrez}, {James},
  {Kuehn}, {Kuhlmann}, {Lahav}, {Li}, {Lima}, {Maia}, {Marshall}, {Menanteau},
  {Miquel}, {Nichol}, {Nugent}, {Ogando}, {Plazas}, {Reil}, {Romer}, {Sako},
  {Sanchez}, {Santiago}, {Scarpine}, {Schindler}, {Schubnell},
  {Sevilla-Noarbe}, {Sheldon}, {Smith}, {Sobreira}, {Suchyta}, {Swanson},
  {Tarle}, {Thomas}, {Walker}, \& {DES Collaboration}}]{diehl2017a}
{Diehl}, H.~T., {Buckley-Geer}, E.~J., {Lindgren}, K.~A., {et~al.} 2017, \apjs,
  232, 15, \dodoi{10.3847/1538-4365/aa8667}

\bibitem[{Einstein(1936)}]{einstein1936a}
Einstein, A. 1936, Science, 84, 506, \dodoi{10.1126/science.84.2188.506}

\bibitem[{{Filipp} {et~al.}(2023){Filipp}, {Shu}, {Pakmor}, {Suyu}, \&
  {Huang}}]{filipp2023a}
{Filipp}, A., {Shu}, Y., {Pakmor}, R., {Suyu}, S.~H., \& {Huang}, X. 2023,
  \aap, 677, A113, \dodoi{10.1051/0004-6361/202346594}

\bibitem[{{Flaugher} {et~al.}(2015){Flaugher}, {Diehl}, {Honscheid}, {Abbott},
  {Alvarez}, {Angstadt}, {Annis}, {Antonik}, {Ballester}, {Beaufore},
  {Bernstein}, {Bernstein}, {Bigelow}, {Bonati}, {Boprie}, {Brooks},
  {Buckley-Geer}, {Campa}, {Cardiel-Sas}, {Castander}, {Castilla}, {Cease},
  {Cela-Ruiz}, {Chappa}, {Chi}, {Cooper}, {da Costa}, {Dede}, {Derylo},
  {DePoy}, {de Vicente}, {Doel}, {Drlica-Wagner}, {Eiting}, {Elliott}, {Emes},
  {Estrada}, {Fausti Neto}, {Finley}, {Flores}, {Frieman}, {Gerdes},
  {Gladders}, {Gregory}, {Gutierrez}, {Hao}, {Holland}, {Holm}, {Huffman},
  {Jackson}, {James}, {Jonas}, {Karcher}, {Karliner}, {Kent}, {Kessler},
  {Kozlovsky}, {Kron}, {Kubik}, {Kuehn}, {Kuhlmann}, {Kuk}, {Lahav}, {Lathrop},
  {Lee}, {Levi}, {Lewis}, {Li}, {Mandrichenko}, {Marshall}, {Martinez},
  {Merritt}, {Miquel}, {Mu{\~n}oz}, {Neilsen}, {Nichol}, {Nord}, {Ogando},
  {Olsen}, {Palaio}, {Patton}, {Peoples}, {Plazas}, {Rauch}, {Reil}, {Rheault},
  {Roe}, {Rogers}, {Roodman}, {Sanchez}, {Scarpine}, {Schindler}, {Schmidt},
  {Schmitt}, {Schubnell}, {Schultz}, {Schurter}, {Scott}, {Serrano}, {Shaw},
  {Smith}, {Soares-Santos}, {Stefanik}, {Stuermer}, {Suchyta}, {Sypniewski},
  {Tarle}, {Thaler}, {Tighe}, {Tran}, {Tucker}, {Walker}, {Wang}, {Watson},
  {Weaverdyck}, {Wester}, {Woods}, {Yanny}, \& {DES
  Collaboration}}]{flaugher2015a}
{Flaugher}, B., {Diehl}, H.~T., {Honscheid}, K., {et~al.} 2015, \aj, 150, 150,
  \dodoi{10.1088/0004-6256/150/5/150}

\bibitem[{{Freedman} {et~al.}(2020){Freedman}, {Madore}, {Hoyt}, {Jang},
  {Beaton}, {Lee}, {Monson}, {Neeley}, \& {Rich}}]{freedman2020a}
{Freedman}, W.~L., {Madore}, B.~F., {Hoyt}, T., {et~al.} 2020, \apj, 891, 57,
  \dodoi{10.3847/1538-4357/ab7339}

\bibitem[{{Gavazzi} {et~al.}(2014){Gavazzi}, {Marshall}, {Treu}, \&
  {Sonnenfeld}}]{gavazzi2014a}
{Gavazzi}, R., {Marshall}, P.~J., {Treu}, T., \& {Sonnenfeld}, A. 2014, \apj,
  785, 144, \dodoi{10.1088/0004-637X/785/2/144}

\bibitem[{{Gilbank} {et~al.}(2008){Gilbank}, {Yee}, {Ellingson}, {Hicks},
  {Gladders}, {Barrientos}, \& {Keeney}}]{gilbank2008a}
{Gilbank}, D.~G., {Yee}, H.~K.~C., {Ellingson}, E., {et~al.} 2008, \apjl, 677,
  L89, \dodoi{10.1086/588138}

\bibitem[{Grillo {et~al.}(2015)Grillo, Suyu, Rosati, Mercurio, Balestra,
  Munari, Nonino, Caminha, Lombardi, De~Lucia, \& et~al.}]{grillo2015a}
Grillo, C., Suyu, S.~H., Rosati, P., {et~al.} 2015, \apj, 800, 38,
  \dodoi{10.1088/0004-637x/800/1/38}

\bibitem[{{Grogin} \& {Narayan}(1996)}]{grogin1996a}
{Grogin}, N.~A., \& {Narayan}, R. 1996, \apj, 464, 92, \dodoi{10.1086/177302}

\bibitem[{{Guy} {et~al.}(2023){Guy}, {Bailey}, {Kremin}, {Alam}, {Alexander},
  {Allende Prieto}, {BenZvi}, {Bolton}, {Brooks}, {Chaussidon}, {Cooper},
  {Dawson}, {de la Macorra}, {Dey}, {Dey}, {Dhungana}, {Eisenstein},
  {Font-Ribera}, {Forero-Romero}, {Gazta{\~n}aga}, {Gontcho A Gontcho},
  {Green}, {Honscheid}, {Ishak}, {Kehoe}, {Kirkby}, {Kisner}, {Koposov}, {Lan},
  {Landriau}, {Le Guillou}, {Levi}, {Magneville}, {Manser}, {Martini},
  {Meisner}, {Miquel}, {Moustakas}, {Myers}, {Newman}, {Nie},
  {Palanque-Delabrouille}, {Percival}, {Poppett}, {Prada}, {Raichoor},
  {Ravoux}, {Ross}, {Schlafly}, {Schlegel}, {Schubnell}, {Sharples},
  {Tarl{\'e}}, {Weaver}, {Y{\'e}che}, {Zhou}, {Zhou}, \& {Zou}}]{guy2023a}
{Guy}, J., {Bailey}, S., {Kremin}, A., {et~al.} 2023, \aj, 165, 144,
  \dodoi{10.3847/1538-3881/acb212}

\bibitem[{{Hennawi} {et~al.}(2008){Hennawi}, {Gladders}, {Oguri}, {Dalal},
  {Koester}, {Natarajan}, {Strauss}, {Inada}, {Kayo}, {Lin}, {Lampeitl},
  {Annis}, {Bahcall}, \& {Schneider}}]{hennawi2008a}
{Hennawi}, J.~F., {Gladders}, M.~D., {Oguri}, M., {et~al.} 2008, \aj, 135, 664,
  \dodoi{10.1088/0004-6256/135/2/664}

\bibitem[{{Hezaveh} {et~al.}(2016){Hezaveh}, {Dalal}, {Marrone}, {Mao},
  {Morningstar}, {Wen}, {Blandford}, {Carlstrom}, {Fassnacht}, {Holder},
  {Kemball}, {Marshall}, {Murray}, {Perreault Levasseur}, {Vieira}, \&
  {Wechsler}}]{hezaveh2016a}
{Hezaveh}, Y.~D., {Dalal}, N., {Marrone}, D.~P., {et~al.} 2016, \apj, 823, 37,
  \dodoi{10.3847/0004-637X/823/1/37}

\bibitem[{{Huang} {et~al.}(2009){Huang}, {Morokuma}, {Fakhouri}, {Aldering},
  {Amanullah}, {Barbary}, {Brodwin}, {Connolly}, {Dawson}, {Doi}, {Faccioli},
  {Fadeyev}, {Fruchter}, {Goldhaber}, {Gladders}, {Hennawi}, {Ihara}, {Jee},
  {Kowalski}, {Konishi}, {Lidman}, {Meyers}, {Moustakas}, {Perlmutter},
  {Rubin}, {Schlegel}, {Spadafora}, {Suzuki}, {Takanashi}, \&
  {Yasuda}}]{huang2009a}
{Huang}, X., {Morokuma}, T., {Fakhouri}, H.~K., {et~al.} 2009, \apj, 707, L12,
  \dodoi{10.1088/0004-637X/707/1/L12}

\bibitem[{Huang {et~al.}(2020)Huang, Storfer, Ravi, Pilon, Domingo, Schlegel,
  Bailey, Dey, Gupta, Herrera, Juneau, Landriau, Lang, Meisner, Moustakas,
  Myers, Schlafly, Valdes, Weaver, Yang, \& Y{\`{e}}che}]{huang2020a}
Huang, X., Storfer, C., Ravi, V., {et~al.} 2020, The Astrophysical Journal,
  894, 78, \dodoi{10.3847/1538-4357/ab7ffb}

\bibitem[{{Huang} {et~al.}(2021){Huang}, {Storfer}, {Gu}, {Ravi}, {Pilon},
  {Sheu}, {Venguswamy}, {Banka}, {Dey}, {Landriau}, {Lang}, {Meisner},
  {Moustakas}, {Myers}, {Sajith}, {Schlafly}, \& {Schlegel}}]{huang2021a}
{Huang}, X., {Storfer}, C., {Gu}, A., {et~al.} 2021, \apj, 909, 27,
  \dodoi{10.3847/1538-4357/abd62b}

\bibitem[{{Huang} {et~al.}(2025){Huang}, {Baltasar}, {Ratier-Werbin},
  {Storfer}, {Sheu}, {Agarwal}, {Tamargo-Arizmendi}, {Schlegel}, {Aguilar},
  {Ahlen}, {Aldering}, {Banka}, {BenZvi}, {Bianchi}, {Bolton}, {Brooks},
  {Cikota}, {Claybaugh}, {de la Macorra}, {Dey}, {Doel}, {Edelstein}, {Filipp},
  {Forero-Romero}, {Gaztanaga}, {Gontcho}, {Gu}, {Gutierrez}, {Honscheid},
  {Jullo}, {Juneau}, {Kehoe}, {Kirkby}, {Kisner}, {Kremin}, {Kwon}, {Lambert},
  {Landriau}, {Lang}, {Le Guillou}, {Liu}, {Meisner}, {Miquel}, {Moustakas},
  {Myers}, {Perlmutter}, {Perez-Rafols}, {Prada}, {Rossi}, {Rubin}, {Sanchez},
  {Schubnell}, {Shu}, {Silver}, {Sprayberry}, {Suzuki}, {Tarle}, {Weaver}, \&
  {Zou}}]{huang2025a}
{Huang}, X., {Baltasar}, S., {Ratier-Werbin}, N., {et~al.} 2025, arXiv
  e-prints, arXiv:2502.03455, \dodoi{10.48550/arXiv.2502.03455}

\bibitem[{{Jacobs} {et~al.}(2019){Jacobs}, {Collett}, {Glazebrook},
  {Buckley-Geer}, {Diehl}, {Lin}, {McCarthy}, {Qin}, {Odden}, {Caso Escudero},
  {Dial}, {Yung}, {Gaitsch}, {Pellico}, {Lindgren}, {Abbott}, {Annis}, {Avila},
  {Brooks}, {Burke}, {Carnero Rosell}, {Carrasco Kind}, {Carretero}, {da
  Costa}, {De Vicente}, {Fosalba}, {Frieman}, {Garc{\'\i}a-Bellido},
  {Gaztanaga}, {Goldstein}, {Gruen}, {Gruendl}, {Gschwend}, {Hollowood},
  {Honscheid}, {Hoyle}, {James}, {Krause}, {Kuropatkin}, {Lahav}, {Lima},
  {Maia}, {Marshall}, {Miquel}, {Plazas}, {Roodman}, {Sanchez}, {Scarpine},
  {Serrano}, {Sevilla-Noarbe}, {Smith}, {Sobreira}, {Suchyta}, {Swanson},
  {Tarle}, {Vikram}, {Walker}, {Zhang}, \& {DES Collaboration}}]{jacobs2019b}
{Jacobs}, C., {Collett}, T., {Glazebrook}, K., {et~al.} 2019, \apjs, 243, 17,
  \dodoi{10.3847/1538-4365/ab26b6}

\bibitem[{{Jaelani} {et~al.}(2020){Jaelani}, {More}, {Oguri}, {Sonnenfeld},
  {Suyu}, {Rusu}, {Wong}, {Chan}, {Kayo}, {Lee}, {Chao}, {Coupon}, {Inoue}, \&
  {Futamase}}]{jaelani2020a}
{Jaelani}, A.~T., {More}, A., {Oguri}, M., {et~al.} 2020, \mnras, 495, 1291,
  \dodoi{10.1093/mnras/staa1062}

\bibitem[{{Jullo} {et~al.}(2010){Jullo}, {Natarajan}, {Kneib}, {D'Aloisio},
  {Limousin}, {Richard}, \& {Schimd}}]{jullo2010a}
{Jullo}, E., {Natarajan}, P., {Kneib}, J.~P., {et~al.} 2010, Science, 329, 924,
  \dodoi{10.1126/science.1185759}

\bibitem[{{Kelly} {et~al.}(2023){Kelly}, {Rodney}, {Treu}, {Birrer}, {Bonvin},
  {Dessart}, {Foley}, {Filippenko}, {Gilman}, {Jha}, {Hjorth}, {Mandel},
  {Millon}, {Pierel}, {Thorp}, {Zitrin}, {Broadhurst}, {Chen}, {Diego},
  {Dressler}, {Graur}, {Jauzac}, {Malkan}, {McCully}, {Oguri}, {Postman},
  {Schmidt}, {Sharon}, {Tucker}, {von der Linden}, \&
  {Wambsganss}}]{kelly2023a}
{Kelly}, P.~L., {Rodney}, S., {Treu}, T., {et~al.} 2023, \apj, 948, 93,
  \dodoi{10.3847/1538-4357/ac4ccb}

\bibitem[{{Kochanek}(1991)}]{kochanek1991a}
{Kochanek}, C.~S. 1991, \apj, 373, 354, \dodoi{10.1086/170057}

\bibitem[{{Koopmans} \& {Treu}(2002)}]{koopmans2002a}
{Koopmans}, L.~V.~E., \& {Treu}, T. 2002, \apjl, 568, L5,
  \dodoi{10.1086/340143}

\bibitem[{{Koopmans} {et~al.}(2006){Koopmans}, {Treu}, {Bolton}, {Burles}, \&
  {Moustakas}}]{koopmans2006a}
{Koopmans}, L. V.~E., {Treu}, T., {Bolton}, A.~S., {Burles}, S., \&
  {Moustakas}, L.~A. 2006, \apj, 649, 599, \dodoi{10.1086/505696}

\bibitem[{{Kubo} {et~al.}(2010){Kubo}, {Allam}, {Drabek}, {Lin}, {Tucker},
  {Buckley-Geer}, {Diehl}, {Soares-Santos}, {Hao}, {Wiesner}, {West}, {Kubik},
  {Annis}, \& {Frieman}}]{kubo2010a}
{Kubo}, J.~M., {Allam}, S.~S., {Drabek}, E., {et~al.} 2010, \apjl, 724, L137,
  \dodoi{10.1088/2041-8205/724/2/L137}

\bibitem[{{Lanusse} {et~al.}(2018){Lanusse}, {Ma}, {Li}, {Collett}, {Li},
  {Ravanbakhsh}, {Mandelbaum}, \& {P{\'o}czos}}]{lanusse2018a}
{Lanusse}, F., {Ma}, Q., {Li}, N., {et~al.} 2018, \mnras, 473, 3895,
  \dodoi{10.1093/mnras/stx1665}

\bibitem[{{Li} {et~al.}(2020){Li}, {Napolitano}, {Tortora}, {Spiniello},
  {Koopmans}, {Huang}, {Roy}, {Vernardos}, {Chatterjee}, {Giblin}, {Getman},
  {Radovich}, {Covone}, \& {Kuijken}}]{li2020a}
{Li}, R., {Napolitano}, N.~R., {Tortora}, C., {et~al.} 2020, \apj, 899, 30,
  \dodoi{10.3847/1538-4357/ab9dfa}

\bibitem[{{Miller} {et~al.}(2024){Miller}, {Doel}, {Gutierrez}, {Besuner},
  {Brooks}, {Gallo}, {Heetderks}, {Jelinsky}, {Kent}, {Lampton}, {Levi},
  {Liang}, {Meisner}, {Sholl}, {Silber}, {Sprayberry}, {Aguilar}, {de la
  Macorra}, {Eisenstein}, {Fanning}, {Font-Ribera}, {Gazta{\~n}aga}, {Gontcho A
  Gontcho}, {Honscheid}, {Jimenez}, {Joyce}, {Kehoe}, {Kisner}, {Kremin},
  {Landriau}, {Le Guillou}, {Magneville}, {Martini}, {Miquel}, {Moustakas},
  {Nie}, {Percival}, {Poppett}, {Prada}, {Rossi}, {Schlegel}, {Schubnell},
  {Seo}, {Sharples}, {Tarl{\'e}}, {Vargas-Maga{\~n}a}, {Zhou}, \& {the DESI
  Collaboration}}]{miller2023a}
{Miller}, T.~N., {Doel}, P., {Gutierrez}, G., {et~al.} 2024, \aj, 168, 95,
  \dodoi{10.3847/1538-3881/ad45fe}

\bibitem[{{More} {et~al.}(2012){More}, {Cabanac}, {More}, {Alard}, {Limousin},
  {Kneib}, {Gavazzi}, \& {Motta}}]{more2012a}
{More}, A., {Cabanac}, R., {More}, S., {et~al.} 2012, \apj, 749, 38,
  \dodoi{10.1088/0004-637X/749/1/38}

\bibitem[{{More} {et~al.}(2016){More}, {Verma}, {Marshall}, {More}, {Baeten},
  {Wilcox}, {Macmillan}, {Cornen}, {Kapadia}, {Parrish}, {Snyder}, {Davis},
  {Gavazzi}, {Lintott}, {Simpson}, {Miller}, {Smith}, {Paget}, {Saha},
  {K{\"u}ng}, \& {Collett}}]{more2016a}
{More}, A., {Verma}, A., {Marshall}, P.~J., {et~al.} 2016, \mnras, 455, 1191,
  \dodoi{10.1093/mnras/stv1965}

\bibitem[{{Moustakas} {et~al.}(2023){Moustakas}, {Scholte}, {Dey}, \&
  {Khederlarian}}]{moustakas2023a}
{Moustakas}, J., {Scholte}, D., {Dey}, B., \& {Khederlarian}, A. 2023,
  {FastSpecFit: Fast spectral synthesis and emission-line fitting of DESI
  spectra}, Astrophysics Source Code Library, record ascl:2308.005

\bibitem[{{Moustakas} {et~al.}(2012){Moustakas}, {Brownstein}, {Fadely},
  {Fassnacht}, {Gavazzi}, {Goodsall}, {Griffith}, {Keeton}, {Kneib},
  {Koekemoer}, {Koopmans}, {Marshall}, {Merten}, {Metcalf}, {Oguri},
  {Papovich}, {Rein}, {Ryan}, {Stewart}, \& {Treu}}]{moustakas2012a}
{Moustakas}, L.~A., {Brownstein}, J., {Fadely}, R., {et~al.} 2012, in American
  Astronomical Society Meeting Abstracts, Vol. 219, American Astronomical
  Society Meeting Abstracts \#219, 146.01

\bibitem[{{Myers} {et~al.}(2023){Myers}, {Moustakas}, {Bailey}, {Weaver},
  {Cooper}, {Forero-Romero}, {Abolfathi}, {Alexander}, {Brooks}, {Chaussidon},
  {Chuang}, {Dawson}, {Dey}, {Dey}, {Dhungana}, {Doel}, {Fanning},
  {Gazta{\~n}aga}, {Gontcho A Gontcho}, {Gonzalez-Morales}, {Hahn},
  {Herrera-Alcantar}, {Honscheid}, {Ishak}, {Karim}, {Kirkby}, {Kisner},
  {Koposov}, {Kremin}, {Lan}, {Landriau}, {Lang}, {Levi}, {Magneville},
  {Napolitano}, {Martini}, {Meisner}, {Newman}, {Palanque-Delabrouille},
  {Percival}, {Poppett}, {Prada}, {Raichoor}, {Ross}, {Schlafly}, {Schlegel},
  {Schubnell}, {Tan}, {Tarle}, {Wilson}, {Y{\`e}che}, {Zhou}, {Zhou}, \&
  {Zou}}]{myers2023a}
{Myers}, A.~D., {Moustakas}, J., {Bailey}, S., {et~al.} 2023, \aj, 165, 50,
  \dodoi{10.3847/1538-3881/aca5f9}

\bibitem[{{O'Donnell} {et~al.}(2022){O'Donnell}, {Wilkinson}, {Diehl},
  {Aros-Bunster}, {Bechtol}, {Birrer}, {Buckley-Geer}, {Carnero Rosell},
  {Carrasco Kind}, {da Costa}, {Gonzalez Lozano}, {Gruendl}, {Hilton}, {Lin},
  {Lindgren}, {Martin}, {Pieres}, {Rykoff}, {Sevilla-Noarbe}, {Sheldon},
  {Sif{\'o}n}, {Tucker}, {Yanny}, {Abbott}, {Aguena}, {Allam},
  {Andrade-Oliveira}, {Annis}, {Bertin}, {Brooks}, {Burke}, {Carretero},
  {Costanzi}, {De Vicente}, {Desai}, {Dietrich}, {Eckert}, {Everett},
  {Ferrero}, {Flaugher}, {Fosalba}, {Frieman}, {Garc{\'\i}a-Bellido},
  {Gaztanaga}, {Gerdes}, {Gruen}, {Gschwend}, {Gill}, {Gutierrez}, {Hinton},
  {Hollowood}, {Honscheid}, {James}, {Jeltema}, {Kuehn}, {Lahav}, {Lima},
  {Maia}, {Marshall}, {Melchior}, {Menanteau}, {Miquel}, {Morgan}, {Nord},
  {Ogando}, {Paz-Chinch{\'o}n}, {Pereira}, {Plazas Malag{\'o}n},
  {Rodriguez-Monroy}, {Romer}, {Roodman}, {Sanchez}, {Scarpine}, {Schubnell},
  {Serrano}, {Smith}, {Suchyta}, {Swanson}, {Tarle}, {Thomas}, {To}, \&
  {Varga}}]{odonnell2022a}
{O'Donnell}, J.~H., {Wilkinson}, R.~D., {Diehl}, H.~T., {et~al.} 2022, \apjs,
  259, 27, \dodoi{10.3847/1538-4365/ac470b}

\bibitem[{{Oguri} {et~al.}(2012){Oguri}, {Bayliss}, {Dahle}, {Sharon},
  {Gladders}, {Natarajan}, {Hennawi}, \& {Koester}}]{oguri2012a}
{Oguri}, M., {Bayliss}, M.~B., {Dahle}, H., {et~al.} 2012, \mnras, 420, 3213,
  \dodoi{10.1111/j.1365-2966.2011.20248.x}

\bibitem[{{Petrillo} {et~al.}(2019){Petrillo}, {Tortora}, {Vernardos},
  {Koopmans}, {Verdoes Kleijn}, {Bilicki}, {Napolitano}, {Chatterjee},
  {Covone}, {Dvornik}, {Erben}, {Getman}, {Giblin}, {Heymans}, {de Jong},
  {Kuijken}, {Schneider}, {Shan}, {Spiniello}, \& {Wright}}]{petrillo2019a}
{Petrillo}, C.~E., {Tortora}, C., {Vernardos}, G., {et~al.} 2019, \mnras, 484,
  3879, \dodoi{10.1093/mnras/stz189}

\bibitem[{{Pierel} \& {Rodney}(2019)}]{pierel2019a}
{Pierel}, J.~D.~R., \& {Rodney}, S. 2019, \apj, 876, 107,
  \dodoi{10.3847/1538-4357/ab164a}

\bibitem[{{Poppett} {et~al.}(2024){Poppett}, {Tyas}, {Aguilar}, {Bebek},
  {Bramall}, {Claybaugh}, {Edelstein}, {Fagrelius}, {Heetderks}, {Jelinsky},
  {Jelinsky}, {Lafever}, {Lambert}, {Lampton}, {Levi}, {Martini}, {Rockosi},
  {Schmoll}, {Sharples}, {Sirk}, {Wishnow}, {Yu}, {Ahlen}, {Bault}, {BenZvi},
  {Brooks}, {Cole}, {de la Macorra}, {Dey}, {Doel}, {Fanning}, {Font-Ribera},
  {Forero-Romero}, {Gazta{\~n}aga}, {Gontcho A Gontcho}, {Gonzalez-Morales},
  {Hahn}, {Honscheid}, {Jimenez}, {Juneau}, {Kirkby}, {Kremin}, {Landriau}, {Le
  Guillou}, {Manera}, {Meisner}, {Miquel}, {Moustakas}, {Mueller},
  {Mu{\~n}oz-Guti{\'e}rrez}, {Myers}, {Nie}, {Niz}, {Palanque-Delabrouille},
  {Percival}, {Prada}, {Rabinowitz}, {Rezaie}, {Rossi}, {Sanchez}, {Schlafly},
  {Schlegel}, {Schubnell}, {Seo}, {Sprayberry}, {Tarl{\'e}},
  {Vargas-Maga{\~n}a}, {Weaver}, \& {Zhou}}]{poppett2024a}
{Poppett}, C., {Tyas}, L., {Aguilar}, J., {et~al.} 2024, \aj, 168, 245,
  \dodoi{10.3847/1538-3881/ad76a4}

\bibitem[{{Riess} {et~al.}(2022){Riess}, {Yuan}, {Macri}, {Scolnic}, {Brout},
  {Casertano}, {Jones}, {Murakami}, {Anand}, {Breuval}, {Brink}, {Filippenko},
  {Hoffmann}, {Jha}, {D'arcy Kenworthy}, {Mackenty}, {Stahl}, \&
  {Zheng}}]{riess2022a}
{Riess}, A.~G., {Yuan}, W., {Macri}, L.~M., {et~al.} 2022, \apjl, 934, L7,
  \dodoi{10.3847/2041-8213/ac5c5b}

\bibitem[{{Rigby} {et~al.}(2018){Rigby}, {Bayliss}, {Sharon}, {Gladders},
  {Chisholm}, {Dahle}, {Johnson}, {Paterno-Mahler}, {Wuyts}, \&
  {Kelson}}]{rigby2018a}
{Rigby}, J.~R., {Bayliss}, M.~B., {Sharon}, K., {et~al.} 2018, \aj, 155, 104,
  \dodoi{10.3847/1538-3881/aaa2ff}

\bibitem[{{Rojas} {et~al.}(2022){Rojas}, {Savary}, {Cl{\'e}ment}, {Maus},
  {Courbin}, {Lemon}, {Chan}, {Vernardos}, {Joseph}, {Ca{\~n}ameras}, \&
  {Galan}}]{rojas2022a}
{Rojas}, K., {Savary}, E., {Cl{\'e}ment}, B., {et~al.} 2022, \aap, 668, A73,
  \dodoi{10.1051/0004-6361/202142119}

\bibitem[{{Romanowsky} \& {Kochanek}(1999)}]{romanowsky1999a}
{Romanowsky}, A.~J., \& {Kochanek}, C.~S. 1999, \apj, 516, 18,
  \dodoi{10.1086/307077}

\bibitem[{{Savary} {et~al.}(2022){Savary}, {Rojas}, {Maus}, {Cl{\'e}ment},
  {Courbin}, {Gavazzi}, {Chan}, {Lemon}, {Vernardos}, {Ca{\~n}ameras},
  {Schuldt}, {Suyu}, {Cuillandre}, {Fabbro}, {Gwyn}, {Hudson}, {Kilbinger},
  {Scott}, \& {Stone}}]{savary2022a}
{Savary}, E., {Rojas}, K., {Maus}, M., {et~al.} 2022, \aap, 666, A1,
  \dodoi{10.1051/0004-6361/202142505}

\bibitem[{{Schlafly} {et~al.}(2023){Schlafly}, {Kirkby}, {Schlegel}, {Myers},
  {Raichoor}, {Dawson}, {Aguilar}, {Allende Prieto}, {Bailey}, {BenZvi},
  {Bermejo-Climent}, {Brooks}, {de la Macorra}, {Dey}, {Doel}, {Fanning},
  {Font-Ribera}, {Forero-Romero}, {Garc{\'\i}a-Bellido}, {Gontcho A Gontcho},
  {Guy}, {Hahn}, {Honscheid}, {Ishak}, {Juneau}, {Kehoe}, {Kisner}, {Kremin},
  {Landriau}, {Lang}, {Lasker}, {Levi}, {Magneville}, {Manser}, {Martini},
  {Meisner}, {Miquel}, {Moustakas}, {Newman}, {Nie}, {Palanque-Delabrouille},
  {Percival}, {Poppett}, {Rockosi}, {Ross}, {Rossi}, {Tarl{\'e}}, {Weaver},
  {Y{\`e}che}, {Zhou}, \& {DESI Collaboration}}]{schlafly2023a}
{Schlafly}, E.~F., {Kirkby}, D., {Schlegel}, D.~J., {et~al.} 2023, \aj, 166,
  259, \dodoi{10.3847/1538-3881/ad0832}

\bibitem[{{Shajib} {et~al.}(2020){Shajib}, {Birrer}, {Treu}, {Agnello},
  {Buckley-Geer}, {Chan}, {Christensen}, {Lemon}, {Lin}, {Millon}, {Poh},
  {Rusu}, {Sluse}, {Spiniello}, {Chen}, {Collett}, {Courbin}, {Fassnacht},
  {Frieman}, {Galan}, {Gilman}, {More}, {Anguita}, {Auger}, {Bonvin},
  {McMahon}, {Meylan}, {Wong}, {Abbott}, {Annis}, {Avila}, {Bechtol}, {Brooks},
  {Brout}, {Burke}, {Carnero Rosell}, {Carrasco Kind}, {Carretero},
  {Castander}, {Costanzi}, {da Costa}, {De Vicente}, {Desai}, {Dietrich},
  {Doel}, {Drlica-Wagner}, {Evrard}, {Finley}, {Flaugher}, {Fosalba},
  {Garc{\'\i}a-Bellido}, {Gerdes}, {Gruen}, {Gruendl}, {Gschwend}, {Gutierrez},
  {Hollowood}, {Honscheid}, {Huterer}, {James}, {Jeltema}, {Krause},
  {Kuropatkin}, {Li}, {Lima}, {MacCrann}, {Maia}, {Marshall}, {Melchior},
  {Miquel}, {Ogando}, {Palmese}, {Paz-Chinch{\'o}n}, {Plazas}, {Romer},
  {Roodman}, {Sako}, {Sanchez}, {Santiago}, {Scarpine}, {Schubnell}, {Scolnic},
  {Serrano}, {Sevilla-Noarbe}, {Smith}, {Soares-Santos}, {Suchyta}, {Tarle},
  {Thomas}, {Walker}, \& {Zhang}}]{shajib2020a}
{Shajib}, A.~J., {Birrer}, S., {Treu}, T., {et~al.} 2020, \mnras, 494, 6072,
  \dodoi{10.1093/mnras/staa828}

\bibitem[{Sharma \& Linder(2022)}]{sharma2022a}
Sharma, D., \& Linder, E.~V. 2022, Journal of Cosmology and Astroparticle
  Physics, 2022, 033, \dodoi{10.1088/1475-7516/2022/07/033}

\bibitem[{{Sheu} {et~al.}(2024){Sheu}, {Huang}, {Cikota}, {Suzuki}, {Palmese},
  {Schlegel}, \& {Storfer}}]{sheu2024a}
{Sheu}, W., {Huang}, X., {Cikota}, A., {et~al.} 2024, \apj, 973, 24,
  \dodoi{10.3847/1538-4357/ad5dad}

\bibitem[{{Sheu} {et~al.}(2023){Sheu}, {Huang}, {Cikota}, {Suzuki}, {Schlegel},
  \& {Storfer}}]{sheu2023a}
---. 2023, \apj, 952, 10, \dodoi{10.3847/1538-4357/acd1e4}

\bibitem[{{Shu} {et~al.}(2016){Shu}, {Bolton}, {Moustakas}, {Stern}, {Dey},
  {Brownstein}, {Burles}, \& {Spinrad}}]{shu2016a}
{Shu}, Y., {Bolton}, A.~S., {Moustakas}, L.~A., {et~al.} 2016, \apj, 820, 43,
  \dodoi{10.3847/0004-637X/820/1/43}

\bibitem[{{Shu} {et~al.}(2012){Shu}, {Bolton}, {Schlegel}, {Dawson}, {Wake},
  {Brownstein}, {Brinkmann}, \& {Weaver}}]{shu2012a}
{Shu}, Y., {Bolton}, A.~S., {Schlegel}, D.~J., {et~al.} 2012, \aj, 143, 90,
  \dodoi{10.1088/0004-6256/143/4/90}

\bibitem[{{Shu} {et~al.}(2022){Shu}, {Ca{\~n}ameras}, {Schuldt}, {Suyu},
  {Taubenberger}, {Inoue}, \& {Jaelani}}]{shu2022a}
{Shu}, Y., {Ca{\~n}ameras}, R., {Schuldt}, S., {et~al.} 2022, \aap, 662, A4,
  \dodoi{10.1051/0004-6361/202243203}

\bibitem[{{Shu} {et~al.}(2017){Shu}, {Brownstein}, {Bolton}, {Koopmans},
  {Treu}, {Montero-Dorta}, {Auger}, {Czoske}, {Gavazzi}, {Marshall}, \&
  {Moustakas}}]{shu2017a}
{Shu}, Y., {Brownstein}, J.~R., {Bolton}, A.~S., {et~al.} 2017, \apj, 851, 48,
  \dodoi{10.3847/1538-4357/aa9794}

\bibitem[{{Smail} {et~al.}(2007){Smail}, {Swinbank}, {Richard}, {Ebeling},
  {Kneib}, {Edge}, {Stark}, {Ellis}, {Dye}, {Smith}, \& {Mullis}}]{smail2007a}
{Smail}, I., {Swinbank}, A.~M., {Richard}, J., {et~al.} 2007, \apjl, 654, L33,
  \dodoi{10.1086/510902}

\bibitem[{{Sonnenfeld} {et~al.}(2013){Sonnenfeld}, {Gavazzi}, {Suyu}, {Treu},
  \& {Marshall}}]{sonnenfeld2013a}
{Sonnenfeld}, A., {Gavazzi}, R., {Suyu}, S.~H., {Treu}, T., \& {Marshall},
  P.~J. 2013, \apj, 777, 97, \dodoi{10.1088/0004-637X/777/2/97}

\bibitem[{{Sonnenfeld} {et~al.}(2019){Sonnenfeld}, {Jaelani}, {Chan}, {More},
  {Suyu}, {Wong}, {Oguri}, \& {Lee}}]{sonnenfeld2019b}
{Sonnenfeld}, A., {Jaelani}, A.~T., {Chan}, J., {et~al.} 2019, \aap, 630, A71,
  \dodoi{10.1051/0004-6361/201935743}

\bibitem[{{Sonnenfeld} {et~al.}(2018){Sonnenfeld}, {Chan}, {Shu}, {More},
  {Oguri}, {Suyu}, {Wong}, {Lee}, {Coupon}, {Yonehara}, {Bolton}, {Jaelani},
  {Tanaka}, {Miyazaki}, \& {Komiyama}}]{sonnenfeld2018a}
{Sonnenfeld}, A., {Chan}, J.~H.~H., {Shu}, Y., {et~al.} 2018, \pasj, 70, S29,
  \dodoi{10.1093/pasj/psx062}

\bibitem[{{Sonnenfeld} {et~al.}(2020){Sonnenfeld}, {Verma}, {More}, {Baeten},
  {Macmillan}, {Wong}, {Chan}, {Jaelani}, {Lee}, {Oguri}, {Rusu}, {Veldthuis},
  {Trouille}, {Marshall}, {Hutchings}, {Allen}, {O'Donnell}, {Cornen}, {Davis},
  {McMaster}, {Lintott}, \& {Miller}}]{sonnenfeld2020a}
{Sonnenfeld}, A., {Verma}, A., {More}, A., {et~al.} 2020, \aap, 642, A148,
  \dodoi{10.1051/0004-6361/202038067}

\bibitem[{Stark {et~al.}(2013)Stark, Auger, Belokurov, Jones, Robertson, Ellis,
  Sand, Moiseev, Eagle, \& Myers}]{stark2013a}
Stark, D.~P., Auger, M., Belokurov, V., {et~al.} 2013, Monthly Notices of the
  Royal Astronomical Society, 436, 1040, \dodoi{10.1093/mnras/stt1624}

\bibitem[{Storfer {et~al.}(2024)Storfer, Huang, Gu, Sheu, Banka, Dey,
  Inchausti~Reyes, Jain, Kwon, Lang, Lee, Meisner, Moustakas, Myers,
  Tabares-Tarquinio, Schlafly, \& Schlegel}]{storfer2024a}
Storfer, C., Huang, X., Gu, A., {et~al.} 2024, The Astrophysical Journal
  Supplement Series, 274, 16, \dodoi{10.3847/1538-4365/ad527e}

\bibitem[{Suyu {et~al.}(2024)Suyu, Goobar, Collett, More, \&
  Vernardos}]{suyu2023a}
Suyu, S.~H., Goobar, A., Collett, T., More, A., \& Vernardos, G. 2024, Space
  Science Reviews, 220, \dodoi{10.1007/s11214-024-01044-7}

\bibitem[{{Suyu} {et~al.}(2017){Suyu}, {Bonvin}, {Courbin}, {Fassnacht},
  {Rusu}, {Sluse}, {Treu}, {Wong}, {Auger}, {Ding}, {Hilbert}, {Marshall},
  {Rumbaugh}, {Sonnenfeld}, {Tewes}, {Tihhonova}, {Agnello}, {Blandford},
  {Chen}, {Collett}, {Koopmans}, {Liao}, {Meylan}, \& {Spiniello}}]{suyu2017a}
{Suyu}, S.~H., {Bonvin}, V., {Courbin}, F., {et~al.} 2017, \mnras, 468, 2590,
  \dodoi{10.1093/mnras/stx483}

\bibitem[{{Talbot} {et~al.}(2021){Talbot}, {Brownstein}, {Dawson}, {Kneib}, \&
  {Bautista}}]{talbot2021a}
{Talbot}, M.~S., {Brownstein}, J.~R., {Dawson}, K.~S., {Kneib}, J.-P., \&
  {Bautista}, J. 2021, \mnras, 502, 4617, \dodoi{10.1093/mnras/stab267}

\bibitem[{{Tanaka} {et~al.}(2016){Tanaka}, {Wong}, {More}, {Dezuka}, {Egami},
  {Oguri}, {Suyu}, {Sonnenfeld}, {Higuchi}, {Komiyama}, {Miyazaki}, {Onoue},
  {Oyamada}, \& {Utsumi}}]{tanaka2016a}
{Tanaka}, M., {Wong}, K.~C., {More}, A., {et~al.} 2016, \apjl, 826, L19,
  \dodoi{10.3847/2041-8205/826/2/L19}

\bibitem[{{Tessore} {et~al.}(2016){Tessore}, {Bellagamba}, \&
  {Metcalf}}]{tessore2016a}
{Tessore}, N., {Bellagamba}, F., \& {Metcalf}, R.~B. 2016, \mnras, 463, 3115,
  \dodoi{10.1093/mnras/stw2212}

\bibitem[{{Tran} {et~al.}(2022){Tran}, {Harshan}, {Glazebrook}, {Keerthi
  Vasan}, {Jones}, {Jacobs}, {Kacprzak}, {Barone}, {Collett}, {Gupta},
  {Henderson}, {Kewley}, {Lopez}, {Nanayakkara}, {Sanders}, \&
  {Sweet}}]{tran2022a}
{Tran}, K.-V.~H., {Harshan}, A., {Glazebrook}, K., {et~al.} 2022, \aj, 164,
  148, \dodoi{10.3847/1538-3881/ac7da2}

\bibitem[{{Treu} \& {Koopmans}(2002)}]{treu2002a}
{Treu}, T., \& {Koopmans}, L.~V.~E. 2002, \mnras, 337, L6,
  \dodoi{10.1046/j.1365-8711.2002.06107.x}

\bibitem[{{Vegetti} {et~al.}(2010){Vegetti}, {Koopmans}, {Bolton}, {Treu}, \&
  {Gavazzi}}]{vegetti2010a}
{Vegetti}, S., {Koopmans}, L.~V.~E., {Bolton}, A., {Treu}, T., \& {Gavazzi}, R.
  2010, \mnras, 408, 1969, \dodoi{10.1111/j.1365-2966.2010.16865.x}

\bibitem[{{Williams} {et~al.}(2004){Williams}, {Olszewski}, {Lesser}, \&
  {Burge}}]{williams2004a}
{Williams}, G.~G., {Olszewski}, E., {Lesser}, M.~P., \& {Burge}, J.~H. 2004, in
  \procspie, Vol. 5492, Ground-based Instrumentation for Astronomy, ed.
  A.~F.~M. {Moorwood} \& M.~{Iye}, 787--798, \dodoi{10.1117/12.552189}

\bibitem[{{Wong} {et~al.}(2022){Wong}, {Chan}, {Chao}, {Jaelani}, {Kayo},
  {Lee}, {More}, \& {Oguri}}]{wong2022a}
{Wong}, K.~C., {Chan}, J. H.~H., {Chao}, D. C.~Y., {et~al.} 2022, \pasj, 74,
  1209, \dodoi{10.1093/pasj/psac065}

\bibitem[{{Wong} {et~al.}(2018){Wong}, {Sonnenfeld}, {Chan}, {Rusu}, {Tanaka},
  {Jaelani}, {Lee}, {More}, {Oguri}, {Suyu}, \& {Komiyama}}]{wong2018a}
{Wong}, K.~C., {Sonnenfeld}, A., {Chan}, J. H.~H., {et~al.} 2018, \apj, 867,
  107, \dodoi{10.3847/1538-4357/aae381}

\bibitem[{{Wong} {et~al.}(2020){Wong}, {Suyu}, {Chen}, {Rusu}, {Millon},
  {Sluse}, {Bonvin}, {Fassnacht}, {Taubenberger}, {Auger}, {Birrer}, {Chan},
  {Courbin}, {Hilbert}, {Tihhonova}, {Treu}, {Agnello}, {Ding}, {Jee},
  {Komatsu}, {Shajib}, {Sonnenfeld}, {Blandford}, {Koopmans}, {Marshall}, \&
  {Meylan}}]{wong2020a}
{Wong}, K.~C., {Suyu}, S.~H., {Chen}, G. C.~F., {et~al.} 2020, \mnras, 498,
  1420, \dodoi{10.1093/mnras/stz3094}

\bibitem[{{Y{\`e}che} {et~al.}(2020){Y{\`e}che}, {Palanque-Delabrouille},
  {Claveau}, {Brooks}, {Chaussidon}, {Davis}, {Dawson}, {Dey}, {Duan},
  {Eftekharzadeh}, {Eisenstein}, {Gazta{\~n}aga}, {Kehoe}, {Landriau}, {Lang},
  {Levi}, {Meisner}, {Myers}, {Newman}, {Poppett}, {Prada}, {Raichoor},
  {Schlegel}, {Schubnell}, {Staten}, {Tarl{\'e}}, \& {Zhou}}]{yeche2020a}
{Y{\`e}che}, C., {Palanque-Delabrouille}, N., {Claveau}, C.-A., {et~al.} 2020,
  Research Notes of the American Astronomical Society, 4, 179,
  \dodoi{10.3847/2515-5172/abc01a}

\bibitem[{{Zhang} {et~al.}(2023){Zhang}, {Manwadkar}, {Gladders}, {Khullar},
  {Dahle}, {Napier}, {Mahler}, {Sharon}, {Matthews Acu{\~n}a}, {Ashmead},
  {Cerny}, {Gonz{\`a}lez}, {Gozman}, {Levine}, {Marohnic}, {Martinez}, {Merz},
  {Pan}, {Sanchez}, {Sierra}, {Sisco}, {Sukay}, {Tavangar}, \&
  {Zaborowski}}]{zhang2023a}
{Zhang}, Y., {Manwadkar}, V., {Gladders}, M.~D., {et~al.} 2023, \apj, 950, 58,
  \dodoi{10.3847/1538-4357/acc9be}

\bibitem[{Zhou {et~al.}(2020)Zhou, Newman, Mao, Meisner, Moustakas, Myers,
  Prakash, Zentner, Brooks, Duan, Landriau, Levi, Prada, \& Tarle}]{zhou2020a}
Zhou, R., Newman, J.~A., Mao, Y.-Y., {et~al.} 2020, Monthly Notices of the
  Royal Astronomical Society, 501, 3309–3331, \dodoi{10.1093/mnras/staa3764}

\bibitem[{Zwicky(1937)}]{zwicky1937a}
Zwicky, F. 1937, Phys. Rev., 51, 290, \dodoi{10.1103/PhysRev.51.290}

\end{thebibliography}
\bibliographystyle{aasjournal}

\appendix

In this Appendix, we provide spectra for the remaining targets in the DESI Strong Lensing secondary target Program not presented in the main text.
Systems that remain candidates are shown in Appendix~\ref{sec:cands} and those are confirmed to be not lenses in Appendix~\ref{sec:nonlenses}.

\section{Systems that Remain Candidates}\label{sec:cands}
In this section, we show all the systems that remain candidates in the order of ascending RA.
These are listed in Table~\ref{tab:master}. 
DESI targeted the sources of some systems but failed to obtain redshifts. 
For these, we will pursue NIR spectroscopy and/or deeper optical spectroscopy (on 8 - 10 m telescopes).

\begin{figure}[h]
\centering
\includegraphics[width=.7\textwidth]{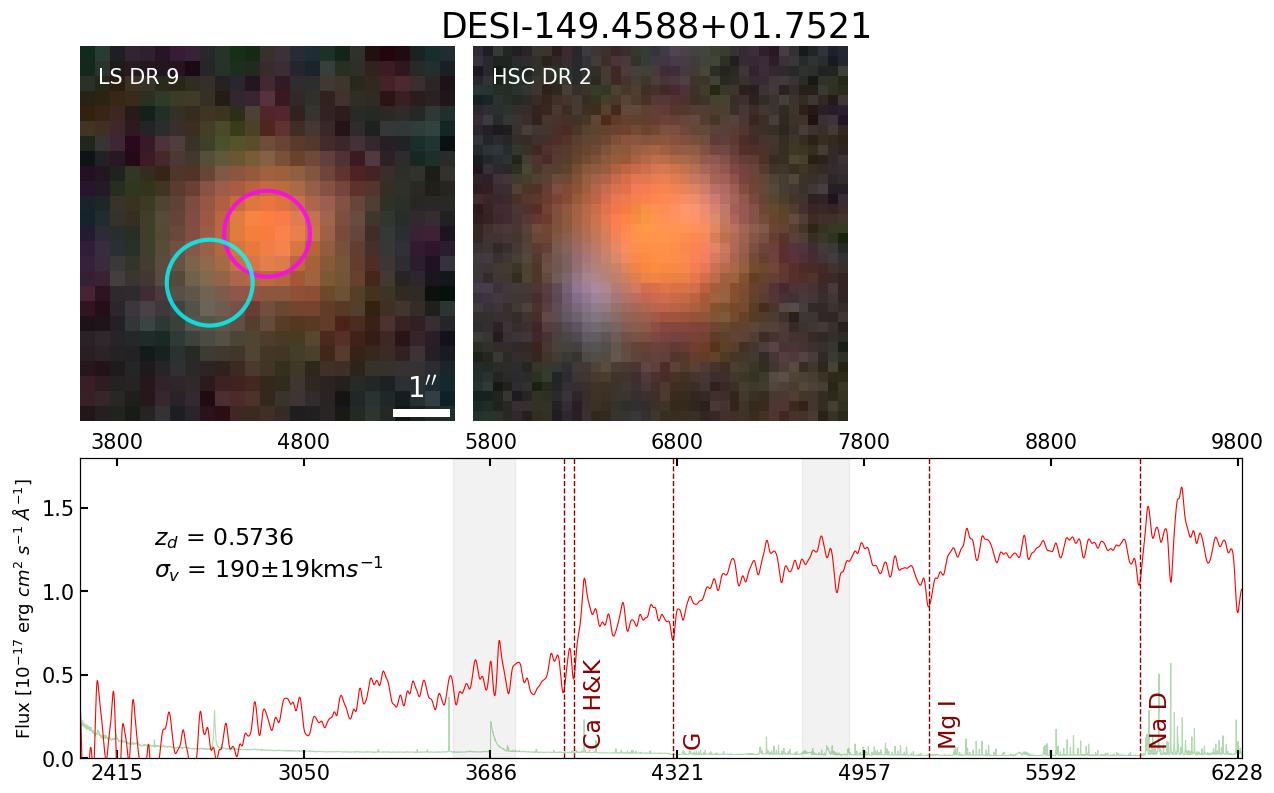}
\caption{DESI-149.4588+01.7521.
For the arrangement of the panels, see Figure~\ref{fig:desi149.6+01}.
}
\label{fig:desi149.4+01}
\end{figure}

\href{https://www.legacysurvey.org/viewer/?ra=149.4588&dec=+01.7521&layer=hsc-dr2&pixscale=0.262&zoom=16}{\emph{DESI-149.4588+01.7521}}\, The putative lens is at 0.5736  (Figure~\ref{fig:desi149.4+01}). 
There are two blue objects with similar colors on opposite sides of the putative lenses. \cwr{DESI observed the brighter of the two but failed to obtain its redshift}.

\begin{figure}[!h]
  \centering
  \includegraphics[width=.7\textwidth]{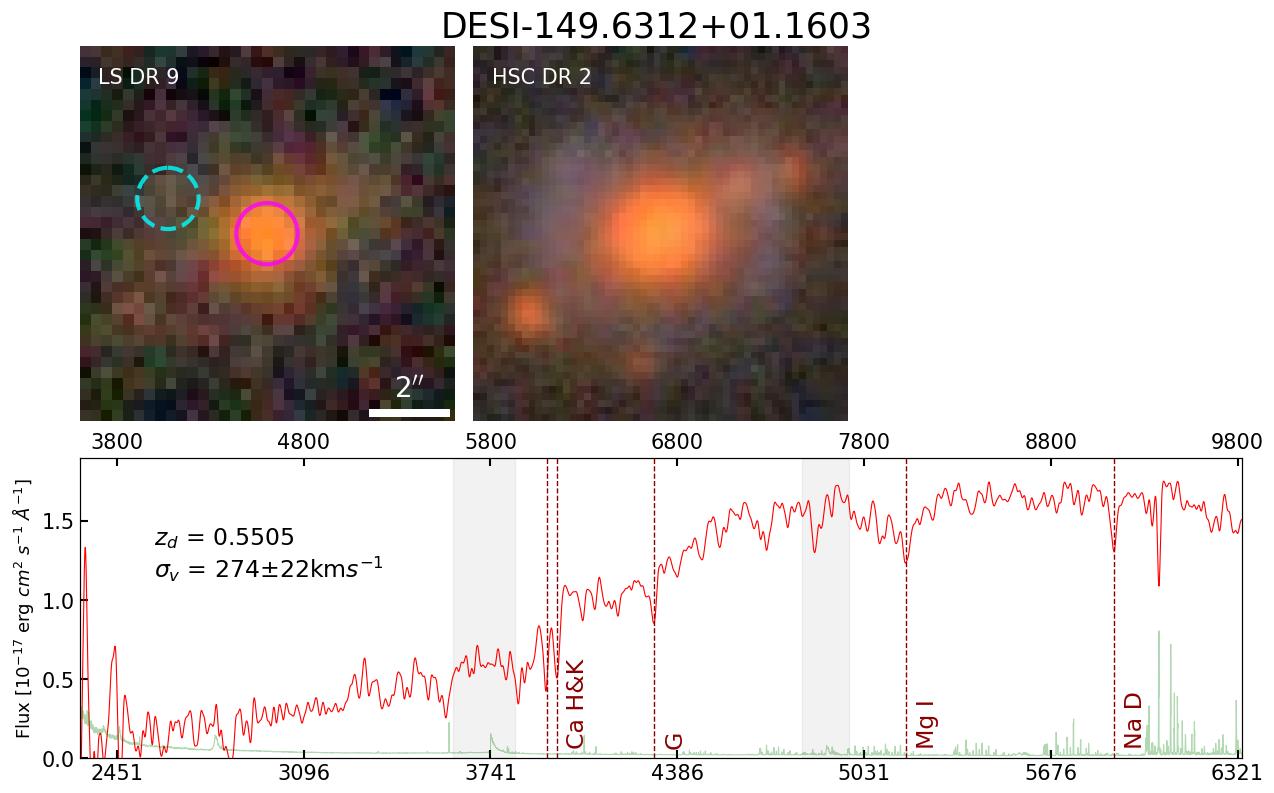}
  \caption{DESI-149.6312+01.1603. 
  The two images shown, from left to right, are from the LS and HSC. The spectrum from the fiber centered on the putative lens is shown in the second row. 
  The gray bands indicate the two dichroics of the DESI spectrograph.
  The 4000 \ang break can clearly be seen.
  We indicate the locations of the most common absorption lines, which for this objects are all detected at high significance, while for other systems, at least Ca H\&K is detected but not necessarily the other lines.}
  \label{fig:desi149.6+01}
\end{figure}

\newpage
\href{https://www.legacysurvey.org/viewer/?ra=149.6312&dec=+01.1603&zoom=16&layer=hsc-dr2}{\emph{DESI-149.6312+01.1603}} \, This system has a bright elliptical galaxy at a redshift of 0.5505 (Figure~\ref{fig:desi149.6+01}).
The \hsc images shows a pair of arc and counter arc, with similar blue colors, 
to the E and W of the elliptical galaxy, respectively.
The brighter arc on the E side  will be observed by DESI.

\begin{figure}[h]
  \centering
  \includegraphics[width=.7\textwidth]{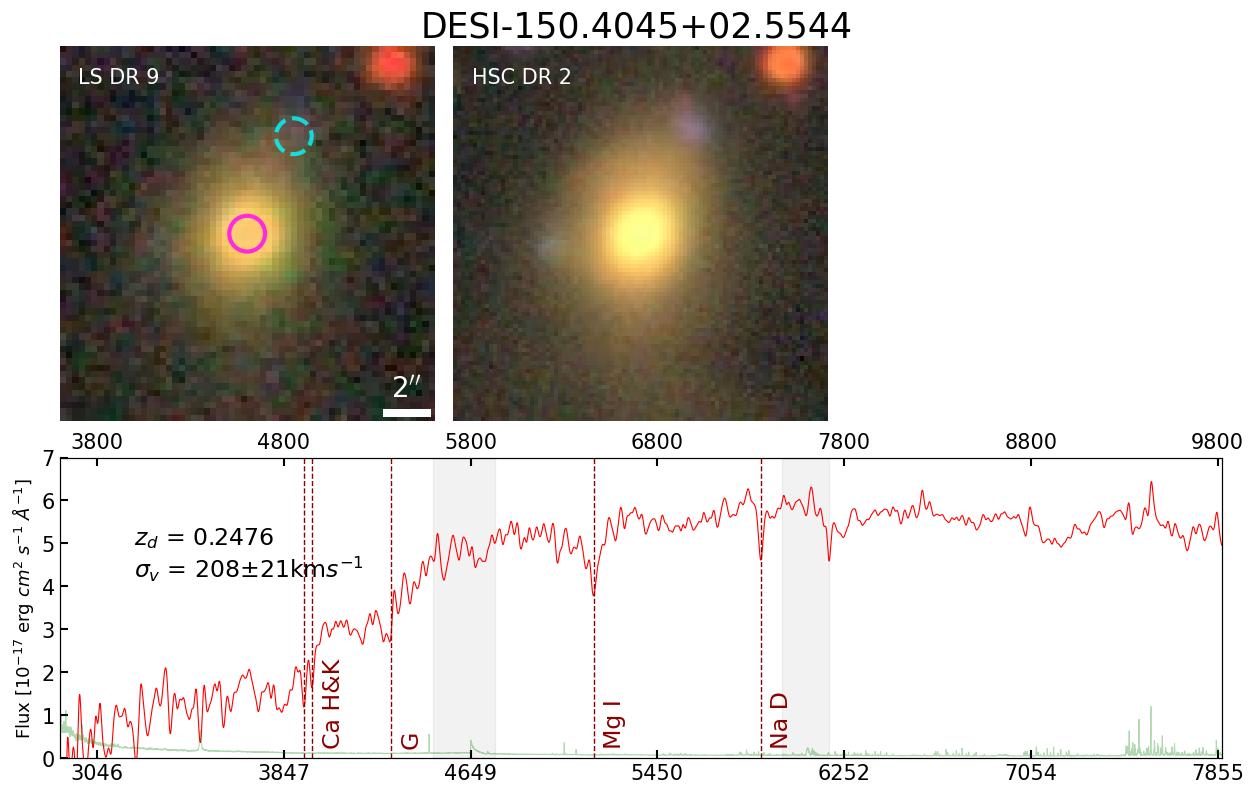}
  \caption{DESI-150.4045+02.5544. 
  For the arrangement of the panels, see Figure~\ref{fig:desi149.6+01}. 
}
  \label{fig:desi150+02}
\end{figure}

\href{https://www.legacysurvey.org/viewer/?ra=150.4045&dec=+02.5544&zoom=16&layer=hsc-dr2}{\emph{DESI-150.4045+02.5544}} \, 
The putative lens of this system 
(Figure~\ref{fig:desi150+02}) is at $z = 0.2476$.
Clearly visible in \hsc (and faintly in \ls) is an arc to the NE, with what appear to be two star formation regions at the ends.

\newpage
\href{https://www.legacysurvey.org/viewer/?ra=151.7664&dec=+02.1430&zoom=16&layer=hsc-dr2}{\emph{DESI-151.7664+02.1430}} \,
The putative lens of this system is at $z = 0.3708$.
There is a blue arc-like object to the SE visible in both \hsc and \ls images.
\begin{figure}[h]
  \centering
  \includegraphics[width=.7\textwidth]{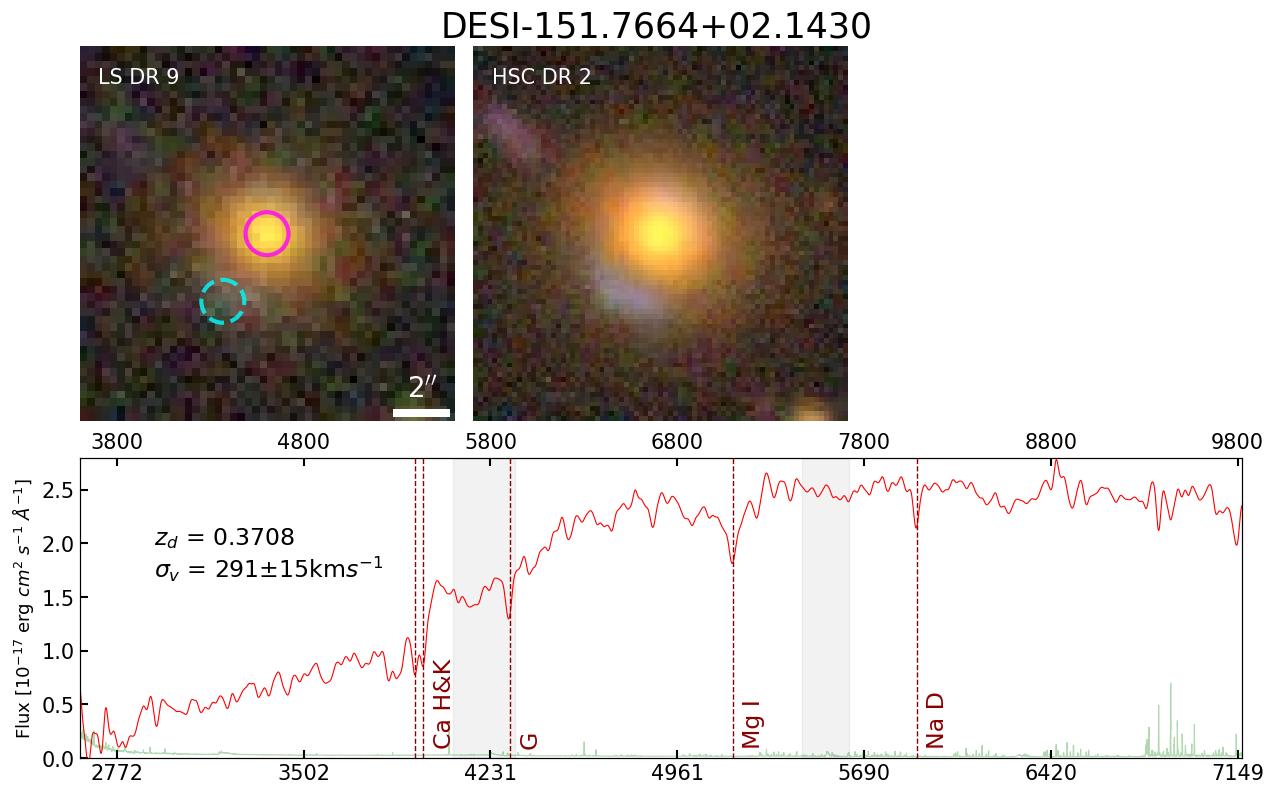}
  \caption{DESI-151.7664+02.1430.
  For the arrangement of the panels, see Figure~\ref{fig:desi149.6+01}.}
  \label{fig:desi151.7}
\end{figure}

\begin{figure}[h]
  \centering
  \includegraphics[width=0.7\textwidth]{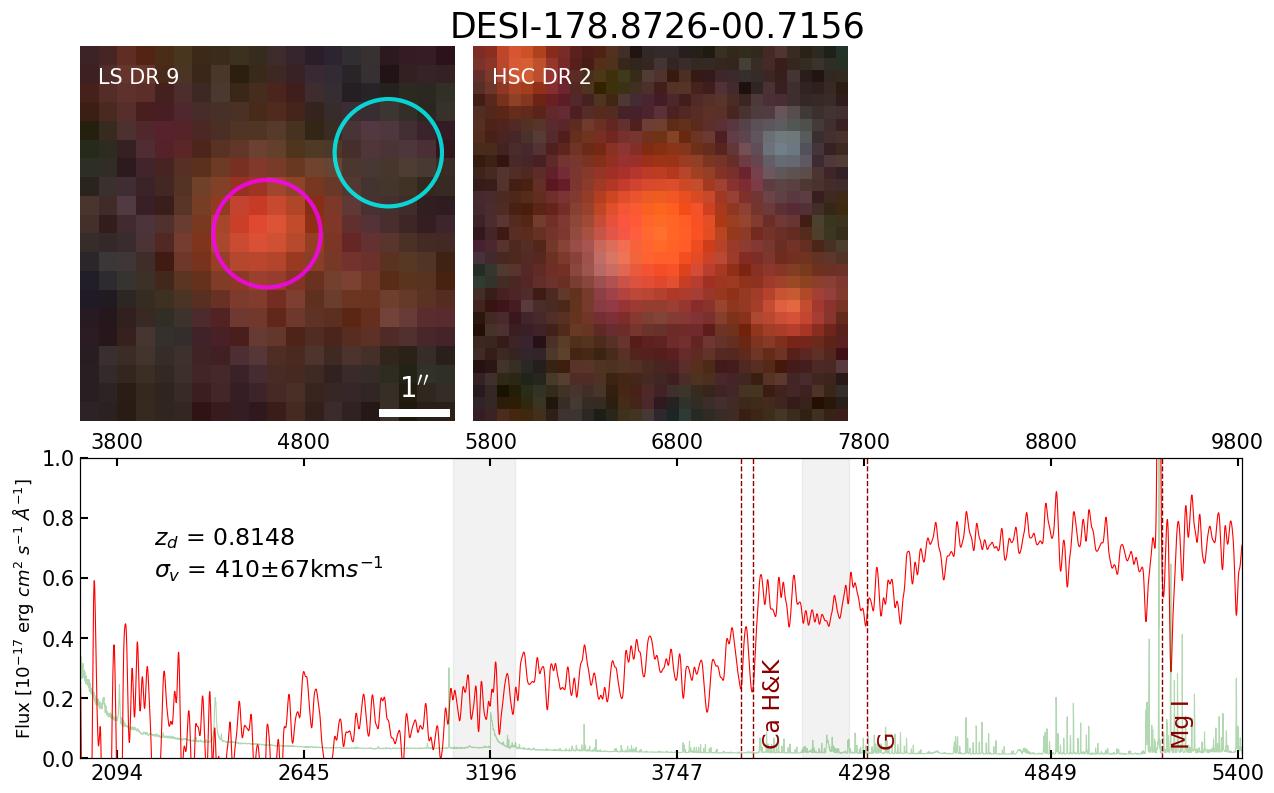}
  \caption{DESI-178.8726-00.7156. 
  For the arrangement of the panels, see Figure~\ref{fig:desi149.6+01}.}
  \label{fig:desi178}
\end{figure}

\href{https://www.legacysurvey.org/viewer/?ra=178.8726&dec=-00.7156&layer=hsc-dr2&pixscale=0.262&zoom=16}{\emph{DESI-178.8726-00.7156}}\, The putative lens is at a high redshift of 0.8148 (Figure~\ref{fig:desi178}).
The fiber for the source is centered on the bright blue object in the HSC cutout. 

\begin{figure}[h]
  \centering
  \includegraphics[width=.7\textwidth]{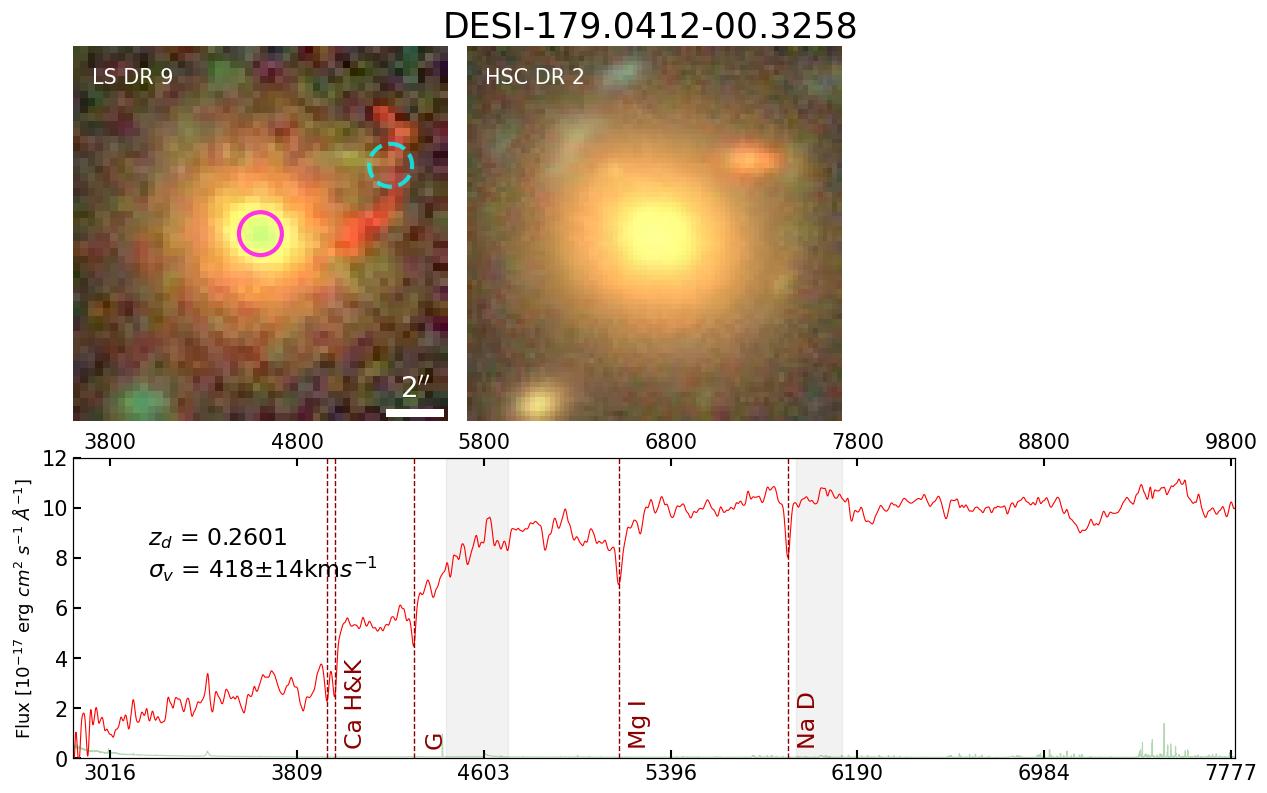}
  \caption{DESI-179.0412-0.3258. 
   For the arrangement of the panels, see Figure~\ref{fig:desi149.6+01}.}
  \label{fig:desi179.0}
\end{figure}

\newpage
\href{https://www.legacysurvey.org/viewer/?ra=179.0412&dec=-0.3258&zoom=16&layer=hsc-dr2}{\emph{DESI-179.0412-00.3258}}\, 
This system has a bright elliptical galaxy at a redshift of $z = 0.2601$ (Figure~\ref{fig:desi179.0}).
It appears to be at the center of a sub-clump in a galaxy cluster.
The putative arc to the NW will be targeted in future DESI observations.

\begin{figure}[h]
  \centering
  \includegraphics[width=.7\textwidth]{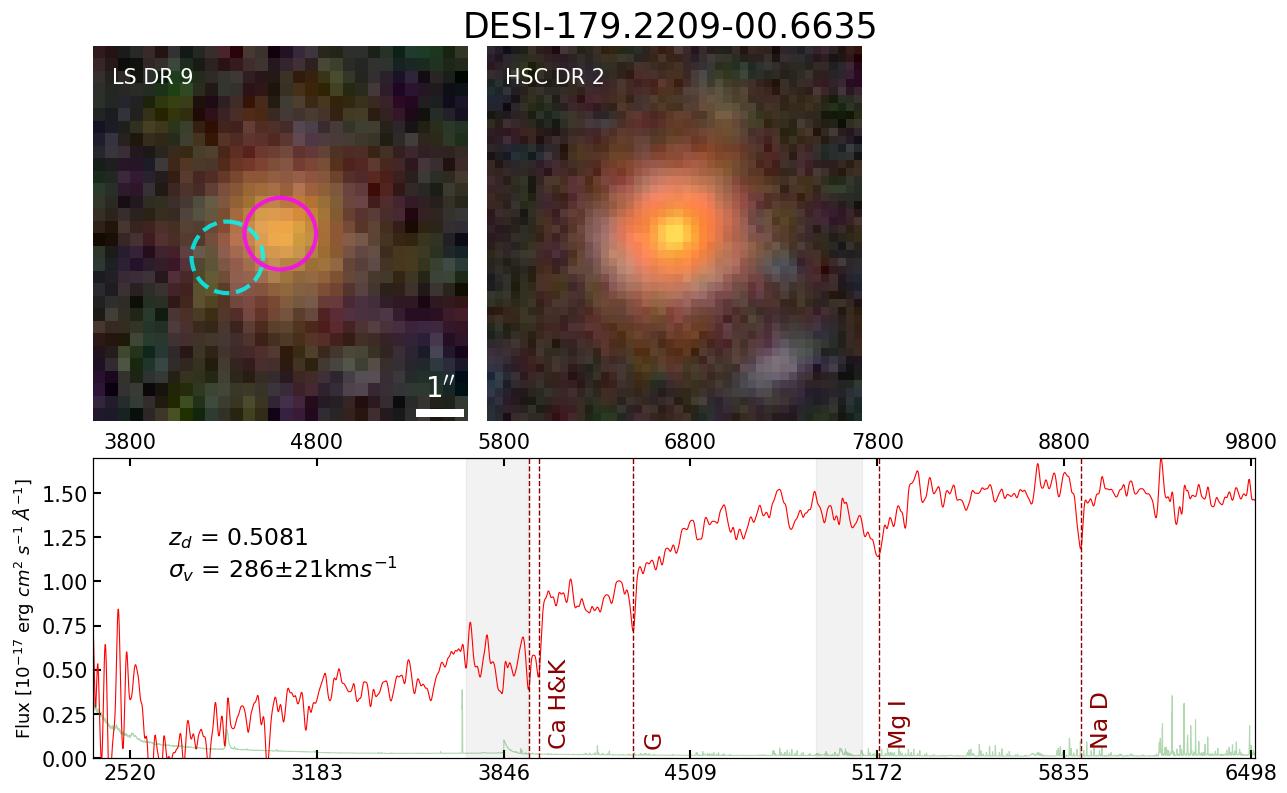}
  \caption{DESI-179.2209-00.6635.
  For the arrangement of the panels, see Figure~\ref{fig:desi149.6+01}.}
  \label{fig:desi179.2-00}
\end{figure}

\href{https://www.legacysurvey.org/viewer/?ra=179.2209&dec=-00.6635&zoom=16&layer=hsc-dr2}{\emph{DESI-179.2209-00.6635}} \,
The putative lens for this system has a redshift of 0.5081 (Figure~\ref{fig:desi179.2-00}).
There appears to an arc and a counterarc, with similar bluish colors, below and above, respectively, the putative lens. 
The putative arc will be targeted in future DESI observations.

\begin{figure}[h]
\centering
\includegraphics[width=.7\textwidth]{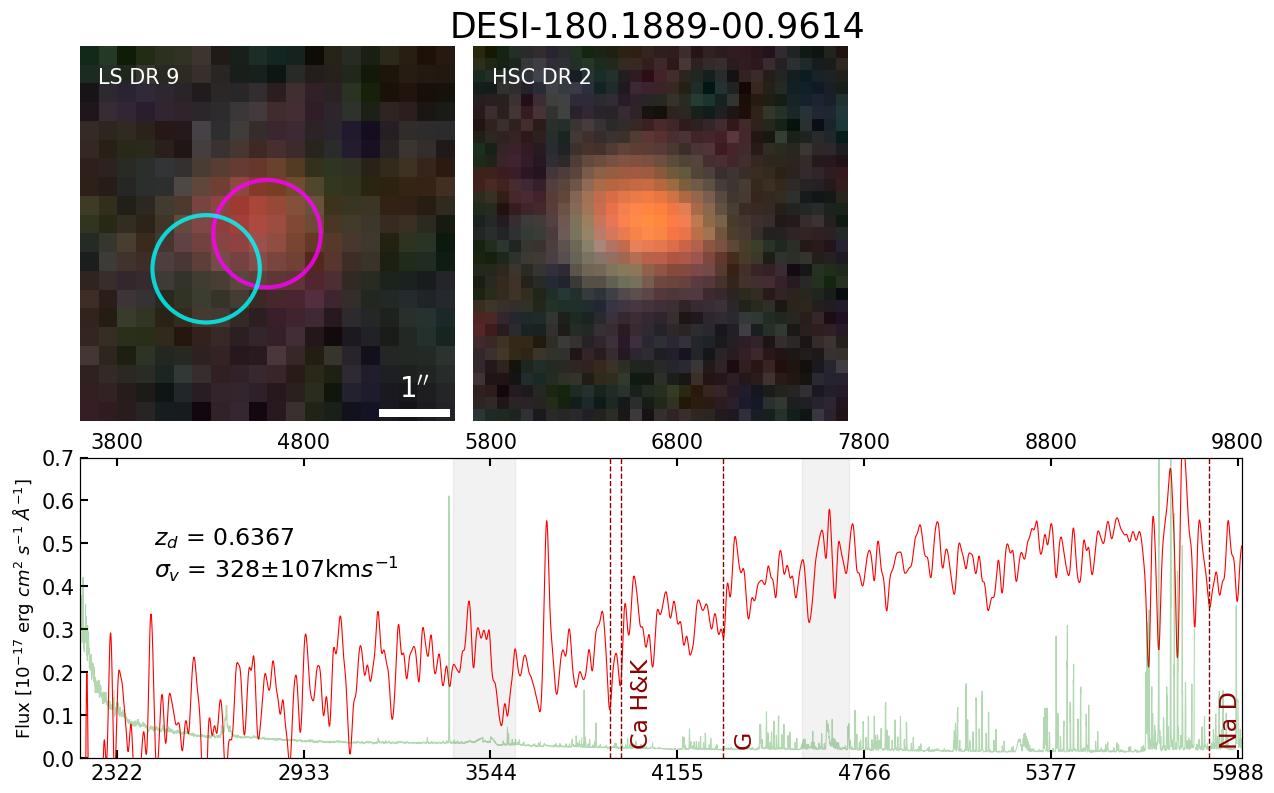}
\caption{DESI-180.1889-00.9614.
For the arrangement of the panels, see Figure~\ref{fig:desi149.6+01}.}
\label{fig:desi180-00}
\end{figure}

\newpage
\href{https://www.legacysurvey.org/viewer/?ra=180.1889&dec=-00.9614&layer=hsc-dr2&pixscale=0.262&zoom=16}{\emph{DESI-180.1889-00.9614}}\, 
The putative lens is at a redshift of 0.6367 (Figure~\ref{fig:desi180-00}).
In \hsc, there is a greenish arc that wraps around the E side of putative lens (with a counterarc on the opposite side).
 \cwr{DESI targeted the brightest part of the arc but failed to obtain its redshift}.

\begin{figure}[h]
  \centering
  \includegraphics[width=.7\textwidth]{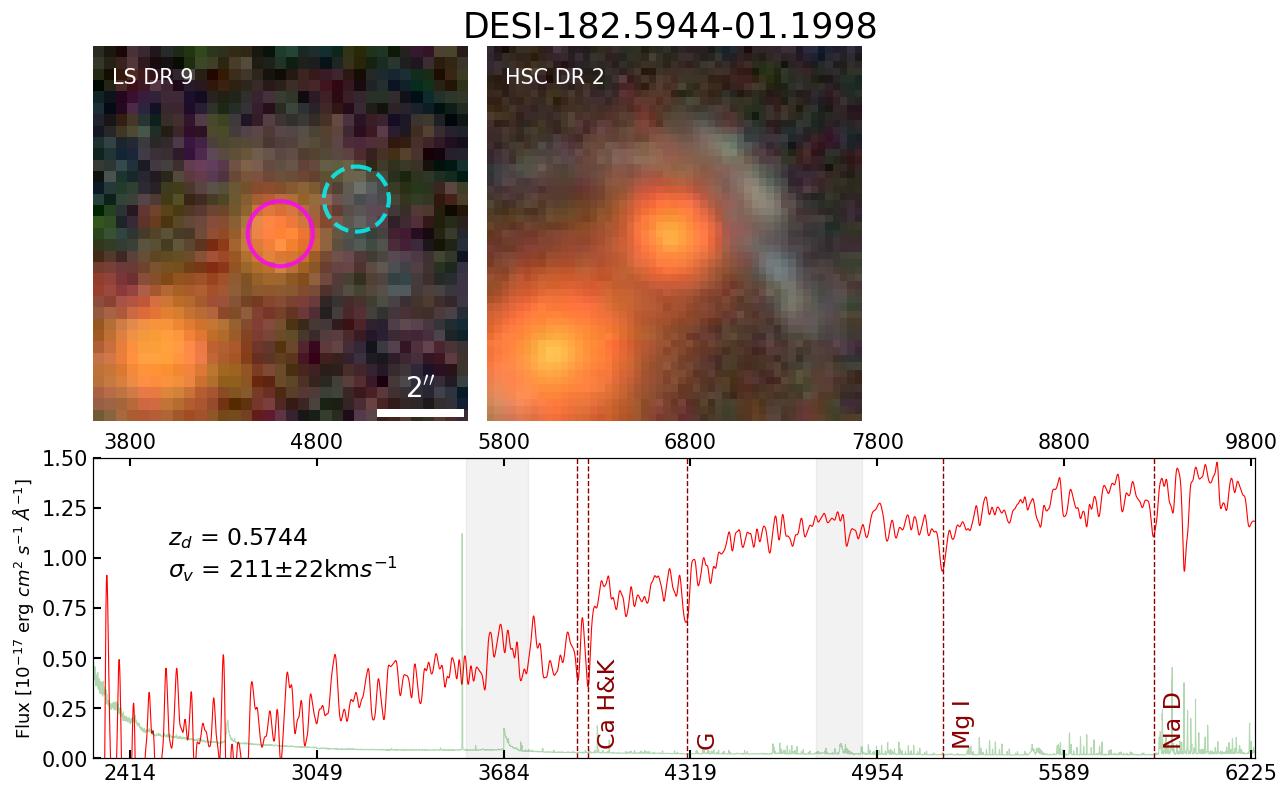}
  \caption{DESI-182.5944-01.1998.
  For the arrangement of the panels, see Figure~\ref{fig:desi149.6+01}.}
  \label{fig:desi182.6-01}
\end{figure}

\href{https://www.legacysurvey.org/viewer/?ra=182.5944&dec=-01.1998&zoom=16&layer=hsc-dr2}{\emph{DESI-182.5944-01.1998}}\, The putative lens has a redshift of 0.5744 (Figure~\ref{fig:desi182.6-01}).
In both the \ls and the \hsc images (more clearly), 
an arc-like blue image can be seen around the putative lens.
This object will be observed by DESI.

\begin{figure}[h]
\centering
\includegraphics[width=.7\textwidth]{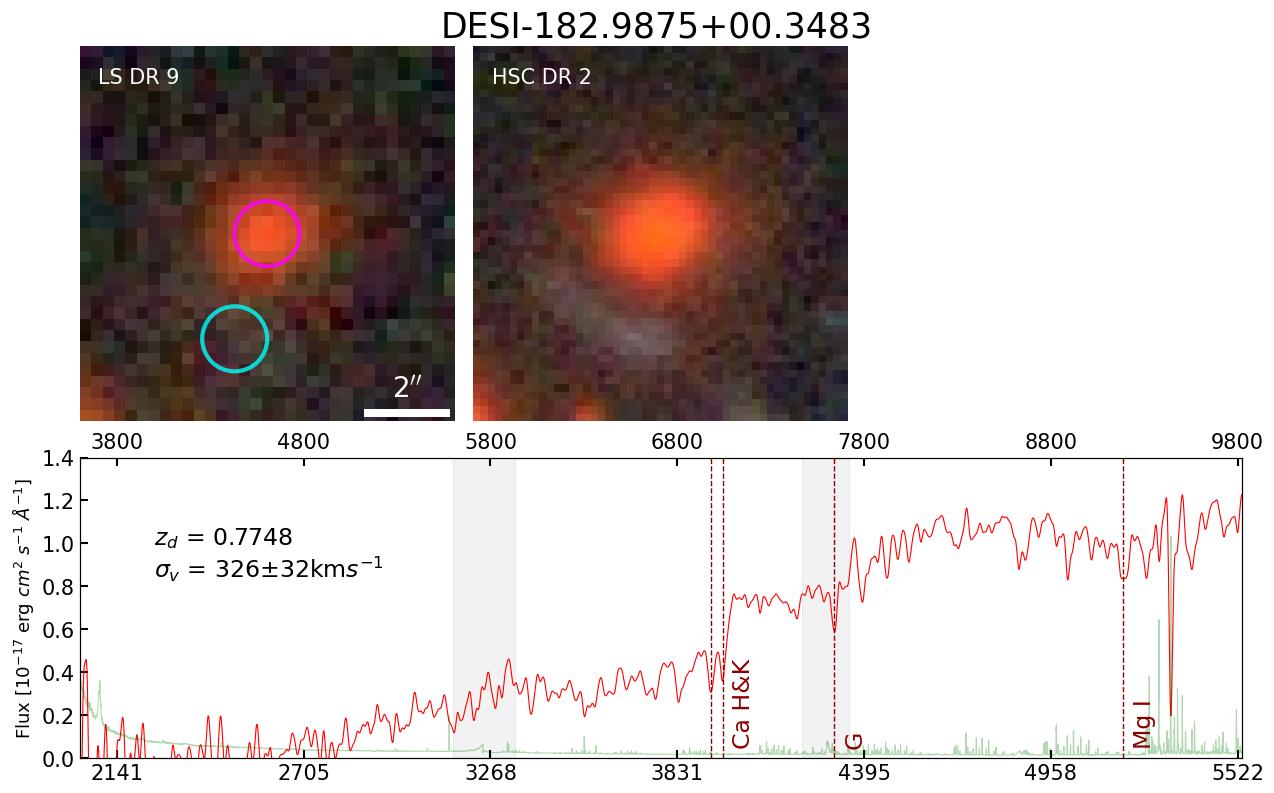}
\caption{DESI-182.9875+00.3483.
For the arrangement of the panels, see Figure~\ref{fig:desi149.6+01}.}
\label{fig:desi182+00}
\end{figure}

\newpage
\href{https://www.legacysurvey.org/viewer/?ra=182.9875&dec=+00.3483&layer=hsc-dr2&pixscale=0.262&zoom=16}{\emph{DESI-182.9875+00.3483}}\,
The putative lens is at a redshift of 0.6367 (Figure~\ref{fig:desi182+00}).
In \hsc, there is a large blue arc-like object to the SE of the putative lens.
 \cwr{DESI targeted the brightest part of the arc but failed to obtain its redshift}.

\begin{figure}[h]
  \centering
  \includegraphics[width=.7\textwidth]{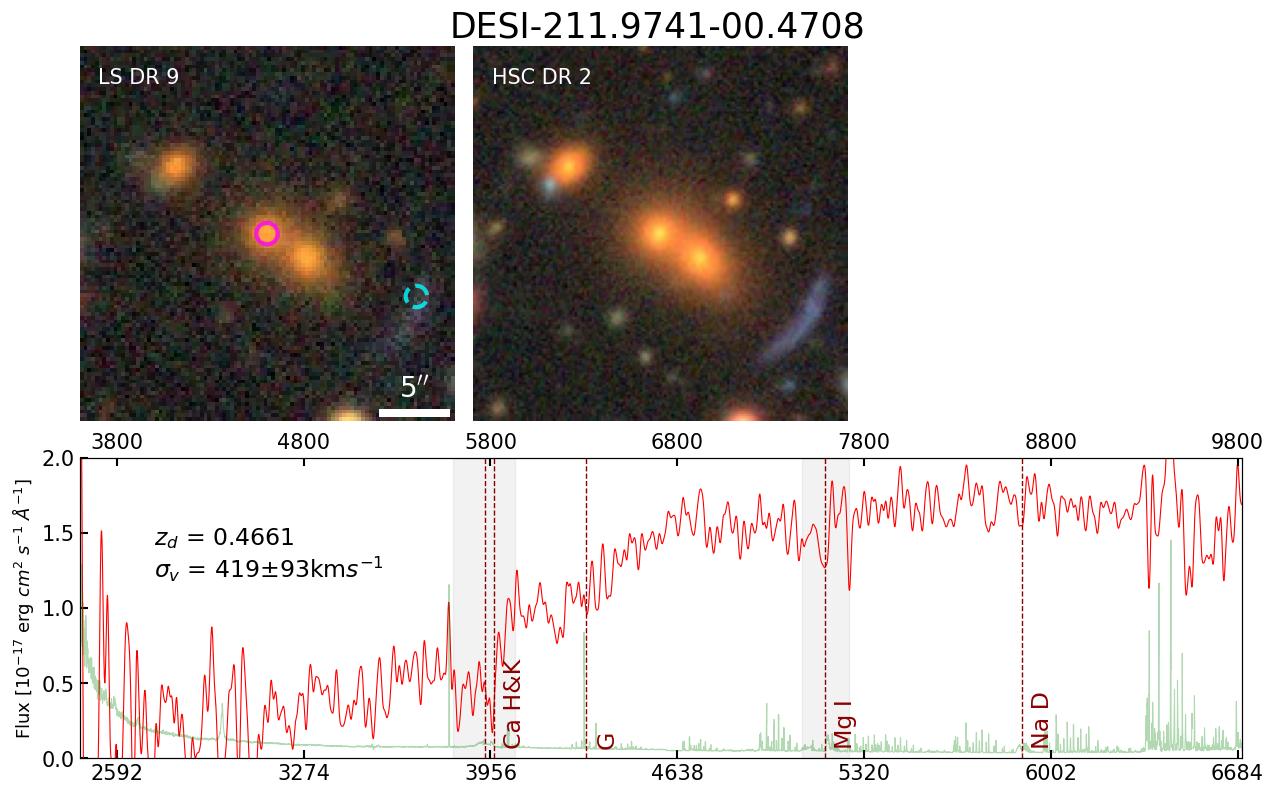}
  \caption{DESI-211.9741-00.4708.For the arrangement of the panels, see Figure~\ref{fig:desi149.6+01}.
}
  \label{fig:211.97-00}
\end{figure}

\href{https://www.legacysurvey.org/viewer/?ra=211.9741&dec=-00.4708&zoom=16&layer=hsc-dr2}{\emph{DESI-211.9741-00.4708}}\, The putative lens  has a redshift of 0.4661 (Figure~\ref{fig:211.97-00}).
DESI will target a prominent blue arc to the SW, visible in both \ls and \hsc images.

\begin{figure}[h]
  \centering
  \includegraphics[width=.7\textwidth]{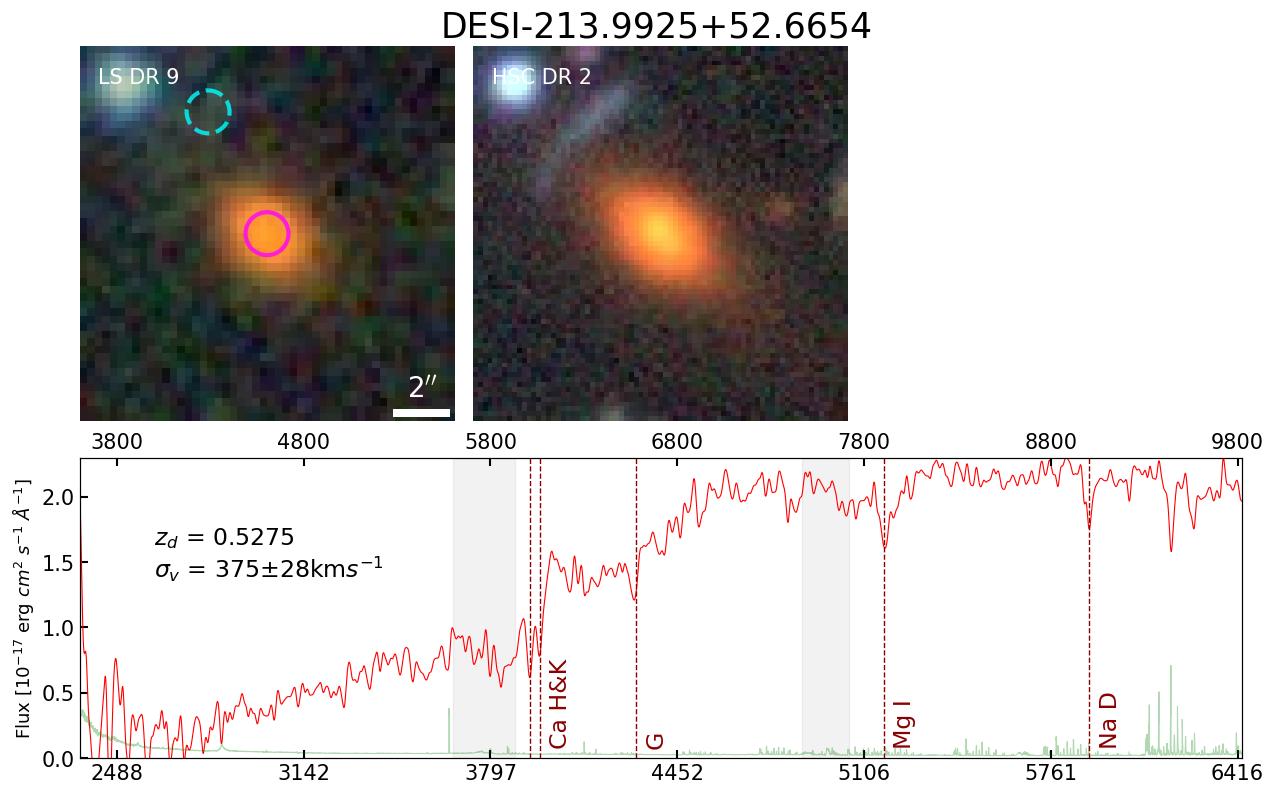}
  \caption{DESI-213.9925+52.6654. For the arrangement of the panels, see Figure~\ref{fig:desi149.6+01}.}
  \label{fig:desi213.99+52}
\end{figure}

\newpage
\href{https://www.legacysurvey.org/viewer/?ra=213.9925&dec=+52.6654&zoom=16&layer=hsc-dr2}{\emph{DESI-213.9925+52.6654}} \,
The putative lens has a redshift of 0.5275 (Figure~\ref{fig:desi213.99+52}).
In the \hsc image, a blue arc can be seen to the NE.
It will be targeted by DESI.

\begin{figure}[h]
  \centering
  \includegraphics[width=.7\textwidth]{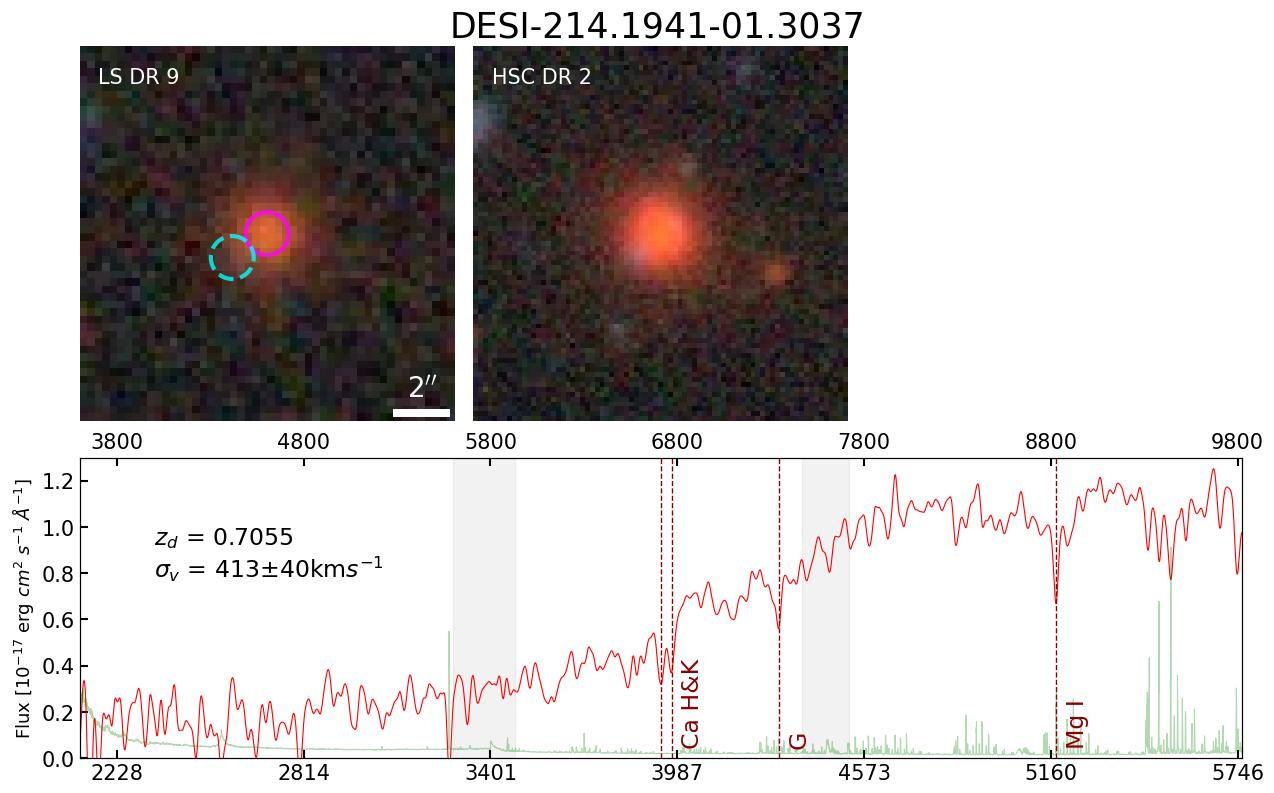}
  \caption{DESI-214.1941-01.3037. 
  For the arrangement of the panels, see Figure~\ref{fig:desi149.6+01}.}
  \label{fig:desi214.1-01}
\end{figure}

\href{https://www.legacysurvey.org/viewer/?ra=214.1941&dec=-01.3037&zoom=16&layer=hsc-dr2}{\emph{DESI-214.1941-01.3037}}\,
The putative lens has a redshift of 0.7055 (Figure~\ref{fig:desi214.1-01}).
From the \hsc image, a pair of faint images, with similar bluish colors,  above and blow putative lens can be seen.
The lower one will be targeted by DESI.

\begin{figure}[!h]
  \centering
  \includegraphics[width=.7\textwidth]{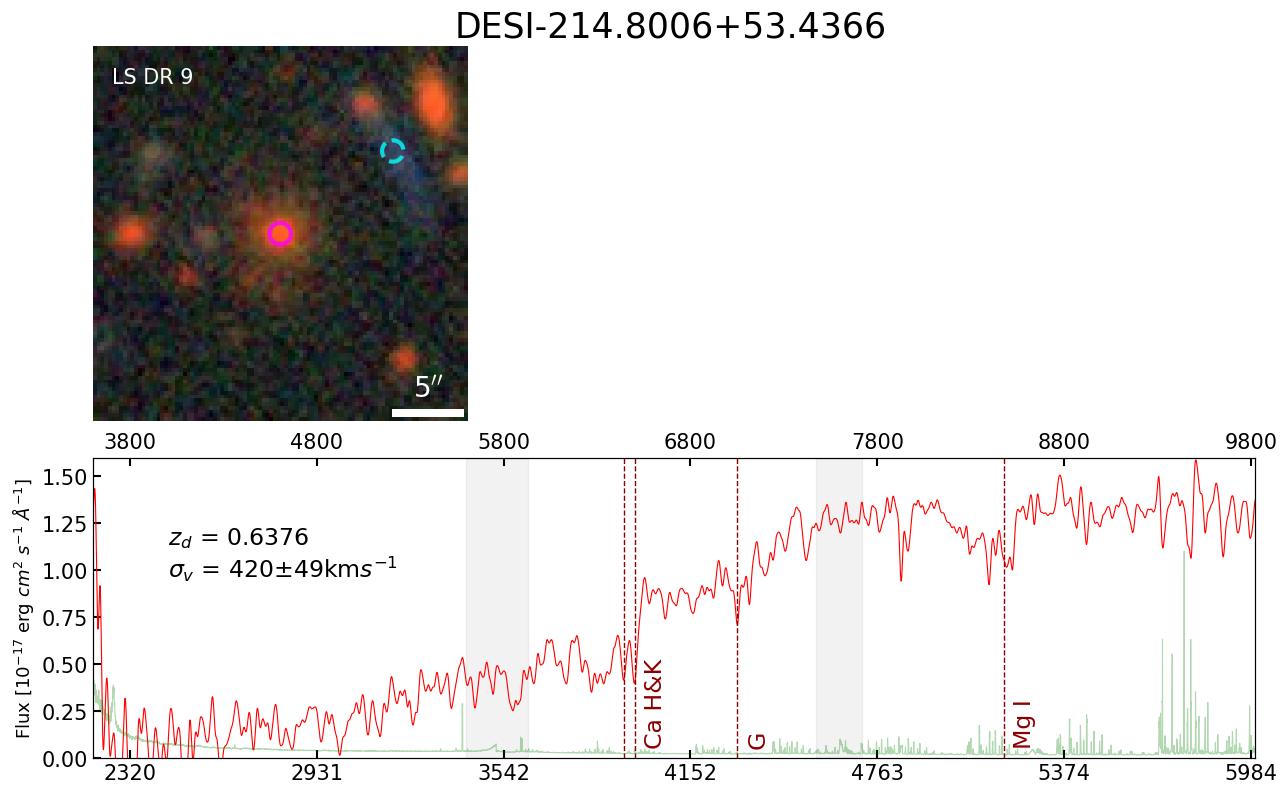}
  \caption{DESI-214.8006+53.4366. For the arrangement of the panels, see Figure~\ref{fig:desi149.6+01}.}
  \label{fig:desi214.8+53}
\end{figure}

\newpage
\href{https://www.legacysurvey.org/viewer/?ra=214.8006&dec=+53.4366&zoom=14&layer=ls-dr9}{\emph{DESI-214.8006+53.4366}}\, The putative lens has a redshift of 0.6376 (Figure~\ref{fig:desi214.8+53}).
DESI will target the blue arc-like object to the NW.

\begin{figure}[!h]
  \centering
  \includegraphics[width=.7\textwidth]{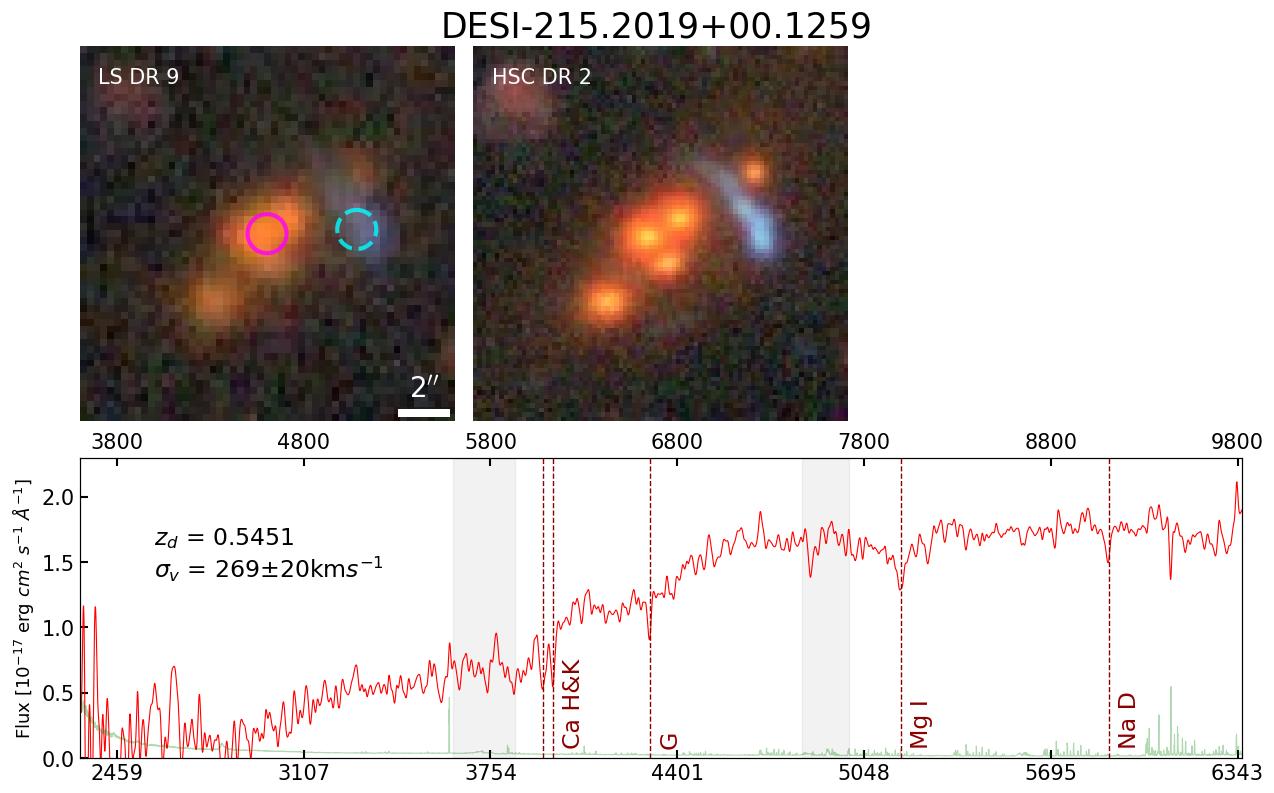}
  \caption{DESI-215.2019+00.1259.
   For the arrangement of the panels, see Figure~\ref{fig:desi149.6+01}.}
  \label{fig:desi215.20+00}
\end{figure}

\href{https://www.legacysurvey.org/viewer/?ra=215.2019&dec=+00.1259&zoom=16&layer=hsc-dr2}{\emph{DESI-215.2019+00.1259}} \, The putative lens appears to be a small group (Figure~\ref{fig:desi215.20+00}).
The brightest galaxy, which is at the center of the cutout image, has a redshift of 0.5451.
In the \hsc image, there is a bright blue arc to its NW, which can be seen in the \ls image as well.
This arc will be observed by DESI.

\begin{figure}[!h]
  \centering
  \includegraphics[width=.7\textwidth]{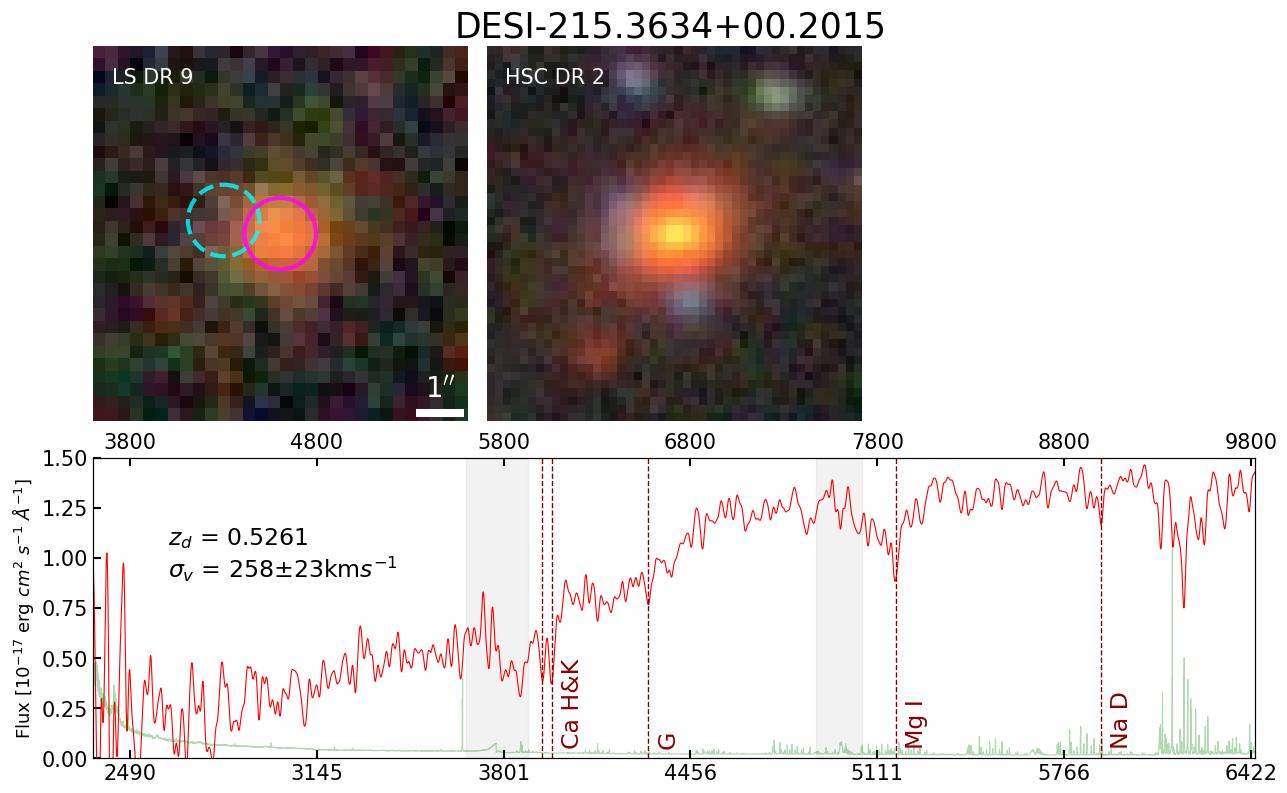}
  \caption{DESI-215.3634+00.2015.
  For the arrangement of the panels, see Figure~\ref{fig:desi149.6+01}.}
  \label{fig:desi215.3+00}
\end{figure}

\newpage
\href{https://www.legacysurvey.org/viewer/?ra=215.3634&dec=+00.2015&zoom=16&layer=hsc-dr2}{\emph{DESI-215.3634+00.2015}} \, The putative lens has a redshift of 0.5261 (Figure~\ref{fig:desi215.3+00}).
From the \hsc image, a pair of images,
with similar blue colors, are to the E --- DESI will target this object --- and SW of the putative lens.
There is a hint of another blue image just above and to the NW of the putative lens.

\begin{figure}[h]
  \centering
  \includegraphics[width=.7\textwidth]{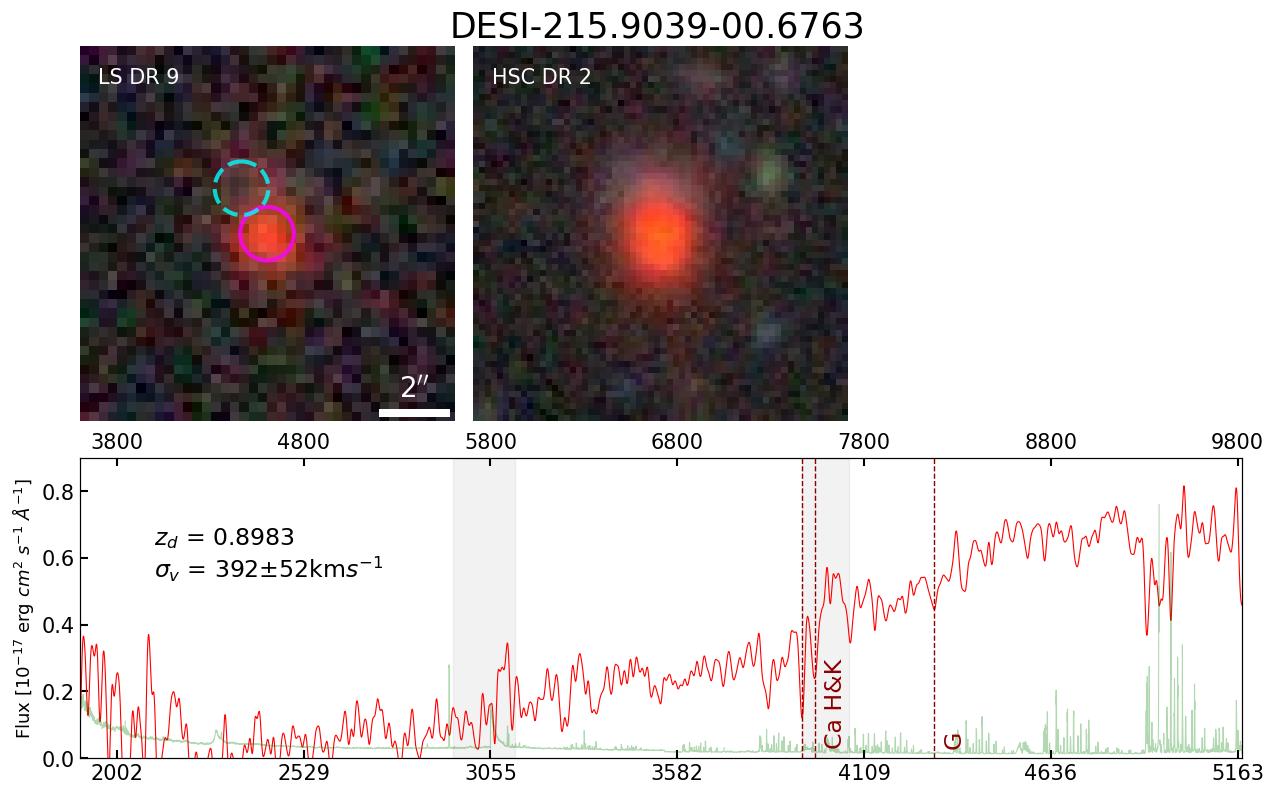}
  \caption{DESI-215.9039-00.6763.
  For the arrangement of the panels, see Figure~\ref{fig:desi149.6+01}.}
  \label{fig:desi215.9-00}
\end{figure}

\href{https://www.legacysurvey.org/viewer/?ra=215.9039&dec=-00.6763&zoom=16&layer=hsc-dr2}{\emph{DESI-215.9039-00.6763}} \,
 In the \hsc image, a very faint blue arc can be seen above the putative lens (Figure~\ref{fig:desi215.9-00}). 
 It will be targeted by DESI.
 The putative lens has a high redshift of 0.8983.

\begin{figure}[h]
\centering
\includegraphics[width=.7\textwidth]{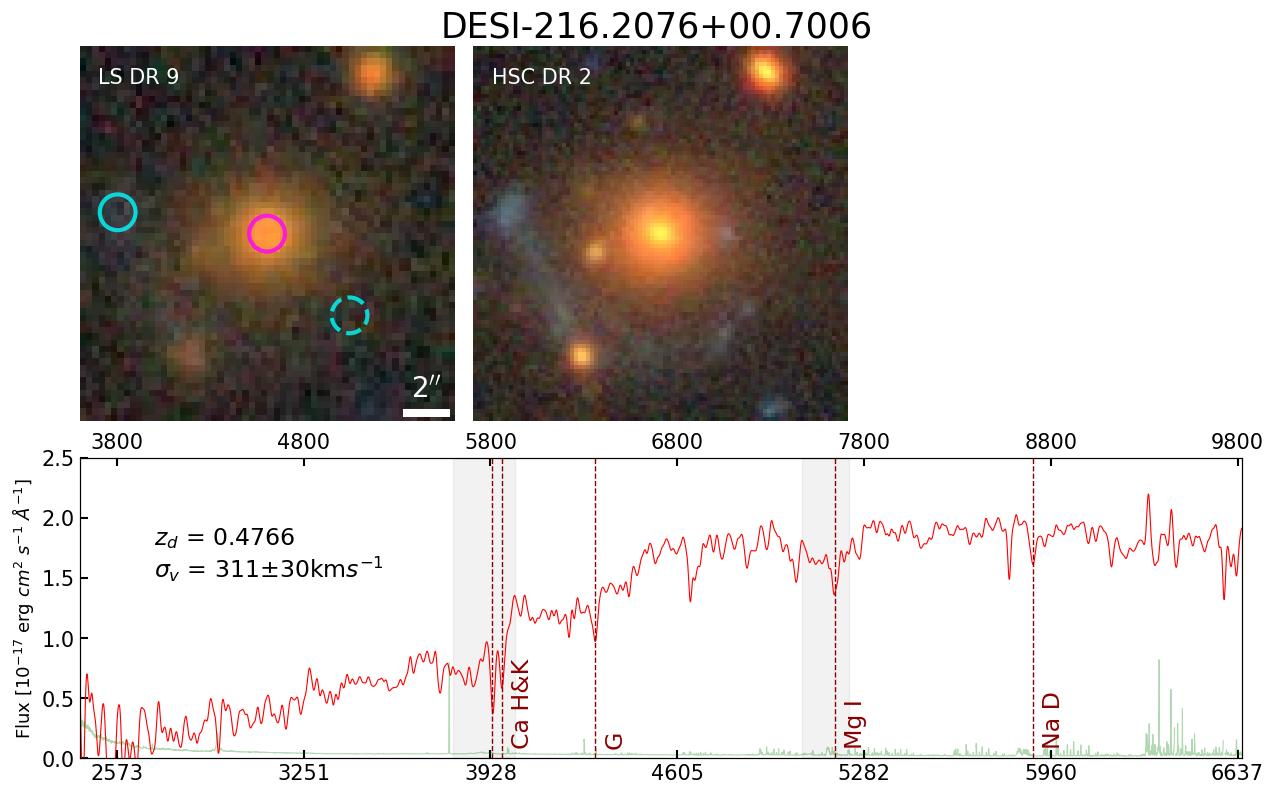}
\caption{DESI-216.2076+00.7006.
For the arrangement of the panels, see Figure~\ref{fig:desi149.6+01}.}
\label{fig:desi216+00}
\end{figure}

\begin{figure}[h]
  \centering
  \includegraphics[width=.7\textwidth]{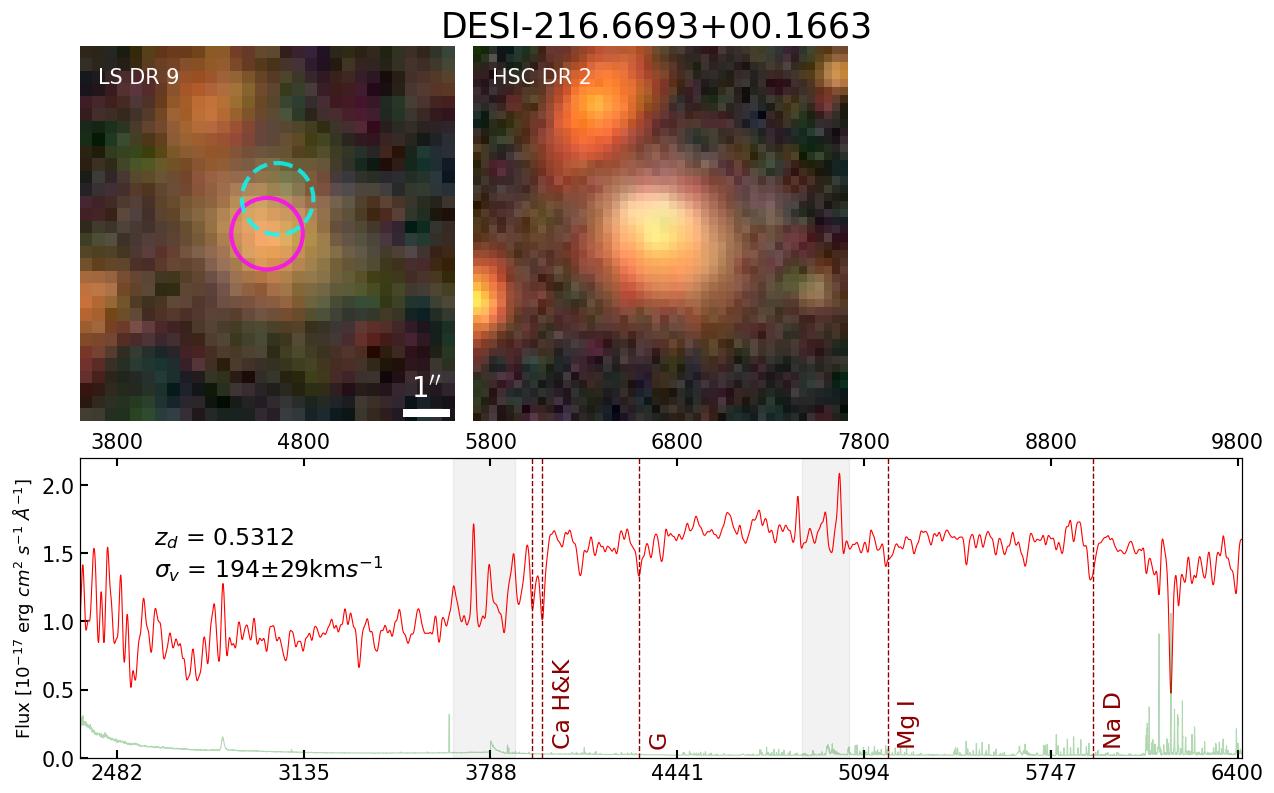}
  \caption{DESI-216.6693+00.1663. For the arrangement of the panels, see Figure~\ref{fig:desi149.6+01}.}
  \label{fig:desi216.7+00}
\end{figure}

\newpage 
\href{https://www.legacysurvey.org/viewer/?ra=216.2076&dec=+00.7006&layer=hsc-dr2&pixscale=0.262&zoom=16}{\emph{DESI-216.2076+00.7006}}\,
The putative lens is at a redshift of 0.4766 (Figure~\ref{fig:desi216+00}).
In \hsc, there are one blue arc-like object and another blue object that may be a ``straight arc''.  
DESI observed the latter \cwr{but failed to obtain its redshift,} and will target former.
\href{https://www.legacysurvey.org/viewer/?ra=216.6693&dec=+00.1663&zoom=16&layer=hsc-dr2}{\emph{DESI-216.6693+00.1663}} \,
The putative lens has a redshift of 0.5312 (Figure~\ref{fig:desi216.7+00}). 
In \hsc, a faint arc-like object can be seen just above the core of the putative lens. 
This object will be targeted by DESI.

\begin{figure}[h]
  \centering
  \includegraphics[width=.7\textwidth]{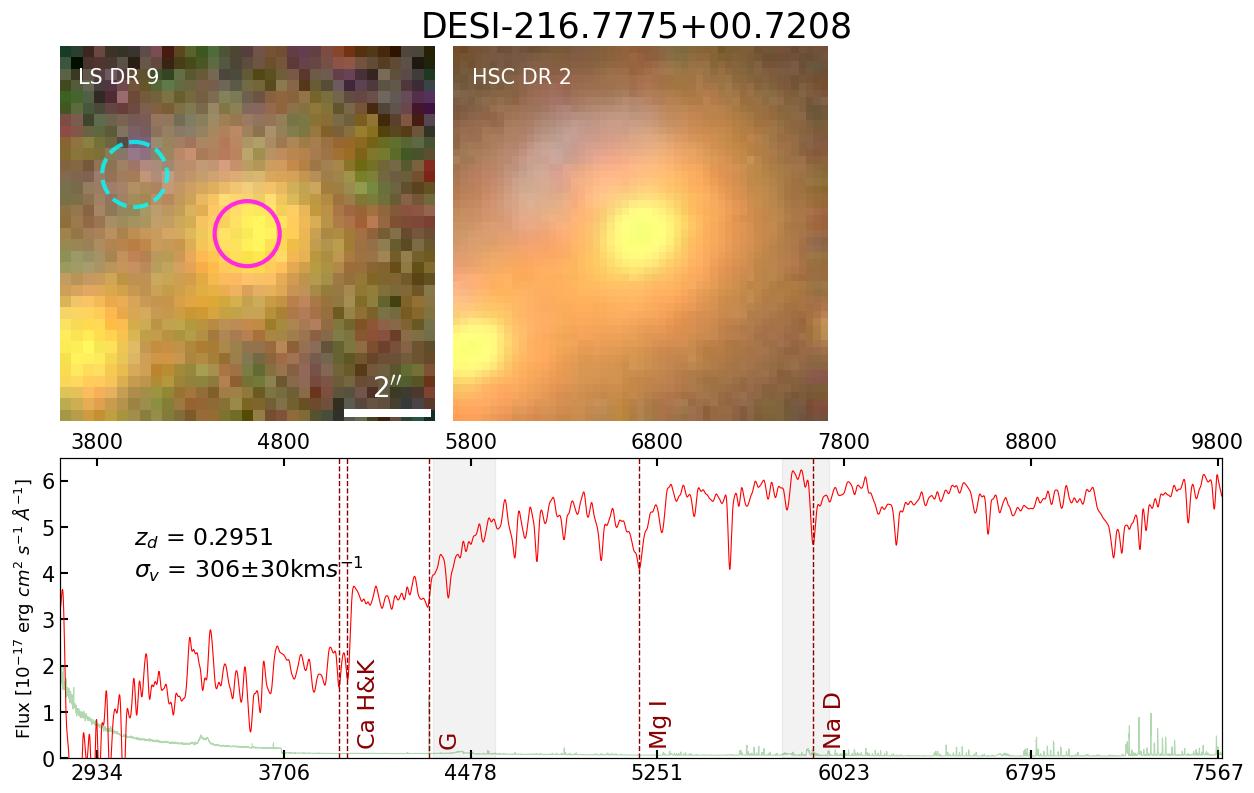}
  \caption{DESI-216.7775+00.7208. 
  For the arrangement of the panels, see Figure~\ref{fig:desi149.6+01}.}
  \label{fig:desi216.8+00}
\end{figure}

\newpage
\href{https://www.legacysurvey.org/viewer/?ra=216.7775&dec=+00.7208&zoom=16&layer=hsc-dr2}{\emph{DESI-216.7775+00.7208}} \,
The putative lens has a redshift of 0.2951 (Figure~\ref{fig:desi216.8+00}).
A long blue arc can be seen in the \hsc image (though not in the \ls image).
It will be observed by DESI.

\begin{figure}[h]
  \centering
  \includegraphics[width=.7\textwidth]{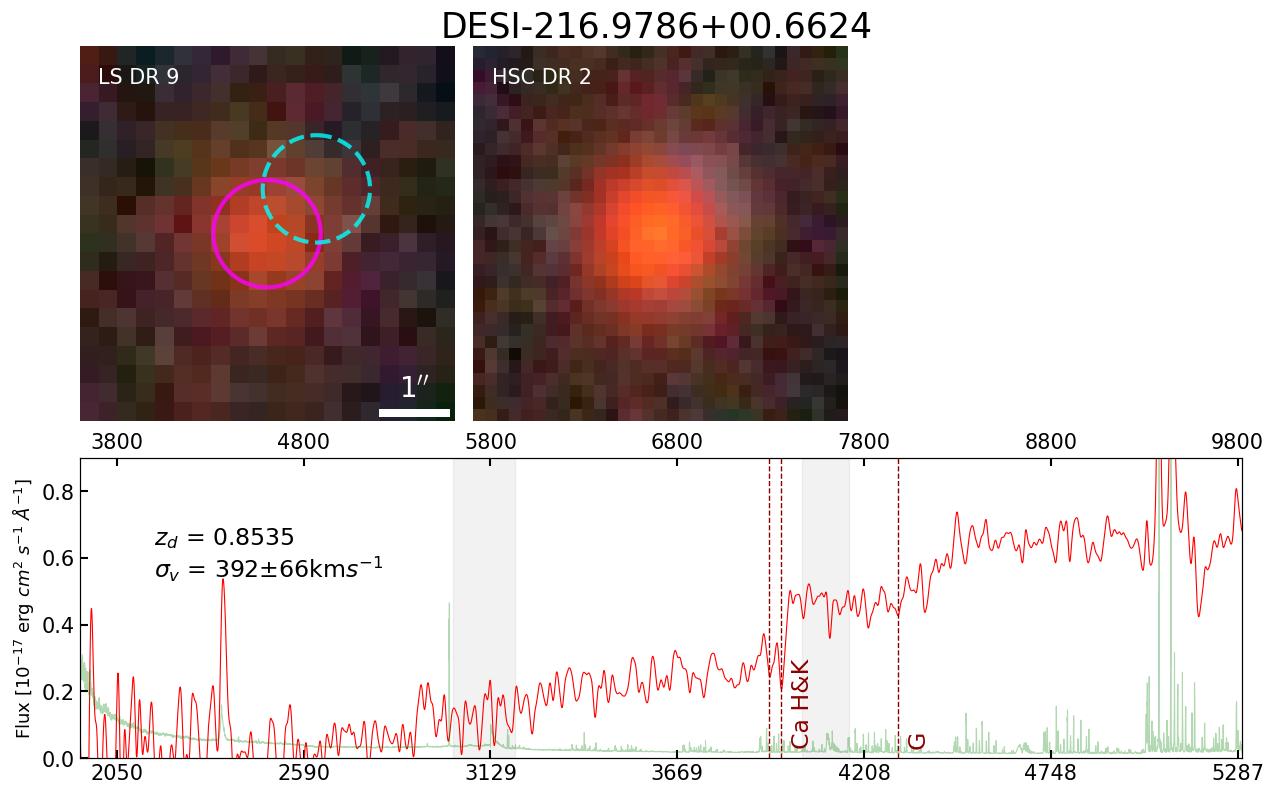}
  \caption{DESI-216.9786+00.6624. For the arrangement of the panels, see Figure~\ref{fig:desi149.6+01}.}
  \label{fig:desi216.9+00}
\end{figure}

\href{https://www.legacysurvey.org/viewer/?ra=216.9786&dec=+00.6624&zoom=16&layer=hsc-dr2}{\emph{DESI-216.9786+00.6624}} \,The putative lens is at a high redshift of 0.8535 (Figure~\ref{fig:desi216.9+00}).
A faint blue arc to the NW of the putative lens can be seen in the \hsc image (with weak evidence of it in the \ls image).
DESI will target this object.
On the other side, to the E of the putative lens, there is another faint blue image.

\begin{figure}[!h]
\centering
\includegraphics[width=.7\textwidth]{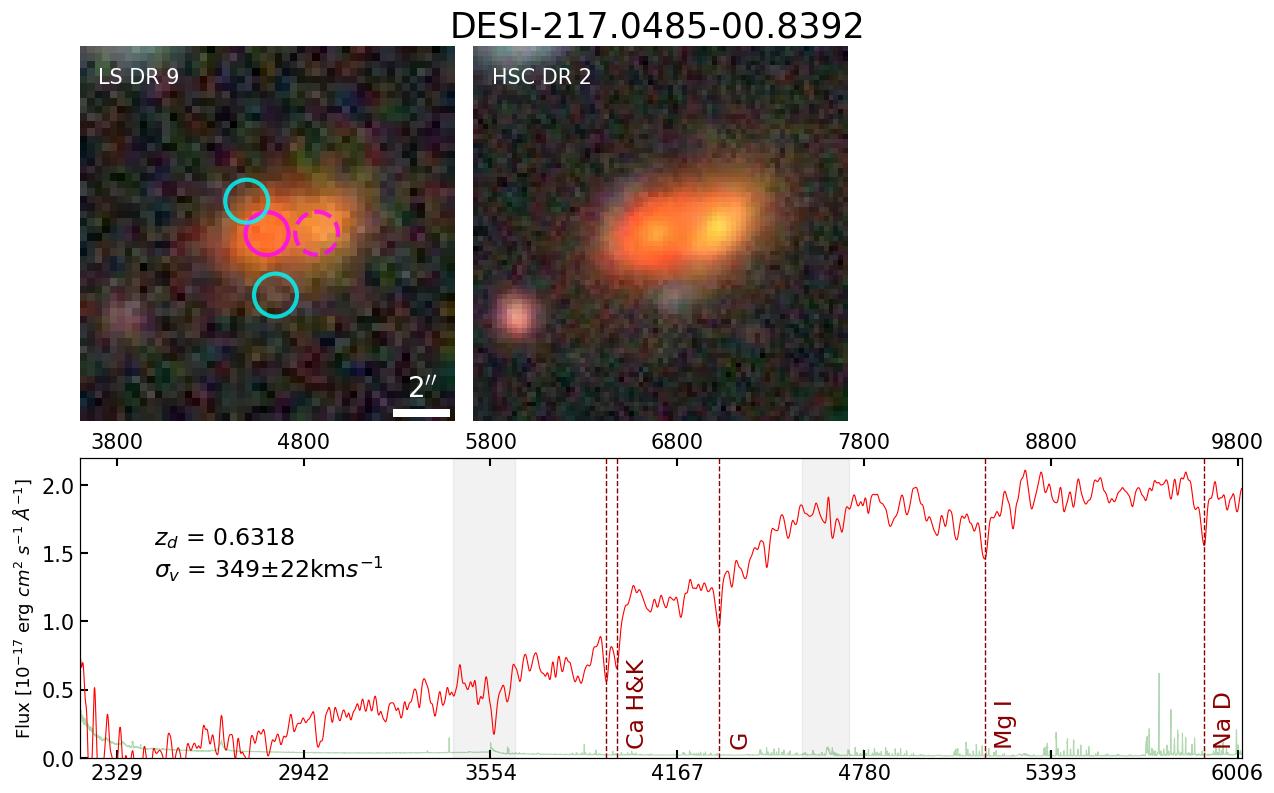}
\caption{DESI-217.0485-00.8392.
For the arrangement of the panels, see Figure~\ref{fig:desi149.6+01}.}
\label{fig:desi217-00}
\end{figure}

\newpage
\href{https://www.legacysurvey.org/viewer/?ra=217.0485&dec=-00.8392&zoom=16&layer=hsc-dr3}{\emph{DESI-217.0485-00.8392}}\,
The putative lens is at a redshift of 0.6318 (Figure~\ref{fig:desi217-00}).
In \hsc, there are two arc-like objects, with similar blue colors, above and below the putative lens.
The DESI \progname targeted the southern object.
DESI also targeted the northern object in its main program.
\cwr{Neither observations succeeded in obtaining redshifts.}
Note that there is another galaxy to the W of the putative lens (red object at the center of the cutout), with a less red color. 
It will be targeted by DESI, as a BGS target.

\begin{figure}[h]
  \centering
  \includegraphics[width=.7\textwidth]{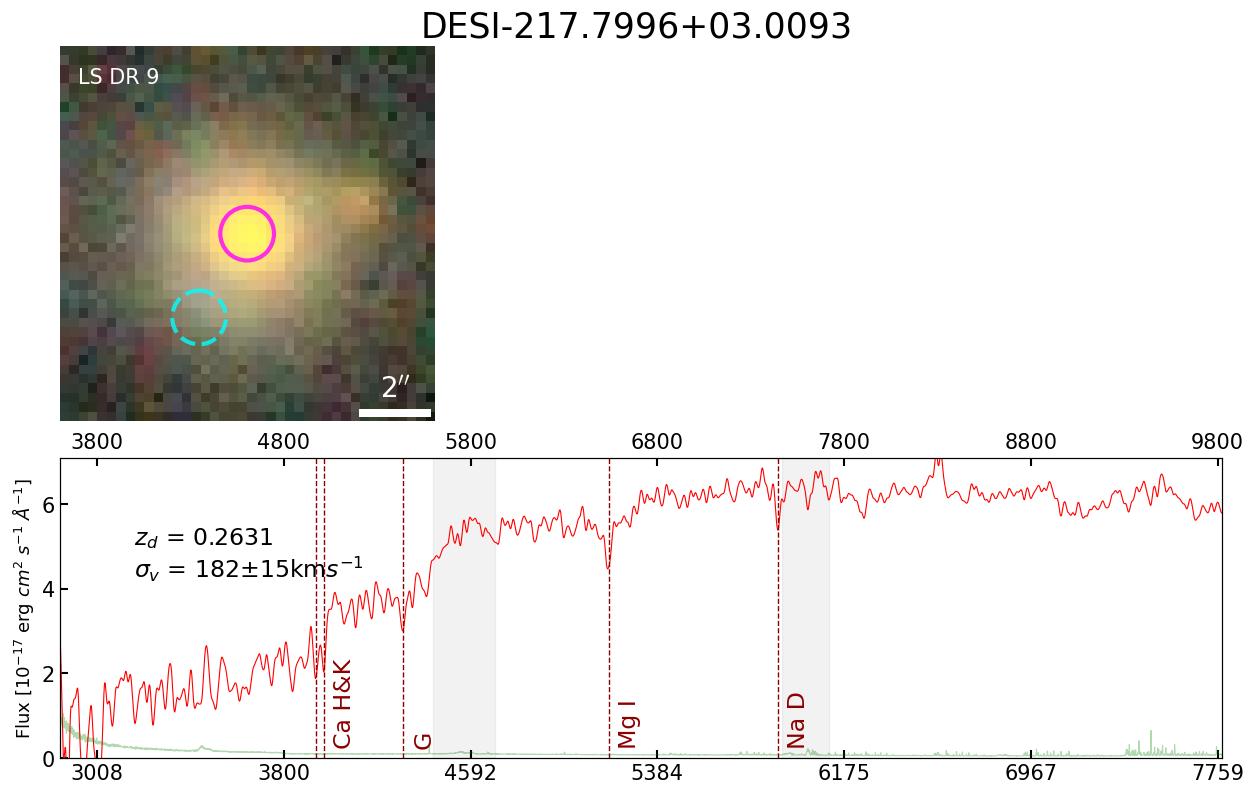}
  \caption{DESI-217.7996+03.0093.
  For the arrangement of the panels, see Figure~\ref{fig:desi149.6+01}.}
  \label{fig:desi217.8+03}
\end{figure}

\href{https://www.legacysurvey.org/viewer/?ra=217.7996&dec=+03.0093&zoom=16&layer=ls-dr9}{\emph{DESI-217.7996+03.0093}}\,
The putative lens is at 0.2631 (Figure~\ref{fig:desi217.8+03}).
There is a blue arc to the SE --- DESI will target this object --- and what appears to be a counterarc on the opposite side.

\begin{figure}[h]
  \centering
  \includegraphics[width=.7\textwidth]{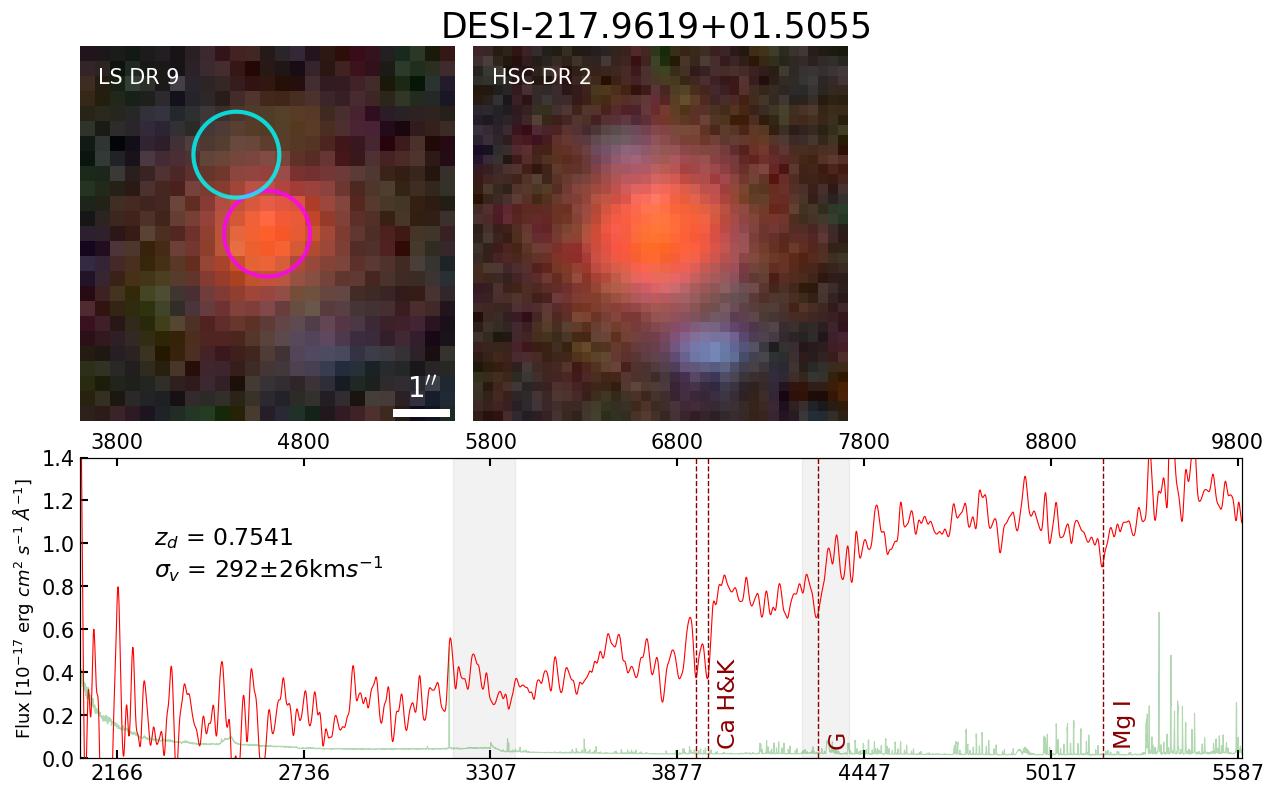}
  \caption{DESI-217.9619+01.5055.
  For the arrangement of the panels, see Figure~\ref{fig:desi149.6+01}.}
  \label{fig:DESI-217+01}
\end{figure}

\newpage
\href{https://www.legacysurvey.org/viewer/?ra=217.9619&dec=+01.5055&zoom=16&layer=hsc-dr3}{\emph{DESI-217.9619+01.5055}}\,
The putative lens is at a high redshift of 0.7541 (Figure~\ref{fig:DESI-217+01}).
In \hsc, there appears to be a faint blue arc that wraps around the putative lens from 11 o'clock to 5 o'clock. 
DESI targeted the brightest part of this arc \cwr{but failed to obtain its redshift}.

\begin{figure}[h]
  \centering
  \includegraphics[width=.7\textwidth]{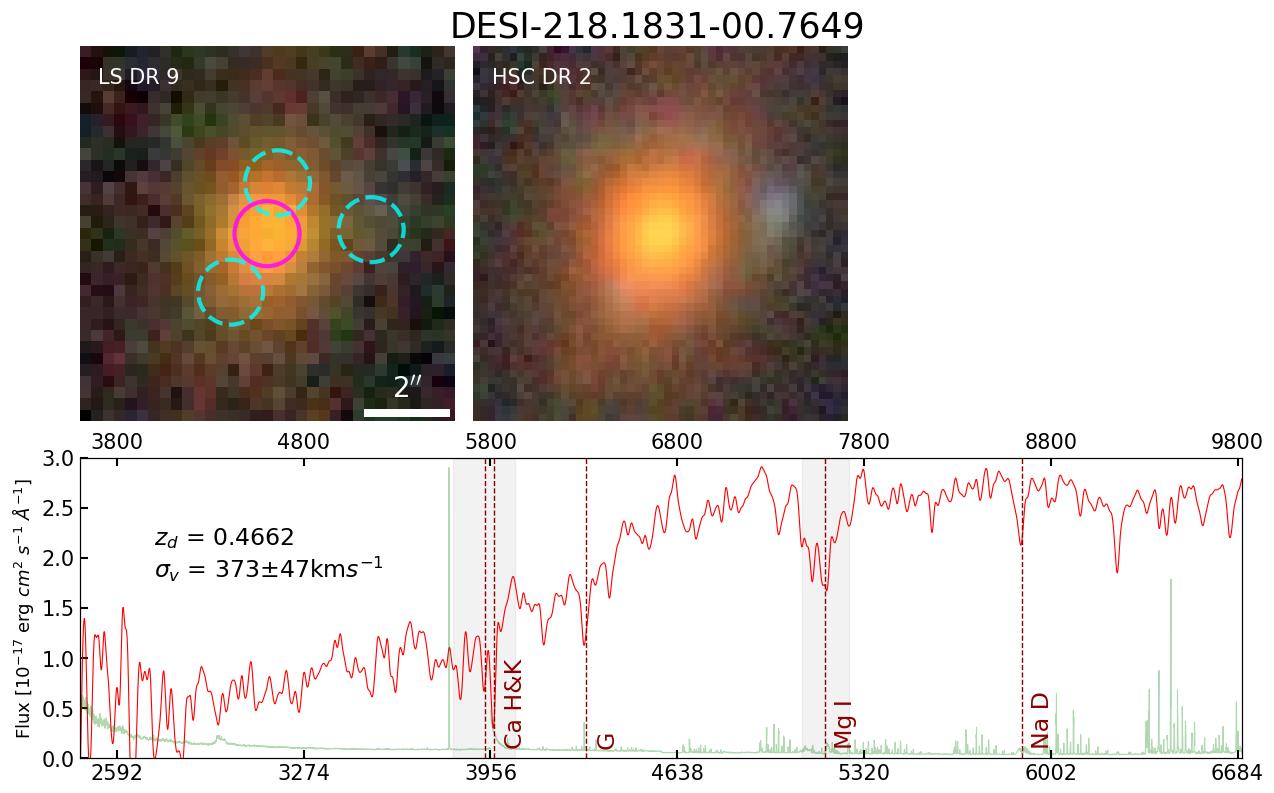}
  \caption{DESI-218.1831-00.7649. 
  For the arrangement of the panels, see Figure~\ref{fig:desi149.6+01}.}
  \label{fig:desi218.1-00}
\end{figure}

\href{https://www.legacysurvey.org/viewer/?ra=218.1831&dec=-00.7649&layer=hsc-dr2&pixscale=0.262&zoom=16}{\emph{DESI-218.1831-00.7649}}\,
The putative lens  is at a redshift of 0.4662 (Figure~\ref{fig:desi218.1-00}).
Three possible lensed arcs can be identified in the \hsc image around the putative lens:
1) a blue object to the W of the putative lens (also visible in the \ls image); 2) a faint greenish arc to the NW; and 3) a possible counterarc with a similar greenish hue to the SE.
The DESI \progname will observe objects 1 and 3.
Another DESI secondary program \citep[\texttt{UNWISE\_BLUE}; see][Table~16]{desi2023a} will target object 2.

\begin{figure}[h]
  \centering
  \includegraphics[width=.7\textwidth]{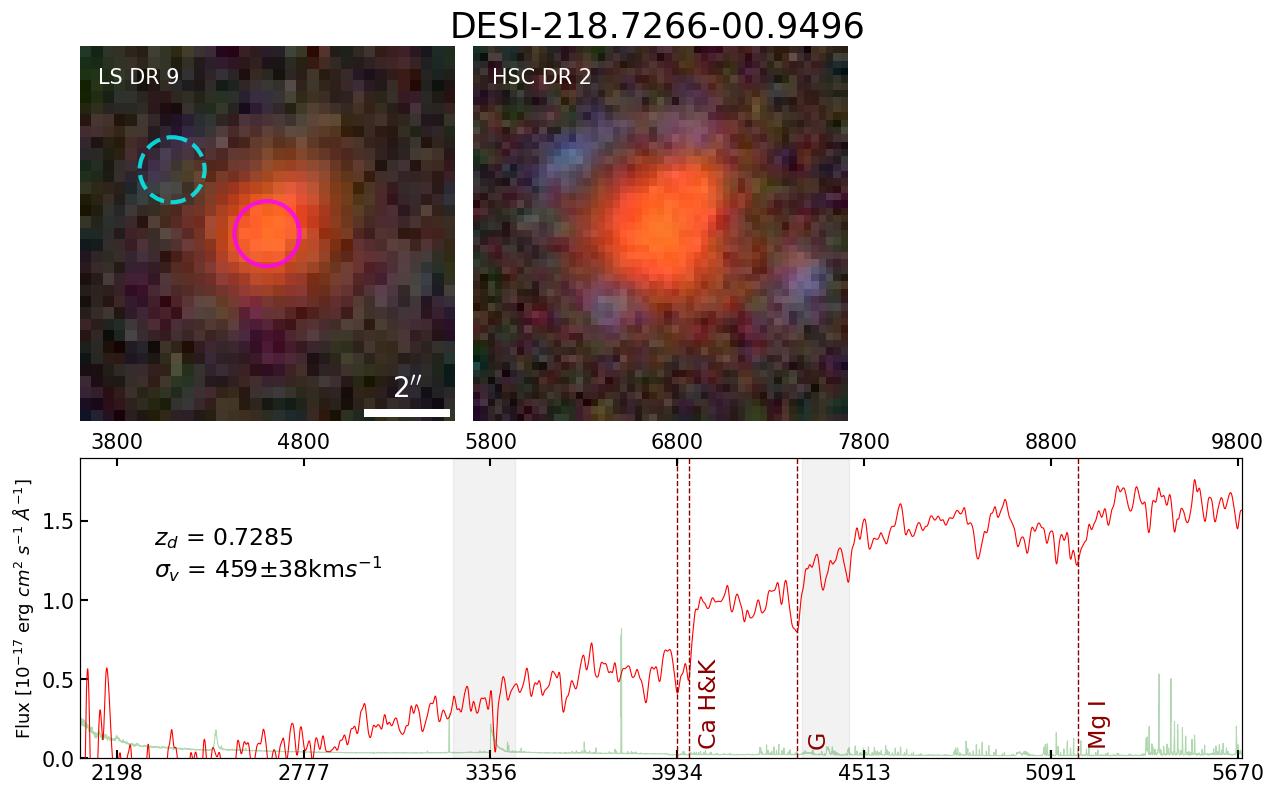}
  \caption{DESI-218.7266-00.9496. 
  For the arrangement of the panels, see Figure~\ref{fig:desi149.6+01}.}
  \label{fig:desi218.7-01}
\end{figure}

\newpage
\href{https://www.legacysurvey.org/viewer/?ra=218.7266&dec=-00.9496&layer=hsc-dr2&pixscale=0.262&zoom=16}{\emph{DESI-218.7266-00.9496}} \, 
The putative lens has a redshift of 0.7285 (Figure~\ref{fig:desi218.7-01}).
The \hsc image shows three images (fainter in the \ls image)  with similar blue colors around the putative lens.
DESI will target the brightest of these (the one to the NE). 

\begin{figure}[h]
  \centering
  \includegraphics[width=.7\textwidth]{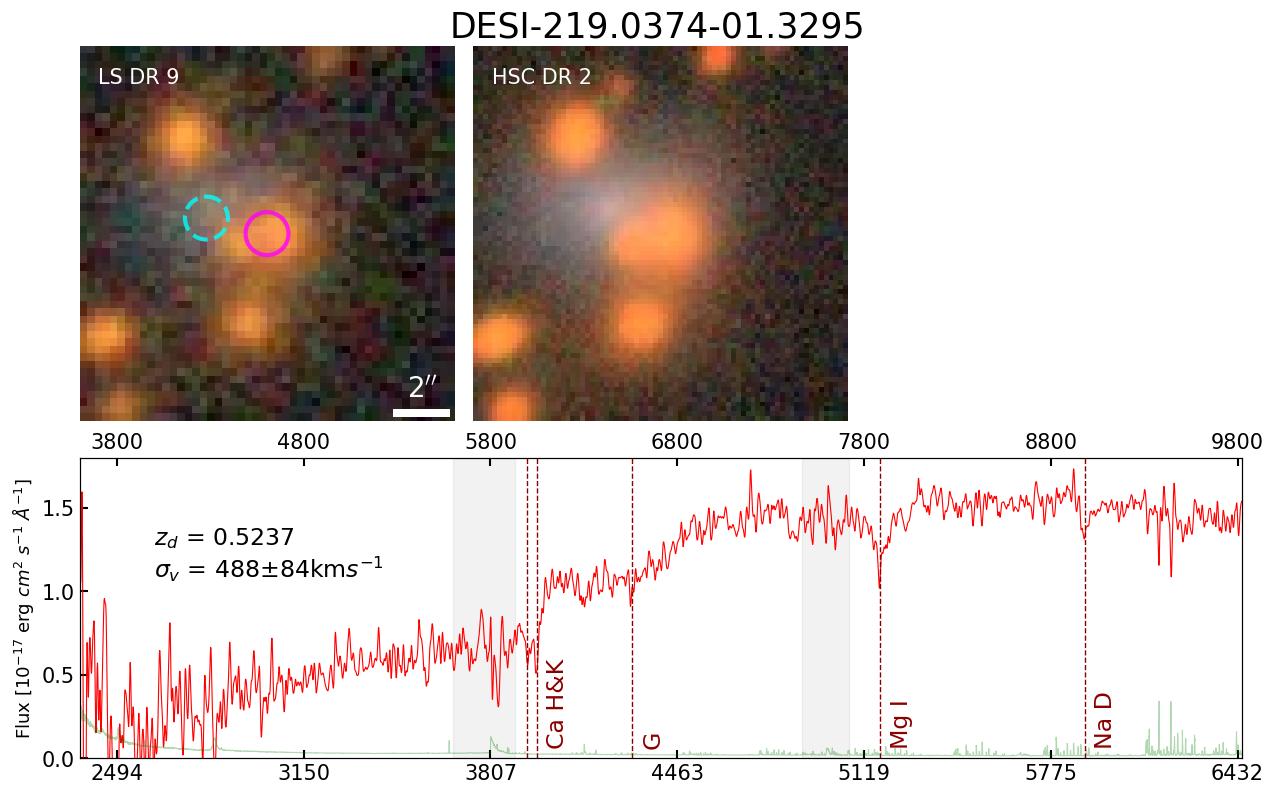}
  \caption{DESI-219.0374-01.3295. 
  For the arrangement of the panels, see Figure~\ref{fig:desi149.6+01}.}
  \label{fig:desi219-01}
\end{figure}

\href{https://www.legacysurvey.org/viewer/?ra=219.0374&dec=-01.3295&layer=hsc-dr2&pixscale=0.262&zoom=16}{\emph{DESI-219.0374-01.3295}} \,
The putative lens has a redshift of 0.5237 (Figure~\ref{fig:desi219-01}).
It appears to be part of a group.
From the \ls and \hsc images, there is a blue image next to one of the group members. DESI will target this object.

\begin{figure}[h]
\centering
\includegraphics[width=.7\textwidth]{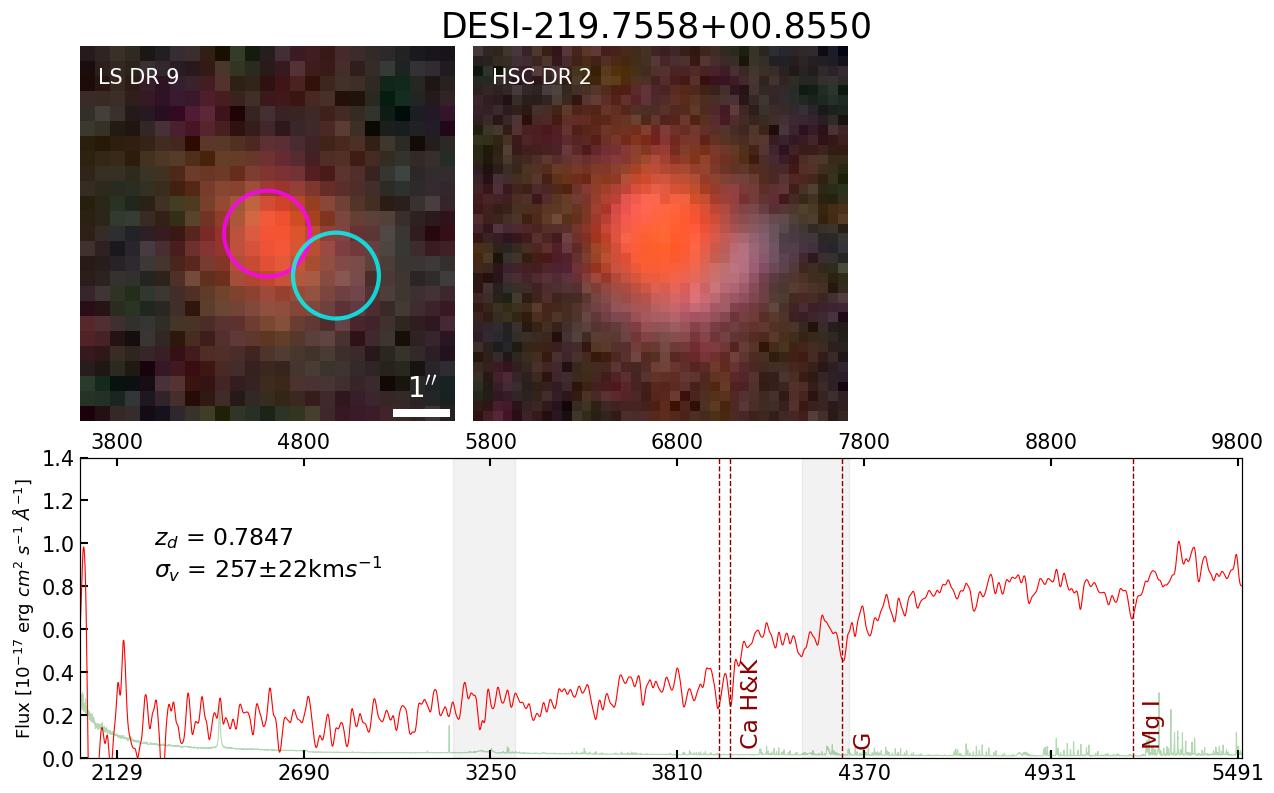}
\caption{DESI-219.7558+00.8550.
For the arrangement of the panels, see Figure~\ref{fig:desi149.6+01}.}
\label{fig:desi219+00}
\end{figure}

\newpage
\href{https://www.legacysurvey.org/viewer/?ra=219.7558&dec=+00.8550&zoom=16&layer=hsc-dr3}{\emph{DESI-219.7558+00.8550}}\,
The putative lens is at a high redshift of 0.7847 (Figure~\ref{fig:desi219+00}).
In \hsc, there is clearly a bluish arc to the SW (faintly visible in the \ls image).  DESI targeted this object \cwr{but failed to obtain its redshift}.

\begin{figure}[h]
  \centering
  \includegraphics[width=.7\textwidth]{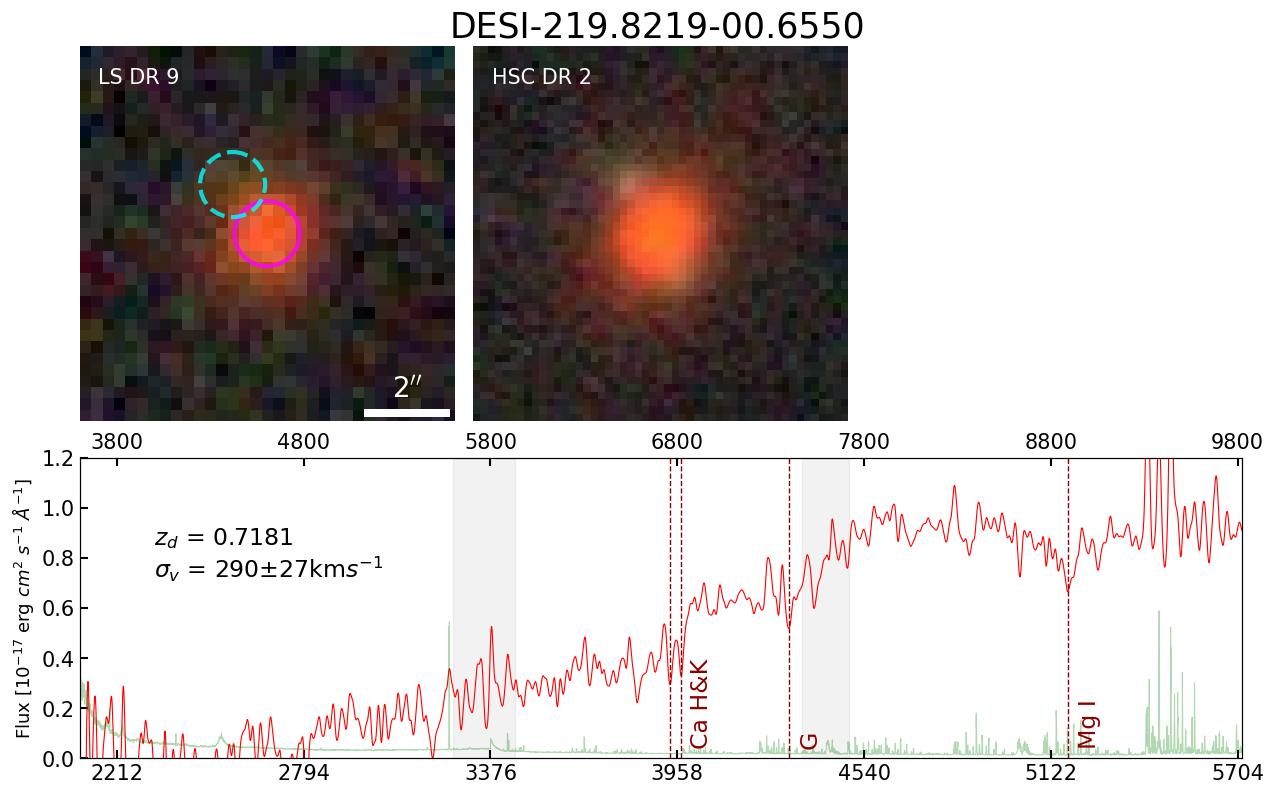}
  \caption{DESI-219.8219-00.6550. For the arrangement of the panels, see Figure~\ref{fig:desi149.6+01}.}
  \label{fig:desi219.8-00}
\end{figure}

 \href{https://www.legacysurvey.org/viewer/?ra=219.8219&dec=-00.6550&layer=hsc-dr2&pixscale=0.262&zoom=16}{\emph{DESI-219.8219-00.6550}} \,
 The putative lens is at z = 0.7181 (Figure~\ref{fig:desi219.8-00}).
There are a pair of putative lensed images with similar blue colors, to the NE and SW of the putative arc. 
The brighter (NE) arc will be observed by DESI.

\begin{figure}[h]
  \centering
  \includegraphics[width=.7\textwidth]{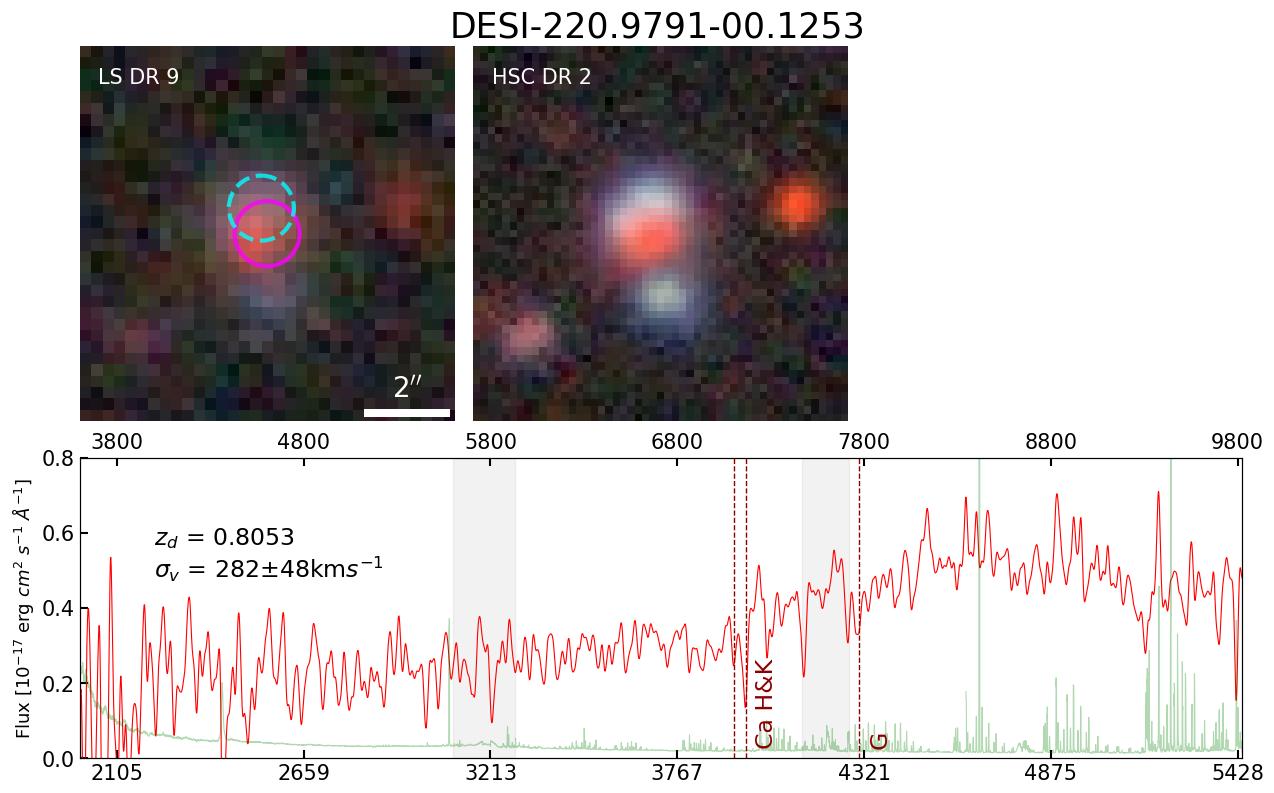}
  \caption{DESI-220.9791-00.1253.
  For the arrangement of the panels, see Figure~\ref{fig:desi149.6+01}.}
  \label{fig:desi220.9-00}
\end{figure}

\newpage
\href{https://www.legacysurvey.org/viewer/?ra=220.9791&dec=-00.1253&layer=hsc-dr2&pixscale=0.262&zoom=16}{\emph{DESI-220.9791-00.1253}}\, The putative lens for  has a high redshift of 0.8053 (Figure~\ref{fig:desi220.9-00}).
From the \hsc image, a pair of arc and counterarc, with similar blue colors,
can clearly be seen (both also visible in the \ls image).
It is worth noting that despite how close the lensed arc to the N is to the lens, 
the DESI spectrum does not appear to show any features belonging to an higher redshift object, possibly indicating the lensed source is at a redshift higher than 1.6.

\begin{figure}[h]
  \centering
  \includegraphics[width=.7\textwidth]{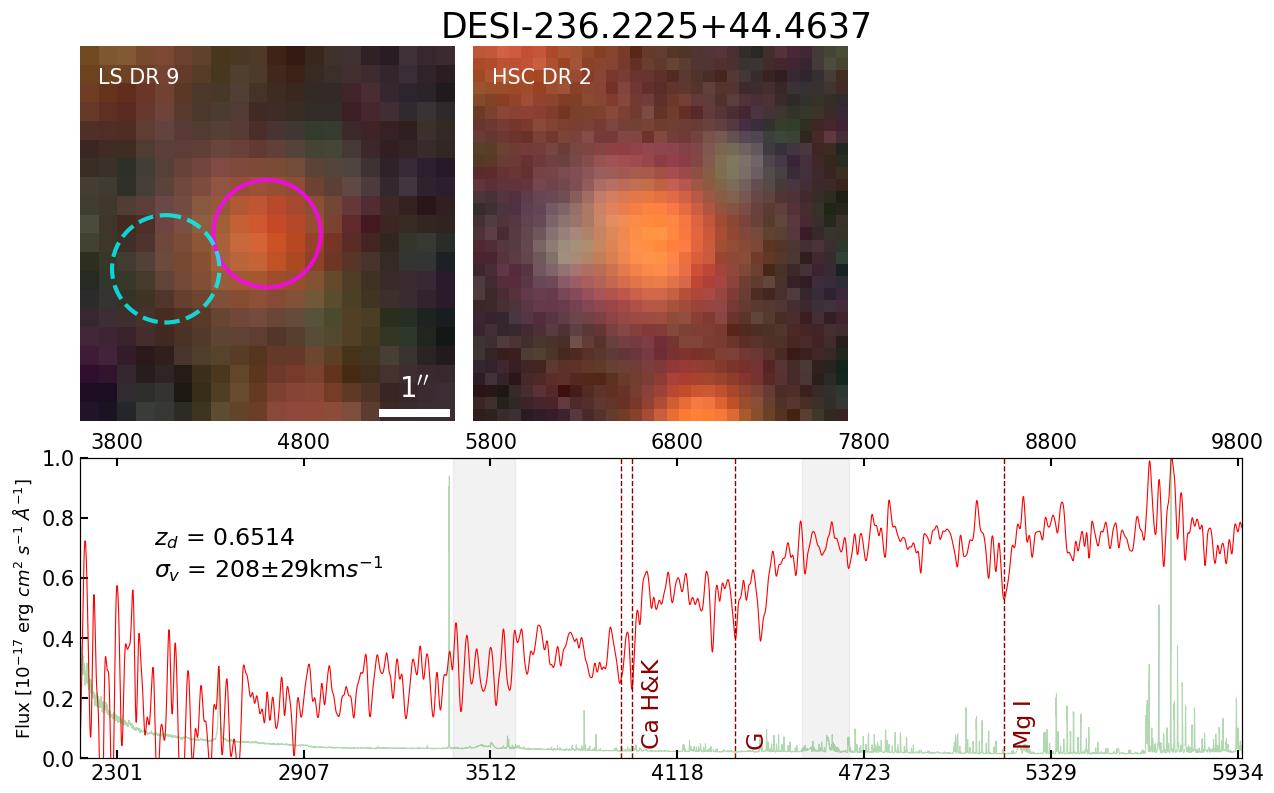}
  \caption{DESI-236.2225+44.4637.
  For the arrangement of the panels, see Figure~\ref{fig:desi149.6+01}.}
  \label{fig:desi236.2+44}
\end{figure}

\href{https://www.legacysurvey.org/viewer/?ra=236.2225&dec=+44.4637&layer=hsc-dr2&pixscale=0.262&zoom=16}{\emph{DESI-236.2225+44.4637}}\, The putative lens has a redshift of 0.6714 (Figure~\ref{fig:desi236.2+44}).
In the \hsc image, there are a pair of image, with similar blue colors, on either side of the putative lens (barely visible in the \ls image).
DESI will target the one to the E.

\begin{figure}[h]
  \centering
  \includegraphics[width=.7\textwidth]{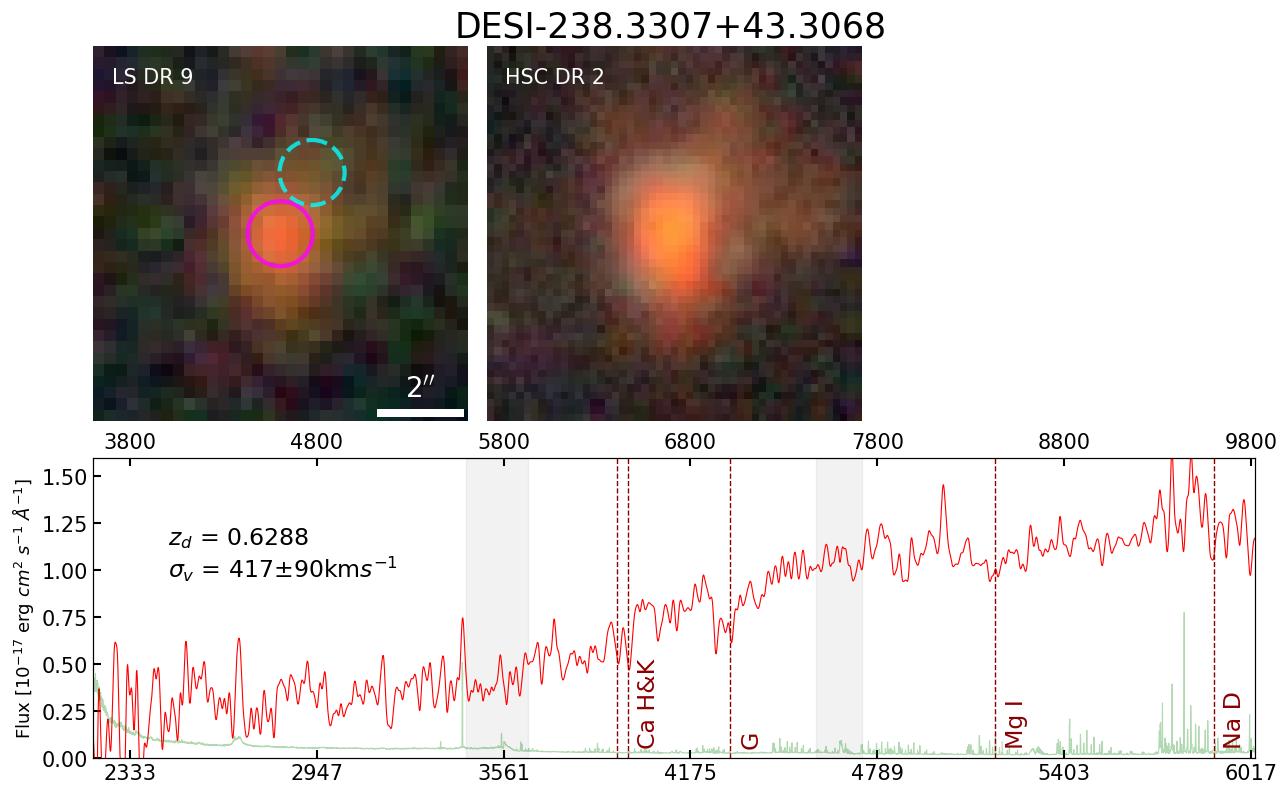}
  \caption{DESI-238.3307+43.3068. For the arrangement of the panels, see Figure~\ref{fig:desi149.6+01}.}
  \label{fig:desi238.3+43}
\end{figure}

\newpage
\href{https://www.legacysurvey.org/viewer/?ra=238.3307&dec=+43.3068&layer=hsc-dr2&pixscale=0.262&zoom=16}{\emph{DESI-238.3307+43.3068}} \, The putative lens has a redshift of 0.6288 (Figure~\ref{fig:desi238.3+43}).
From the \hsc image, a blue arc can be seen to the north, just N of the putative lens.
This arc will be targeted by DESI.
In addition, there is a pair of small blue images to the SE and SW, with similar blue colors.

\begin{figure}[h]
  \centering
  \includegraphics[width=.7\textwidth]{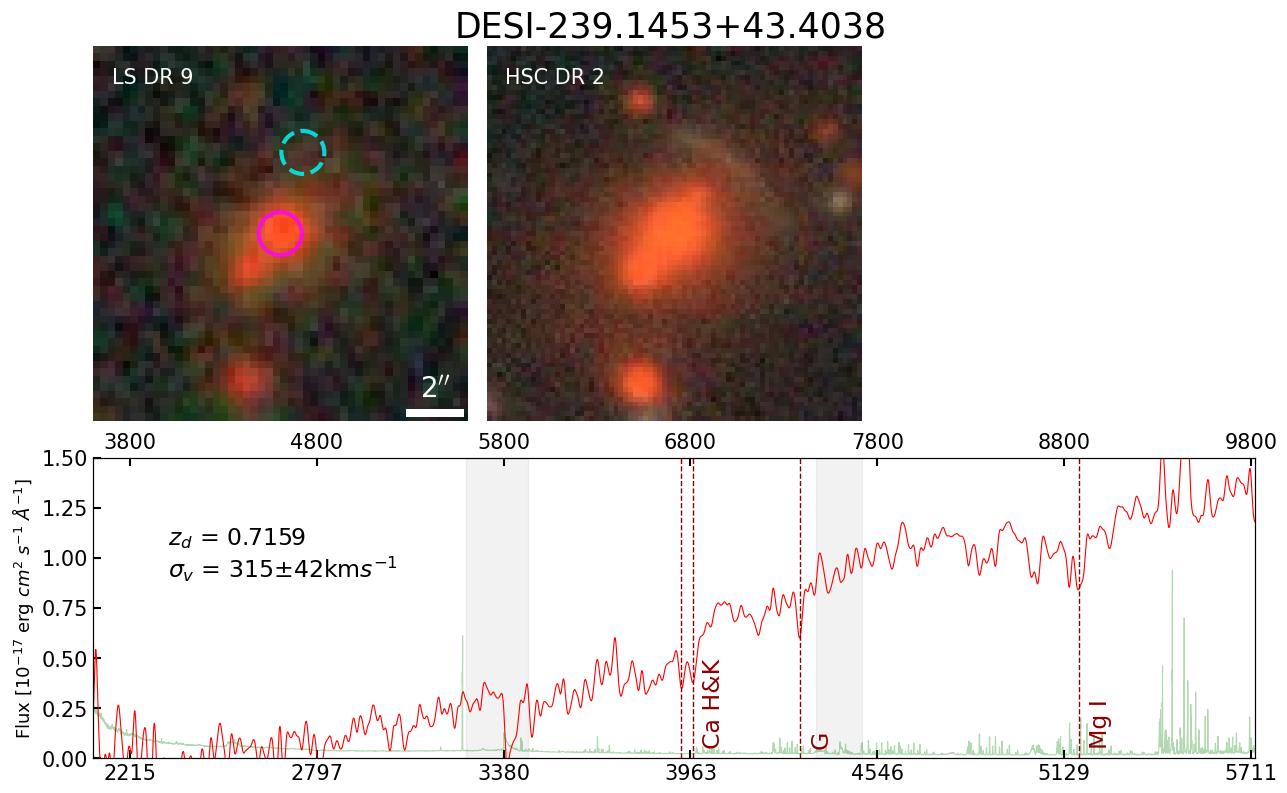}
  \caption{DESI-239.1453+43.4038. For the arrangement of the panels, see Figure~\ref{fig:desi149.6+01}.}
  \label{fig:desi239.1+43}
\end{figure}

\href{https://www.legacysurvey.org/viewer/?ra=239.1453&dec=+43.4038&layer=hsc-dr2&pixscale=0.262&zoom=16}{\emph{DESI-239.1453+43.4038}}\, The putative lens has a redshift of 0.7159 (Figure~\ref{fig:desi239.1+43}).
A long, bluish arc can be seen in the \hsc image (and faintly in the \ls image).
It will be targeted by DESI.

\newpage

\begin{figure}[!h]
\centering
\includegraphics[width=.7\textwidth]{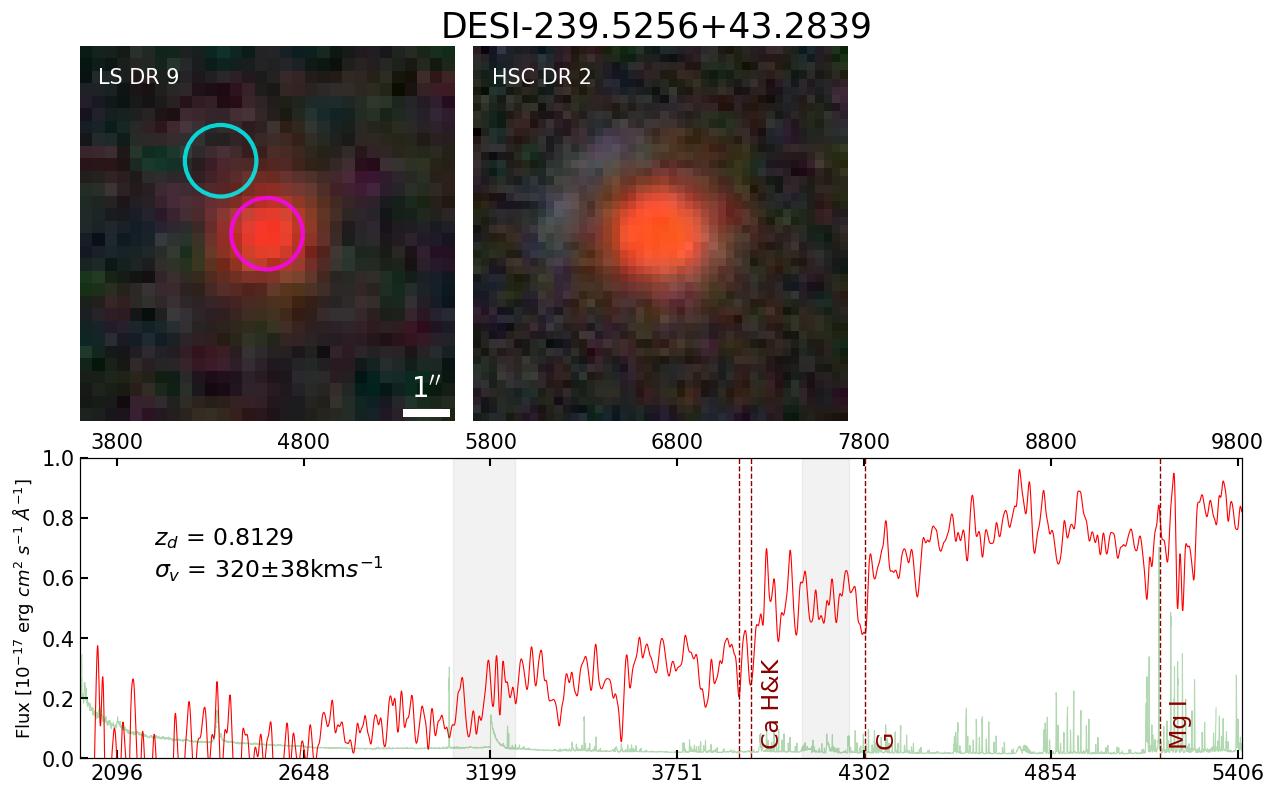}
\caption{DESI-239.5256+43.2839.
For the arrangement of the panels, see Figure~\ref{fig:desi149.6+01}.}
\label{fig:desi239+43}
\end{figure}

\href{https://www.legacysurvey.org/viewer/?ra= 239.5256&dec=+43.2839&layer=hsc-dr2&pixscale=0.262&zoom=16}{\emph{DESI-239.5256+43.2839}}\,
The putative lens is at a high redshift of 0.8129 (Figure~\ref{fig:desi239+43}).
In \hsc, there is clearly a blue arc to the NE (there is a hint of it in the \ls image). 
DESI targeted this object \cwr{but failed to obtain its redshift}.

\begin{figure}[h]
  \centering
  \includegraphics[width=.7\textwidth]{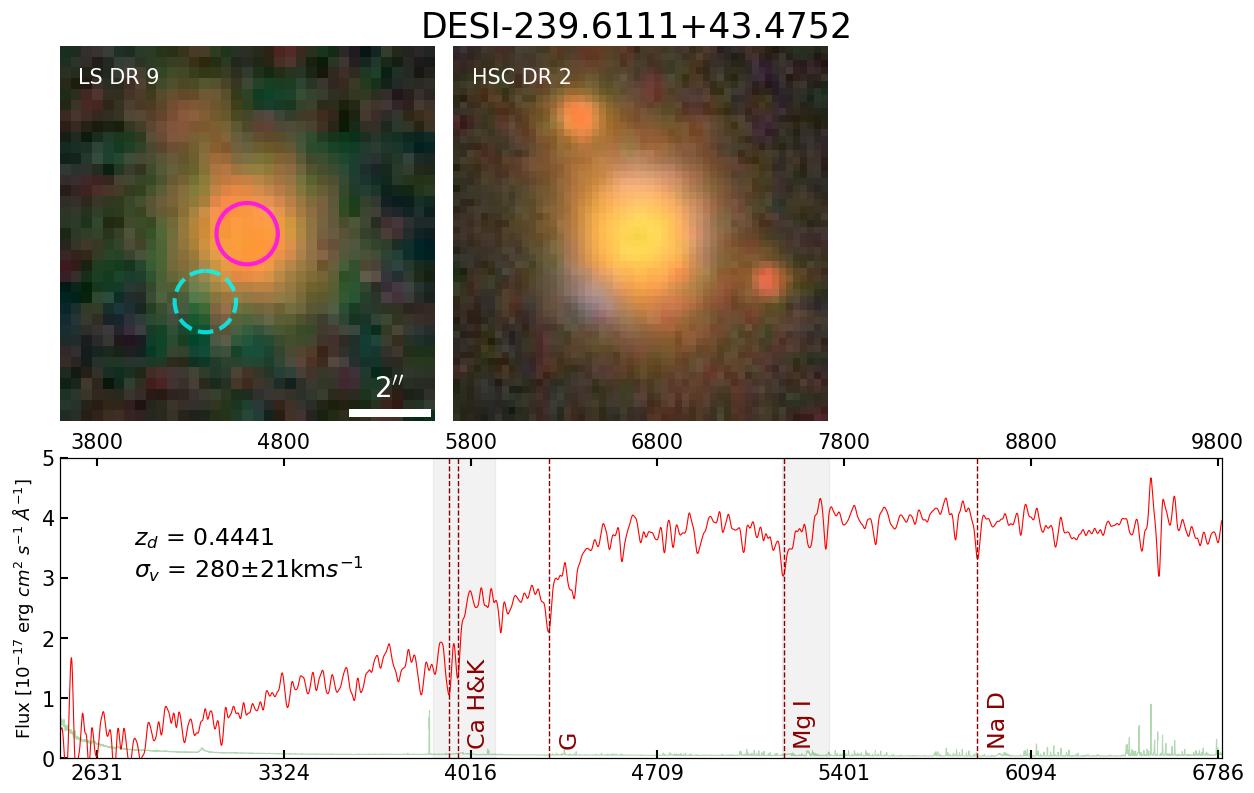}
  \caption{DESI-239.6111+43.4752. For the arrangement of the panels, see Figure~\ref{fig:desi149.6+01}.}
  \label{fig:desi239.6+43}
\end{figure}

\href{https://www.legacysurvey.org/viewer/?ra=239.6111&dec=+43.4752&layer=hsc-dr2&pixscale=0.262&zoom=16}{\emph{DESI-239.6111+43.4752}}\, The putative lens has a redshift of 0.4441 (Figure~\ref{fig:desi239.6+43}).
From the \hsc image, a small blue image can be seen to the SE, which DESI will target, with a hint of a counter image on the other side.

\newpage

\begin{figure}[h]
  \centering
  \includegraphics[width=.7\textwidth]{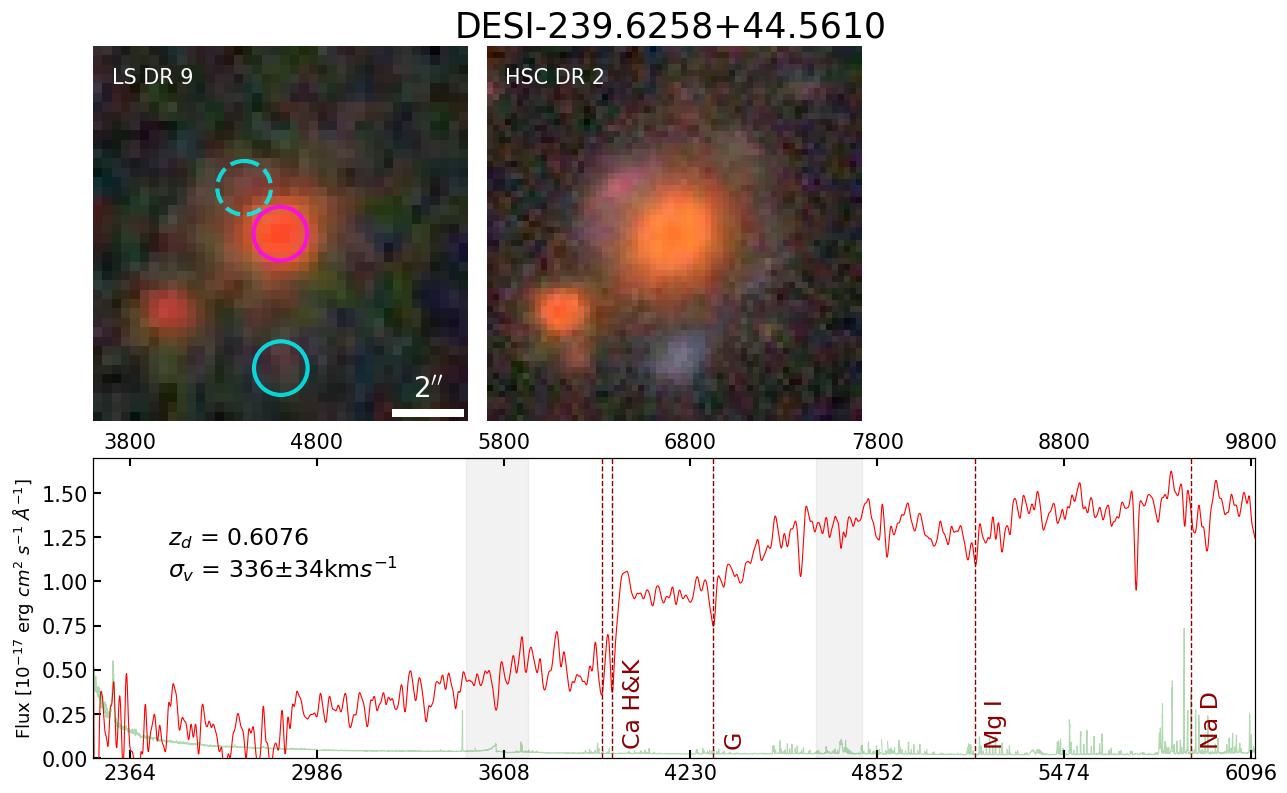}
  \caption{DESI-239.6258+44.5610. For the arrangement of the panels, see Figure~\ref{fig:desi149.6+01}.}
  \label{fig:desi239+44}
\end{figure}

\href{https://www.legacysurvey.org/viewer/?ra=239.6258&dec=+44.5610&layer=hsc-dr2&pixscale=0.262&zoom=16}{\emph{DESI-239.6258+44.5610}}\, The putative lens has a redshift of 0.6076 (Figure~\ref{fig:desi239+44}).
In the \hsc image, there is a bright purplish arc-like object to the NE. 
It will be targeted by DESI.
A faint blue arc-like object can be also seen to the W of the putative lens.
Finally the blue object to the S was observed as a DESI main target (ELG) and is in DESI~EDR, but the redshift is inconclusive.

\begin{figure}[h]
\centering
\includegraphics[width=.7\textwidth]{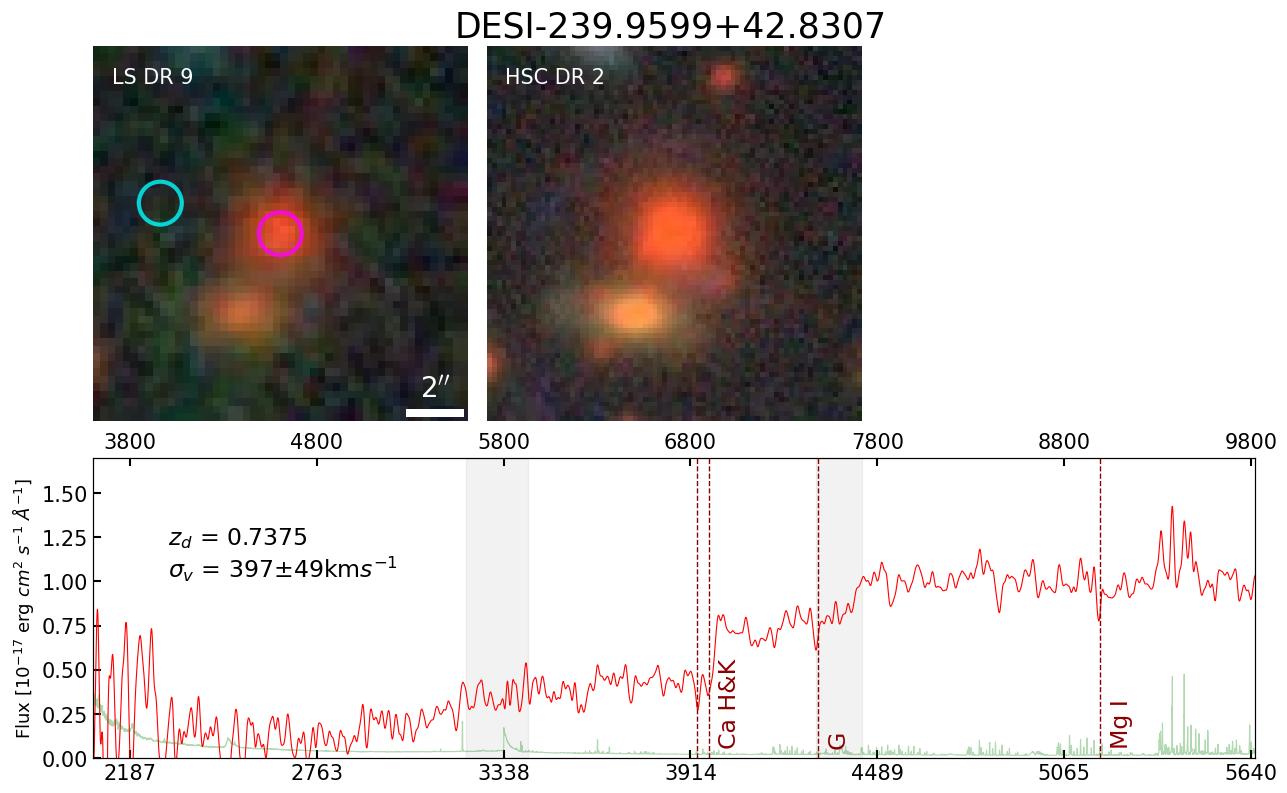}
\caption{DESI-239.9599+42.8307. 
For the arrangement of the panels, see Figure~\ref{fig:desi149.6+01}.
}
\label{fig:desi239+42}
\end{figure}

\href{https://www.legacysurvey.org/viewer/?ra=239.9599&dec=+42.8307&layer=hsc-dr2&pixscale=0.262&zoom=16}{\emph{DESI-239.9599+42.8307}} \,
The putative lens is at $z = 0.7375$ (Figure~\ref{fig:desi239+42}).
Though the ZWARN value is 2048, VI indicates that \rr best-fit redshift is correct.
DESI targeted a blue, faint arc-like object to the E \cwr{but failed to obtain its redshift.}

\begin{figure}[!h]
\centering
\includegraphics[width=.7\textwidth]{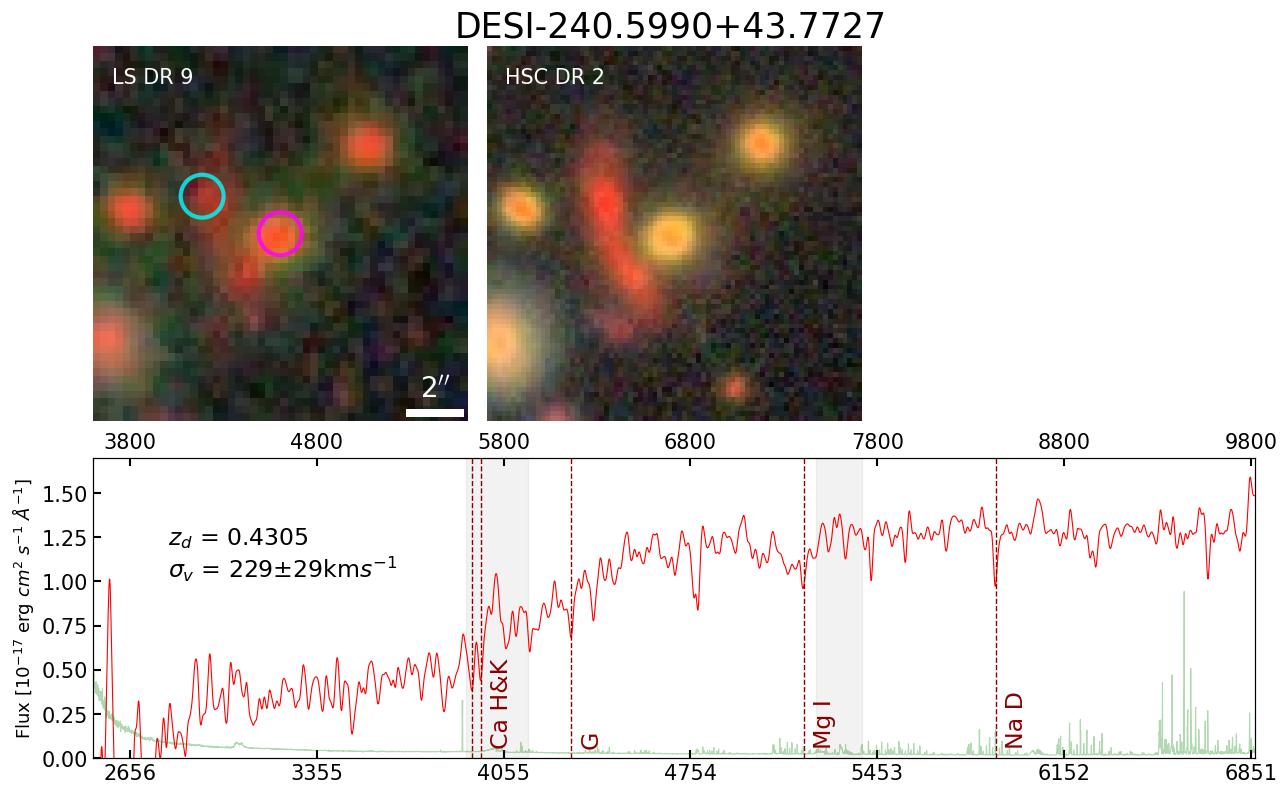}
\caption{DESI-240.5990+43.7727.
For the arrangement of the panels, see Figure~\ref{fig:desi149.6+01}.}
\label{fig:desi240+43.7727}
\end{figure}

\newpage
\href{https://www.legacysurvey.org/viewer/?ra=240.5990&dec=+43.7727&layer=hsc-dr2&pixscale=0.262&zoom=16}{\emph{DESI-240.5990+43.7727}}\, The putative lens is at a redshift of 0.4305 (Figure~\ref{fig:desi240+43.7727}).
In \hsc, there is a red arc to the E (visible also in the \ls image).  
DESI will target the brightest part of this object.

\begin{figure}[h]
  \centering
  \includegraphics[width=.7\textwidth]{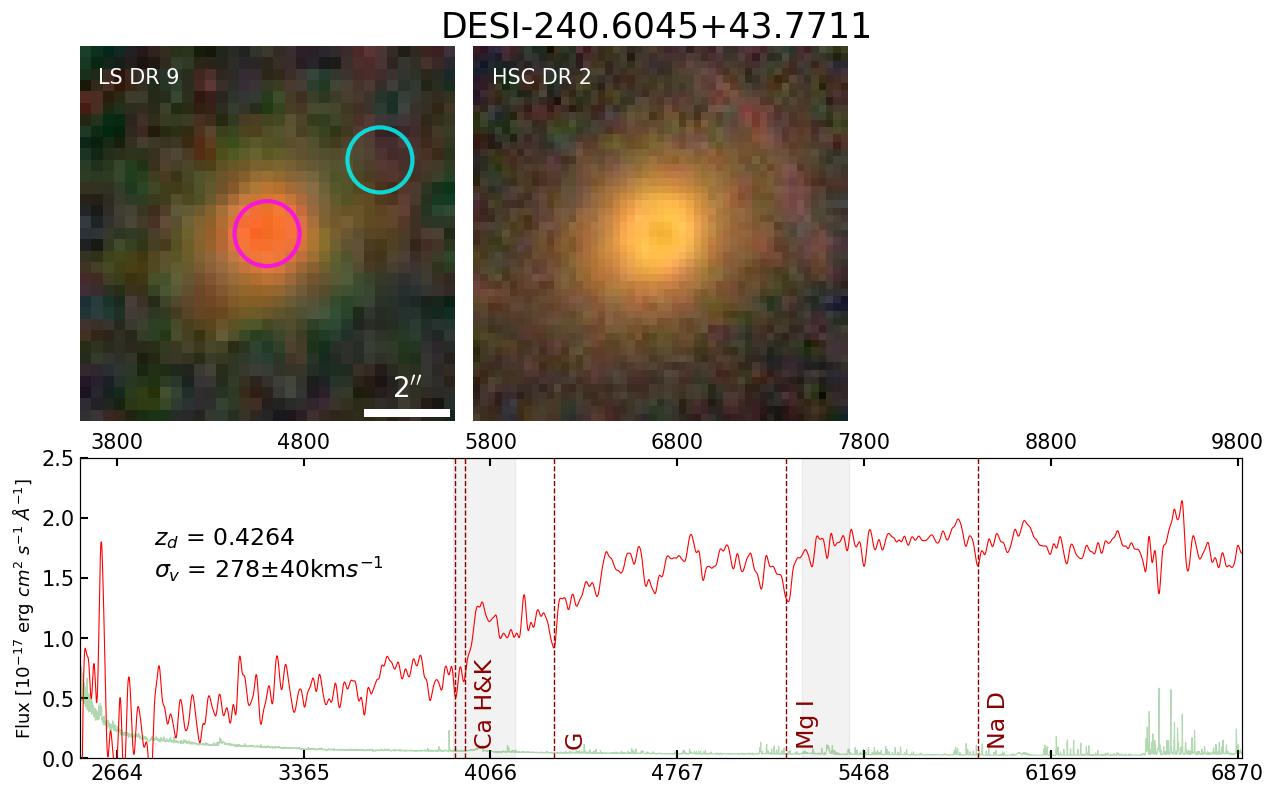}
  \caption{DESI-240.6045+43.7711.
  For the arrangement of panels see Figure~\ref{fig:desi149.6+01}.
  }
  \label{fig:desi240.6+43}
\end{figure}

\href{https://www.legacysurvey.org/viewer/?ra=240.6045&dec=+43.7711&layer=hsc-dr2&pixscale=0.262&zoom=16}{\emph{DESI-240.6045+43.7711}}\, The putative lens is at a redshift of 0.4264 (Figure~\ref{fig:desi240.6+43}).
In \hsc, there is a giant red arc to the NW (faintly visible in the \ls image).  
DESI targeted this object \cwr{but failed to obtain its redshift}.

\newpage

\begin{figure}[h]
  \centering
  \includegraphics[width=.7\textwidth]{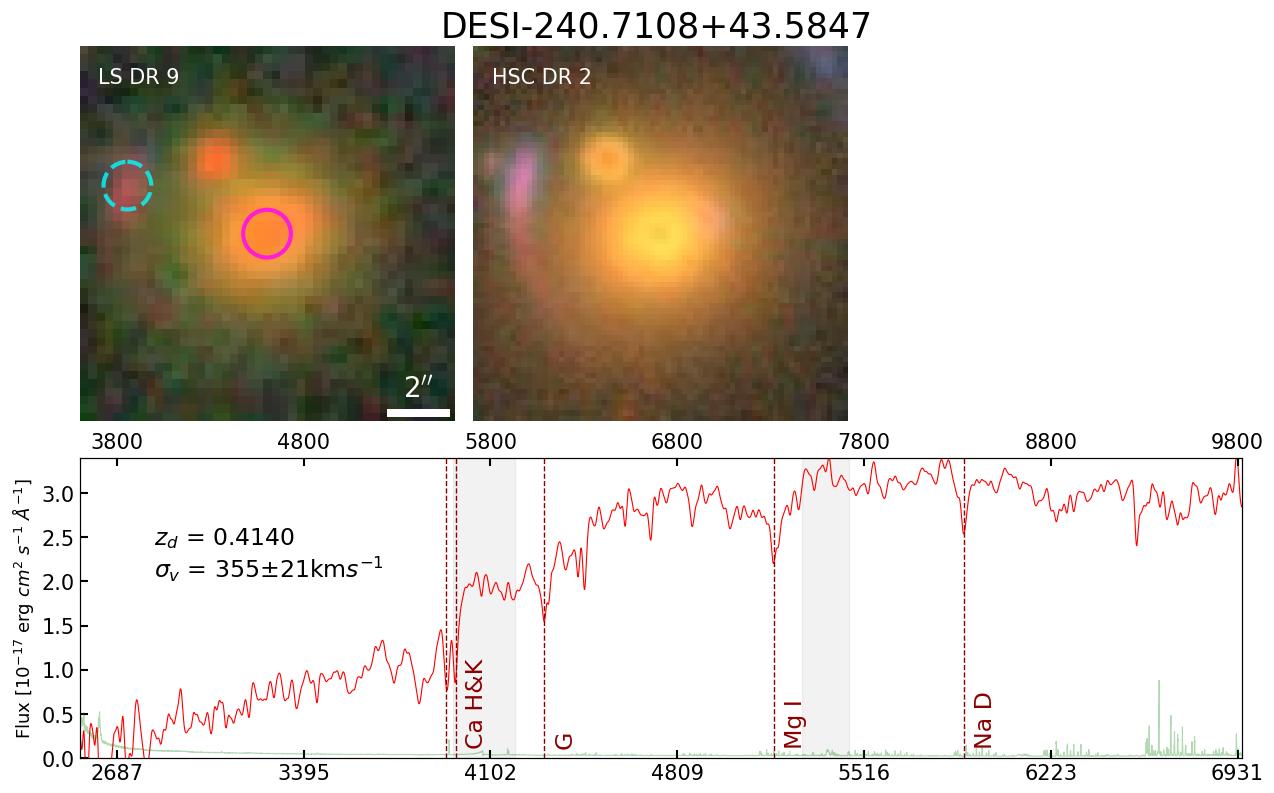}
  \caption{DESI-240.7108+43.5847. For the arrangement of the panels, see Figure~\ref{fig:desi149.6+01}.}
  \label{fig:desi240+43}
\end{figure}

\href{https://www.legacysurvey.org/viewer/?ra=240.7108&dec=+43.5847&layer=hsc-dr2&pixscale=0.262&zoom=16}{\emph{DESI-240.7108+43.5847}}\, The putative lens has a redshift of 0.4140 (Figure~\ref{fig:desi240+43}).
In the \hsc image, a large purplish arc can be seen to the E (also visible in the \ls image).
This object will be targeted by DESI.

\begin{figure}[h]
\centering
\includegraphics[width=.7\textwidth]{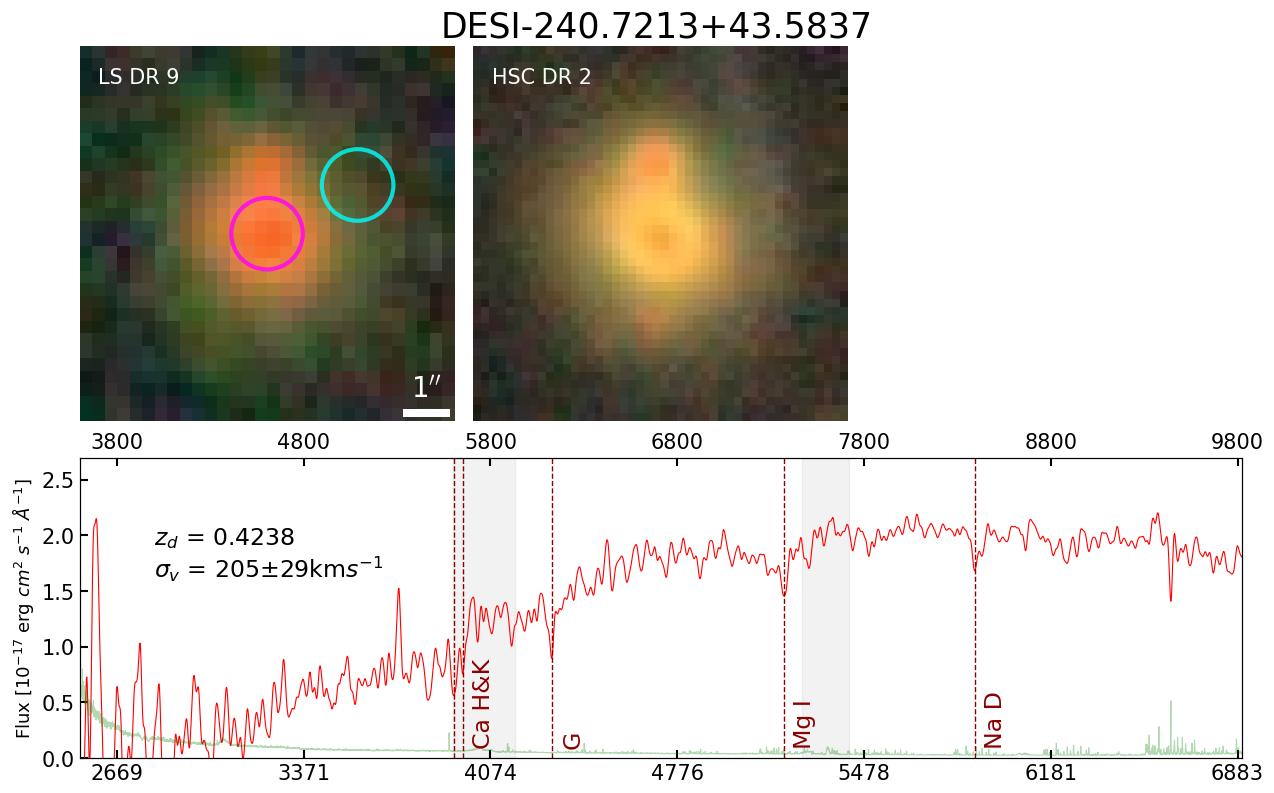}
\caption{DESI-240.7213+43.5837.
For the arrangement of the panels, see Figure~\ref{fig:desi149.6+01}.}
\label{fig:desi240.7+43}
\end{figure}

\href{https://www.legacysurvey.org/viewer/?ra=240.7213&dec=+43.5873&layer=hsc-dr2&pixscale=0.262&zoom=16}{\emph{DESI-240.7213+43.5873}}\,
The putative lens is at a redshift of 0.4238 (Figure~\ref{fig:desi240.7+43}).
In \hsc, there is a faint arc-like object to the NW.
DESI targeted this object \cwr{but failed to obtain its redshift}.

\begin{figure}[h]
  \centering
  \includegraphics[width=.7\textwidth]{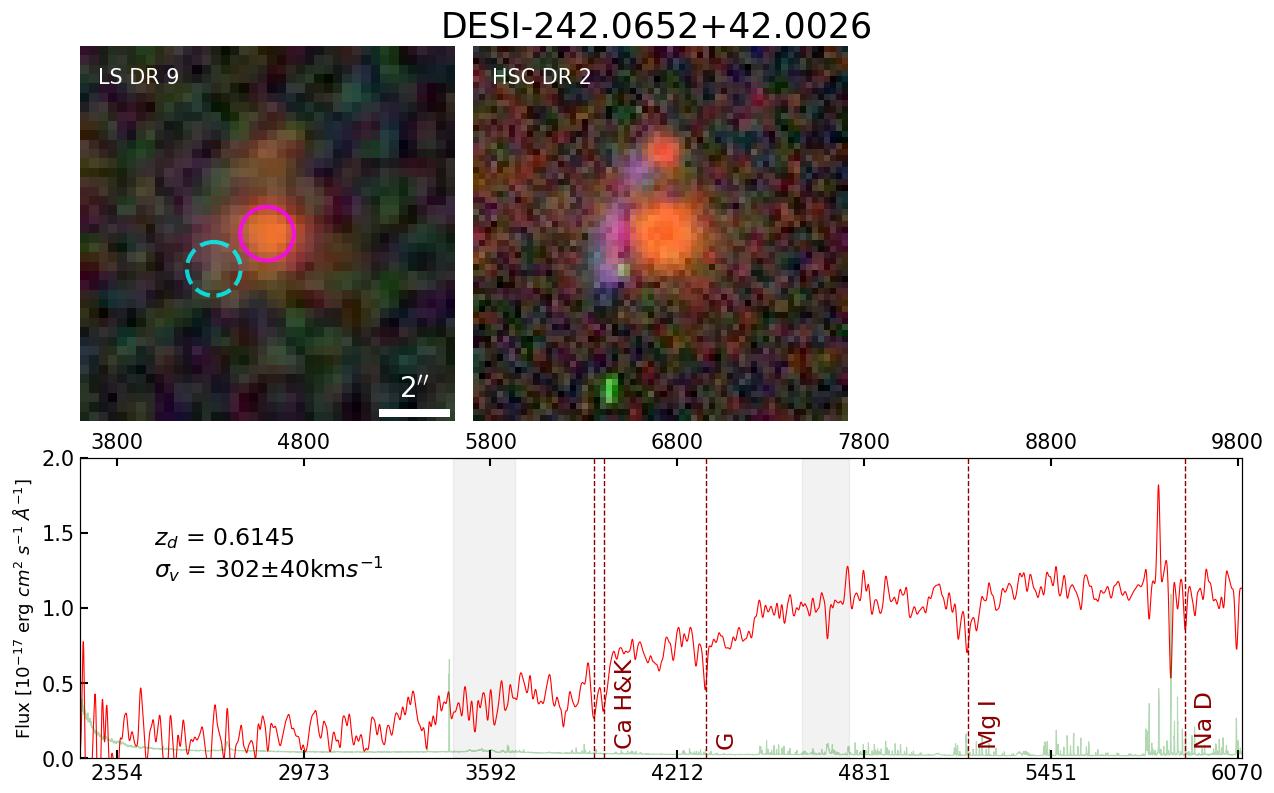}
  \caption{DESI-242.0652+42.0026.
  For the arrangement of the panels, see Figure~\ref{fig:desi149.6+01}.}
  \label{fig:desi242+42}
\end{figure}

\newpage
\href{https://www.legacysurvey.org/viewer/?ra=242.0652&dec=+42.0026&layer=hsc-dr2&pixscale=0.262&zoom=16}{\emph{DESI-242.0652+42.0026}}\, The putative lens is at $z = 0.6145$ (Figure~\ref{fig:desi242+42}).
The \ls image shows an blue arc to the NE, which can also be seen in the \hsc image.
This object will be targeted by DESI.

\begin{figure}[h]
\centering
\includegraphics[width=.7\textwidth]{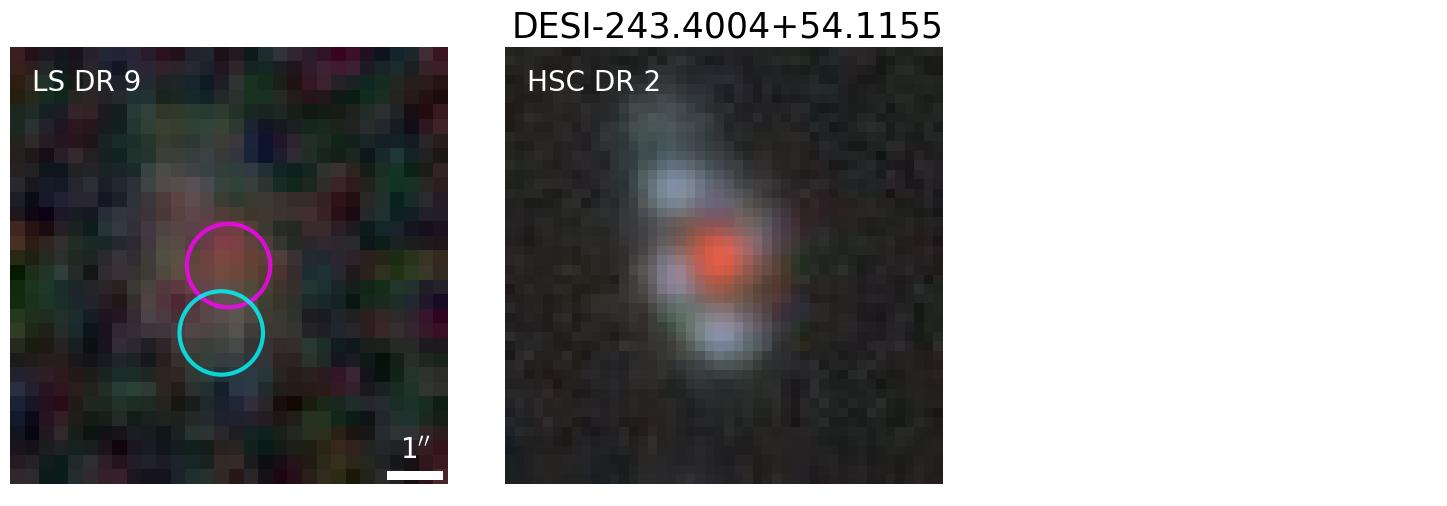}
\caption{DESI-243.4004+54.1155.
 We only show the \ls (left) and \hsc (right) images because the putative lens was observed by DESI but the redshift was inconclusive, whereas the putative source has not yet been observed by DESI. 
}
\label{fig:desi243+54}
\end{figure}

\href{https://www.legacysurvey.org/viewer/?ra=243.4004&dec=+54.1155&layer=hsc-dr2&pixscale=0.262&zoom=16}{\emph{DESI-243.4004+54.1155}}\, 
Given the way in which four blue images with very similar colors are arranged around a red galaxy, 
this is almost certainly a strong lensing system (Figure~\ref{fig:desi243+54}).
DESI will target the brightest of these four.
This is the only case where the putative lens was observed by DESI, 
but the redshift was inconclusive. 
The photo-$z$ for the putative lens is $0.974 \pm 0.148$. 
Given that photometric redshifts for elliptical galaxies, which this appears to be one, are generally reliable, 
this is almost certainly another high redshift lensing galaxy (see also DESI\cwr{~J}212.9021-01.0377 in \S~\ref{sec:confirmed-lenses}).
It is possible that deeper optical spectroscopic observation will reveal clear features (e.g., Ca H\&K and the 4000 \ang break) to determine its redshift. 
But it may be the case that NIR spectroscopy is necessary, if the redshift is too high.

\begin{figure}[h]
  \centering
  \includegraphics[width=.7\textwidth]{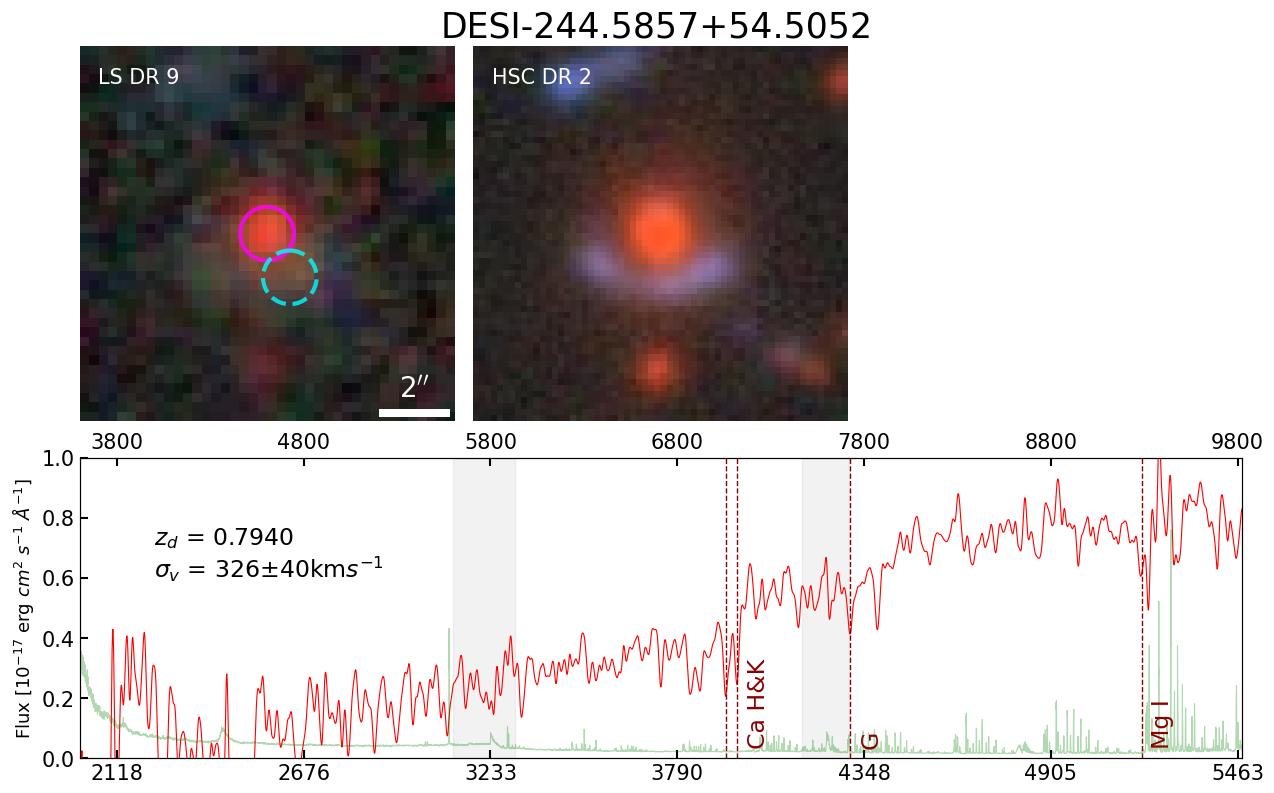}
  \caption{DESI-244.5857+54.5052. For the arrangement of the panels, see Figure~\ref{fig:desi149.6+01}.}
  \label{fig:desi244.5+54}
\end{figure}

\newpage
\href{https://www.legacysurvey.org/viewer/?ra=244.5857&dec=+54.5052&layer=hsc-dr2&pixscale=0.262&zoom=16}{\emph{DESI-244.5857+54.5052}}\, The putative lens has a high redshift of 0.7940 (Figure~\ref{fig:desi244.5+54}).
DESI will target the bright blue arc  to the S in the \hsc image (faintly visible in the \ls image).

\begin{figure}[h]
  \centering
  \includegraphics[width=.7\textwidth]{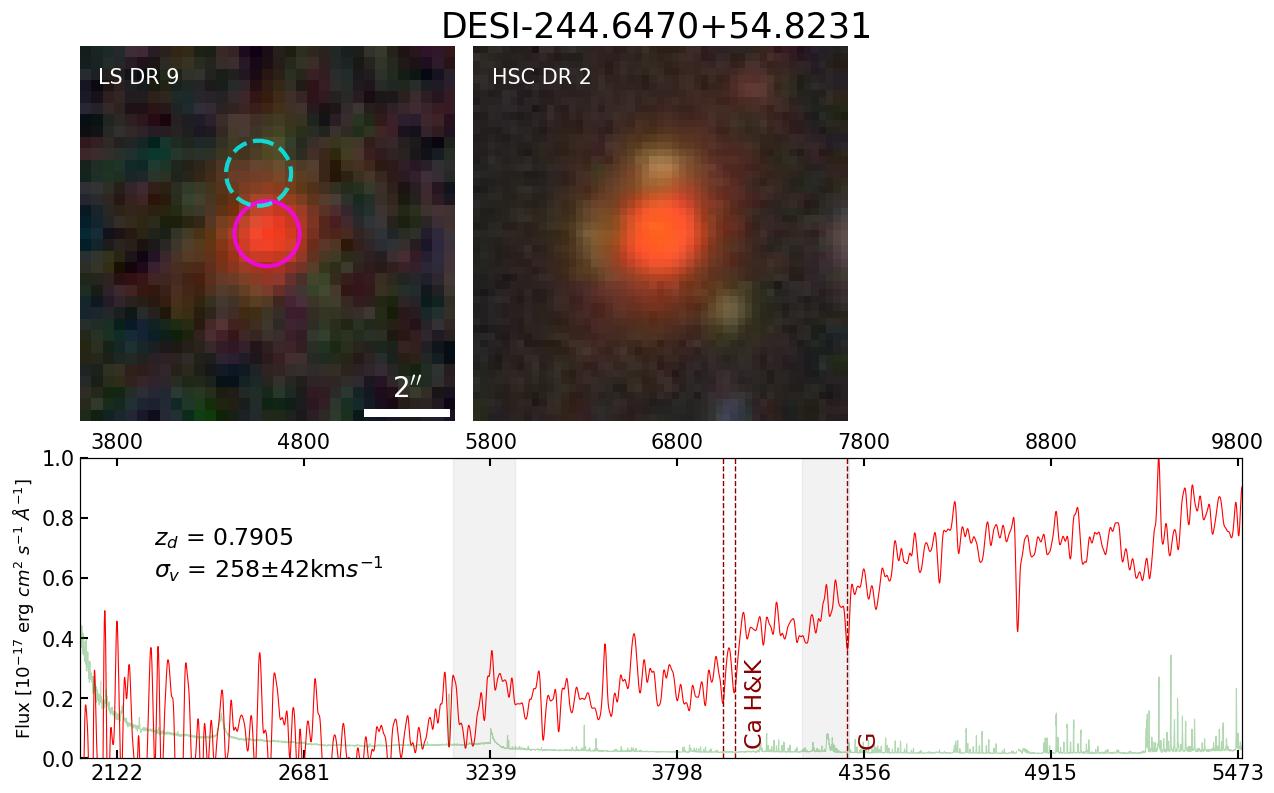}
  \caption{DESI-244.6470+54.8231. For the arrangement of the panels, see Figure~\ref{fig:desi149.6+01}.}
  \label{fig:desi244.6+54}
\end{figure}

\href{https://www.legacysurvey.org/viewer/?ra=244.6470&dec=+54.8231&layer=hsc-dr2&pixscale=0.262&zoom=16}{\emph{DESI-244.6470+54.8231}}\,  The putative lens has a high redshift of 0.7905 (Figure~\ref{fig:desi244.6+54}).
In the \hsc image, 
two greenish images can be seen to the N and NE forming an arc, with a possible counterarc to the SW.
The brightest of these three will be targeted by DESI.
(There is a hint of these images in the \ls cutout.)

\begin{figure}[h]
  \centering
  \includegraphics[width=.7\textwidth]{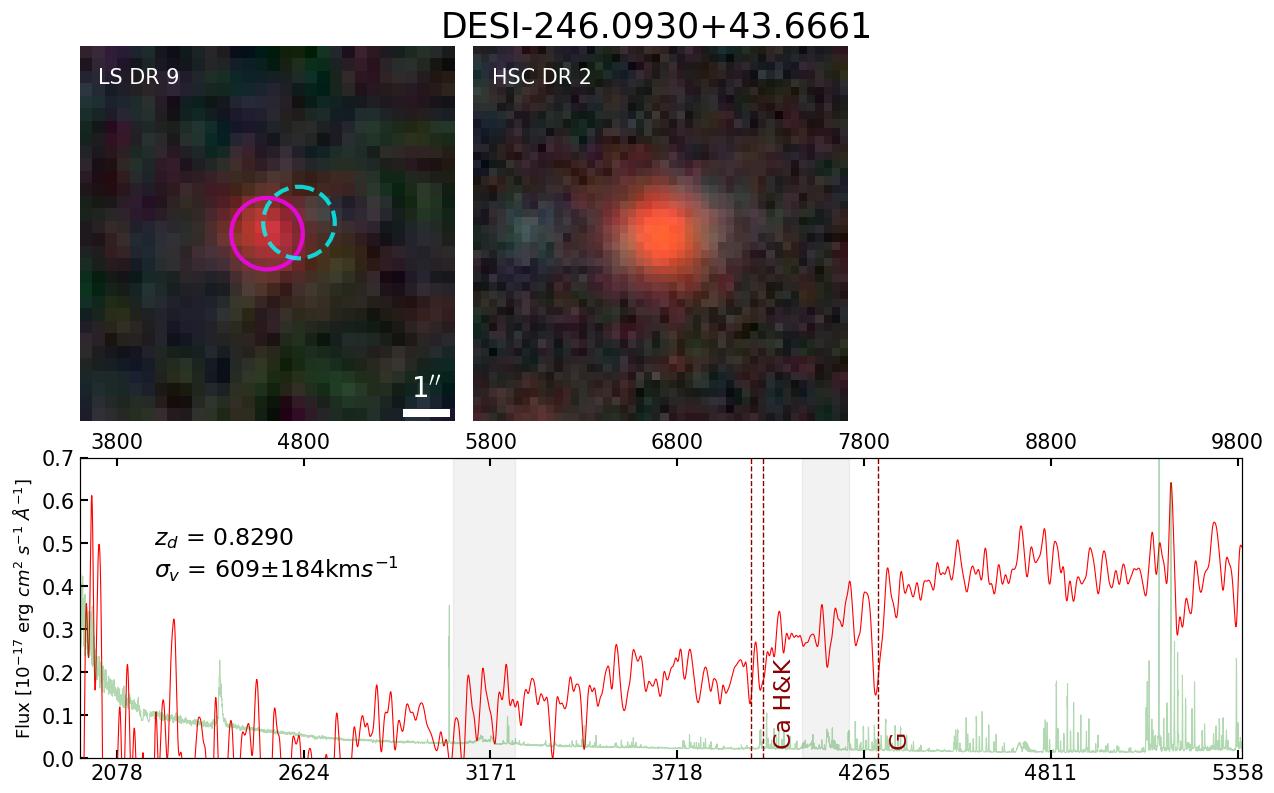}
  \caption{DESI-246.0903+43.6661. For the arrangement of the panels, see Figure~\ref{fig:desi149.6+01}.}
  \label{fig:desi246+43}
\end{figure}

\newpage
\href{https://www.legacysurvey.org/viewer/?ra=246.0903&dec=+43.6661&layer=hsc-dr2&pixscale=0.262&zoom=16}{\emph{DESI-246.0903+43.6661}}\,The putative lens has a high redshift of 0.8290 (Figure~\ref{fig:desi246+43}).
In the \hsc image, a pair of blue images can be seen on the E and W sides of the putative lens (the one on the W side barely visible in the \ls image).
The brighter of these two (on the W side) will be targeted by DESI. 

\begin{figure}[h]
  \centering
  \includegraphics[width=.7\textwidth]{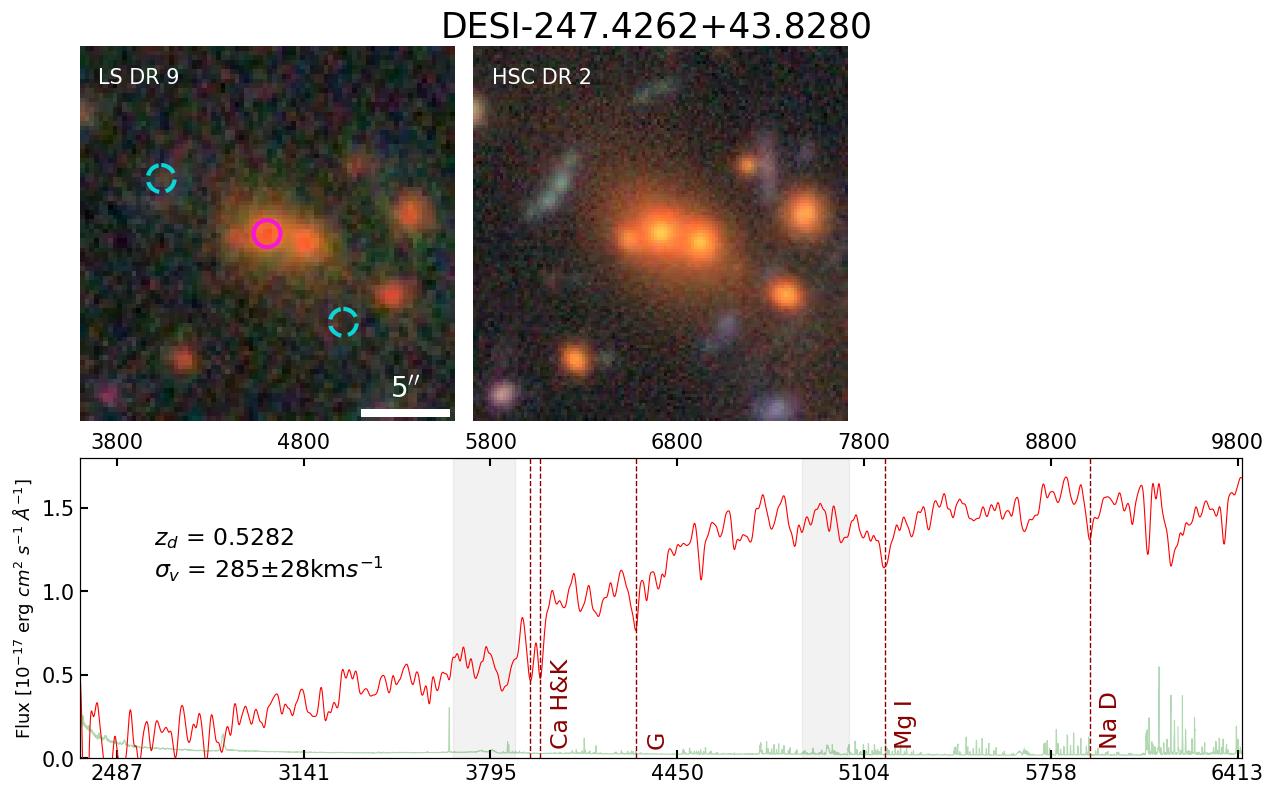}
  \caption{DESI-247.4262+43.8280. 
  For the arrangement of the panels, see Figure~\ref{fig:desi149.6+01}.}
  \label{fig:desi247+43}
\end{figure}

\href{https://www.legacysurvey.org/viewer/?ra=247.4262&dec=+43.8280&layer=hsc-dr2&pixscale=0.262&zoom=16}{\emph{DESI-247.4262+43.8280}}\, The putative lens has a redshift of 0.5282 (Figure~\ref{fig:desi247+43}).
There are two prominent arcs to the NE and SW of the putative lens, with different colors. 
Both will be targeted by DESI.

\begin{figure}
  \centering
  \includegraphics[width=.7\textwidth]{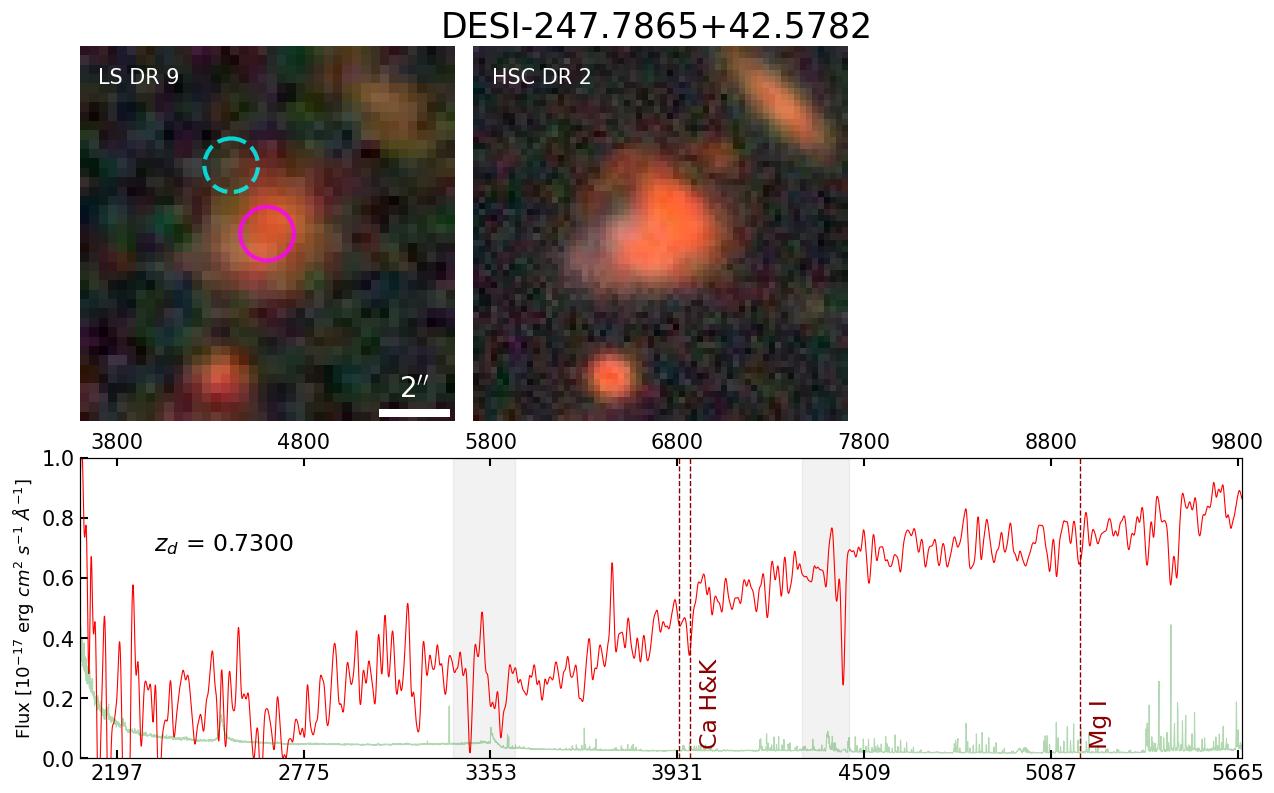}
  \caption{DESI-247.7865+42.5782. For the arrangement of the panels, see Figure~\ref{fig:desi149.6+01}. 
  }
  \label{fig:desi247+42}
\end{figure}

\href{https://www.legacysurvey.org/viewer/?ra=247.7865&dec=+42.5782&layer=hsc-dr2&pixscale=0.262&zoom=16}{\emph{DESI-247.7865+42.5782}}\, The putative lens  is at 0.7300 (Figure~\ref{fig:desi247+42}). 
The lens appears to be a galaxy group. 
The fiber is on one of the two brightest galaxies.
There are two reddish arc-like objects to the N of the putative lenses. 
The one slightest to the E will be targeted by DESI.

\newpage
\section{Confirmed Non-lenses}\label{sec:nonlenses}
In this section, we present the four confirmed non-lenses, 
For these, a fiber centered on a putative lens is labeled as ``lens target", 
and a fibers centered on a putative lensed source, ``src target''.
Clearly the redshifts for each system show that these are not strong lensing systems.
The spectra of these systems are presented below.
Even in the HSC image they all appear to be plausible lens candidates.



These four fall into two subcategories.
For the first two (DESI-150.2022+01.653 and DESI-217-2090+51.6467), a low redshift blue galaxy happens to have an arc-like appearance.
For each of the next two (DESI-218.87813+00.3034 and DESI-219.9227+0.507), the putative source and lens have very similar redshifts.
For these systems, we arrange each of the figures in the same way as those that are confirmed lenses (see Figure~\ref{fig:eg-lens-fig}).
For the figures in this section, 
the second row and third row show the spectra from the fiber centered on the \cwr{``lens target''} and \cwr{``src target"}, respectively.
The \rr pipeline redshifts are shown in the respective panels, confirming that these are not nonlenses.

\begin{figure}
  \centering
  \includegraphics[width=.7\textwidth]{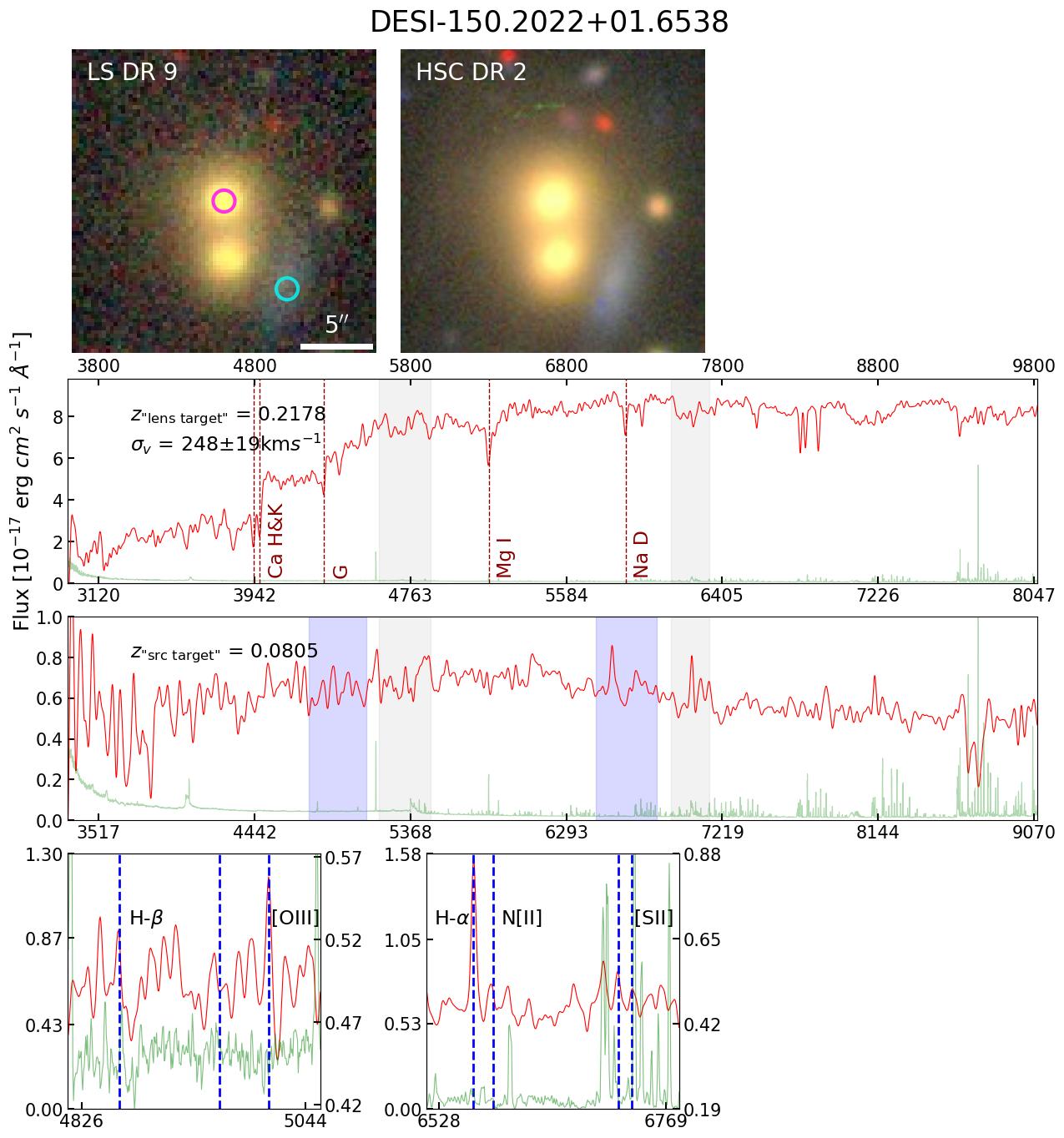}
  \caption{DESI-150.2022+01.6538.
  Even though this is a confirmed non-lens, the arrangement of panels is the same as for Figure~\ref{fig:eg-lens-fig}.
  Except that here the spectra are from fibers targeting the putative lens (``lens target'') and putative source (``source target'').}
  \label{fig:desi150+01}
\end{figure}

\href{https://www.legacysurvey.org/viewer/?ra=150.2022&dec=+01.6538&layer=hsc-dr2&pixscale=0.262&zoom=16}{\emph{DESI-150.2022+01.6538}}\, 
There appears to have a blue elongated object that arcs toward two massive elliptical galaxies (Figure~\ref{fig:desi150+01}).
But this blue object turns out to be a foreground galaxy.
DESI\cwr{~J}183.0990-01.5510 (Figure~\ref{fig:desi183-01}) has a similar appearance at least in the \ls image and it is confirmed by DESI as a lensing system (see \S~\ref{sec:discuss} for more discussion on this comparison).

\begin{figure}
  \centering
  \includegraphics[width=.7\textwidth]{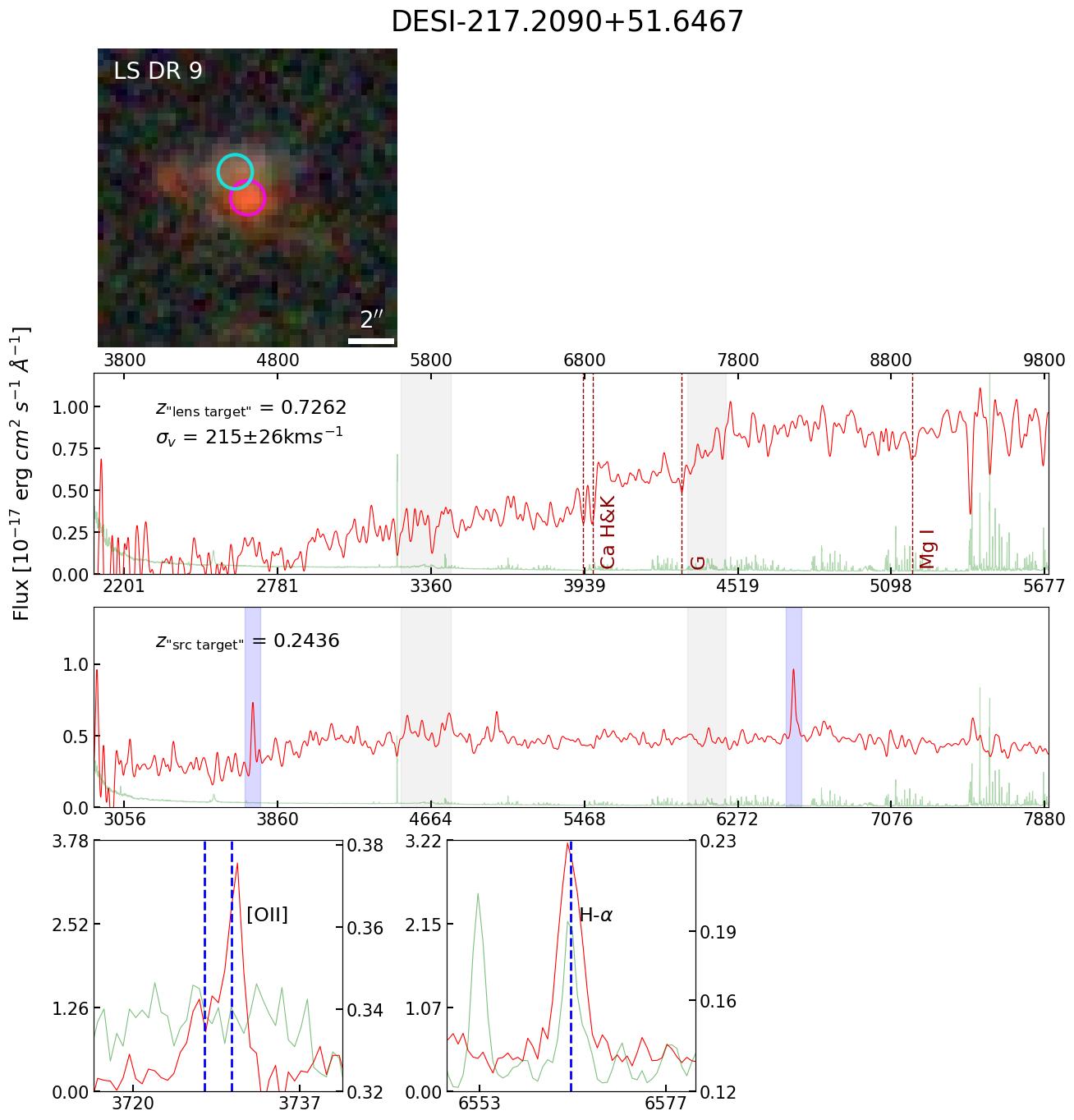}
  \caption{DESI-217.2090+51.6467. For the arrangement of panels, see Figure~\ref{fig:desi150+01}.}
  \label{fig:desi217+51}
\end{figure}

\newpage
In \href{https://www.legacysurvey.org/viewer/?ra=217.2090&dec=+51.6467&layer=ls-dr10-grz&pixscale=0.262&zoom=16}{\emph{DESI-217.2090+51.6467}}\, There appears to be a faint blue arc-like object next to an elliptical galaxy (Figure~\ref{fig:desi217+51}). 
Based on the \ls image, it is reasonable to consider this a lensing candidate. 
But this blue object turns out to be a foreground galaxy.

\begin{figure}
  \centering
  \includegraphics[width=.7\textwidth]{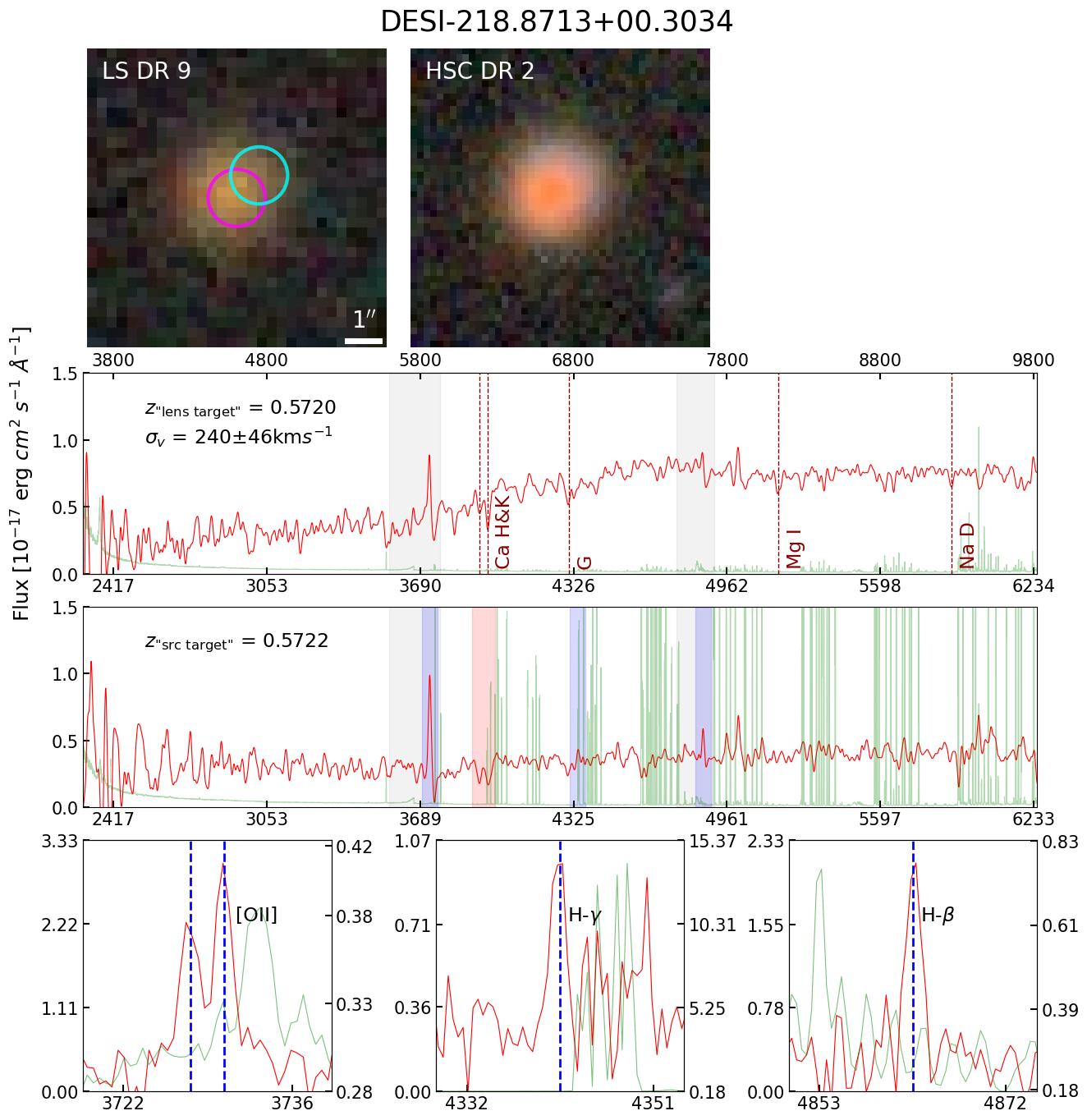}
  \caption{
  DESI-218.87813+00.3034. For the arrangement of panels, see Figure~\ref{fig:desi150+01}.
From the \hsc image, this seems a plausible lens candidate.
But DESI spectra reveal it to be a likely spiral galaxy.
The \oii emission appears in both the spectra of the ``lens target" and ``source target", 
and so do the most prominent absorption lines, Ca~H\&K (highlighted with a red band in the latter).
Given the proximity of the two fibers, it is not clear whether these features are indeed present in both the orange ``blob'' and the blue arc-like feature, 
but they could be.
In addition to the \oii zoom-in from the ``source target" spectrum,
we also show the zoom-in's for the weaker \hb and \hc emission lines.
The totality of the spectral evidence strong suggests that the orange blob (the bulge) and blue arc-like feature (spiral arm) are parts of the same galaxy.
}\label{fig:desi218+00}
\end{figure}

\newpage
\href{https://www.legacysurvey.org/viewer/?ra=218.87813&dec=+00.3034&layer=hsc-dr2-grz&pixscale=0.262&zoom=16}{\emph{DESI-218.87813+00.3034}}\, 
From the appearance of the imaging data, even with the resolution and depth of \hsc (Figure~\ref{fig:desi218+00}), 
it is reasonable to consider it a lens candidate.
The ``orange blob'' has the appearance of an elliptical galaxy and around it, the blue arc-like feature has the appearance of a strongly lensed background galaxy.
DESI fibers centered on the putative lens and arc yielded very similar redshifts, 0.5722 and 0.5770, respectively.
This small difference is comparable to the uncertainty of the DESI redshift measurement (but may indicate a small line-of-sight rotation speed). 
This suggests that the two fibers are probing two different parts of the same galaxy, probably a spiral galaxy,
with one centered on the red core and the other on the blue spiral arms.

\begin{figure}
\centering
\includegraphics[width=.7\textwidth]{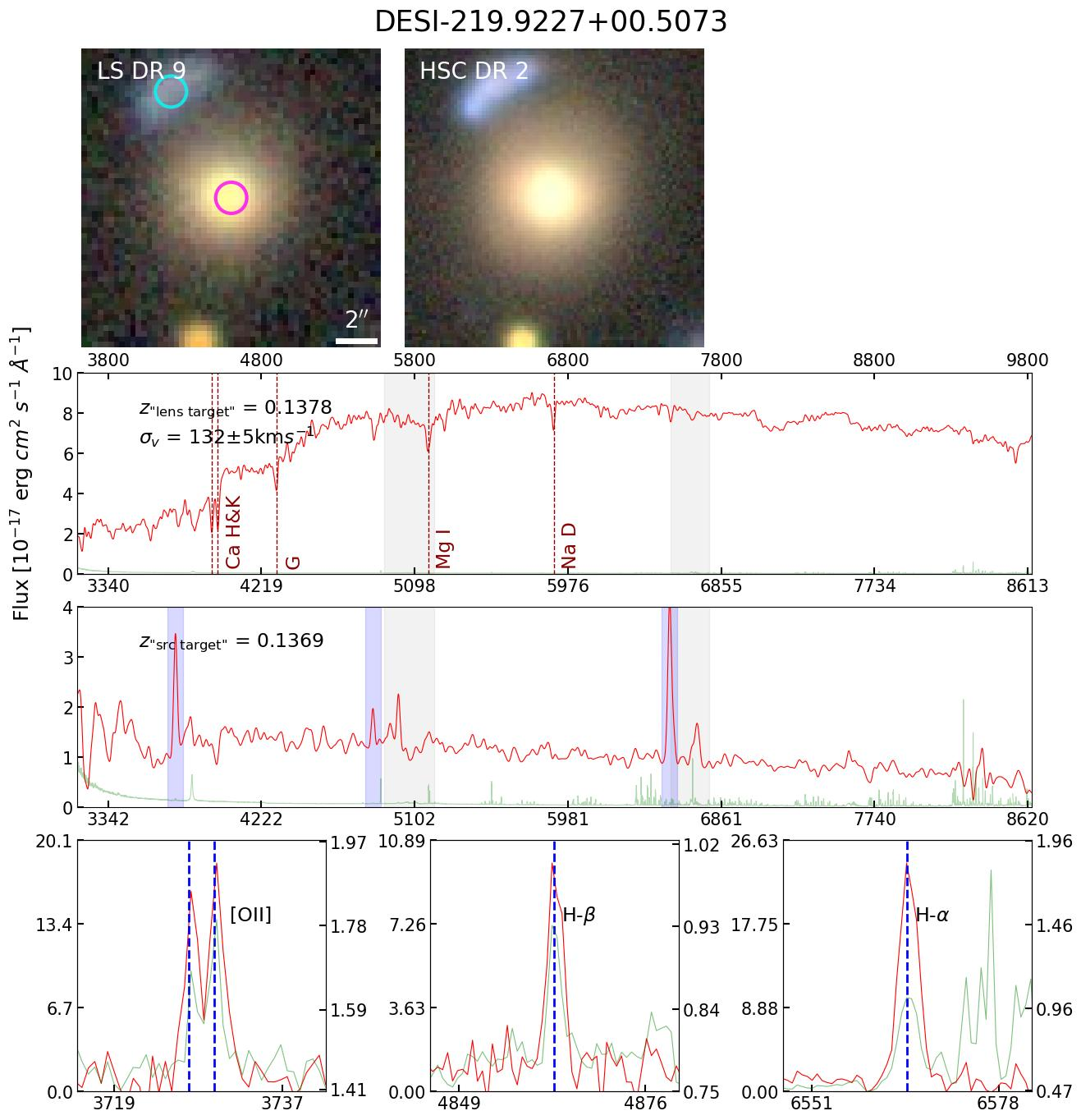}
\caption{DESI-219.9227+00.5073. 
For the arrangement of panels, see Figure~\ref{fig:desi150+01}. 
}
\label{fig:desi219.9+00}
\end{figure}

\newpage
\href{https://www.legacysurvey.org/viewer/?ra=219.9227&dec=+0.5073&layer=hsc-dr2&pixscale=0.262&zoom=16}{\emph{DESI-219.9227+0.5073}}\,
This seems to be a plausible lensing candidate, with a blue elongated object that appears to arc toward an elliptical galaxy.
The DESI spectra (Figure~\ref{fig:desi219.9+00}) reveal that it is not a lensing system.
The redshift difference between the two galaxies  is $\sim 0.001$.
It is likely that they are in the same galaxy group.
From the \hsc image, 
the blue galaxy thought to be an arc candidate appears to be an edge-on spiral galaxy with a reddish core and blue spiral arms.
Based on the \ls image alone, the evidence for this interpretation is much weaker. 


\end{document}